%% file: main.tex
\documentclass[PhD]{iitddiss}
\usepackage{rotating,tabularx}

\usepackage{t1enc}
\usepackage[a4paper]{geometry}
\usepackage{graphicx,wrapfig,lipsum}
\usepackage{amsmath} % easier math formulae, align, subequations \ldots
\usepackage[english]{babel}
\usepackage[utf8]{inputenc,xcolor}
\usepackage{fancyhdr}
\usepackage{comment}
\usepackage[colorlinks=true,citecolor=blue,urlcolor=blue,linkcolor=blue]{hyperref}
\usepackage{adjustbox}
\usepackage{multirow}%\linespread{1.2}
\allowdisplaybreaks
\usepackage{lscape}
\usepackage{amsmath}

	\usepackage{booktabs}

\usepackage{caption}
\usepackage{physics}
\usepackage{mathrsfs}

\usepackage[caption=false]{subfig}
\usepackage{verbatim}
\usepackage{float}

\usepackage[backend=biber,
            uniquename=init,
            giveninits,natbib=true,
            hyperref=true,
            url=false,
            isbn=false,
            backref=true,
            style=phys,
            citereset=chapter,
            biblabel=brackets,
            articletitle=false,
            maxcitenames=5,
            maxbibnames=5,
            block=none,
            refsection=chapter,
            pageranges=false, 
            defernumbers=true]{biblatex}
            
\newcommand*{\mkandothers}{\mkbibemph}
\renewbibmacro*{name:andothers}{%
  \ifboolexpr{
    test {\ifnumequal{\value{listcount}}{\value{liststop}}}
    and
    test \ifmorenames
  }
    {\ifnumgreater{\value{liststop}}{1}
       {\finalandcomma}
       {}%
     \andothersdelim\bibstring[\mkandothers]{andothers}}
    {}}

\usepackage{epigraph} 

\usepackage{csquotes}

\usepackage{subfiles}
\newcommand{\ols}[1]{\mskip.4\thinmuskip\overline{\mskip-.4\thinmuskip {#1} \mskip-.4\thinmuskip}\mskip.4\thinmuskip} % overline short

\makeatletter
\def\thickhrulefill{\leavevmode \leaders \hrule height 1ex \hfill \kern \z@}
\def\@makechapterhead#1{%
  \vspace*{10\p@}%
  {\parindent \z@ \centering \reset@font
        \thickhrulefill\quad
        \scshape \@chapapp{} \thechapter
        \quad \thickhrulefill
        \par\nobreak
        \vspace*{10\p@}%
        \interlinepenalty\@M
        \hrule
        \vspace*{10\p@}%
        \Huge \bfseries #1\par\nobreak
        \par
        \vspace*{10\p@}%
        \hrule
    \vskip 100\p@
  }}
\def\@makeschapterhead#1{%
  \vspace*{10\p@}%
  {\parindent \z@ \centering \reset@font
        \thickhrulefill
        \par\nobreak
        \vspace*{10\p@}%
        \interlinepenalty\@M
        \hrule
        \vspace*{10\p@}%
        \Huge \bfseries #1\par\nobreak
        \par
        \vspace*{10\p@}%
        \hrule
    \vskip 100\p@
  }}

\begin{document}

\pagenumbering{roman}
% Non-Content pages
\input{titlepage}%Title page
\input{nonindex/dedication}%dedication page
\input{nonindex/certificate}% Certificate
\input{nonindex/acknowledgements}% Acknowledgements
\subfile{\cwd/nonindex/abstract}% Abstract
%%%%%%%%%%%%%%%%%%%%%%%%%%%%%%%%%%%%%%%%%%%%%%%%%%%%%%%%%%%%%%%%%
% Table of contents etc.
\begin{singlespace}
\tableofcontents
\thispagestyle{empty}

\listoftables
\addcontentsline{toc}{chapter}{LIST OF TABLES}
\listoffigures
\addcontentsline{toc}{chapter}{LIST OF FIGURES}
\end{singlespace}

%%%%%%%%%%%%%%%%%%%%%%%%%%%%%%%%%%%%%%%%%%%%%%%%%%%%%%%%%%%%
% List of papers
\listofpapers
\addcontentsline{toc}{chapter}{LIST OF PUBLICATIONS}
\textbf{In International Journals}
%%%%%%%%%%%%%%%%%%%%%%%%%%%
\begin{enumerate}
\item  
 ``Magnetised neutron star crust within effective relativistic mean-field model''\\ 
 \textbf{Vishal Parmar}, H. C. Das, M. K. Sharma, and S. K. Patra \\
 \href{https://link.aps.org/doi/10.1103/PhysRevD.107.043022}{Phys. Rev. D, \textbf{107}, 043022 (2023)}
  
  %\emph{\textbf{Impact Factor:} 3.043}
  
\item  
 ``Pasta properties of the neutron star within effective relativistic mean-field model''\\ 
 \textbf{Vishal Parmar}, H. C. Das, Ankit Kumar, Ankit Kumar, M. K. Sharma, P. Arumugam and S. K. Patra \\
 \href{https://doi.org/10.1103/PhysRevD.106.023031}{Phys. Rev. D, \textbf{106}, 023031 (2022)}
  
  %\emph{\textbf{Impact Factor:} 3.043}
  
  \item
  ``Crustal properties of a neutron star within an effective relativistic mean-field model''\\
   \textbf{Vishal Parmar}, H. C. Das, M. K. Sharma, and S. K. Patra \\
   \href{https://doi.org/10.1103/PhysRevD.105.043017}{Phys. Rev. D, \textbf{105}, 043017 (2022)}
  
  %\emph{\textbf{Impact Factor:} 1.659}

  \item
  ``Properties of hot finite nuclei and associated correlations with infinite nuclear matter''\\
   \textbf{Vishal Parmar}, M. K. Sharma, and S. K. Patra \\
   \href{https://doi.org/10.1103/PhysRevC.105.024316}{Phys. Rev. C, \textbf{105}, 024316 (2022)}

%  \emph{\textbf{Impact Factor:} 3.296}

  \item
  ``Thermal effects in hot and dilute homogeneous asymmetric nuclear matter''\\
   \textbf{Vishal Parmar}, M. K. Sharma, and S. K. Patra \\
   \href{https://doi.org/10.1103/PhysRevC.103.055817}{    Phys. Rev. C, \textbf{103}, 055817 (2021)}

%    \emph{\textbf{Impact Factor:} 2.145}

  \item
  ``Critical properties of symmetric nuclear matter in
low-density regime using effective-relativistic mean
field formalism''\\
  \textbf{Vishal Parmar}, M. K. Sharma, and S. K. Patra \\
  \href{https://doi.org/10.1088/1361-6471/abc864}{J. Phys. G: Nucl. Part. Phys, \textbf{48}, 025108 (2021)}

 %   \emph{\textbf{Impact Factor:} 3.296}

%%%%%%%%%%%%%%%%%%%%%%%  
\end{enumerate}
%%%%%%%%%%%%%%%%%%%%%%%%%%
\textbf{In National and International Conferences/Symposiums}
%%%%%%%%%%%%%%%%%%%%%%%%%%%

 \begin{enumerate}
     \item  \textbf{Vishal Parmar}, H. C. Das, Ankit Kumar, Manoj K. Sharma and  S. K. Patra, \emph{Sensitivity of crustal properties of neutron star on the parameters of compressible liquid drop model},  Accepted in XXV DAE-BRNS High Energy Physics (HEP) Symposium  held in IISER Mohali, India. 
     
     \item  \textbf{Vishal Parmar},  Manoj K. Sharma and  S. K. Patra, \emph{Role of symmetry energy on the inner crust composition of neutron star},  \href{http://sympnp.org/proceedings/66/C32.pdf}{Proceedings of the DAE Symp. on Nucl. Phys. 66, 786 (2022)}

    \item  \textbf{Vishal Parmar}, H. C. Das, Manoj K. Sharma and  S. K. Patra, \emph{Effect of Landau quantization of the electron on neutron star crust within effective relativistic mean-field model}, \href{https://indico.cern.ch/event/921532/book-of-abstracts.pdf}{IWARA2022 - 10th International Workshop on Astronomy and Relativistic Astrophysics, Contribution ID: 47}
    
        \item  \textbf{Vishal Parmar}, Manoj K. Sharma and  S. K. Patra, \emph{Limiting temperature of nuclei within effective relativistic mean-field theory}, \href{https://indico.cern.ch/event/1012633/contributions/4478244/}{LXXI International conference “NUCLEUS – 2021”, Contribution ID: 100 }
    
    \item  \textbf{Vishal Parmar}, Manoj K. Sharma and  S. K. Patra, \emph{Density fluctuation near the critical points in symmetric nuclear matter}, \href{https://inspirehep.net/files/15a4c752acae3f8b2467924e31483c81}{Proceedings of the DAE Symp. on Nucl. Phys. 65, 579 (2021)}

             \item  \textbf{Vishal Parmar}, Manoj K. Sharma and  S. K. Patra, \emph{Thermodynamics of phase transition in asymmetric nuclear matter within relativistic mean field framework}, \href{http://sliet.ac.in/wp-content/uploads/2019/10/contents-details.pdf}{Punjab Science Congress held at SLIET Punjab - February 2020}         
   
     \item  \textbf{Vishal Parmar}, Manoj K. Sharma and  S. K. Patra, \emph{Critical parameters of liquid-gas phase transition in symmetric nuclear matter}, \href{http://sympnp.org/proceedings/64/A52.pdf}{Proceedings of the DAE Symp. on Nucl. Phys. 64, 158 (2019)}

    \item  \textbf{Vishal Parmar}, Manoj K. Sharma and  S. K. Patra, \emph{Stability condition in hot symmetric nuclear matter}, \href{https://inspirehep.net/files/15a4c752acae3f8b2467924e31483c81}{Proceedings of the DAE Symp. on Nucl. Phys. 64, 160 (2019)}

\end{enumerate}
 
%%%%%%%%%%%%%%%%%%%%%%%%%%%%%%%%%

%%%%%%%%%%%%%%%%%%%%%%%%%%%%%%%%%%%%%%%%%%%%%%%%%%%%%%%%%%%%%%%%%%%%%%
\input{nonindex/abrv}% Abbreviations
%\input{nonindex/notations}% Notations
%%%%%%%%%%%%%%%%%%%%%%%%%%%%%%%%%%%%%%%%%%%%%%%%%%%%%%%%%%%%%%%%%%%%%%
\pagebreak
\clearpage

% The main text will follow from this point so set the page numbering
% to arabic from here on.
\pagenumbering{arabic}

%%%%%%%%%%%%%%%%%%%%%%%%%%%%%%%%%%%%%%%%%%%%%%%
\setlength{\parskip}{0pt} % 1ex plus 0.5ex minus 0.2ex}

\subfile{\cwd/Chapter_1/intro.tex}
\subfile{\cwd/Chapter_2/CHAP2.tex}
\subfile{\cwd/Chapter_3/CHAP3.tex}
\subfile{\cwd/Chapter_4/CHAP4.tex}
\subfile{\cwd/Chapter_5/CHAP5.tex}
\subfile{\cwd/Chapter_6/CHAP6.tex}
\subfile{\cwd/Chapter_7/CHAP7.tex}

\subfile{\cwd/Summary/conclusion.tex}

%%%%%%%%%%%%%%%%%%%%%%%%%%%%%%%%%%%%%%%%%%%%%%

%Appendices.
%\appendix
%\chapter{A SAMPLE APPENDIX}

%Just put in text as you would into any chapter with sections and
%whatnot.  Thats the end of it.

%%%%%%%%%%%%%%%%%%%%%%%%%%%%%%%%%%%%%%%%%%%%%%%%%%%%%%%%%%%%
% Bibliography.

%\begin{singlespace}
%  \bibliography{refs}
%\end{singlespace}

%%%%%%%%%%%%%%%%%%%%%%%%%%%
\end{document}

%% file: titlepage.tex
\begin{titlepage}
\setstretch{1.5}
\thispagestyle{empty}
   \begin{center}
       \vspace*{1cm}
       \textbf{\large Exploration of Nuclear Matter Properties
and Related Thermodynamical Aspects}%Thesis Title

       \vspace{0.5cm}
       \textit{\large A THESIS}
        % Thesis Subtitle %Thesis Subtitle
       
       \vspace{0.10cm}
       Submitted for the partial fulfillment of requirement for the award of the degree of
       
       \vspace{0.20cm}
       \textit{\large DOCTOR OF PHILOSOPHY}
       
       \vspace{0.20CM}
       By
       
       \textbf{\LARGE Vishal Parmar\\ \small(Reg. No: 901912001)}%Author Name
       
       \vspace{0.20cm}
       \textit{\large Under the supervision of}
       
\begin{minipage}[t]{7cm}
\flushleft
\textbf{Dr. Manoj K Sharma}\\
Professor\\
TIET Patiala
\end{minipage}
\hfill
\begin{minipage}[t]{7cm}
\flushright
\textbf{Dr. S K Patra}\\
Professor\\
IOP Bhubaneswar
\end{minipage} 
\newline
\centering{\includegraphics[scale=1.5]{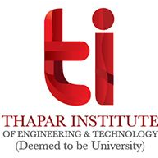}}
\vspace*{0.8cm}        
\\SCHOOL OF PHYSICS AND MATERIALS SCIENCE\\
          THAPAR INSTITUTE OF ENGINEERING AND TECHNOLOGY, PATIALA - 147004\\
          PUNJAB, INDIA\\
          AUGUST 2023
   \end{center}
\end{titlepage}

%% file: nonindex/dedication.tex
\dedication

%\vspace*{0.5in}

%\begin{flushright}
%    \begin{itshape}
%            \begin{LARGE}
%                Dedicated to My Mother,  \\
%	            My Strongest Support.
%            \end{LARGE}
%    \end{itshape}
%\end{flushright}
\begin{center}
    \thispagestyle{empty}
    %\vspace*{2in}
    \begin{itshape}
            \begin{LARGE}

                To Maa, Di \& Bhai  \\
                My Strongest Support
                \rule{\linewidth}{1pt}

            \end{LARGE}
    \end{itshape}
    \vspace*{\fill}
\end{center}

\vspace{1in}
\begin{center}
\centering{\includegraphics[scale=1.1]{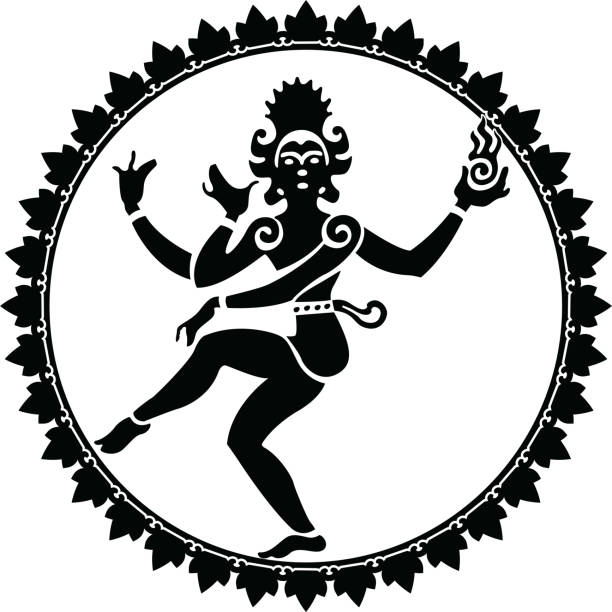}}\\
\textbf{\textit{The cosmic dance  of shiva}}
\end{center}

\begin{center}
    \thispagestyle{empty}
    %\vspace*{1in}
    \begin{itshape}
            %\begin{LARGE}
   
              ``The source of all movement,\\
                Shiva's dance,\\
                Gives rhythm to the universe.\\
                He dances in evil places, 
                In sacred,\\
                He creates and preserves, 
                Destroys and releases.\\
                We are part of this dance, 
                This eternal rhythm,\\
                And woe to us if, Blinded by illusions,\\
                We detach ourselves,\\
                From the dancing cosmos,\\
                This universal harmony''\\
                
                \vspace{0.5in}
                  
                \textbf{\emph{Ruth Peel}}
            \rule{\linewidth}{1pt} 
            %\end{LARGE}
    \end{itshape}
    \vspace*{\fill}
\end{center}

\clearpage

%% file: nonindex/certificate.tex
\certificate

\vspace*{0.5in}

\noindent This is to certify that the thesis titled ``\textbf{Exploration of Nuclear Matter Properties
and Related Thermodynamical Aspects}'', submitted by \textbf{Mr. Vishal Parmar}, for the fulfillment of the requirements for the award of Degree of Doctor of Philosophy in the School of Physics and Materials Science, Thapar Institute of Engineering and Technology, Patiala, is a record of the candidate's own work carried out by him under my supervision. The matter presented in this thesis has not been submitted in part or full for the award of any degree in any other university or institute.\\
\vspace{2.0cm}
%%%%%%%%%%%%%%%%%%%%%%%%%%%
\begin{center}
    \textbf{Supervisors}
\end{center}
\vspace{2.0cm}

\begin{minipage}[t]{8cm}
\flushleft

\textbf{Manoj K Sharma}\\
Professor\\
School of Physics Materials Science\\
TIET, Patiala - 147004\\
Punjab, INDIA\\
Date:.......................\\
Place: Patiala

\end{minipage}
\hfill
\begin{minipage}[t]{8cm}
\flushright

\textbf{S K Patra}\\
Professor\\
Institute of Physics\\
Bhubaneswar - 751005\\
Odisha, INDIA \\
Date:.......................\\
Place: Bhubaneswar

\end{minipage}

%\begin{singlespacing}
%\hspace*{-0.25in}
%\parbox{2.5in}{
%\noindent {\bf Manoj Kumar Sharma} \\
%\noindent 
%Professor \\
%\noindent 
%School of Physics and Materials Science\\
%\noindent TIET, Patiala \\
%}
%\hspace*{1.0in}
%\end{singlespacing}
%\vspace*{0.25in}
%\noindent Place: Patiala\\
%Date: 

%% file: nonindex/acknowledgements.tex
\acknowledgements

In a letter to Robert Hooke in 1675, Isaac Newton made his most famous statement: “If I have seen further, it is by standing on the shoulders of Giants.”  Similarly, this thesis has been possible with the help of various individuals. I would like to express my deepest gratitude and appreciation to everyone who has supported and encouraged me throughout my academic journey. 

Firstly, I would like to thank my supervisors, \textit{Prof. Manoj K Sharma} and \textit{Prof. S K Patra}, for their guidance, expertise, and endless support throughout my research. Their valuable feedback and insights have helped me shape my ideas and push the boundaries of my research. I am grateful for their patience, dedication, and willingness to go the extra mile to ensure my success. Moreover, I am indebted to them, who provided invaluable academic guidance and imparted important life lessons that have shaped me personally and professionally.

I would also like to extend my gratitude to the SPMS family, including \textit{Prof. Kulvir Singh}, HOD SPMS, \textit{respected faculty members, Doctoral Committee ( Dr. S. Jana, Dr. Sunil Devi, Dr. Meenakshi Rana)} and \textit{staff} for providing me with the resources and opportunities needed to conduct my research. Their support has been crucial in my academic development, and I am fortunate to have had them as a mentor. I am also deeply indebted to the \textit{Honourable Director, TIET}  and \textit{DoRDC, TIET} for their constant support and encouragement. Their leadership has inspired me to aim for excellence and push beyond my limits. I am grateful for their vision and commitment to academic excellence. I want to thank \textit{Dr Debabrata Deb}, who has been instrumental in learning various aspects of research and life in the last few years.  I extend my heartfelt gratitude to \textit{Dr. Debarati Chatterjee} and \textit{Prof Francesca Gulminelli} for their gracious acceptance to review my thesis and for providing invaluable suggestions that greatly enhanced its scientific accuracy.

I am incredibly grateful to \textit{my mother, sister,} and \textit{brother} for the unparallel sacrifices they have made throughout their lives to support me in completing my thesis. Their unwavering love, encouragement, and selflessness have been a constant source of inspiration and motivation for me. Without their sacrifices and support, I would not be where I am today, on the brink of achieving a significant milestone in my academic journey.  Their dedication and sacrifices will always be remembered and cherished, and I owe them a debt of gratitude that can never be repaid. I also thank Shivangi, my friend and my sister-in-law, for all the love and help she has bestowed on me. She has been an inspiration. Thank you, \textit{Maa, Di, Bhai} and \textit{Shivangi} for all that you have done for me.

I am also indebted to my friend \textit{Shilpa}, who has been a pillar of strength and a true companion during this arduous journey. Her unwavering support and cheerful disposition have made this journey easy and memorable. I would like to express my heartfelt gratitude to my dear friends \textit{Shivani Di, Harshit, Tamo, Nirmal Da, Ayushi, Ankur} and \textit{Ashish} for their unwavering support and encouragement throughout my PhD journey. Their friendship has been a constant source of comfort and inspiration, and their insightful feedback and guidance have been invaluable to me. I am also thankful to \textit{Neha di, Kanishka di, Aman di, Gurjit di} and \textit{Ishita di} for helping me during my early days of PhD. You were the best seniors one can even hope for.  My heartfelt thanks to \textit{Bharti, Nitin, Kundan and Jagroop} for being wonderful friends. Their encouragement  have been significant in this journey.  I would like to express my sincere appreciation and respect to all of them for their contributions to my academic and personal growth.  I am thankful to my juniors \textit{Ashutosh, Chahat, Nishu, Shubh} and \textit{Diksha} for always being there and helping me on countless occasions. I feel fortunate to have met them and to have learned so much from their experiences.  I am  thankful to IOPB family members \textit{Harish, Ankit} and \textit{Jeet} for their guidance, support, and collaborative efforts which have helped me to navigate the complex challenges of research and have contributed significantly to my success. I am grateful for their camaraderie and for being an integral part of my PhD journey. 

Last but not least, I would like to extend my sincere gratitude to the various researchers who have helped me on numerous occasions during my PhD research. I am particularly grateful to \textit{Thomas}, who provided me with invaluable assistance in understanding the CLDM code, which was crucial to the success of my research. Furthermore, I must acknowledge the groundbreaking work of the genius minds in the field that has been instrumental in my understanding of the subject matter of my PhD. Their contributions to the field have been a source of inspiration and motivation for me. Their work has laid the foundation for my research, and I am grateful for the knowledge and insights I have gained from their publications. I would like to express my sincere appreciation and respect to all those researchers who have contributed to my research in one way or another. I owe my knowledge and experiences to a combination of imitation and personal encounters. Every person I have met has played a significant role in shaping who I am today. I extend my heartfelt appreciation  to all those who have been part of my life,  whether closely or distantly related,  for being the faculty of my life.

\vspace{.5in}
\begin{flushright}

    \textbf{Vishal Parmar}

\end{flushright}

%% file: nonindex/abstract.tex
% Abstract
\abstract
The present thesis aims to understand the properties of nuclear matter over a wide range of temperatures, density, isospin-asymmetry, pressure and magnetic field. The precise knowledge of the effect of these variables on the nuclear matter is of prime importance due to their relevance in various nuclear physics and astrophysical phenomena such as heavy-ion reactions, nuclear multifragmentation, the neutron star, supernovae etc. The effective relativistic mean field model (E-RMF) is used for the nuclear interaction to achieve these objectives. The nuclear matter is investigated in three forms: infinite nuclear matter, finite nuclei and the neutron star. In the \textbf{ \textit{first}} part, the infinite nuclear matter is investigated to estimate the critical properties in the context of liquid-gas phase transition, finite temperature effects and modifications in the equation of state (EoS) due to temperature. Extending the infinite matter, the \textbf{\textit{second}} part of the thesis studies various properties of hot nuclei and their dependence on the EoS. Possible correlations among the critical properties with respect to a hot nucleus and infinite nuclear matter are discussed in this part. The \textbf{\textit{third}}  part of the thesis aims to study the neutron star interior with a main focus on the neutron star crust. The neutron star crust is investigated for the unmagnetised and magnetised matter, in context to pulsars and magnetars. The thesis is divided into eight chapters which are briefly described below. 
\\
\textbf{Chapter \ref{chap:intro}} discusses the essential concepts and definitions used in the  thesis with an appropriate literature review. It begins with the description of the nuclear matter phase diagram advocating the importance of precise knowledge of nuclear interaction to describe various associated phenomena. The liquid-gas phase transition in the nuclear matter in the low-density regime is discussed for both infinite and finite nuclear matter. Next, the neutron star, a prominent aspect of the phase diagram, is discussed in detail with emphasis on its interior structure. The importance of EoS in nuclear physics is discussed with  available constraints on various nuclear matter observables. Finally, the E-RMF framework is discussed, citing its suitability and success in describing various nuclear matter properties. 
\\
In \textbf{Chapter \ref{methodology}}, the methodology is discussed. It starts with  a brief description of E-RMF theory which is then extended to the finite temperature regime. The finite temperature effects, such as phase transition in symmetric/asymmetric nuclear matter and the effects of temperature on EoS, are discussed. The Gibbs and Maxwell phase rules are discussed for the description of liquid-gas phase transition in nuclear matter. Extending the infinite nuclear matter to a finite nucleus, a simplistic liquid drop model is considered with the aprropriate density dependence in various coefficients. The free energy of the nucleus in terms of the liquid drop model is formulated to study a hot finite nucleus and associated phenomena. The relevance of various properties such as excitation energy, level density, limiting temperature is discussed in context of nuclear dimensions. Finally, the outer crust using the Baym-Pethick-Sutherland (BPS)  model and inner crust using compressible liquid drop model (CLDM) are discussed to understand  the internal structure of a neutron star. Global properties of a neutron star, such as mass, radius, the moment of inertia, crust mass etc., are also described.
\\
In \textbf{Chapter \ref{chap3}}, various finite temperature properties of isospin symmetric and asymmetric nuclear matter over a wide range of density and pressure are investigated.  The E-RMF formalism employing the  FSUGarnet, IOPB-I and G3 forces, along with one of the most used NL3 parameter sets, are used in the finite temperature limit realising their narrow range of bulk matter properties at zero temperature. The liquid-gas phase transition in context to symmetric and asymmetric matter is discussed. The binodal and spinodal diagrams in reference to the symmetric and asymmetric matter are estimated due to their relevance in neutron star and supernovae physics. The effects of the temperature of the EoS and various nuclear matter observables, such as symmetry energy, are worked out. The thermal properties of state variables in context to their importance in supernovae matter is also discussed. 
\\
In \textbf{Chapter \ref{chap4}}, thermal properties of hot nuclei are investigated within E-RMF formalism. The free energy of a nucleus is estimated by using temperature and density-dependent parameters of the liquid-drop model. The surface free energy is parametrised using two approaches based on the sharp interface of the liquid-gaseous phase and the semi-classical Seyler-Blanchard interaction. Various properties, such as limiting temperature, Excitation energy, level density, fissility etc., are evaluated for a wide atomic mass range. Since the calculations are inevitably model dependent, a detailed correlation matrix analysis is worked out to account for  large deviations in the magnitude of critical parameters among various E-RMF sets.
\\
In \textbf{Chapter \ref{chap5}}, using the E-RMF model, the crustal properties of the neutron star are investigated. The unified equations of state (EoS) are constructed using recently developed E-RMF parameter sets, such as FSUGarnet, IOPB-I, and G3. The outer crust composition is determined using the atomic mass evaluation 2020 data along with the available Hartree-Fock-Bogoliubov mass models  for neutron-rich nuclei. The structure of the inner crust is estimated by performing the compressible liquid drop model calculations using the same E-RMF functional as that for the uniform nuclear matter in the liquid core. Various neutron star properties such as mass-radius ($M-R$) relation, the moment of inertia ($MI$), the fractional crustal moment of inertia ($I_{crust}/I$), mass ($M_{crust})$ and thickness ($l_{crust}$) of the crust are calculated with three unified EoSs.  The crustal properties are found to be sensitive to the density-dependent symmetry energy and slope parameter advocating the importance of the unified treatment of neutron star EoS. The three unified EoSs, IOPB-I-U, FSUGarnet-U, and G3-U, reproduced the observational data obtained with different pulsars, NICER, and glitch activity and are found suitable for further description of the structure of the neutron star. 
\\
Extending above analysis, \textbf{Chapter \ref{chap6}} investigates the properties of pasta structures and their influence on the neutron star observables employing the E-RMF model. The compressible liquid drop model is used to incorporate the finite size effects, considering the possibility of nonspherical structures in the inner crust. The unified equation of states are constructed for several E-RMF parameters to study various properties such as pasta mass and thickness in the neutron star's crust. The majority of the pasta properties are sensitive to the symmetry energy in the subsaturation density region. Using the results from Monte Carlo simulations, the shear modulus of the crust in the context of quasiperiodic oscillations from soft gamma-ray repeaters  and the frequency of fundamental torsional oscillation mode in the inner crust is estimated. Global properties of the neutron star such as mass-radius profile, the moment of inertia, crustal mass, crustal thickness, and fractional crustal moment of inertia etc. are worked out. The results are consistent with various observational and theoretical constraints. 
\\
In \textbf{Chapter \ref{chap7}}, the crustal properties of a  neutron star are investigated within E-RMF framework in the presence of magnetic field strength $\sim 10^{17}$G. The equilibrium composition of the outer crust is calculated by minimizing the Gibbs free energy using the most recent atomic mass evaluations. The magnetic field significantly affects the equation of state (EoS) and the properties of the outer crust, such as neutron drip density, pressure, melting temperature etc. For the inner crust,  the compressible liquid drop model is used for the first time to study the crustal properties in a magnetic environment. The inner crust properties, such as mass and charge number distribution, isospin asymmetry, cluster density, etc., show typical quantum oscillations (De Haas–van Alphen effect) sensitive to the magnetic field's strength. The density-dependent symmetry energy influences the magnetic inner crust like the field-free case. The primary aim here is to study the probable modifications  in the pasta structures  and it is observed that their mass and thickness changes by $\sim 10-15 \%$  depending upon the magnetic field strength. The fundamental torsional oscillation mode frequency is investigated for the magnetized crust in the context of quasiperiodic oscillations (QPO) in soft gamma repeaters. The magnetic field strength considered in this work influences only the EoS of outer and shallow regions of the inner crust, which results in no significant change in global neutron star properties. However, the outer crust mass and its moment of inertia increase considerably with increase in  magnetic field strength. 
\\
Finally, \textbf{Chapter \ref{chap8}} summarises important results of the  thesis and the possible future scopes are outlined here.\\

%\noindent \textcolor{red}{KEYWORDS: \hspace*{0.5em} \parbox[t]{4.4in}{\LaTeX ; Thesis;
%  Style files; Format.}}

%\vspace*{24pt}

%\textcolor{red}{The formatting is as (as far as the author is aware) per the current
%institute guidelines.}

\pagebreak

%% file: nonindex/abrv.tex
% Abbreviations
\abbreviations

\noindent
\begin{tabbing}
xxxxxxxxxxx \= xxxxxxxxxxxxxxxxxxxxxxxxxxxxxxxxxxxxxxxxxxxxxxxx \kill
%\textbf{TIET}   \> Thapar Institute of Engineering and Technology, Patiala \\
\textbf{ANM} \> Asymmetric nuclear matter\\
\textbf{AME} \> Atomic mass evaluations\\
\textbf{BCC} \> Body-centered cubic\\
\textbf{CCSN} \>  Core-collapse supernova\\
\textbf{CLDM} \> Compressible liquid drop model\\
\textbf{CC} \> Core-crust \\
\textbf{E-RMF} \> Effective relativistic mean field\\
\textbf{EoS} \> Equation of state\\
\textbf{ETF} \> Extended Thomas-Fermi\\
\textbf{FMI} \> Fractional moment of inertia\\
\textbf{HF} \> Hartree-Fock\\
\textbf{HFB} \> Hartree-Fock-Bogoliubov\\
\textbf{HIC} \> Heavy-ion collision\\
\textbf{LGPT} \> Liquid gas phase transition\\
\textbf{NSE} \>  Nuclear symmetry energy\\
\textbf{NS} \> Neutron star \\
\textbf{PNM} \> Pure neutron matter\\
\textbf{QPOs} \>  Quasiperiodic oscillations\\
\textbf{RMF} \> Relativistic mean field  \\
\textbf{SNA}   \>  Single-nucleus approximation\\
\textbf{SGRs} \>  Soft gamma repeaters \\
\textbf{SNM} \>  Symmetric nucelar matter \\
\textbf{TF}    \> Thomas-Fermi\\
\textbf{TOV}   \>  Tolman-Oppenheimer-Volkoff\\
\textbf{WS} \> Wigner–Seitz\\
\end{tabbing}

\pagebreak

%% file: Chapter_1/intro.tex
% Introduction.
%\baselineskip=1.3\baselineskip
\chapter{Introduction}
\label{chap:intro}
 \setlength{\epigraphwidth}{6in}
 %\begin{center}
 %   \epigraph{``It was= quite the mos=t incredible event that has= ever happened to me in my life. It was= almos=t as= incredible as if you fired a 15-inch s=hell at a piece of tis=s=ue paper, and it came back and hit you."}{ \emph{Ernes=t Rutherford on the back-s=cattering effect of metal foil on alpha-particles=}}
     
 %\end{center}

%\epigraph{With all reserve we advance the view that a supernova represents the transition of an ordinary star into a neutron star, consisting mainly of neutrons. Such a star may possess a very small radius and an extremely high density. As neutrons can be packed much more closely than ordinary nuclei and electrons, the gravitational packing energy in a cold neutron star may become very large, and under certain conditions may far exceed the ordinary nuclear packing fractions}

%%%%%%%%%%%%%%%%%%%%%%%%%
%THE DOCUMENT BEGINS HERE
%%%%%%%%%%%%%%%%%%%%%%%%%c

\section{Prologue}

Rutherford's alpha-scattering experiment discovered the atomic nucleus over a century ago and revolutionised our understanding of the subatomic regime. The idea of the nucleus at the centre of an atom was as extraordinary as the model of the universe by Nicolaus Copernicus 400 years earlier, which placed the sun at the centre rather than the earth. These monumental discoveries in science not only questioned the conventional philosophical foundations but forced the mankind to think beyond the visible range of perception.   While the prepositions of Copernicus were validated and modified in subsequent years, investigating the atomic nucleus posed an uphill task owing to its size,  lack of available fundamental theories and behaviours unconventional to scientific understanding. Schrodinger's quantum picture in subsequent years, which was far less intuitive than the conventional classical mechanics, helped to unravel the intricacies of the nucleus.   Various experimental landmarks in the 19$^{th}$ century helped our understanding of the nucleus, such as the seminal discovery of the neutron by James Chadwick \cite{Chadwick_1932}, the splitting of a nucleus by  John Cockroft and Ernest Walton \cite{Walton_1932}, nuclear fission by Otto Hahn and Lise Meitner \cite{Hahn_1937}, etc. On the theoretical front, Hideki Yukawa,  proposed the strong force for the first time, which binds the nucleons in an atomic nucleus \cite{yukawa1935proc}. The semi-empirical formula by Weizsacker and the groundbreaking nuclear shell model by Maria Goeppert Mayer \cite{Mayer_1948} made the picture of a nucleus more transparent.

In 1934, two years after the discovery of neutrons, astronomers Baade and Zwicky \cite{Baade254, Baade259}, in their pioneering work, coined the term “supernova” and hypothesized the existence of neutron stars, which Hewish \textit{et al.} discovered in 1968 \cite{Hewish_1968}. After a supernova explosion (Type II, Ib or Ic \cite{Turatto2003}), the core of a massive star collapses and forms a neutron star which is mostly comprised of neutrons (also known as a gigantic nucleus \cite{Lifshitz_1969}). Both a neutron star and an atomic nucleus are governed by  the strong force, one of the four fundamental forces,  while having huge difference in their size. For a typical neutron star, the radius is $\approx$ 10 km in contrast to 10 fm of the radius of a typical nucleus. The density of the neutron star core, where the maximum  mass is concentrated, becomes $8-10$ times the  nuclear saturation density. These contrasting features of a neutron star and an atomic nucleus prompted the researchers to investigate the structure and dynamics of a neutron star and an atomic nucleus in a parallel fashion. While the main thrust of the nuclear physicists is to understand the structure of the nucleus and related aspects, such as  nuclear potential, nuclear reactions, search for superheavy nuclei, fusion, fission etc., the nuclear astrophysicists are focused on  the global properties of the neutron star, consistent with the observations and understanding of supernovae  mechanisms.

Unlike an atomic nucleus that can be investigated in a terrestrial laboratory, a neutron star is an elusive astrophysical object. Only indirect estimations of its properties, such as mass-radius profile, structure, composition etc., is possible \cite{Lattimer_2012}. It requires various kinds of telescopes in a wide range of the electromagnetic spectrum, both ground-based and on-board, to infer the neutron star properties. Additionally, analysis of the neutrino emission from supernovae explosions in neutrino observatories across the globe provides essential information on the neutron star properties  \cite{burgio2020equation}. The astrophysical observations on neutron stars are often correlated with the properties of atomic nuclei, which serve as crucial tools for better understanding of nuclear matter. Since neutron stars and supernovae explosion are asymmetric systems, the properties such as neutron skin thickness of highly asymmetric nuclei such as $^{208}$Pb and $^{132}$Sn are correlated with the radius of the neutron star with low mass \cite{STEINER2005325}. These similarities between the two highly contrasting nuclear systems allow researchers to develop and test nuclear theories and experiments in the unknown and unconventional regimes.

%The main ingredient to study the structure and dynamics of the neutron star, supernovae explosion and the atomic nucleus is the nuclear equation of state (EoS) \cite{KARSCH2011136, BURGIO2021103879, Miller_2021}.
The nuclear equation of state (EoS) plays a central role in determining the static properties of a cold-catalyzed neutron star and is  vital in describing phenomena such as supernovae explosions and the atomic nucleus \cite{KARSCH2011136, BURGIO2021103879, Miller_2021}. Theoretical description of the nuclear EoS in conventional and unconventional regimes is crucial for understanding of nuclear matter in its various forms. While the low-dense EoS dictates the nuclear phenomena such as the structure of the nuclei, nuclear reaction, supernovae matter, etc., the EoS at high density is essential for describing neutron star interiors, Quark-gluon plasma, etc. It is desirable that a nuclear EoS is capable of describing all these phenomena, which is, however, a very challenging task due to the scarcity of the model-independent nuclear interaction calculations within the framework of quantum chromodynamics (QCD). As a remedy, a nuclear EoS is first developed by fitting various nuclear models on the available data on finite nuclei, such as binding energy, charge radius, heavy-ion collision (HIC) etc. This EoS is then used to predict the neutron star's global properties and the supernovae's composition. In the last two decades, an enormous amount of effort has been made to deduce various nuclear models based on microscopic and phenomenological approaches. There is a continuous attempt to make these models as realistic as possible, based on the constraints on nuclear EoS \cite{huth2022constraining}.
 
The first detection of gravitational waves (GW) in 2015 by  LIGO and Virgo Collaboration revolutionised nuclear and astrophysics and unfolded a new era of science \cite{Abbott_2016, Abbott_2017, Abbott_2018}. With the constant improvement in nuclear theory and instrumentation, including third-generation GW observatories like the Einstein Telescope (ET) and Cosmic Explorer (CE \cite{gupta2022determining}), the investigation of nuclear matter under various conditions has recently gained momentum. The aim is to have stringent constraints on EoS, a more exact picture of neutron star interior, developing unified EoS to describe nuclear matter etc. This thesis is a work in the same direction and aims to investigate the properties of nuclear matter over a wide range of density, temperature, isospin asymmetry and magnetic field.

\section{ Nuclear matter and its phases}
\begin{figure}[htb!]
    \centering
    \includegraphics[scale=0.6]{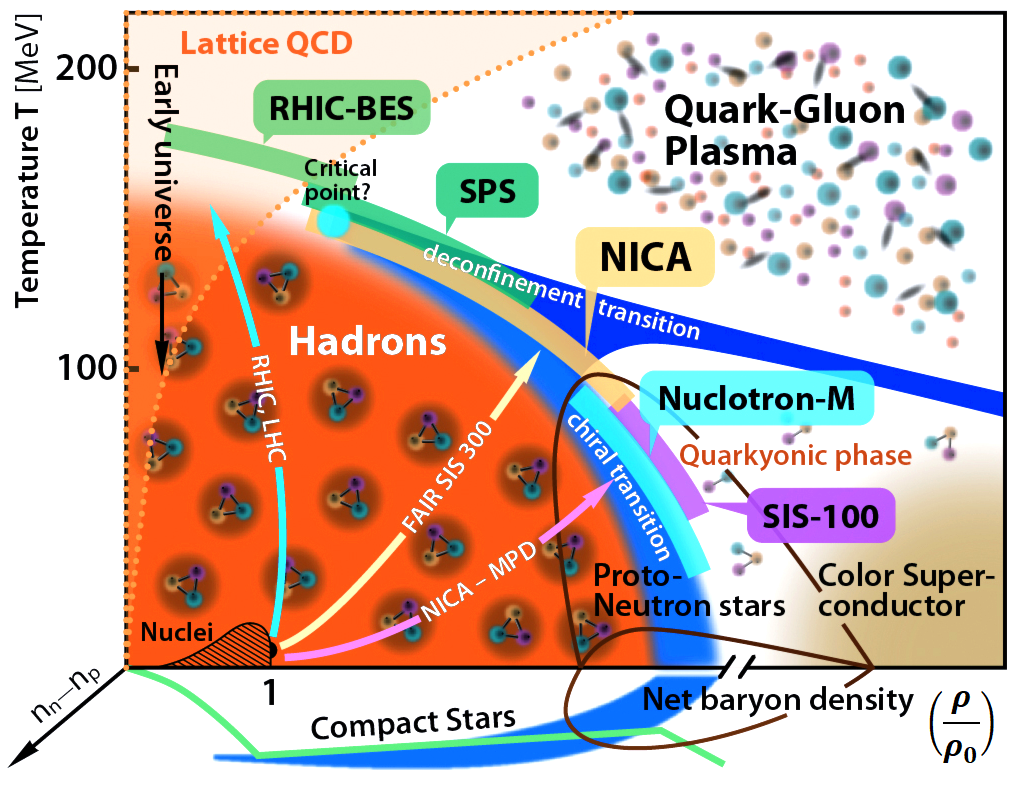}
    \caption{Nuclear matter phase diagram  as a function of baryon density and temperature \cite{universe4030052}. Various experimental set ups point towards the applicable regions in the phase diagram.}
    \label{fig:nuclear_matter_phases}
\end{figure}

After about $10 \mu $S of the Big Bang,  the quark-hadron transition resulted in the formation of neutrons and protons when the temperature was about a  trillion degrees \cite{Olive_1991}. These particles, also known as nucleons, constitute the ``nuclear matter" in today's universe. Nucleons inside an atomic nucleus are bound by the residual strong or nuclear force. As we go inside the nucleus, the nucleons are colorless and the bound states of quarks governed by the strong force mediated by massless gluons. One of the astonishing universality in the laws of nature is the resemblance between the strong and the molecular force. Molecular force is of van der Waals type, and strong force behaves similarly, albeit on a different energy and length scale. Therefore, the nuclear matter phase diagram resembles the phase diagram of water which has been confirmed by various phenomenological and microscopic effective nuclear interactions \cite{RIOS201058, DUCOIN2007403, Typel_2010}. One can see in Fig. \ref{fig:nuclear_matter_phases} that the phase diagram of nuclear matter is quite involved, which is drawn in the Temperature-Baryon density ($T-\rho$ )  plane.  

At T $\sim$ 0 axis, one sees two significant entities: atomic nuclei and the neutron star or protoneutron star. For various isospin asymmetries, the atomic nuclei exist only in a small region of the phase diagram at low densities. In contrast, the neutron star is situated at extreme density and low temperatures. The effective degree of freedom in atomic nuclei and the neutron star are the hadrons. At higher temperature, the quarks and gluons become the degrees of freedom and form an exotic state of matter known as Quark-gluon plasma (QGP). In QGP, the quarks become deconfined, i.e. they are allowed to exist in the free state. The existence of the QGP originates from the property of asymptotic freedom in the strong interaction \cite{Gross_1973}. QGP is also expected to exist in the interiors of neutron star with mass exceeding 2$M_\odot$ \cite{Beni_2015}. The QGP and Hadron phase are separated by the crossover region, which ultimately makes a  first-order phase transition at low and zero density \cite{Guenther_2020}. At extremely high densities, some fascinating phenomena, such as the color superconducting phase, is expected to exist \cite{RISCHKE2004197}. Finally the $\rho_b \sim$0 or $\mu_b=0$ line resembles the early universe and is a crucial part of phase diagram.

In the last few decades, continuous efforts have been made to investigate the phase diagram of nuclear matter using various theoretical and experimental techniques. While Lattice quantum chromodynamics (LQCD) explains the low-density region or the region near zero chemical potential, most of the regions in phase diagram are examined in various state-of-the-art heavy ion collision (HIC) experiments across the globe. Among these, the  Large Hadron Collider (LHC) and Relativistic Heavy Ion Collider (RHIC) aim to understand the crossover regions of hadron gas (HG) and QGP, while the Super Proton Synchrotron (SPS) experiment tries to understand and estimate the phase transition line and critical points of HG and QGP. Researchers have faced a challenging endeavor in identifying the critical points at both $\mu_b=0$ and $\mu_b \ne 0$. While the LQCD ab-initio calculations can only be applied to $\mu_b=0$ cross-over thermodynamics \cite{STEINBRECHER2019847}, the exploration of the remaining phase diagram involves two approaches: i) extrapolating the low $\mu_b$ region, and ii) utilizing effective QCD models like pNJL for finite $\mu_b$ scenarios \cite{Ratti2007}. With more and more accurate LQCD calculations using supercomputers, the overall conclusion is that the crossover region exists at $T_c(\mu_b=0)=154\pm 9$ $MeV$ \cite{Bazavov_2017}. At the same time, the critical point is not expected to occur for $\mu_b/T \le2$ and $T/T_c (\mu_b=0) \ge 2$ \cite{Bazavov_2014}. The accurate estimation of this chiral critical point is highly desirable to understand the fundamental mechanisms associated with the strong interaction and, consequently, the understanding of the early universe.

\subsection{Liquid gas phase transition in nuclear matter}

\begin{figure}
    \centering
    \includegraphics[scale=0.25]{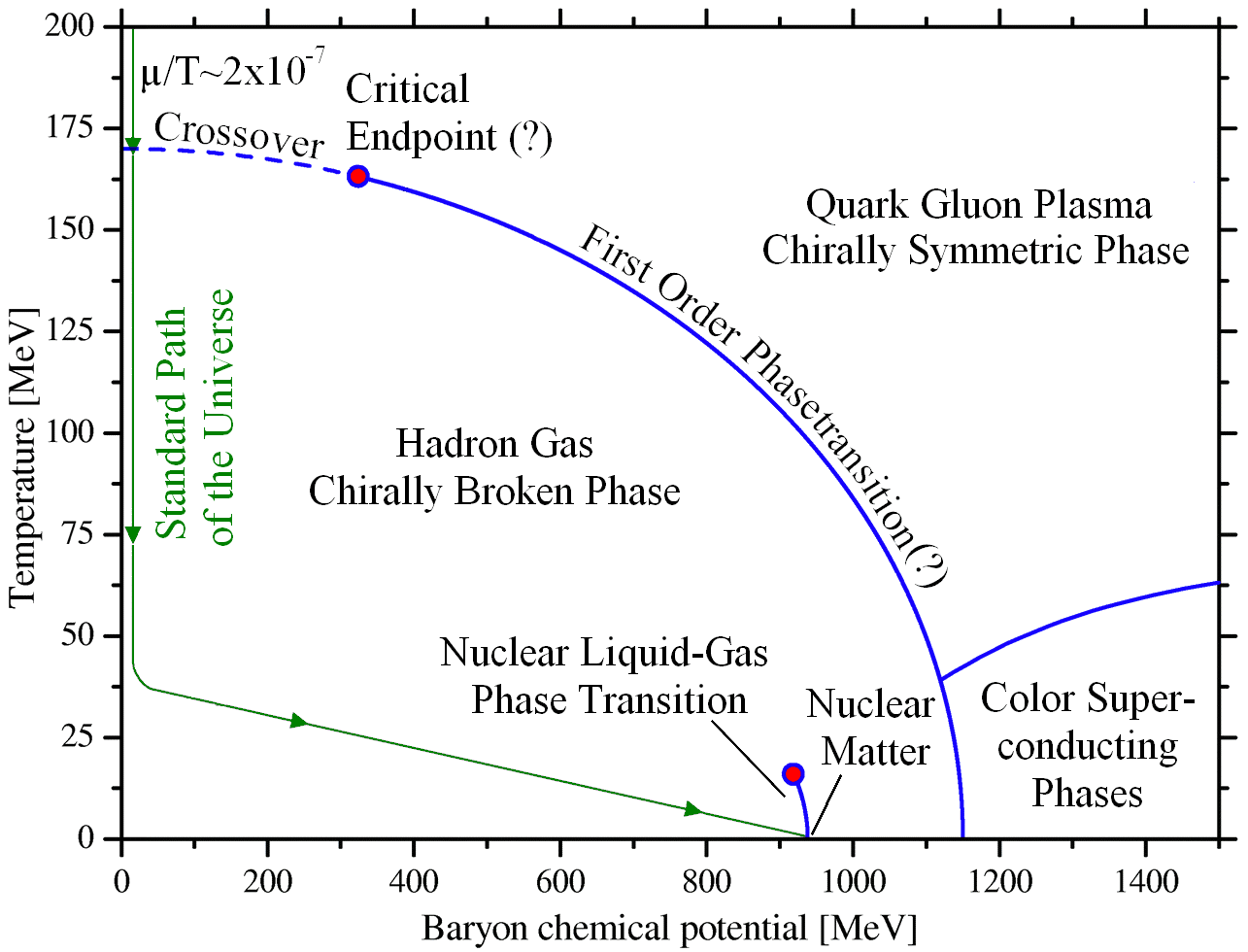}
    \caption{Various phase boundary, critical points and accepted path of the universe \cite{Boeckel_2010} in the nuclear matter phase diagram. (Adopted from \cite{universe4030052}) }
    \label{fig:lgpt_diag}
\end{figure}

Out of various fascinating phenomena in nuclear matter phase diagram, the nuclear liquid gas phase transition (LGPT) is one of the  intriguing and vital mechanism. While the order of the chiral phase transition remains a subject of debate \cite{Cuteri2021}, it is established that the nuclear LGPT  undergoes a first-order phase transition. To put it in perspective, Fig. \ref{fig:lgpt_diag} shows various phase boundaries, critical points and the accepted path of the universe \cite{Boeckel_2010}, in the nuclear matter phase diagram. The phenomenon of LGPT in both infinite nuclear matter and finite nuclei is an important feature of heavy-ion-induced reactions (HIR) \cite{PhysRevLett.72.3321, tlimexpdata, WU1997385}. In these reactions, the participating hot nuclei undergo multi-fragmentation after the initial dynamic stage of compression upon reaching the sub-saturation density ($\approx 0.2\rho_0$) \cite{KARNAUKHOV200691}. In this sub-saturation density region, the properties of nuclei are modified \cite{PhysRevLett.94.162701, PhysRevC.95.061601, BORDERIE201982} which are very essential for the understanding of the thermodynamics of hot nuclei, and the medium in which they are created. The knowledge of the nuclear matter in the sub-saturation region is also important in context to the core-collapse supernovae \cite{Liebendorfer_2005}, neutron star crust and giant astrophysical explosions where nuclear matter minimizes its energy by forming clusters at temperature $\approx 4$ $MeV$ \cite{HOROWITZ200655}.

The $\gamma$ ray emission is the dominating process in the nucleus at low excitation energy, where nuclear levels are well resolved. As excitation energy increases slightly, nuclear energy levels are substantially modified. The single-particle energy states become degenerate and nuclear shells start melting, leading to a continuum spectrum after a  temperature usually known as shell melting temperature T$_m \approx 1-2$ MeV \cite{Quddus_2018}. Further increase in temperature leads to nucleon emission, which is generally studied within the framework of Hauser-Feshbach theory \cite{Hauser_1952}. On further heating, the nucleon evaporation turns violent, and at a certain limiting temperature, T$_{l}$, a new decay channel known as multi-fragmentation becomes dominant. This T$_{l}$ was found to be $\approx 5.6$ $MeV$ for the mass region A $\approx$ 90 in the  experiment  using the ALADIN forward spectrometer at the heavy ion synchrotron (SIS) at Gesellschaft für Schwerionenforschung (GSI) Germany, which was established to conduct research on and with heavy-ion accelerators \cite{PhysRevLett.102.152701}.   Nuclear multi-fragmentation occurs in the spinodal or phase instability boundary region in the nuclear matter phase diagram \cite{vishalsymmetric}. The nucleus, which resembles a hot liquid drop, expands because of thermal pressure and moves to the spinodal region where a nucleon gas surrounds it. As the spinodal is the region of instability, the nucleus explodes violently, and the process is known as multi-fragmentation which  occur in a low density configuration \cite{karnau2003, BORDERIE201982}.

There have been several qualitative attempts to study the limiting temperature of nuclei in terms of Coulomb instability, where the  EoS of infinite matter is taken from  various  theoretical frameworks such as Skyrme effective NN interaction \cite{PhysRevC.44.2505, PhysRevC.69.014602}, microscopic EoS such as Friedman and Pandharipande \cite{fp}, finite temperature relativistic Dirac-Brueckner, chiral perturbation theory \cite{microscopic, m1, NICOTRA2005118}, EoS considering the degeneracy of the Fermi system \cite{PhysRevC.39.169}, relativistic calculations using quantum hadrodynamics, Thomas-Fermi approach \cite{PhysRevC.47.2001, PhysRevC.49.3228, PhysRevC.55.R1641},  Gogny interactions \cite{PhysRevC.54.1137} and  chiral symmetry model \cite{PhysRevC.59.3292}.  Some calculations  have been carried out by analyzing the plateau in caloric curve obtained from various experimental observations \cite{PhysRevC.65.034618}. The caloric curve of nuclei depicts the relationship between temperature and the excitation energy of the system. It exhibits a distinctive plateau-like behaviour at a specific temperature, akin to the phenomenon of Latent heat observed in the caloric curve of water. These calculations  give a qualitative picture of $T_{l}$ and it is seen that $T_{l}$ is  model dependent and hence needs to be investigated for appropriate outcome. The plateau observed in the caloric curve can offer insights into the limiting temperature of hot nuclei, whereas the critical points for infinite nuclear matter are determined by extrapolating data from the multifragmentation reaction on finite nuclei \cite{Elliott_2013}. However, this method has various limitations due to factors such as the Coulomb interactions, angular momentum, isospin effects, finite size,  and secondary decay of excited fragments. Additionally, it is important to note that finite nuclei can only be excited up to a certain temperature beyond which Coulomb effects, combined with the reduction  in  the surface tension, lead to their thermal dissociation. In the experiments to understand the phase transition in nuclear matter \cite{Panagiotou_1984, Li_1994, Bonard_2002, Zagrebaev_2003, Elliott_2013}, the critical temperature $T_c$ is hardly constrained. There is a large uncertainty in the value of $T_c$ among these experiments.
Moreover, the model dependence in these experimental calculations arises inevitably. At the theoretical front, there have been numerous attempts to estimate critical temperature using non-relativistic approaches such as   Hartree-Fock theory \cite{BONCHE1981496, Hua_2004}, Skyrme interaction, Thomas-Fermi model \cite{Bliss_2020} and Gogny interaction \cite{HEYER1988465} etc. Several calculations have also been done in the relativistic domain using relativistic mean-field framework \cite{Malheiro_1998}, and the critical temperature is found in the range $14.2-16.1$ $MeV$. In recent years, various ab-initio or fully microscopic calculations such as self-consistent Green’s functions, Monte Carlo method, etc., of the nuclear matter equation of state (EoS) at the finite temperature have been performed \cite{HOLT201335}. These calculations yield the critical temperature ($T_c$) of infinite symmetric nuclear matter as $T_c=15.80^{+0.32}_{-1.60}$ $MeV$ \cite{Lu_2020}.

\section{Neutron star}
Astrophysical objects known as neutron stars are among the most compact and densest. They are the remnant collapsed core after supernovae explosions  that mark the evolutionary end-points of giant stars with mass $8-20\ M_\odot$ \cite{Couch_2017}. During their lifetime, these stars burn their core through nuclear fusion, which forms heavy elements up to the iron, after which the fusion stops. As a result, the star exhibits an onion-like structure in the ﬁnal stage, with no more possible energy production \cite{BETHE1979487}. As the core mass keeps increasing up to (1.44 $M_\odot$, the Chandrashekar mass limit) through accretion, the electron degeneracy pressure can no longer sustain the gravitational pull, triggering the core collapse, which is known as the core-collapse supernova (CCSN) explosion \cite{JANKA200738, couch2017mechanism, ccsn}. The CCSN forces the electrons and protons to  form the neutrons and neutrinos. While the neutrinos escape the contracting core, neutrons keep coming closer providing the necessary neutron degeneracy pressure (along with the nuclear interaction) to sustain if the core is less than maximum possible neutron star mass \cite{Ozel_2016, Woosley2005, Lattimer_2001}. For larger masses, this pressure is not enough to counter gravity, and it ultimately collapses to become a stellar black hole. The resulting neutron star (or the protoneutron star (PNS)) possesses a temperature of $\sim 10^{10}$ $K$, which cools down by the emission of neutrinos for which the neutron star is transparent and photons. The PNS is catalyzed or in the ground state when it reaches a temperature of $\sim 10^{8}$ $K$.

%%%%%%%%%%%%%%%%%%%%%%%%%%%%%%%%%%%%%%%%%%%%%%%%%%%%%%%%%%%%%%%%%
\subsection{Neutron star structure}
%%%%%%%%%%%%%%%%%%%%%%%%%%%%%%%%%%%%%%%%%%%%%%%%%%%%%%%%%%%%%%%%%
\begin{figure}
    \centering
    \includegraphics[scale=0.5]{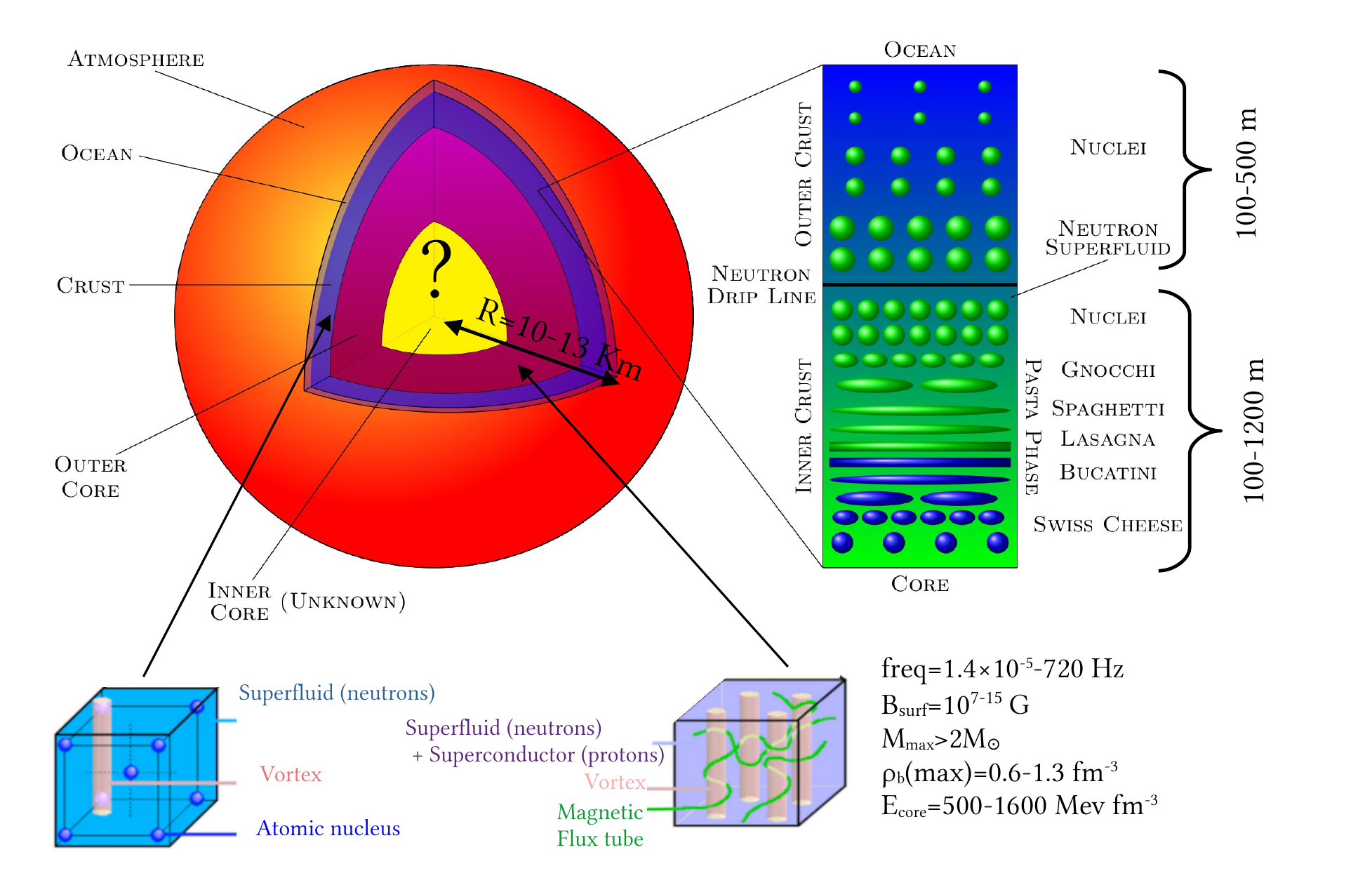}
    \caption{A schematic diagram of a neutron star with its surface and interior. Various components and data are referred from \cite{Dany_2006, chamel2008physics, Jason_2006, nstar_web}}
    \label{fig:ns_interior}
\end{figure}
%%%%%%%%%%%%%%%%%%%%%%%%%%%%%%%%%%%%%%%%%%%%%%%%%%%%%%%%%%%%%%%%%

The internal structure of a typical cold nonaccreting neutron star can be divided into three distinct parts below its thin atmosphere:  two concentric inhomogeneous outer and inner crusts followed by a dense homogeneous liquid core \cite{Haensel_2008, Haensel_2009, BKS_2015}. The neutron star remains in complete thermodynamic equilibrium against all possible interactions and in the lowest energy state as per the cold catalyzed matter hypothesis \cite{chamel2008physics}. A schematic diagram of a neutron star with its surface and interior is shown in Fig. \ref{fig:ns_interior}. The outermost layer consists of  layers namely atmosphere and ocean/envelope, which have combined thickness of few meters only. In this region the density is below $10^{-10} fm^{-3}$ and the electrons are still bounded to the nuclei. One can use generalised Thomas-Fermi (TF) theory to calculate the properties of this thin layer \cite{Feynman_1949, Haenel_2007}. Thermal electromagnetic spectrum  especially soft $X-$ ray \cite{Haenel_2007} from this layer provide valuable information on surface temperature and mass-radius profile of the star. Beneath the thin atmosphere, lie the two concentric inhomogeneous outer and inner crusts followed by a dense homogeneous liquid core \cite{Haensel_2008, Haensel_2009, BKS_2015}. These layers of the neutron star account for majority of the structural and dynamical aspects of the star and are discussed below.

\subsubsection{Outer crust}
The composition of the outer crust is not only significant for the neutron star but also crucial in describing white dwarfs which require the same theoretical tools such as dense plasmas energetics \cite{VANHORN19903}. Globally charge neutral, the "Outer crust" is composed of fully ionised atoms arranged in a body-centered cubic (BCC) lattice and immersed in a sea of electrons.  The BCC lattice is prefered in the crust due to its lowest lattice energy. The electron density is uniform and the charge screening effects are negligibly small due to screening length being larger than the lattice spacing \cite{Watanabe_2003, Pethick_1995}. 
As the depth of the stars increases, there is a noticeable increase in the presence of neutron-rich nuclei resulting from electron capture. This trend continues until the inner crust is reached, at which point high-density conditions cause the neutrons to drip out of nuclei \cite{Chamel_2016, Roca_2008, Pearson_2018}.  Using the variational formalism developed by Baym-Pethick-Sutherland (BPS) \cite{BPS_1971}, the composition of the outer crust is computed from a density of $10^{-10}$ fm$^{-3}$ up to the point of neutron drip onset. It considers that the ensemble of heavy nuclei may be represented by a single nucleus commonly known as the single-nucleus approximation (SNA) \cite{Shen_2011}, thus giving a unique configuration for given thermodynamic conditions. However, it is important to note that the SNA   is applicable only to cold, non-accreting neutron stars. In scenarios like core collapse supernovae, where the temperature exceeds 0.5 MeV, the SNA  becomes inadequate. This is because statistical mechanics dictates that at finite temperatures, there is a mixing of various microstates. To address such cases, modern approaches rely on an extended concept called Nuclear Statistical Equilibrium (NSE) \cite{Gulminelli_2015, RADUTA2019252}. In the NSE framework, the distribution of nuclear clusters over all possible mass numbers is taken into account and self-consistently obtained under conditions of statistical equilibrium \cite{Blinnikov_2011}. Although the SNA  remains reasonably accurate when estimating the thermodynamic properties of matter \cite{Burrows_1984}, it becomes less reliable for dynamical processes that heavily rely on reaction rates of specific nuclei \cite{JUODAGALVIS2010454}. In the calculation of outer crust, the only uncertain input is atomic mass evaluations.  Therefore, the accurate and precise determination of the outer crust composition stems from our ability to measure the masses of nuclei in terrestrial laboratories \cite{Huang_2021}.   As the mass evaluations are not possible for highly neutron-rich nuclei in the laboratory,  the need to use a mass model become inevitable. Various mass models such as those using the Hartree-Fock-Bogoliubov (HFB) theory with the Skyrme   \cite{Samyn_2002} and Gogny interaction \cite{sym13091613}, finite range liquid drop model (FRLDM) \cite{MOLLER20161}, BCPM \cite{BKS_2015},  etc. have been used in the literature to estimate the outer crust structure. Recently machine learning algorithms have also been developed for this purpose \cite{Murarka_2022}.

\subsubsection{Inner crust}

The ocurrence of neutron drip indicates the beginning of the inner crust, which has an intricate structure making it a challenging problem. The inner crust is marked by the assembly of the clusters formed by neutrons and protons along with the unbound neutrons making the neutron gas. The system is neutralized by the electron gas, which is distributed uniformly over the cluster and neutron gas \cite{Margaritis_2021, Oyamatsu_2007}.  Various calculations hint that the neutron in the inner crust have superfluid nature which is confirmed by the glitch behaviour in the neutron star \cite{Prix_2002}. This gives rise to the  entrainment effect which reduces the density of the free neutrons \cite{Chamel_2012, Martin_2016}. As the density increases, the size of the cluster in the inner crust increases, and at a density called transition density, the inhomogeneities disappear, and we enter the liquid core of the star. The estimation of the composition of the inner crust is limited by the inevitability of using an empirical mass model because of our inability to measure the mass excess of highly neutron-rich nuclei in a neutron gas background. At low densities, the clusters are at a sufficient distance from each other and are expected to be spherical in shape \cite{Dinh_2021}. However, at high densities, i.e., near the crust-core transition density, the system becomes ``frustrated'' due to the competition between the Coulomb and nuclear interactions \cite{Avancini_2008, Avancini_2009}. The frustration leads to the system arranging itself into various exotic geometries commonly known as "nuclear pasta" \cite{Ravenhall_1983,  Lorenz_1993, Maruyama_1998}. These configurations of nuclear pasta are related to the complex terrestrial fluids such as glassy system \cite{Newton_2022} rather than a solid and have a variety of responses towards the mechanical stimuli \cite{PETHICK19987, watanabe2007dynamical}. 

Although there exists no direct and robust observational evidence of nuclear pasta, various tantalizing observations indicate its existence \cite{Horowitz_2014,  Pons_2013, Horowitz_2015}. Numerous  theoretical attempts  based on molecular dynamics simulations \cite{Lin_2020, li2021tasting, watanabe2007dynamical}, compressible liquid-drop models \cite{Carreau_2019, Newton_2021}, Thomas-Fermi method \cite{Haensel_2008, Furtado_2021, bcpm} and nuclear density functional theory \cite{schuetrumpf2016clustering} point towards the possibility of the pasta structures near the crust-core transition. The amount of these structures in the inner crust plays pivotal role in the explanation of  various neutron star mechanisms such as crust cooling \cite{Horowitz_2015, Ootes_2018}, spin period \cite{Pons_2013}, quasiperiodic oscillation in giant flares \cite{Steiner_2009}, transport \cite{rezzolla2018physics}, shattering of the crust \cite{Tsang_2012} etc.  The discovery of quasiperiodic oscillations (QPOs) in soft gamma repeaters (SGRs), which are related to the torsional vibrations of the neutron star crust, enables us to put constraints on the thickness and mass of the pasta structure and quadrupole ellipticity sustainable by the crust \cite{Gearheart_2011}. Theoretically, this is achieved by new approaches to nuclear models in the form of Bayesian inference \cite{Newton_2021, Carreau_2019, Balliet_2021}   and establishing possible correlations between parameters and crust properties through systematic surveys of models \cite{Newton_2013, Zhang_2018, Horowitz_2001, Oyamatsu_2007}. These approaches of nuclear models are essential to account for the upgradation of data on the nuclear matter  observables with improved quantity and fidelity. However, one must take a simplistic energy density functional (EDF) to account for the computational requirements.  

Different treatments of inner crust are available such as microscopic calculation pioneered by Negele and Vautherin \cite{NEGELE1973298} using the microscopic Hartree-Fock approach and subsequently modified by Baldo {\it et al.} \cite{Baldo_2006}, and Onsi {\it et al.} \cite{Onsi_2008} which uses the extended Thomas-Fermi (ETF) formulation.   Microscopic calculations that explicitly consider the quantum nature are highly accurate, but they face challenges due to the extensive computational power required for large-scale statistical analyses like Bayesian analysis. Additionally, extending these calculations to multi-component systems, which are necessary for finite temperature calculations, is not a straightforward task.
On the other hand, classical formalism such as the compressible liquid drop model (CLDM) \cite{Mackie_1977, Newton_2021} is computationally economical and can be easily extended to the finite temperature regime. The CLDM model is modified from the conventional semiempirical model by Baym-Bethe-Pethick \cite{Baym_1971}, which incorporated the compressibility of nuclear matter, negative lattice Coulomb energy, and the suppression of surface tension by the neutron gas. The results of CLDM are known to be at par with those of ETF, and TF calculations \cite{Newton_2012}. It should be noted that the CLDM requires that the same functional be used for the calculations of bulk as well as the finite size contributions. The CLDM is recently applied in the work of Refs. \cite{Carreau_2019, Carreau_2020, carreau2020modeling} where the energy-density functional is taken in the form of meta-modeling, a technique developed to mimic the original relativistic or nonrelativistic functional using the isoscalar and the isovector energy of the EoS \cite{Margueron_2018} and for the Bayesian inference of neutron star crust properties \cite{Newton_2021}.  The meta-modeling reduces the computational difficulties when studying the statistical properties such as Bayesian inference to constrain the EoS.

\subsubsection{Core}

%As one moves deeper into the neutron star's crust, the nuclei can not exit for densities ($> 0.1-0.12$) $fm^{-3}$, and the matter becomes a homogenous mixture of neutrons, protons and electrons. This part of the neutron star is called the core. The core is further divided into the outer and inner cores. In the outer core, the ground state composition consists of a mixture of neutrons, protons, electrons and muons under $\beta-$ equilibrium. The neutrons are believed to be superfluid while the protons act as superconductors \cite{Prix_2002}. The properties of the core and structure can be calculated reliably using nuclear many-body theories that have been applied to investigate the microscopic nuclear matter in recent times. On the other hand, the exact composition of the inner core is still unknown. There have been a plethora of predictions regarding the presence of hyperons, Kaon or pion condensate in this region of the neutron stars. Various studies also point towards a possible phase transition to form a deconfined quark plasma state \cite{Huang_2022}. 

As one delves deeper into the crust of a neutron star, nuclei are unable to exist beyond densities of ($> 0.1-0.12$) $fm^{-3}$, causing the matter to transform into a homogeneous mixture of protons, neutrons, and electrons. This region is known as the core and is further classified into inner and outer cores. The ground state composition of the outer core is believed to consist of a mixture of neutrons, protons, electrons, and muons under $\beta-$ equilibrium, where neutrons are superfluid and protons act as superconductors \cite{Prix_2002}. The core's properties and structure can be reliably determined using nuclear many-body theories, which have been recently applied to investigate the microscopic nuclear matter. However, the exact composition of the inner core remains unknown. There are several predictions regarding the presence of hyperons, Kaon, or pion condensates in this region \cite{Thapa_2021, Thapa_2020}, and some studies suggest the possibility of a phase transition leading to a deconfined quark plasma state \cite{Huang_2022, Mallick_2021, Prasad_2022}.

\subsection{Magnetar and Pulsar}

In a recent breakthrough, astronomers detected an extremely bright radio burst from the Galactic magnetar SGR 1935+2154 \cite{chime2020bright, bochenek2020fast}, which confirmed that the gamma-ray bursts (GRBs) originate from the magnetars at cosmological distances \cite{kirsten2021detection}. Magnetars are the family of neutron stars with an extremely intense magnetic field ($\ge 10^{15}$ $G$) known for their observed quiescent in a wide range of the electromagnetic spectrum that includes $X-$  ray and $\gamma-$ ray in the form of powerful bursting emissions \cite{Kaspi_2017, Konstantinos_2016}. The origin of the colossal magnetic field in the magnetar is still controvertible; however, a common hypothesis is that strong dynamo effects caused due to the initial spin period are responsible for such an extreme environment \cite{Thompson_1995, Vink_2006}. The activities of the magnetars are principally associated with the crustal motion, which twists its magnetosphere \cite{Beloborodov_2014, Thompson_2002}. Currently, researchers have a catalogue of only 30 magnetars%\footnote{\url{https://www.physics.mcgill.ca/~pulsar/magnetar/main.html}}
, which primarily originate from  Soft Gamma-ray Repeaters (SGRs)  and Anomalous X-ray Pulsars (AXPs). Both SGRs and AXPs are associated with the core-collapse supernovae (CCSN), hinting that they are relatively younger stars \cite{Zhang_2000}. These emit powerful $X-$ ray with the difference that SGRs occasionally emit $\gamma-$ ray bursts. 

In literature, most of the observed neutron stars are pulsars, which convert the star's rotational energy into periodic multi-wavelength radiations \cite{Kaspi_2010, Baiko_2017}. The pulsars are rapidly rotating objects with the highest measured frequency of 716 Hz \cite{Jason_2006} and are generally characterised by their spin periods, i.e. the time lapsed between two successive pulses. These are also characterised by a strong magnetic field ($10^{12}-10^{14}$ G) \cite{lorimer2008binary}, which channel streams of particles through their two magnetic poles. These accelerated particles give rise to intensely energetic beams of light. The pulsed emission, which is in the radio frequency band, is the direct way of measuring the rotation of the crust using the pulsar timing technique \cite{Basu_2018, Strappers_2011}. By measuring the time of arrival of the pulse, one can estimate the crust's rotational speed and glitch activity. The glitches are produced due to the sudden spin-ups in the radio pulsars. This is because the angular momentum transfers from the superfluid component of the stellar interior to the solid crust. Therefore, there is a change of moment of inertia from the superfluid to the rest of the star. As compared to the magnetars, currently there are more than 3359 catalogued pulsars %\footnote{\url{https://www.atnf.csiro.au/research/pulsar/psrcat/}}
 \cite{Manchester_2005}. Magnetars, along with pulsars, provide extraordinary opportunities to develop and test theories or models to describe and explain the wide range of associated observational phenomena \cite{Turolla_2015, mereghetti2015magnetars}, such as gamma-ray bursts (GMR), fast radio bursts (FRB), $x-$ ray outbursts, etc. 

\section{Equation of state}
The properties of a system under specific physical conditions can be described by a thermodynamic relation between state variables known as an equation of state (EoS). EoS is one of the preeminent aspects of any nuclear system. Its theoretical determination is highly desirable to estimate the properties of various phases of nuclear matter. Nuclear EoS plays a paramount role in the modelling of static neutron star while it is one of the  required inputs in various phenomena such as the Heavy ion collision (HIC) reaction mechanism \cite{KARSCH2011136}, neutron star structure \cite{BURGIO2021103879}, core-collapse supernova explosions (CCSN) \cite{Rosswog_2015, shibata2011coalescence}, neutron star merger and associated stellar nucleosynthesis \cite{Pons_1999}, structural and reactional aspect of finite nuclei \cite{burrello2019symmetry} etc.

The HIC experiments as shown in Fig. \ref{fig:nuclear_matter_phases} provide us with crucial information on the EoS over a wide range of pressure, temperature, and neutron-proton asymmetry etc. with density slightly above the nuclear saturation density $\rho_0 \sim 0.16$ $ fm^{-3}$ \cite{FU2008359, BURGIO2021103879}. For the dense EoS relevant to the neutron star, one relies on various astrophysical observations related to the neutron star and gravitational waves, such as data from Neutron Star Interior Composition Explorer (NICER) \cite{Miller_2021, Miller_2019}, VIRGO, LIGO collaboration \cite{Abbott_2016, Abbott_2017, Abbott_2018} etc. Moreover, the typical time scale of the HIC is different from the neutron star, which hinders the weak processes crucial for the neutron star's relaxation, impeding a direct link between the two fields. Furthermore, HIC observations do not directly measure the energy functional and require a transport theory (which is still not well-established) to extract information on the EoS \cite{huth2022constraining}.

In the past few decades, there were significant attempts to understand the nuclear EoS theoretically using various baryon-baryon interaction models. The meson theory of  Yukawa in 1935 was the starting point of such developments. As quantum chromodynamics (QCD) has emerged as the leading theory of strong interaction, it has become increasingly necessary to derive nuclear Hamiltonians from QCD. However, this poses a mammoth task to the researchers \cite{Rujula_2019} and prompts them to look for the alternatives such as chiral effective field theory ($_\mathcal{X}$EFT) \cite{WEINBERG1990288}, phenomenological models \cite{Machleidt_2001}, resonating group models \cite{FUJIWARA2007439} etc. The primary aim of these nuclear interaction models is to have a simplified understanding of nuclear interaction without compromising the fundamental properties such as symmetries arising in the QCD. Based on these interaction models, numerous theoretical frameworks for understanding various nuclear systems (many body systems) have been developed. These approaches are broadly classified into two categories: \textit{Ab-initio} and Phenomenological approaches. 

The \textit{Ab-initio} approaches which involve many body methods such as variational methods \cite{RevModPhys.51.821}, Dirac–Brueckner–Hartree –Fock (DBHF) framework \cite{Baldo_2007},  Self-consistent Green’s function (SCGF) approach \cite{Dickhoff_2008}, Chiral EFT approach \cite{Entem_2003},  Quantum Monte Carlo (QMC) framework \cite{carlson2015quantum} etc., 
are based on bare two- and three-nucleon interactions that can reproduce the nucleon scattering data and bound few-nucleon systems. Although efficient, this approach is limited to dense matter (such as neutorn star core) besides being computationally expensive \cite{BURGIO2021103879}. On the other hand, phenomenological approaches are based on effective interactions and rely on EDF, which relates the energy of the system to various parameters such as density, kinetic and potential energy etc. These models are fitted to various finite nuclei properties. The primary challenge with this approach lies in extrapolating to high-density conditions since these models are fitted based on the ground-state properties of finite nuclei. However, the phenomenological approach is simple compared to the \textit{Ab-initio} method and it can describe the otherwise complex many-body nuclear system. The phenomenological approach is generally used in two domains; relativistic such as Relativistic mean-field (RMF) models based on quantum hadrodynamics \cite{Serot_1986, Serot_1997}, and non-relativistic, which include Skyrme forces, Gogny interactions \cite{HEYER1988465, decharge1980hartree}, etc. The EDF is constructed to reproduce the available data of finite nuclei and extrapolated to the case of the neutron star giving rise to uncertainties at high densities. Recently, a new approach in the form of Taylor series expansion around the nuclear saturation density, such as metamodel, has also been developed \cite{Margueron_2018, carreau2020modeling}. However, the behaviour of its convergence is not yet fully established \cite{Gil_2019}.

The determination of EoS is not only important in its ground state, but its behaviour at finite temperatures is equally significant. It plays the determining role in describing the CCSN and neutron star merger. Nuclear EoS at finite temperatures determines the transition state of both CCSN and neutron star binaries \cite{Baiotti_2017, stergioulas2003rotating}. While the zero-temperature EoS has sufficient constraints, it is not straightforward to establish a link between finite-temperature EoS and its experimental and observational counterparts. The main feature of finite temperature is the critical behaviour of nuclear matter, as discussed in the above section. The critical temperature  is found to have a significant deviation among various experimental and theoretical calculations. This fact points towards our limited understanding of the finite temperature behaviour of nuclear matter and the need to adopt a holistic approach to understanding it. In literature, there are few EoS available which include the effect of temperature over a wide range of density and isospin asymmetry \cite{SHEN1998435, HEMPEL2010210, TOGASHI201778, refId0}. The effect of temperature on the neutron star properties is also debatable, as few models suggest an increase in the maximum mass of neutron star  with temperature. At the same time, some of them hint towards the reduction in the maximum  mass and  neutron star stability with temperature \cite{BURGIO2021103879}.

\subsection{Constraints}
Even after continuous efforts by the scientific community, we are still forced to use various interaction models to understand nuclear phenomena. The persistent non-viability of  \textit{Ab-initio} QCD calculations of nuclear matter results in the deviations among adopted alternatives. Therefore, one has to test a given EoS against the available experimental and astrophysical observational data, known as constraints, to give reliable results in the unknown territories. Independent of the interaction models, the energy of the nuclear matter can be expanded as Taylor series around saturation density in terms of isoscalar (such as binding energy, incompressibility, skewness, etc.)  and isovector (such as symmetry energy, slope parameter etc.) parameters. The major thrust of the nuclear physics community  working in the field of neutron star and associated physics is to put stringent constraints on these nuclear matter observables to minimize the uncertainties in the various estimations for which an experimental or astrophysical observation is unavailable. 

Among various isoscalar observables, incompressibility ($K_0$) plays a crucial role in deciding the stiffness of the EoS. An EoS with larger $K_0$ at the saturation yields stiffer behaviour and, consequently, corresponds to a larger neutron star mass. In experiments, the value of $K_0$ can be extracted from the isoscalar giant monopole resonances (ISGMR) in heavy nuclei such as lead ($Pb$). The currently accepted value of $K_0$ is 240$\pm$ 20 MeV \cite{Col_2004}. Unlike isoscalar observables, the isovector observables are known to be uncertain and are limited by our understanding of nuclear force, especially its spin and isospin content. The isovector parameter symmetry energy and its higher order derivatives, such as slope parameter, symmetry incompressibility etc., can be deduced from isobaric analog state, kaon and pion production in HIC, isospin diffusion measurement, neutron skin thickness of the heavy nuclei etc. While the symmetry energy ($J$) is found to be $\approx$ 30 MeV ($29<J<32.7$ \cite{Lattimer_2013}, $30.2<J<33.7$ \cite{Danielewicz_2014} MeV), the slope parameter ($L$) is constrained to ($40.5<L<61.9$ \cite{Lattimer_2013},  $35<L<70$ \cite{Danielewicz_2014} MeV).  Recently, the nuclear symmetry energy constraints are also computed by analyzing the data from resonant shattering flare (RSF) from the binary neutron star merger and related gravitational wave signals \cite{Neil_2023}. The second or higher-order derivatives of the symmetry energy are still inadequately understood and lack sufficient constraints. In recent times, various theoretical attempts in the form of a survey of models and Bayesian inference techniques have also been made to determine the constraints on nuclear matter observables \cite{Xie_2019, Xu_2022,Malik_2022}. These calculations yield different ranges of the nuclear matter observables and hence  more astrophysical and experimental data is required to test their future validation.

A  nuclear EoS should be able to describe both finite nuclei and infinite nuclear matter. It should determine the structural property of atomic nuclei, neutron star and CCSN at par with the experimental calculations and astrophysical observations. Therefore, in addition to HIC data on finite nuclei, the astrophysical constraints on the EoS become equally essential. Constraints on the neutron star mass, radius, and its internal structure are one of the most sought after in nuclear astrophysics. The neutron star mass can be deduced from the binary system and CCSN. Various techniques such as gravitational waves (GW) measurements, pulsar timing, $x-$ ray measurments are used to deduce neutron star mass. The massive pulsar such as PSR J0740+6620 \cite{Cromartie_2019} ,   PSR J0348+0432 \cite{Antoniadis_2013} and PSR J1614–2230 \cite{Demorest_2010} estimate that the maximum neutron star mass should be greater than $2~M_\odot$ (($M = 2.14_{-0.09}^{+0.10} \ M_\odot$) \cite{Cromartie_2019}). The radius constraints are  given by Riley {\it et al.} \cite{Riley_2019}, Miller {\it et al.} \cite{Miller_2019},  and PSR J0030+0451 with X-ray Multi-Mirror Newton for canonical star with $R_{1.4} = 12.35 \pm 0.75$ km \cite{Miller_2021}. Recently, the largest neutron star mass was measured for pulsar PSR J0952-0607 as $M = 2.35 \pm 0.17 \ M_\odot$ \cite{Romani_2022}. Furthermore, the tidal deformability provide crucial information on the dense matter EoS and  helps to constrain the neutron rich matter at extreme density  \cite{BAIOTTI2019103714}. The GW170817 event \cite{Abbott_2017, Abbott_2018} provides an upper limit on the tidal deformability with the accepted canonical tidal deformability ($\Lambda_{1.4}=190^{+390}_{-120}$). Apart from these, various astrophysical observations such as gravitational redshift, magnetic fields measurements, pulsar llitches, surface temperatures, rotational periods etc. are used to provide valuable information on the neutron star internal structure, and they also help to constrain the nuclear EoS \cite{BURGIO2021103879}. 

\section{Effective relativistic mean field model}
%Quantum chromodynamics (QCD) is a local, relativistic, and fundamental gauge-invariant theory that describes the interactions between colored quarks and gluons, which are responsible for the strong force.  However, it can not be used to describe hadron matter as it suffers from nonperturbative properties. At large length or low energy scales, quarks cluster into colourless hadrons and therefore, QCD prediction, if at all available in future, will be inefficient and cumbersome  \cite{Serot2004}. On the other hand, Quantum hadrodynamics (QHD) is an effective theory of strong interaction at low energy. QHD is a Lorent-covariant effective field theory of nuclear many-body problems \cite{Serot_1992}. In QHD, the nucleons are considered relativistic four-component Dirac spinors and the NN interaction is mediated by the exchange of mesons. Based on QHD, relativistic mean field theory (RMF) provides an excellent description of nuclear matter and has been applied to a wide range of nuclear physics problems such as nuclear structure \cite{Patra_2002}, nuclear reaction \cite{Rana_2022}, neutron star structure \cite{Rather_2021, Ankit_2020, Das_2020} etc. The RMF theory is basically a  relativistic Hartree or Hartree-Fock approximation of one-boson exchange (OBE) interaction \cite{Kumar_2018}. In OBE, various isovector ($\pi$, $\rho$, $\delta$) and isoscalar ($\eta$, $\omega$) mesons serve as mediating particles. Among these, $\pi$ and $\eta$ are pseudo-scalar mesons and do not show ground state parity.  

Although Quantum Chromodynamics (QCD) is the fundamental, local, and relativistic gauge-invariant theory that describes the strong interaction between colored quarks and gluons, it is not applicable for studying hadron matter due to its nonperturbative properties. At low energy or large length scales, quarks bind together to form colorless hadrons, making QCD predictions, even in the future, inefficient and cumbersome \cite{Serot2004}. An alternative approach is Quantum Hadrodynamics (QHD), an effective theory of the strong interaction at low energy. QHD is a Lorentz-covariant effective field theory that addresses nuclear many-body problems \cite{Serot_1992}. In QHD, nucleons are treated as relativistic four-component Dirac spinors, and meson exchange mediates the NN interaction. Relativistic Mean Field (RMF) theory, based on QHD, provides an accurate description of nuclear matter and has been utilized in various nuclear physics applications such as nuclear structure \cite{Patra_2002}, nuclear reactions \cite{Rana_2022}, and neutron star structure \cite{Rather_2021, Ankit_2020, Das_2020}. The RMF theory is essentially a relativistic Hartree or Hartree-Fock approximation of the one-boson exchange (OBE) interaction \cite{Kumar_2018}, where several isovector ($\pi$, $\rho$, $\delta$) and isoscalar ($\eta$, $\omega$) mesons serve as mediators. However, $\pi$ and $\eta$ are pseudo-scalar mesons and lack ground state parity.

The RMF theory has proven to be an efficient method for describing the bulk matter and spin-orbit properties of nuclei across the Periodic Table. This success can be attributed to the presence of large scalar and vector mean fields, which reflect relativistic interaction effects under normal conditions (e.g., in finite nuclei). Another notable advantage of the RMF theory is that it naturally incorporates important aspects of the nuclear force, including spin-orbit interaction, that are typically added in a more ad-hoc manner in non-relativistic frameworks, as discussed in \cite{Patra_2002}. The RMF model begins with a Lagrangian density that takes into account contributions from nucleons, free mesons, and interactions between mesons and nucleons. The interaction in the RMF theory depends on various couplings that are determined based on available data from finite nuclei and astrophysical observations. Following the suggestion of Duerr and Teller \cite{Johnson_1955, Duerr_1956}, the first successful model was developed by including only $\sigma$ and $\omega$ mesons, known as the Walecka model \cite{WALECKA1974491}. However, this model is not favored due to its high incompressibility of $K_0 \approx 550$ MeV for infinite nuclear matter at saturation, and its stiff equation of state (EoS). To account for the incompressibility and finite nuclei, cubic and quartic nonlinearities of the $\sigma$ meson were added, resulting in improved models. Renormalizability constraints led to the addition of scalar-vector and vector-vector interactions, and subsequent parameter sets such as NL1, NL2, and NL-SH were introduced \cite{Reinhard_1989}. These models included the $\rho$ meson to account for isospin asymmetry and the self-couplings of the $\sigma$ meson to reduce the incompressibility. However, these models still produce a stiff EoS at supernormal densities.

Based on the effective field theory (EFT) motivated relativistic mean-field (E-RMF) approach, Furnstahl,  Serot,  and  Tang obtained the G1 and G2 parameter sets \cite{Frun_1997}. Here,  they neglected the idea of renormalizability and included all possible couplings up to the fourth-order  of expansion in the fields, which include scalar-vector and vector-vector self-interactions in addition to tensor couplings. The E-RMF Lagrangian contains all    the    non-renormalizable   couplings   consistent   with   the   underlying symmetries of QCD. It is also well established that the effective Lagrangian with meson couplings up to the fourth order is a good approximation to predict finite nuclei and nuclear matter observables satisfactorily \cite{FURNSTAHL1997441, MULLER1996508, Del_2001}. The ambiguity in expansion is avoided by using the concept of naturalness \cite{van2020naturalness}. The  contributions  of each term in effective Lagrangian is   determined by counting  powers  in  the  expansion  parameters. To achieve a certain level of accuracy, the Lagrangian is truncated by expanding fields in terms of the mass of nucleons. For the truncation to be valid, the coupling constants must display naturalness and none of them should be omitted without additional symmetry arguments, as noted in \cite{Furn_1987}. In literature, there are hundreds of parameter sets based on the RMF or E-RMF framework. However, only few have been able to satisfy the continuously updating constraints on the nuclear matter EoS \cite{Dutra_2014}. In an attempt to construct E-RMF models that are consistent with recent observational and experimental data, parameter sets FSUGarnet and IOPB-I are designed \cite{Kumar_2018, Chen_2015}. Recently it is observed that at high densities, $\delta$ meson significantly influences the highly asymmetric nuclear matter \cite{Singh_2013, g3}. To incorporate the interaction of $\delta$ meson, the parameter set G3 is constructed \cite{g3}, which has been successfully applied for the description of dense nuclear matter \cite{Ankit_2020, Das_2020}.

\section{Motivation and objective}

With the advancement in the nuclear theory, HIC instrumentation and observational data related to the neutron star and CCNS, there is a need to understand the properties of nuclear matter in various environmental conditions. In recent years, the origin, structure, and dynamics of neutron stars have played a central role in multimessenger and gravitational-wave astronomy \cite{Abbott_2017, Abbott_2018}. It has provided us  the opportunity to understand the behaviour of fundamental forces in extreme environmental situations. With the available multifaceted data from various astrophysical observations,  we now can better constrain the neutron star observables such as mass, radius, tidal deformation, etc., and the behaviour of the equation of state (EoS) over a wide density range \cite{Cromartie_2019, Antoniadis_2013, Demorest_2010}. The "crust" has drawn particular attention among various layers of its internal structure because of its complexity and importance in multiple astrophysical phenomena. Since a neutron star's crust contains nuclear matter at subsaturation density, it acts as the most advanced cosmic laboratory where theory can be confronted with neutron star observations. 

The CCNS, which is the cause of the birth of a neutron star,  is nature's one of the brightest optical display where million-year life of a giant star ($M > 8M\odot$) is put to an end violently and abruptly within fractions of a second \cite{couch2017mechanism, ccsn}. The exact mechanism of collapse explosion is still not well understood even after several decades of thorough investigations. In recent years, such explosion has been studied using several ab-initio core-collapse simulations where the hydrodynamics equations are solved numerically \cite{Muller2010, ccsnsimulation}.  These simulations estimate that the explosion energy of $\approx$ 10$^{51}$ erg is attained within the time scale of $\ge$ 1s  \cite{Sawada_2019}.  The temperature of the matter rises to 20 MeV and the density of the bounce can vary up to two times the nuclear saturation density.  The short time scale of collapse does not allow the matter to reach $\beta$ equilibrium and calculations needs extensive tables of the EoS in different conditions of density, temperature and asymmetry\cite{Alam2017,Nishizaki1994}.
In neutron star crust, the density of the matter is low ($\rho$<$\rho_0=0.15$$fm^{-3}$) and gives rise to the clusterisation of the matter due to the compitition between Coulomb and nuclear force.

In view of the above, the thesis aims to investigate the properties of nuclear matter over a wide range of parameters such as temperature, pressure, density and magnetic field. Such conditions are observed in various terrestrial laboratories, such as HIC experiments and astrophysical phenomena, including CCNS and various kind of neutron stars. Investigation of nuclear matter in such circumstances provides us with much-needed information to fine-tune the nuclear models, which are inevitable to have better idea of nuclear interaction. The principle aim of the thesis is threefold: a) investigation of various nuclear matter properties and to deduce the unified equation of state for hadronic and stellar matter systems to study the relevant structural and thermodynamical properties; b) study of the model dependency of the results and c) making the possible predictions for future validation. To achieve this, it begins with studying an ideal infinite nuclear matter. Properties of infinite nuclear matter at zero and finite temperature limit are essential to describe both finite and infinite nuclear matter and their phase diagrams. The principle aim is to investigate the critical phenomena in isospin symmetric and asymmetric systems and to work out their model dependence. Since the finite temperature properties of nuclear matter are poorly constrained, they play a seminal role in the finite temperature EoS which is an essential aspect of CCNS simulations. The properties of a hot nucleus are discussed in the next part of the thesis. Due to the Coulomb instability, the limiting temperature of the nucleus decreases sharply as compared to the critical temperature of infinite matter. This part of the thesis investigates the impact of this reduction and the role of surface and Coloumb energy. A precise knowledge of the limiting temperature in a nucleus is essential to describe the multifragmentation process in HIC experiments. The last and most crucial part of the present thesis comprehensively studies the neutron star crust structure, one of the most exotic nuclear systems that gives rise to various astrophysical scenarios. Since the majority of neutron stars are observed as pulsars, crustal properties are investigated at zero and finite magnetic field strengths. Various data obtained in the thesis are publically available for the possible use in the future neutron star and supernova simulations.  

\section{Organization of the thesis}
This thesis aims to explore the nuclear matter properties, especially in the low-density regime under various circumstances crucial to our understanding of nuclear interactions. It attempts to study three distinct features of nuclear matter, i.e. infinite nuclear matter and associated thermal properties (chapter 3), hot finite nuclei (chapter 4) and the structure of the neutron star crust (chapter 5, 6 and 7). A brief description of each chapter is as follows: 

\textbf{Chapter \ref{methodology}} of methodology discusses the detailed theoretical framework adopted in this thesis. The E-RMF model is discussed briefly for nuclear interactions, giving essential formulations at zero and finite temperatures. The finite temperature effects, such as phase transition in symmetric/asymmetric nuclear matter and the effects of temperature on EoS, are discussed. The free energy of the nucleus in terms of the liquid drop model is formulated to study a hot finite nucleus and associated phenomena. Finally, it discusses the outer crust (Baym-Pethick-Sutherland (BPS)  model and inner crust (compressible liquid drop model (CLDM)) to study the internal structure of a neutron star. The characteristics that describe a neutron star including its radius, mass, moment of inertia, and crust thickness, crust mass, etc. are also analysed.

In the \textit{\textbf{first}} part of the thesis, the \textbf{Chapter \ref{chap3}} examines the effect of temperature on the EoS. It presents a comprehensive analysis of hot and dilute isospin-symmetric and asymmetric nuclear matter employing the temperature-dependent effective-relativistic mean-field theory (E-RMF). The chapter has two sections. In the first section, the ideal homogeneous symmetric nuclear matter is investigated under the influence of temperature. The goal is to study the liquid-gas phase transition and estimate the critical temperature, pressure and density. In the second section, the influence of temperature on the equation of state (EoS) for densities, temperature and asymmetry, which are relevant for astrophysical simulations such as supernovae explosion and neutron star crust, is given. Using various E-RMF forces, the extent of model dependency of the calculations is investigated. 

In the  \textit{\textbf{second}} part, \textbf{Chapter \ref{chap4}} introduces the Coloumb instability in the infinite nuclear matter, which is the case in HIC experiments. It focuses on the study of various thermal characteristics of nuclei in view of the saturation and critical behaviour of infinite nuclear matter. The density-dependent parameters of the liquid-drop model are used to model the free energy of a hot nucleus. Surface free energy is calculated using two approaches based on the sharp interface of the liquid-gaseous phase and the semi-classical Seyler-Blanchard interaction. The main aim of this chapter is to investigate the limiting temperature of nuclei and its dependence on parameters such as EoS, size, etc.

The  \textit{\textbf{third}} and final part of this thesis studies the crustal properties of the neutron star and attempts to make various predictions on associated phenomena. \textbf{Chapters \ref{chap5} and \ref{chap6}} examine the crust structure of a cold-catalyzed nonaccreting neutron star. The outer crust composition is estimated using the recent atomic mass evaluations and Hartree-Fock-Bogoliubov nuclear mass models. The compressible liquid drop model is utilized for the inner crust, employing the E-RMF  framework. While \textbf{Chapter \ref{chap5}} considers the spherical cluster in the inner crust, \textbf{Chapter \ref{chap6}} studies the existence of non-spherical shapes or nuclear pasta. Since most of the observed neutron stars are pulsars or magnetars, it becomes essential to understand the properties of magnetised neutron star crust. \textbf{Chapter \ref{chap7}} aims to study the influence of the magnetic field on the crustal properties of the neutron star and its various implication on the astrophysical observations. 

Finally, Chapter \ref{chap8} provides a summary and future prospects of the work carried out in this thesis.

%\printbibliography
%\begin{singlespace}
\addcontentsline{toc}{section}{Bibliography}
\printbibliography
%\end{singlespace}

%% file: Chapter_2/CHAP2.tex
\chapter{\label{methodology} Methodology}

\section{Effective relativistic mean-field model}
The Effective relativistic mean-field model (E-RMF) model has been employed in this study to characterize nuclear interactions. Motivated by effective field theory and inspired by the relativistic mean field model (RMF), the E-RMF approach is reliable and in line with Quantum chromodynamics symmetries. Additionally, it addresses the renormalization challenge associated with the RMF theory \cite{Frun_1997}. Over the last few years, this method has been utilized to address various nuclear physics problems \cite{MULLER_1996, Wang_2000, DelPairing_2001, Kumar_2020, Das_2020, Das_2021}. The effective Lagrangian for E-RMF incorporates interactions among various mesons, including $\sigma$, $\omega$, $\rho$, $\delta$, and photon, and is written as \cite{Patra_2002, Kumar_2017, Kumar_2018, DasBig_2020}.

%%%%%%%%%%%%%%%%%%%%%%%%%%%%%%%%%%%%%%%%%%%%%%%%%%%%%%%%%%%%%%%%%%%%%%%%%%%%%%%%%%%%%
\begin{equation}
\begin{aligned}
\label{rmftlagrangian}
\mathcal{E}=&\psi^{\dagger}\Big(i\alpha.\grad+\beta[M-\Phi -\tau_3D ]+W +\frac{1}{2}\tau_3R +\frac{1+\tau_3}{2} A  -\frac{i\beta \alpha }{2M}(f_\omega \grad W \\&
+\frac{1}{2}f_\rho \tau_3 \grad R )\Big)\psi +  \qty(\frac{1}{2}+\frac{k_3\Phi }{3! M}+\frac{k_4}{4!}\frac{\Phi^2 }{M^2})\frac{m^2_s}{g^2_s}\Phi ^2  
+\frac{1}{2g^2_s}\qty\Big(1+\alpha_1\frac{\Phi }{M})(\grad \Phi )^2\\&
-\frac{1}{2g^2_\omega}\qty\Big(1+\alpha_2\frac{\Phi }{M})(\grad W )^2-\frac{1}{2}\qty\Big(1+\eta_1\frac{\Phi }{M}+\frac{\eta_2}{2}\frac{\Phi^2 }{M^2})
\frac{m^2_\omega}{g^2_\omega}W^2 -\Lambda_\omega R^2 W^2 \\&
-\frac{1}{2}\qty\Big(1+\eta_\rho\frac{\Phi }{M})\frac{m^2_\rho}{g^2_\rho}R^2  +\frac{1}{2g^2_\delta}(\grad D )^2+\frac{1}{2}\frac{m^2_\delta}{g^2_\delta}(D )^2
-\frac{1}{2e^2}(\grad A^2 )^2\\&
 -\frac{\zeta_0}{4!}\frac{1}{g^2_\omega}W ^4-\frac{1}{2g^2_\rho}(\grad R )^2.
\end{aligned}
\end{equation} 
%%%%%%%%%%%%%%%%%%%%%%%%%%%%%%%%%%%%%%%%%%%%%%%%%%%%%%%%%%%%%%%%%%%%%%%%%%%%%%%%%%%%%
The fields $\Phi $, $W $, $R $, $D $, and $A $ correspond to the $\sigma$, $\omega$, $\rho$, $\delta$, mesons  and photon, respectively. The corresponding coupling constants are $g_s$, $g_{\omega}$, $g_{\rho}$, $g_{\delta}$ and $\frac{e^2}{4\pi }$ with respective masses as $m_s$, $m_{\omega}$, $m_{\rho}$ and $m_{\delta}$. The  field equations for nucleons and mesons can be solved by applying variational  principle within the  mean field approximations \cite{Kumar_2017, Kumar_2018}. In the present work, the  homogeneous nuclear matter is considered within the mean-field approximation. By utilizing the Lagrange multiplier technique, which represents the energy eigenvalue of the Dirac equation employing the normalization condition $\sum_{\alpha}\psi^{\dagger}\psi_{\alpha}=1$ \cite{Furn_1987, Furnstahl_1997}, one can derive the single-particle energy for the nucleons. The baryon and  scalar density can be written as \cite{Kumar_2018}
\begin{equation}
\rho_b=\sum_q\frac{2}{(2\pi)^3}\int_0^{k_q} d^3k \hspace{1cm}(q=n,p),
\end{equation}
\begin{equation}
\rho_s=\sum_q\frac{2}{(2\pi)^3}\int_0^{k_q} d^3k \frac{M^*_q}{k_q^2+M^{*^2}_q} \hspace{1cm}(q=n,p),
\end{equation}
The effective mass of the nucleons, denoted as $M^*_q$, due to their motion in the mean potential generated by the mesons can be expressed as follows \cite{Kumar_2018}
\begin{equation}
\begin{aligned}
\label{effmass}
&M^*_q& = &M-\Phi \pm D , \\
%&M^*_n& = &M-\Phi +D .
\end{aligned}
\end{equation}
Lastly, the zeroth component of  energy-momentum tensor
%%%%
\begin{equation}
\label{set}
T_{\mu\nu}=\partial^\nu\phi(x)\frac{\partial\mathcal{E}}{\partial\partial_\mu \phi(x)}-\eta^{\nu\mu}\mathcal{E},
\end{equation}
%%%%
denoted as $T_{00}$, yields the energy density, while its third component, represented as $T_{ii}$, gives us the pressure density which are written as,%
\begin{equation}
\begin{aligned}
E=&\sum_{p,n}\frac{\gamma}{(2\pi)^3}\int_0^{k_f}\dd[3]{k}E^*_{p,n}(k)+\qty\Big(\frac{1}{2}+\frac{k_3\Phi}{3! M}+\frac{k_4}{4!}\frac{\Phi^2}{M^2})\frac{m^2_s}{g^2_s}\Phi^2
-\frac{1}{2}\qty\Big(1+\eta_1\frac{\Phi}{M}+\frac{\eta_2}{2}\frac{\Phi^2}{M^2})\\
&\frac{m^2_\omega}{g^2_\omega}W^2-\frac{\zeta_0}{4!}\frac{1}{g^2_\omega}W^4+\frac{1}{2}\rho_3R-\frac{1}{2}\qty\Big(1+
\eta_\rho\frac{\Phi}{M})\frac{m^2_\rho}{g^2_\rho}R^2-\Lambda_\omega(R^2W^2)+\frac{1}{2}\frac{m^2_\delta}{g^2_\delta}(D)^2+
\rho W.\nonumber
\end{aligned}
\end{equation}
\begin{equation}
\begin{aligned}
P=&\sum_{p,n}\frac{\gamma}{3(2\pi)^3}\int_0^{k_f}\dd[3]{k}\frac{k^2}{E^*_{p,n}(k)}-
\qty\Big(\frac{1}{2}+\frac{k_3\Phi}{3! M}+\frac{k_4}{4!}\frac{\Phi^2}{M^2})\frac{m^2_s}{g^2_s}\Phi^2
+\frac{1}{2}\qty\Big(1+\eta_1\frac{\Phi}{M}+\frac{\eta_2}{2}\frac{\Phi^2}{M^2})\\
&\frac{m^2_\omega}{g^2_\omega}W^2+\frac{\zeta_0}{4!}\frac{1}{g^2_\omega}W^4+\frac{1}{2}\qty\Big(1+\eta_\rho\frac{\Phi}{M})\frac{m^2_\rho}{g^2_\rho}R^2
+\Lambda_\omega(R^2W^2)-\frac{1}{2}\frac{m^2_\delta}{g^2_\delta}(D)^2.
\end{aligned}
\end{equation}
%%%%%%%%%%%%%%
In above equations, $\gamma$ represents the spin-isospin degeneracy where  $\gamma$ = 2 denotes the pure neutron matter (PNM) and $\gamma$=4 represent symmetric nuclear matter (SNM). The details regarding the equation of motion, chemical potential, nuclear matter properties are readily available in the literature  and can be found in \cite{Kumar_2018, Das_2019, Dutra_2016}. The E-RMF Lagrangian contains a range of couplings that each serve an important role in providing the model with the flexibility necessary to account for a variety of phenomena related to nuclear matter. The $\zeta_0$ (self-coupling of isoscalar-vector $\omega$ meson ($W  ^4$)) and the self-coupling of $\sigma$ meson ($k_3$, $k_4$) help soften the equation of state. The quartic-order cross-coupling of $\rho$ and $\omega$ mesons, denoted by $\Lambda_\omega$, plays a critical role in regulating the density dependence of symmetry energy and aiding in achieving better agreement with neutron skin thickness data. Moreover, $\Lambda_\omega$ provides flexibility in fitting spherical nuclei without compromising the ability to adjust the neutron skin thickness of $^{208}$Pb across a wide range \cite{Tapas_2005}. Additionally, the cross-couplings of $\sigma-\omega$ and $\sigma-\rho$ mesons, represented by $\eta_1, \eta_2,$ and $\eta_\rho$, influence the surface properties of finite nuclei, while the $\delta$ meson has a softening effect on the symmetry energy at subsaturation densities and stiffens the equation of state at high densities \cite{KUBIS1997191, Singh_2014}.

%%%%%%%%%%%%%%%%%%%%%%%%%%%%%%%%%%%%%%%%%%%%%%%%%%%%%%%%%%%%%%%%%%%%%%%%%%
\section{Nuclear matter thermodynamics}
As dicussed in chapter \ref{chap:intro}, the \textit{first} and \textit{second} part of this thesis focuses on the finite temperature characteristics of infinite and finite nuclear matter. To achieve this, one needs a model with covariant thermodynamic consistency. It should be able to compute both the  thermodynamic (pressure, energy, entropy) and dynamical (transport coefficients, viscosity) properties of nuclear matter. The E-RMF formalism  provides us with one such solution \cite{Furnstahl_1990}. The E-RMF model can be extended to finite temperature in a straightforward manner. As a starting point, For the symmetric nuclear matter, the thermodynamic potential $\Omega$ in the grand canonical ensemble is expressed using principles of statistical mechanics as

\begin{equation}
\Omega=-k_B\text{T}\ln Z
\end{equation}
where
\begin{equation}
Z=Tr\qty{\exp(-\beta(\vu{H}-\mu\vu{B}))}, 
\end{equation}
is the grand canonical partition function where $\vu{H}$ represents the Hamiltonian operator, $\vu{B}$ is the mean field baryon number operator, and $k_B$ is the Boltzmann constant. The fundamental connections between the thermodynamic potential ($\Omega$), entropy (S), chemical potential ($\mu$), and temperature (T) can be expressed as follows:

\begin{equation}
\Omega=-pV=E-TS-\mu B
\end{equation}
and 
\begin{equation}
\dd{\Omega}=-S\dd{T}-p\dd{V}-B\dd{\mu}.
\end{equation}
The mean values of baryon and scalar densities can be computed through ensemble averaging, and can be expressed as shown in \cite{Serot2004, Muller1995}.
\begin{equation}
\rho_b=\frac{1}{V}\qty\Big[\pdv{\Omega}{\mu}]=\sum_{q}\frac{\gamma}{(2\pi)^3}\int \dd[3]k[n_k(T)-\bar{n}_k(T)] 
\end{equation}
and 
\begin{equation}
\rho_s=\frac{1}{V}\qty\Big[\pdv{\Omega}{M}]=\sum_{q}\frac{\gamma}{(2\pi)^3}
\int \dd[3]k\frac{m^*_\alpha}{\sqrt{k^2_\alpha+M^*_\alpha}}[n_k(T)+\bar{n}_k(T)],
\end{equation}
The quantities $n_k(T)$ and $\bar{n}_k(T)$ represent the occupation numbers of baryons and antibaryons, respectively, and are determined by the finite temperature Fermi distribution function as\\
\begin{equation}
\label{fdf1}
n_k(T)=\frac{1}{1+\exp\qty(\frac{(E^*(k)-\nu)}{T})}
\end{equation}
and 
\begin{equation}
\label{fdf2}
\bar{n}_k(T)=\frac{1}{1+\exp\qty(\frac{(E^*(k)+\nu)}{T})},
\end{equation}
where  E$^*$ is $\sqrt{k^2 + {M^*}^2}$.  $\nu$ is defined as the effective chemical potential for proton and neutron in Eqs. (\ref{fdf1}) and (\ref{fdf2}) and  is written as 
\begin{equation}
\begin{aligned}
\label{effcpot}
\nu_q&=& \mu - W  \pm\frac{1}{2}R  \hspace{1cm}(q=n,p).
%\nu_n&=& \mu - W -\frac{1}{2}R .
\end{aligned}
\end{equation}
The energy and pressure densities retain the same form as in the case of zero temperature, except for the alteration in the contribution from neutrons and protons, which becomes: \cite{pwang,Serot2004, Muller1995}
\begin{equation}
\begin{aligned}
&E_{q}=\sum_{q}\frac{2}{(2\pi)^3}\int\dd[3]{k}E^*_{q}(k)[n_k(T)+\bar{n}_k(T)]
\end{aligned}
\end{equation}

\begin{equation}
\begin{aligned}
&P_{q}=\sum_{q}\frac{2}{3(2\pi)^3}\int\dd[3]{k}\frac{k^2}{E^*_{q}(k)}[n_k(T)+\bar{n}_k(T)]
\end{aligned}
\end{equation}
The entropy density, denoted by $s$ can be conveniently determined, and takes a form identical to that of a non-interacting gas. This quantity is expressed as \cite{Alam_2016},
 \begin{equation}
\label{entropyeq}
s_i=-2\sum_{i}\int \frac{\dd^3k}{(2\pi)^3} [n_{k} \ln n_{k} + (1-n_{k})\ln(1-n_{k})+(n_{k}\leftrightarrow \bar n_{k})].
\end{equation}
The free energy of the system then can be written as
\begin{equation}
\mathcal{F}=E-TS.
\end{equation}

%%%%%%%%%%%%%%%%%%%%%%%%%%%%%%%%%%%%%%%%%%%%%%%%%%%%%%%%%%%%%%%%%%%%%%%%%%%%%%%%%%%%%%%%%%%
\section{Liquid-gas phase transition}
\subsection{Symmetric nuclear matter}
The nuclear matter undergoes first-order liquid-gas phase transition (LGPT) under the influence of temperature analogous to the water molecule. It was observed that in  the lowest energy state, nucleons show liquid-like characteristics with density $\rho_0 \approx$ 0.15 fm$^{-3}$ \cite{Walecka_74}. As the temperature increases, the nuclear liquid evaporates and undergoes the liquid-gas phase transition \cite{Muller_1995}. Below certain maximum temperature T$_c$, the liquid and gas phase remain in phase coexistence  which terminates at temperature greater than $T_c$. In symmetric nuclear matter, the LGPT reduces to a conventional one component phase tranistion problem.  The definining feature of any first-order phase transition is the involvement of ``Latent heat”. After obtaining EoS in pressure and energy, it is straightforward to calculate latent heat using the Clausius-Clapeyron equation
\begin{equation}
\label{classiousclapron}
    L = T \Big(\frac{1}{\rho_g}-\frac{1}{\rho_l}\Big)\Big(\frac{dP}{dT}\Big)_{coex}.
\end{equation}
Here the $\frac{dP}{dT}$ is along the coexistence curve which is determined from the Maxwell construction. $\rho_g$ and $\rho_l$ are the densities corresponding to gaseous and liquid phase, respectively. The Gibbs conditions of  phase transition along the isotherms are then written as  \cite{Furnstahl_1990}
\begin{equation}
   \label{chmeicalpotentialeqi}
        \mu_q^{'}(\rho^{'},T) = \mu_q^{''}(\rho^{''},T),\;  (q=n,p),
   \end{equation} 
   \begin{equation}
   \label{pressureeqi}
       P'(\rho',T) = P''(\rho'',T).
   \end{equation}
Here two phases are represented by prime and a double-prime. Alternatively, the latent heat can be estimated from \cite{Carbone_2011}
\begin{equation}
    L = T(s_g-s_l), \; (s=S/\rho_b),
\end{equation}   
which is the amount of heat required to take one particle from the ordered phase to disordered phase at a constant temperature, pressure and chemical potential. The related thermodynamic variables such as free energy ($F=E-TS$), specific heat ($C_V$), isothermal compressibility ($K_T=-\frac{1}{V}\frac{\partial V}{\partial P}$) can be calculated using standard thermodynamic relations \cite{greiner2012thermodynamics}.

%%%%%%%%%%%%%%%%%%%%%%%%%%%%%%%%%%%%%%%%%%%%%%%%%%%%%%%%%%%%%%%%%%%%%%%%%%%%%%%
\subsection{Asymmetric nuclear matter}
 In the context of asymmetric nuclear matter, there are two conserved quantities to consider: the baryon number ($\rho_b=\rho_p+\rho_n$) and the isospin number ($I_3=I_p+I_n$). Consequently, it is necessary to treat this system as a binary mixture. The condition for stability against phase separation is that the free energy of a single phase must be lower than the free energy of all multi-phase configurations. This criterion is expressed as follows \cite{Muller1995}:
\begin{equation}
\label{stability_con1}
\mathcal{F}(T,\rho_i)<(1-\lambda)\mathcal{F}(T,\rho_i^{'})+\lambda \mathcal{F}(T,\rho_i^{''}),
\end{equation}
with 
\begin{equation}
\label{stability_con2}
\rho_i=(1-\lambda)\rho_i^{'}+ \lambda \rho_i^{''},
\end{equation}
In this expression, the two phases are labeled as prime and double prime, and $\lambda$ represents the volume fraction. Mathematically speaking, the criterion for stability implies that the free energy density must exhibit convex behaviour with respect to the densities. This convexity condition guarantees not only stability against phase separation into two phases, but also against separation into any number of phases. Therefore, it follows that the symmetric matrix:
\begin{equation}
\label{symmetricmatrix}
\mathcal{F}_{ij}=\Big(\frac{\partial^2 \mathcal{F}}{\partial\rho_i\partial\rho_j}\Big)_T
\end{equation}
is positive \cite{Muller1995, Avancini2006, KAUR2016133, Alam_2016}. This results in mechanical and diffusive stability conditions as \cite{Muller1995}
\begin{equation}
\label{instabilitycondition}
    \frac{\partial P}{\partial \rho_b}\bigg|_{T,\alpha} > 0 \quad \textrm{and} \quad
    \frac{\partial \mu_p}{\partial \alpha}\bigg|_{T,P} < 0
\end{equation}
If either of the stability conditions is not satisfied, then it is energetically more favorable for the system to exist in the form of two distinct phases. At the critical points, the pressure, density and temperature are written as $P_c, \rho_c$ and $T_c$. For asymmetric nuclear matter, they are calculated by finding an inflation point at chemical potential isobars as
\begin{equation}
    \frac{\partial \mu_q}{ \partial \alpha}\bigg|_{T=T_c}=\frac{\partial^2 \mu_q}{ \partial \alpha^2}\bigg|_{T=T_c}=0
\end{equation}

%%%%%%%%%%%%%%%%%%%%%%%%%%%%%%%%%%%%%%%%%%%%%%%%%%%%%%%%%%%%%%%%%%%%%%%%%%%%%%%%%%%%%%%%%%%%%%%%

\section{Finite temperature nuclear matter observables}
The free energy density of nuclear matter can be expressed as a parabolic function of the isospin asymmetry parameter, denoted by $\alpha=\frac{\rho_n-\rho_p}{\rho_n+\rho_p}$. \cite{bednarek2020forms, tsang2009}
\begin{equation}
\mathcal{F}(\rho,\alpha,T)=\mathcal{F}(\rho,\alpha=0,T)+\mathcal{F}_{sym}(\rho,T)\alpha^2
\end{equation}
where $F_{sym}(\rho,T)\alpha^2$ is the free symmetry energy content per
nucleon of the system and $F(\rho,\alpha=0,T)$ is the free energy per
nucleon of symmetric ($\alpha=0$) nuclear matter. The free symmetric energy using the empirical parabolic approximation then can be written as \cite{Ankit_2020}
\begin{equation}
    \mathcal{F}_{sym}(\rho,T)=  \frac{\mathcal{F}(\rho,T,\alpha=1)}{\rho}-\frac{\mathcal{F}(\rho,T,\alpha=0)}{\rho}.
    \label{freesymenergy}
\end{equation}
The free symmetric energy can be represented by a Taylor series expansion centered at the saturation density $\rho_0$, as \cite{Kumar_2018},
\begin{equation}
\begin{aligned}
\label{fsym}
    \mathcal{F}_{sym}(\rho,T)=    F_{sym}(\rho_0,T)+ L_{sym}\eta + \frac{K_{sym}}{2!}\eta^2+\frac{Q_{sym}}{3!}\eta^3+O(\eta^4),
\end{aligned}
\end{equation}
where $\eta=\frac{\rho-\rho_0}{3\rho_0}$ and $L_{sym}$, $K_{sym}$ and $Q_{sym}$ are the slope parameter, curvature parameter and skewness parameter which are written as \cite{Raduta2021, Ankit_2020}
\begin{equation}
    \begin{aligned}
        L_{sym}&=&3\rho\frac{\partial F_{sym}(\rho,T)}{\partial \rho};\hspace{.5cm}
        K_{sym}&=&9\rho^2\frac{\partial^2 F_{sym}(\rho,T)}{\partial \rho^2};\hspace{.5cm}
        Q_{sym}&=&27\rho^3\frac{\partial^3 F_{sym}(\rho,T)}{\partial \rho^3}.
    \end{aligned}    
\end{equation}
In order to examine the influence of finite temperature, thermal contribution of various state variables is required, which refers to the difference between their values at T=0 and T$\neq$0 for a particular thermodynamic function $\mathcal{X}$ \cite{Constantinou2014, Constantinou2015},
\begin{equation}
    \mathcal{X}=\mathcal{X}(\rho,\alpha,T)-\mathcal{X}(\rho,\alpha,0)
\end{equation}
The thermal energy, thermal pressure, thermal free energy density and thermal index  are then written as \cite{Raduta2021}
\begin{equation}
\label{thermalpart}
    \begin{aligned}
    &E_{th}=E(\alpha,T)-E(\alpha,0)\\
    &P_{th}=P(\alpha,T)-P(\alpha,0)\\
    &F_{th}=F(\alpha,T)-E(\alpha,0)\\
    &\Gamma_{th}=1+\frac{P_{th}}{E_{th}}\\
    \end{aligned}
\end{equation}
Thermal contributions to the free symmetry energy is given by
\begin{equation}
F_{sym,th}=F_{sym}(\alpha,T)-E_{sym}(\alpha,0).
\end{equation}
Furthermore, one can define isothermal incompressibility of nuclear matter at finite temperature T and asymmetry $\alpha$ as \cite{haddad2003}
\begin{equation}
\label{isothermal}
    K^T(\alpha,T)=9\bigg(\rho_b^2\frac{\partial^2 F}{\partial \rho_b^2}\bigg)\bigg|_{\rho_b^T(\alpha,T)} 
\end{equation}
Here, $\rho_b^T$ is the density where free energy has its minimum. The isentropic incompressibility at entropy S and asymmetry $\alpha$ which is an important quantity in supernova collapse is written as \cite{haddad2003}
\begin{equation}
\label{isentropic}
    K^S(\alpha,S)=9\bigg(\rho_b^2\frac{\partial^2 E}{\partial \rho_b^2}\bigg)\bigg|_{\rho_b^T(\alpha,S)}
\end{equation}

%%%%%%%%%%%%%%%%%%%%%%%%%%%%%%%%%%%%%%%%%%%%%%%%%%%%%%%%%%%%%%%%%%%%%%%%%%%%%%%%%%%%%%%%%%%%%%%
\section{\label{limtemp} Limiting temperature of a nucleus}

Unlike infinite matter, a hot nucleus can only sustain  a maximum limiting temperature $T_{l}$ after which, it ceases to exist due to mechanical instabilities. This  results  in a  nuclear multifragmentation reaction \cite{BALDO199413}. To study the  $T_{l}$ of a nucleus and associated properties, it is crucial to describe the free energy of the nucleus. 
%\subsubsection{\label{freeenergyformalism} From infinite matter to finite nuclei}
Considering a nucleus to be a liquid drop and resorting to the conventional liquid-drop model to  define the free energy of the drop with given mass number $A$, proton number $Z$, and neutron number $N$, the free energy can be written as,  
\begin{eqnarray}
\label{freeenergy}
F_A(\rho,T)=&\mathcal{F}_v(\rho,T)A+\mathcal{F}_{corr}(\rho,T).
\end{eqnarray}
Here $\mathcal{F}_v(\rho, T)$ represents the free energy of infinite symmetric nuclear  matter (SNM) corresponding to the volume. $\mathcal{F}_{corr}$ denotes the finite size correction  to symmetric nuclear matter  written as \cite{gt, RAVENHALL1983571, BLAIZOT1980171} 
\begin{eqnarray}
\begin{aligned}
\label{freecorr}
\mathcal{F}_{corr}(\rho,T)=&f_{surf}(\rho,T)4 \pi R^2+f_{sym}(\rho,T)\frac{(N-Z)^2}{A}+f_{Col}.
\end{aligned}
\end{eqnarray}
Here $R$  is defined as
\begin{equation}
    R=\qty(\frac{3 A}{4 \pi \rho(T)})^{1/3},
\end{equation}
which is the radius of the drop. The coefficient of free surface energy (FSE) ($f_{surf}(\rho, T)$) is a crucial parameter that introduces the surface and is assumed to be factorized and density-dependent \cite{RAVENHALL1983571}. This is written as 
\begin{eqnarray}
\begin{aligned}
\label{surffreeenergy}
f_{surf}(\rho,T)=\alpha_{surf}(\rho_0,T=0)\mathcal{D}(\rho)\mathcal{Y}(T).
\end{aligned}
\end{eqnarray}
The surface energy coefficient $\alpha_s(\rho_0,T)$ is defined at the saturation density ($\rho_0$) of infinite symmetric nuclear matter (SNM). As the density of liquid evolves, the surface energy gets modified. Therefore, the density dependence  is written as \cite{BLAIZOT1980171}
\begin{equation}
    \mathcal{D}(\rho)=1-\frac{\mathcal{K_\rho}}{2}\qty(\frac{\rho-\rho_0}{\rho_0})^2.
\end{equation}
The temperature dependence of the coefficient of FSE is another significant parameter that ensures that the surface tension vanishes above a certain temperature $T_c$. For the calculation of $T_l$, two parametrizations of the temperature dependence of surface energy are used which are widely adopted in the calculation of multi-fragmentation in nuclei and structure of neutron star crust.  The first expression is taken from \cite{ravenhall} which takes into account the plane sharp interface between liquid and gaseous phase of nuclear matter in equilibrium. It is written as 
\begin{eqnarray}
\label{s1}
\mathcal{Y}(T)&=&\qty(\frac{T_c^2-T^2}{T_c^2+T^2})^\frac{5}{4}.
\end{eqnarray}
The second expression is derived from the semiclassical modified Seyler-Blanchard interaction and takes the form \cite{seyler} as
\begin{eqnarray}
\label{s2}
\mathcal{Y}(T)&=&\qty(1+1.5\frac{T}{T_c})\qty(1-\frac{T}{T_c})^{\frac{3}{2}}.
\end{eqnarray}
In these expressions, $T_c$ is the critical temperature of LGPT in infinite SNM. $\alpha_s(\rho_0,T)$ is taken as 1.15 MeV fm$^{-2}$ and $\mathcal{K_\rho}$ is a dimensionless constant taken to be 5.0 as prescribed in \cite{jnde}. The  coefficient of free symmetry energy (FSYE) ($f_{sym}(\rho, T)$) which depend on the mass number of liquid drop is written as
\begin{equation}
    f_{sym}(\rho,T)=\alpha_{sym}(\rho,T=0)\mathcal{G}(T)\qty(\frac{\rho}{\rho_0})^\gamma.
\end{equation}
Here, $\alpha_{sym}(\rho,T=0)$ is further defined as
\begin{equation}
    \alpha_{sym}(\rho,T=0)=\frac{J}{1+ \mathcal{C}A^{-1/3}},
\end{equation}
where $J$ denotes the symmetry energy of  SNM at $T=0$ and is taken as $31$ MeV. Here, the symmetry energy $J$ is taken as a constant while the density dependence of the $J$ is introduced in a parmetric form $(\frac{\rho}{\rho_0})^\gamma$. With this, the only input in the Eq. (\ref{freeenergy}) is the energy of the symmetric nuclear matter. The coefficient $\mathcal{C}$ takes care of the mass dependence of $J$ and is taken to be 2.4 \cite{jnde}. The dependence of $f_{sym}(\rho, T)$ on the temperature is ensured using the function $\mathcal{G}(T)$ in line with the infinite matter calculations that suggest that free FSYE increases with temperature \cite{vishalasymmetric}. It  is taken in a schematic form as per \cite{gt}
\begin{equation}
    \mathcal{G}(T)=(1+\mathcal{X}_1T+\mathcal{X}_2T^2+\mathcal{X}_4T^4),
\end{equation}
where $\mathcal{X}_1=-0.00848$, $\mathcal{X}_2=0.00201$, $\mathcal{X}_4=0.0000147$ with dimensions as relevant power of unit of temperature. The density dependence is ensured with the $\gamma=0.69$ in congruence with the experimental observations \cite{gamma}. The free Coulomb energy (FCE) which is otherwise absent in the infinite matter is responsible for the Coulomb instability of the liquid drop. It is taken as \cite{SAUER1976221}
\begin{equation}
    f_{Col}=\frac{3}{5}\frac{Z^2e^2}{R}\qty(1-\frac{5}{2}\qty(\frac{b}{R})^2),
\end{equation}
where b is the surface thickness which is also a temperature-dependent quantity taken as
\begin{equation}    
    b \approx 0.72(1+0.009T^2).
\end{equation}
The ratio $\frac{b}{R}$ increases with temperature resulting in the reduction of Coulomb free energy in addition to that arising from the expansion of bulk matter. The exchange term is not included in Coulomb free energy due to its low contribution. Pairing and shell corrections are also omitted because they become insignificant for temperature $>1-2$ MeV due to shell melting.

%%%%%%%%%%%%%%%%%%%%%%%%%%%%%%%%%%%%%%%%%%%%%%%%%%%%%%%%%%%%%%%%%%%%%%%%%%%%%%%%%%%%%%%%%%	
\subsection{\label{eenergy} Excitation energy, level density and  fissility parameter}
The binding energy $E(T)$ of a liquid-drop with given A and Z can be found by minimizing Eq. (\ref{freeenergy}) to obtain the density of a nucleus at a particular temperature. The excitation energy then attain a simple form as $E^*(T)=E(T)-E(T=0)$, which essentially signifies the difference of binding energy of ground level to that at any given temperature.  Here the energy can be determined from the relation
\begin{equation}
    E(T)=\mathcal{F}(T)+TS.
\end{equation}
The inter-relationship between  excitation energy, entropy and temperature, which determine the level density parameter (a) is written as \cite{baym2008landau}
\begin{equation}
\label{lebeldensityeq}
E^*=aT^2, \hspace{0.5cm} S=2aT, \hspace{0.5cm} S^2=4aE^*.
\end{equation}
In a heavy nucleus, the competition between Coulomb and surface energy determines the fissility of the nucleus: As the ratio increases, the fission barrier decreases proportionally. The fissility parameter  is given by dimensionless parameter $x(T)$ which is defined as \cite{SAUER1976221}
\begin{equation}
\label{fissilityeq}
    x(T)=\frac{\mathcal{F}^0_{Col}}{2\mathcal{F}^0_{s}},
\end{equation}
here superscript signifies the spherical drop. The fission barrier or potential energy of deformed drop  is expressed  as \cite{SAUER1976221}
 \begin{equation} 
\label{fbeq}
    \mathcal{B}_f(T)=((B_s-1)+2x(T)(B_c-1)).
\end{equation} 
Here, $B_s$ and $B_c$ are the surface and Coulomb energy at saddle point in the units of surface and Coulomb free energy, respectively. Values of $B_s$ and $B_c$ can be determined from \cite{nix} where these values are tabulated against fissility parameter $x(T)$.

%%%%%%%%%%%%%%%%%%%%%%%%%%%%%%%%%%%%%%%%%%%%%%%%%%%%%%%%%%%%%%%%%%%%%%%%%%%%%%%%%%%%%%%%%%
\subsection{\label{limtformalism} Limiting temperature}
The most important aspect of the thermodynamics of a finite nucleus is its multi fragmentation which can be explained in terms of liquid-gas phase transition. 
Nucleus is considered to be a spherical drop of liquid surrounded by a gas of nucleons under the assumptions  that the hot nucleus at a temperature T is surrounded by homogeneous gas of symmetric nuclear matter in a complete mechanical and chemical thermodynamic equilibrium with no exchange of particles.  A sharply defined surface separates the liquid and gaseous phase and there is no interaction between nucleons in the gaseous and liquid phase, so the gas remains unchanged (without Coulomb effect). These approximations then lead us to the following modified phase equilibrium condition similar to the infinite matter case.
\begin{subequations}
\label{coexcondition}
\begin{eqnarray}
&P_0^g(\rho^g,T)=P^l_0(\rho^l,T)+\delta P^l ,\\
&\mu_{p0}^g(\rho^g,T)=\mu_{p0}^l(\rho^l,T)+\delta \mu_p^l.
\end{eqnarray}
\end{subequations}
Here, 0 in the subscript refers to the bulk matter conditions, and $\delta P^l$ and $\delta \mu_p^l$ are the pressure and chemical potential corrections which are given as \cite{BANDYOPADHYAY19901}
\begin{subequations}
\label{correction}
\begin{eqnarray}
&\delta P^l=-\rho^2\big(\frac{\partial \mathcal{F}_{corr}}{\partial \rho}\big)\big|_{T,N,Z},\\
&\delta \mu_p^l=\big(\frac{\partial \mathcal{F}_{corr}}{\partial Z}\big)\big|_{T,N,\rho}.
\end{eqnarray}
\end{subequations}
The, $\mathcal{F}_{corr}$ is defined in Eq \ref{freeenergy}. The expressions for other thermodynamical quantities such as critical temperature (T$_c$), flash temperature (T$_f$), etc. can be found in \cite{vishalasymmetric, vishalsymmetric}. The external nucleon gas also defines the stability of a hot nuclear liquid drop. In this context, we define the lifetime of a hot drop by using the concept of statistical average and assuming neutron emission to be the dominant process. Neglecting the energy dependency of the cross-section, the lifetime of a hot nucleus is written as \cite{BONCHE1984278} 
\begin{equation}
\label{timeeq}
    \frac{1}{\tau}=4 \pi \gamma \frac{1}{h^3} 2 m (kT)^2 \sigma \exp{\frac{\mu_n}{kT}},
\end{equation}
where  $\sigma$ is taken to be geometric  cross section.

%%%%%%%%%%%%%%%%%%%%%%%%%%%%%%%%%%%%%%%%%%%%%%%%%%%%%%%%%%%%%%%%%%%%%%%%%%%%%%%%%%%%%%%%%%
\section{Neutron star structure}
Nuclear astrophysics revolves around the fundamental aspects of neutron stars, including their structural and dynamic properties. The interiors of the neutron star offer a plethora of structural aspects and associated astrophysical phenomena. Present day research is predominantly focused on exploring these phenomena using nuclear theories developed for terrestrial nuclei. Accordingly, the \textit{third} part of the thesis is dedicated to examining the properties of the neutron star crust, and the ensuing section outlines the formalism employed to characterize the crust.

\subsection{Outer crust}
To determine the composition of the outer crust, which spans from a density of $10^{-10}$ $fm$$^{-3}$ until the inception of neutron drip, the Baym-Pethick-Sutherland (BPS) variational formalism is employed \cite{BPS_1971}. It considers that the ensemble of heavy nuclei may be expressed by a single nucleus, the single-nucleus approximation \cite{Shen_2011}, thus giving a unique configuration for given thermodynamic conditions. In the outer crust, the energy of Wigner–Seitz (WS) cell at a given baryon density ($\rho_b$) ensuring the charge neutrality reads as \cite{Haensel_2008}
%%%%
\begin{equation}
    \label{wsenergy}
    E(A,Z,\rho_b)_{WS}= E(A,Z)_N + E_L+E_{zp}+E_e,
\end{equation}
%%%%
where $E(A, Z)_N=M(A, Z)$ is the rest mass energy of nucleus with atomic number $Z$ and mass number $A$. $E_L$ and $E_{zp}$ correspond to static-lattice and zero-point energy, which are written as \cite{carreau2020modeling}
%%%%
\begin{equation}
    \begin{aligned}
    &E_L=-C_M\frac{(Ze)^2}{R_N}; \ \ R_N=\left(\frac{3}{4 \pi} \rho_N\right)^{1/3},\\
    &E_{zp}=\frac{3}{2}\hbar\omega_pu.
    \end{aligned}
\end{equation}
%%%%
Here, $C_M=0.895929255682$ is the Madelung constant, $u=0.51138$ is a constant for a BCC lattice \cite{Chamel_2016} and $\omega_p$ is the plasma frequency. $\rho_N$ is the neutron density. $E_e=\mathcal{E}_e V_{WS}$ is the energy of the surrounding relativistic electron gas. $V_{WS}$ is the volume of the WS cell. At a fixed pressure, the Gibbs free energy \cite{BPS_1971}
%%%%
\begin{equation}
    G(A,Z,P)=\frac{\mathcal{E}_{WS}+P}{\rho_b},
\end{equation}
%%%%
is minimized to estimate the equilibrium value of the mass number $A$ and charge $Z$, where $\mathcal{E}_{WS}=E_{WS}/V_{WS}$ is the energy density of WS cell and $\rho_b=A/V_{WS}$ is the baryon density. The advantage of taking pressure as an independent variable is that it increases monotonically while moving from the surface to the core. Thus discontinuity in density suggests the transition from one layer of the nucleus to another.  One also gets rid of the Maxwell construction \cite{Vishal_2021_jpg} to determine the transition pressure from one nucleus to another. The pressure can be calculated using the first law of thermodynamics as \cite{Carreau_2020} 
%%%%
\begin{equation}
    P=\rho_b^2\frac{\partial \mathcal{E}_{WS}/\rho_b}{\partial \rho_b}.
\end{equation}
%%%%
Nucleons exert no pressure in the outer crust, and the total pressure can be written using Eq. (\ref{wsenergy}) as
%%%%
\begin{equation}
\label{ocpress}
    P=\frac{1}{3}E_L\rho_N +\frac{1}{2}E_{zp}\rho_N+P_e.
\end{equation}
%%%%
The Gibbs free energy to minimize thus becomes \cite{bcpm, Carreau_2020}
%%%%
\begin{equation}
\label{eq:gibbsminimization}
    G(A,Z,P)=\frac{M(A,Z)}{A}+\frac{4}{3}\frac{E_L}{A}+\frac{1}{2}\frac{E_{zp}}{A}+\frac{Z}{A}\mu_e,
\end{equation}
%%%%
where $\mu_e$ is the electron chemical potential. The only input in the calculation of outer crust is the nuclear masses that can be taken from experiments \cite{Huang_2021} which are available for $I=(N-Z)/A \leq 0.3$. For the nuclear mass of more neutron-rich nuclei, available theoretical mass tables \cite{Samyn_2002} can be used. The outer crust extends to the density where the chemical potential of neutrons exceeds its rest mass-energy. The neutron chemical potential utilizing the  condition of $\beta-$equilibrium ($\mu_n=\mu_p+\mu_e$) can be simply written as
%%%%
\begin{equation}
    \mu_n=G.
\end{equation}

%%%%%%%%%%%%%%%%%%%%%%%%%%%%%%%%%%%%%%%%%%%%%%%%%%%%%%%%%%%%%%%%%%%%%%%%%%%%%%%%%%%%%%%%%%
\subsection{Inner crust: Compressible liquid drop model (CLDM)}
As one moves deeper into the crust, the neutrons become less and less bound. At the transition density, the neutrons drip out of the nuclei and start filling the continuous energy spectrum. The dripped neutrons stay confined in the WS cell due to the large gravitational pressure. %Since no such system can be produced in terrestrial laboratories as neutrons evaporate, the inner crust will inevitably become model-dependent. 
In the inner crust, the WS consists of a cluster surrounded by ultrarelativistic electron gas and ambient neutron gas. The energy of this cluster can be written as \cite{Pearson_2018, BPS_1971}
%%%%
\begin{equation}
\label{wsenergyic}
E_{WS}=M_i(A,Z)+E_e+V_{WS}(\mathcal{E}_g+\rho_gM_n),
\end{equation}
%%%%
where $M_i(A,Z)$ is the mass of the cluster written as
%%%%
\begin{equation}
    \label{clustermass}
    M_i(A,Z)=(A-Z)M_n + Z M_p + E_{cl}-V_{cl}(\mathcal{E}_g+\rho_g M_n),
\end{equation}
%%%%
where $M_n$ and $M_p$ are the masses of neutron and proton, respectively. ${\cal E}_g$, and $\rho_g$ are the energy density and density of the neutron gas, respectively. The energy of the cluster within the Compressible liquid drop model (CLDM) framework reads
%%%%
\begin{equation}
    \label{ecluster}
    E_{cl}=E_{bulk}(\rho_0,I)A+E_{surf}+E_{curv}+E_{coul},
\end{equation}
%%%%
where $E_{surf}$, $E_{curv}$, and $E_{coul}$ are surface, curvature and Coulomb energy, respectively. In WS approximation, the Coulomb energy, which consists of lattice and finite-size correction, is written as \cite{carreau2020modeling}
%%%%
\begin{equation}
    E_{col}=\frac{3}{20}\frac{e^2}{r_0}\eta_{col}A^{5/3}(1-I)^2,
\end{equation}
%%%%
with
%%%%
\begin{equation}
    \eta_{col}=1-\frac{3}{2}\lambda^{1/3}+\frac{1}{2}\lambda,
\end{equation}
%%%%
where $\lambda=\rho_e/\rho_{0,p}$ is the volume fraction with $\rho_{0,p}$ and $\rho_e$ are the proton and electron density inside the cluster, respectively. Considering cluster to be spherical, the surface energy is defined as
%%%%
\begin{equation}
  E_{surf} = 4\pi R_0^2A^{2/3}\sigma(I),\label{eq:esurf}
\end{equation}
%%%%
where $R_0 = (4\pi \rho_0/3)^{-1/3}$ is related to the cluster density $\rho_0$, and
$\sigma(I)$ is the nuclear surface tension that depends on the isospin asymmetry of the cluster.  In this work, the parametrization of surface tension proposed by Ravenhall \textit{{\it {\it et al.}}}~\cite{Ravenhall1983} which is obtained by fitting Thomas-Fermi and Hartree-Fock numerical values is used and is written as,
%%%%
\begin{equation}
  \sigma(I) = \sigma_0\frac{2^{p+1} + b_s}{Y_p^{-p} + b_s + (1 -
  Y_p)^{-p}},\label{eq:sigma}
\end{equation}
%%%%
where, $\sigma_0,p,b_s$ are the free parameters and $Y_p$ is the proton fraction inside the cluster. This functional is widely employed in calculating the structure of neutron star crusts, and thus, it is utilized here to maintain consistency with the existing literature \cite{carreau2020modeling, Newton_2012}. It's worth noting that this particular functional form of surface energy differs from the one used for calculating the limiting temperature in Section \ref{limtemp}. Similar to surface energy, the curvature energy plays an important part in describing the surface and is written as \cite{Newton_2012} 
%%%%
\begin{equation}
  E_{curv} = 8\pi r_0A^{1/3}\sigma_c.\label{eq:ecurv}
\end{equation}
%%%%
Here $\sigma_c$ is the curvature tension related to the surface tension
$\sigma$ as \cite{carreau2020modeling, Newton_2012},
%%%%
\begin{equation}
  \sigma_c =
  \sigma\frac{\sigma_{0,c}}{\sigma_0}\alpha(\beta-Y_p),\label{eq:sigmac}
\end{equation}
%%%%
with $\alpha=5.5$. $\sigma_{0,c}, \beta$,  $\sigma_0$ and $b_s$  are the parameters which needs to be fitted for a given EoS with the available experimental AME2020 mass table \cite{Huang_2021} at a fixed value of $p$. The equilibrium composition of inhomogeneous matter in the inner crust is obtained by minimizing the energy of WS cell per unit volume at a given baryon density ($\rho_b=\rho_n+\rho_p$), where $\rho_n$ and $\rho_p$ represent the neutron and proton density, respectively. For the minimization,  the variational method used in \cite{Carreau_2019,Newton_2012} is applied where the Lagrange multipliers technique is used so that the auxiliary function to be minimized reads as \cite{Carreau_2019, Carreau_2020}
%%%%
\begin{equation}
    \label{eq:auxillaryfunctio}
    \mathscr{F}(A,I,\rho_0,\rho_g,\rho_p)=\frac{E_{WS}}{V_{WS}}-\mu_b\rho_b,
\end{equation}
%%%%
where $\mu_b$ is the baryonic chemical potential given by \cite{Carreau_2019}
%%%%
\begin{equation}
    \label{baryonicchempot}
    \mu_b=\frac{2\rho_0\rho_p}{\rho_0(1-I)-2\rho_p}\frac{\partial (E_{cl}/A)}{\partial \rho_g}+\frac{d \mathcal{E}_g}{d\rho_g}.
\end{equation}
%%%%
The chemical and mechanical equilibrium along with the Baym virial theorem then transmute to the following set of coupled differential equations  \cite{Carreau_2020},
%%%%
\begin{subequations}
\label{diffeq}
\begin{equation}
  \frac{\partial (E_{cl}/A)}{\partial A} = 0,\label{eq:ic1}
\end{equation}
\begin{equation}
  \frac{\rho_0^2}{A}\frac{\partial E_{cl}}{\partial \rho_0} = P_g,\label{eq:ic2}
\end{equation}
\begin{equation}
  \frac{E_{cl}}{A} + \frac{1-I}{A}\frac{\partial E_{cl}}{\partial I} +
  \frac{P_g}{\rho_0} = \mu_g,\label{eq:ic3}
\end{equation}
\begin{equation}
  \frac{2}{A}\left(\frac{\partial E_{cl}}{\partial I} -
  \frac{\rho_p}{1-I}\frac{\partial E_{cl}}{\partial \rho_p}\right)  = \mu_e(\rho_p),
  \label{eq:ic4}
\end{equation}
\end{subequations}
where $P_g$ is the gas pressure. For the complete derivation of Eqs. (\ref{diffeq}), one can see ref. \cite{carreau2020modeling} The four differential equations (\ref{diffeq}) are solved simultaneously to estimate the equilibrium composition in the inner crust.  The energy density for the homogeneous nuclear matter entering Eq. (\ref{ecluster}) and neutron gas  is determined by employing the E-RMF theory. 

%%%%%%%%%%%%%%%%%%%%%%%%%%%%%%%%%%%%%%%%%%%%%%%%%%%%%%%%%%%%%%%%%%%%%%%%%%%%%%%%%%%%%%%%%%
\subsection{Liquid core}
%%%%%%
As the density increases, the transition from inner solid crust to outer liquid core takes place. In the outer core, the energy density of homogeneous matter is written as
%%%
\begin{equation}
    \label{coreenergy}
    {E}_{core}={E}_{B}(\rho_b,\alpha)+{E}_{e}(\rho_e)+{E}_{\mu}(\rho_\mu),
\end{equation}
%%%%
where $B$ stands for  baryon. The population of baryons and leptons are calculated by the constraints of $\beta-$equilibrium and charge neutrality as  \cite{NKGb_1997, Hcdas_2020,Hcdas_2021}
%%%%
\begin{subequations}
\begin{equation}
\mu_n=\mu_p+\mu_e,  \ \ \mu_e=\mu_\mu.
\end{equation}
\begin{equation}
    \rho_p=\rho_e+\rho_\mu,
\end{equation}
\end{subequations}
%%%%
where $\mu_{p,n,e,\mu}$ are the chemical potential of the proton, neutron electron, and muon in the homogeneous phase, respectively. The crust-core transition from the crust side occurs when the energy density of the WS cell in the inner crust exceeds the energy density of the liquid core. It can be written as
%%%%%
\begin{equation}
    \label{eq:cctransition}
    {E}_{WS}({\rho_t})={E}_{npe\mu}(\rho_{\rho_t}),
\end{equation}
where $\rho_t$ is the density at the crust-core tranistion.

%%%%%%%%%%%%%%%%%%%%%%%%%%%%%%%%%%%%%%%%%%%%%%%%%%%%%%%%%%%%%%%%%%%%%%%%%%%%%%%%%%%%%%%%%%
\subsection{CLDM for nuclear pasta}
\label{cldmform}
\begin{figure}
    \centering
    \includegraphics[scale=0.7]{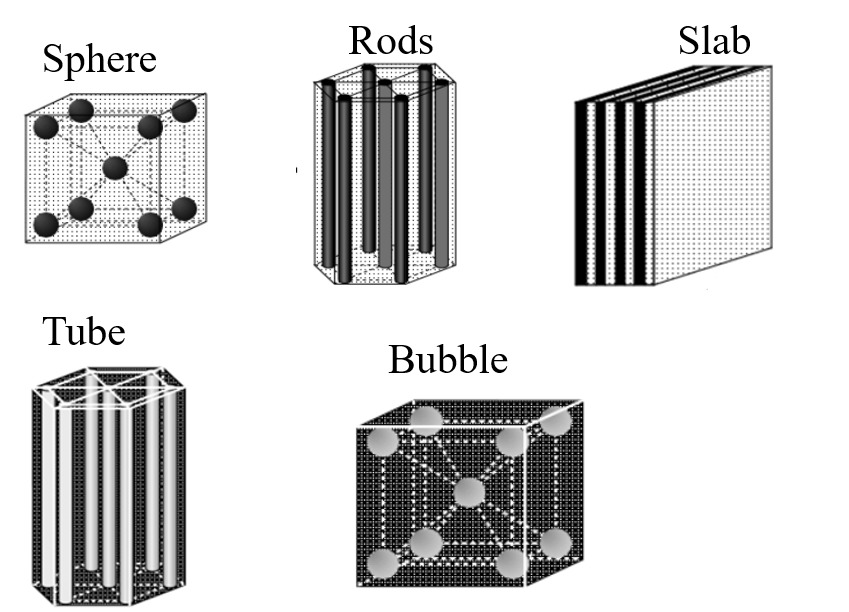}
    \caption{A schematic diagram of various nuclear pasta structure in the inner crust of neutron star.}
    \label{fig:pasta_struct}
\end{figure}

The CLDM formulation originally proposed by Baym, Bethe, Pethick (BBP) \cite{Baym_1971} assumes a repeating unit cell of volume $V_{WS}$ in which clustered structure ``pasta'' resides, immersed in a uniform neutron gas of density $\rho_g$. The system is neutralized by a homogeneous ultra-relativistic electron gas of density $\rho_e$. Using the Wigner-Seitz (WS) approximation, the energy of the system in the inner crust of a neutron star can then be written as \cite{Newton_2013},
%%%%%%%%%%%%%
\begin{align}
    E(r_c, y_p,\rho, \rho_n)&=f(u)\left[E_{\rm bulk}(\rho_b, y_p)\right]
    \nonumber\\
    &
    +E_{\rm bulk}(\rho_g,0)\left[1-f(u)\right]
    \nonumber\\
    &
    +E_{\rm surf}+E_{\rm curv}+E_{\rm coul}+E_e.
    \label{eq:energy_sys}
\end{align}
%%%%%%%%%%%%%
The WS cell's radius (or half-width in the case of planar geometry) is denoted as $r_c$, while $y_p$ represents the proton fraction, and $\rho$ and $\rho_n$ represent the baryon density of the charged nuclear component and the density of neutron gas, respectively. The cluster is characterised by a density  $\rho_i$ and a volume fraction of $u$ which is written as \cite{Newton_2012, Dinh_2021}
%%%%%%%%%%%%%%%%
\begin{equation}
    u=\begin{cases}
       (\rho-\rho_g)/(\rho_i-\rho_g) \, \, \, \text{for clusters},\\ (\rho_i-\rho)/(\rho_i-\rho_g) \, \, \, \, \text{for holes}.
      \end{cases}
\end{equation}
%%%%%%%%%%%%%%
The function $f(u)$ is defined as 
%%%%%%%%%%%%%%%%
\begin{equation}
    f(u) = \begin{cases}
           u  \, \, \, \, \,  \text{for clusters}, \\
           1-u \, \, \, \text{for holes.}
          \end{cases}
\end{equation}
%%%%%%%%%%%%%%
Pasta structure only affects the finite size effects, which can be expressed analytically as a function of the dimension of the pasta structure. This work  consider the three canonical geometries, namely spherical, cylindrical, and planar, defined by a dimensionality parameter $d = 3, 2, 1,$ respectively. They are represented as in Fig.\ \ref{fig:pasta_struct}  The  finite size corrections are defined along the same lines as in \cite{Newton_2013, Dinh_2021}. The surface and curvature energies are written as \cite{Newton_2013, Dinh_2021},
%%%%%%%%%%%%%%%%%%%

\begin{equation}
    E_{\rm surf}+E_{\rm curv}=\frac{u d}{r_N}\left( \sigma_s +\frac{(d-1)\sigma_c}{r_N}\right),
\end{equation}
%%%%%%%%%%%%%%%%%%%
where $r_N$ is the radius/half-width of the cluster/hole and   $\sigma_s$ and $\sigma_c$ are the dimension independent surface and curvature tension defined in Eqs. (\ref{eq:sigma}) and (\ref{eq:sigmac}). Finally, the Coulomb energy reads as \cite{Dinh_2021} 
%%%%%%%%%%%%%%%%%%%
\begin{equation}
    E_{\rm coul}=2\pi (e\,y_p\,n_i\,r_N)^2\,u\,\eta_d(u),
\end{equation}
%%%%%%%%%%%%%%%%%%%
where e is the elementary charge and $\eta_d(u)$ is associated with the pasta structure as \cite{Dinh_2021, Newton_2013}
%%%%%%%%%%%%%%%%%%%
\begin{equation}
    \eta_d(u)=\frac{1}{d+2}\Big[\frac{2}{d-2}\Big(1-\frac{du^{1-\frac{2}{d}}}{2}\Big)+u\Big]
\end{equation}
%%%%%%%%%%%%%%%%%%%
for $d=1$ and 3 whereas for $d=2$ it reads as,
%%%%%%%%%%%%%%%%%%%
\begin{equation}
    \eta_d(u)=\frac{1}{4}\Big[\log (\frac{1}{u}) + u -1\Big].
\end{equation}
For a given baryon density, the equilibrium composition of a WS cell is obtained by minimizing the energy per unit volume using the variational method. To obtain the most stable pasta structure at a given baryon density,  first the composition of a spherical nucleus is calculated. Then keeping this composition fixed, the radius or half-width of five different pasta structures, namely, sphere, rod, plate, tube, and bubble are computed. The equilibrium phase is then the one that minimizes the total energy of the system. This method is an approximation to the variational method and yield similar behaviour as compared to the complete  variational solution \cite{Dinh_2021}.

%%%%%%%%%%%%%%%%%%%%%%%%%%%%%%%%%%%%%%%%%%%%%%%%%%%%%%%%%%%%%%%%%%%%%%%%%%%%%%%%%%%%%%%%%%
\section{Neutron star observables}
%The metric corresponds to static, spherically symmetric stars is in the form of
%5%%
%\begin{eqnarray}
%ds^2= -e^{2\nu }dt^2+e^{2\lambda }dr^2+r^2d\theta^2+r^2sin^2\theta d\phi^2,
%\label{eq:metric}
%\end{eqnarray}
%where $r$, $\theta$ and $\phi$ are the coordinates. $\nu $, $\lambda $ are the metric potential given as  \citep{Krastev_2008}
%%%%
%\begin{eqnarray}
%e^{2\lambda } = [1-\gamma ]^{-1},
%\end{eqnarray}
%%%%5
%\begin{eqnarray}
%e^{2\nu }&=&e^{-2\lambda } = [1-\gamma ], \qquad r>R_{star}
%\end{eqnarray}
%%%%%
%with
%\begin{equation}
%\gamma =\left\{
%\begin{array}{l l}
%\frac{2m }{r}, & \quad \mbox{if $r<R_{star}$}\\\\
%\frac{2M}{R}, & \quad \mbox{if $r>R_{star}$}
%\end{array}
%\right.
%\end{equation}
%%%%
To determine the macroscopic characteristics of a stationary star, such as its mass ($M$) and radius ($R$), one can solve the Tolman-Oppenheimer-Volkoff equations given as  \cite{TOV1, TOV2}
%%%%%
\begin{eqnarray}
\frac{dP }{dr}= - \frac{[P(r)+{{E}}(r)][m(r)+4\pi r^3 P(r)]}{r[r-2m(r)]},
\label{eq:pr}
\end{eqnarray}
and 
\begin{eqnarray}
\frac{dm(r)}{dr}=4\pi r^2 {{E}}(r).
\label{eq:mr}
\end{eqnarray}
%%%%
Here $P(r)$ and ${E}$(r) is the pressure and energy density, respectively, $r$ is the radius of star defined within the Schwarzschild metric 
\begin{eqnarray}
ds^2= -e^{2\nu(r)}dt^2+e^{2\lambda(r)}dr^2+r^2d\theta^2+r^2sin^2\theta d\phi^2,
\label{eq:metric}
\end{eqnarray}
where, $r$, $\theta$ and $\phi$ corresponds to the coordinates while $\nu(r)$, $\lambda(r)$ are the metric potential \citep{Krastev_2008}. To obtain the $M$ and $R$ of the star, the coupled differential equations are solved with the boundary conditions $r=0, P = P_c$ and $r=R, P = P_0$ at certain central density.
To calculate the moment of inertia of the neutron star, the metric of slowly, uniformly rotating NS is given by \citep{Stergioulas_2003} 
%%%%
\begin{eqnarray}
ds^2= -e^{2\nu}dt^2+e^{2\psi}(d\phi - \omega dt^2)+e^{2\alpha}(r^2d\theta^2+ d\phi^2),
\end{eqnarray}
%%%%
%The moment of inertia (MI) of the NS is calculated in the Refs.  \cite{Stergioulas_2003,Jha_2008,Sharma_2009,Friedmanstergioulas_2013,Paschalidis_2017,Quddus_2019,Kol cccccccccccccccccccccccccccccccccccccccccccccccccccccccccccccccciogiannis_2020}. 
The moment of inertia (MI) of the uniformly rotating neutron star with angular frequency $\omega$ is can be expressed as  \cite{Hartle_1967,Lattimer_2000,Worley_2008}
%%%%
\begin{equation}
I \approx \frac{8\pi}{3}\int_{0}^{R}\ dr \ ({\cal E(r)}+P(r))\  e^{-\phi(r)}\Big[1-\frac{2m(r)}{r}\Big]^{-1}\frac{\Bar{\omega}}{\Omega}\ r^4,
\label{eq:moi}
\end{equation}
%%%%
where variable $\Bar{\omega}$ refers to the angular velocity of drag experienced by a uniformly rotating star. The the boundary conditions satisfying $\Bar{\omega}$ are 
%%%%
\begin{equation}
\Bar{\omega}(r=R)=1-\frac{2I}{R^3},\qquad \frac{d\Bar{\omega}}{dr}\Big|_{r=0}=0 .
\label{eq:omegabar}
\end{equation}
%%%%%%%%%%%%%%%%%
Finally the crustal moment of inertia can be calculated using the Eq. (\ref{eq:moi}) from transition radius ($R_c$) to the surface of the star ($R$) and is given by \cite{Fattoyev_2010, Basu_2018} 
%%%%%%%%%%%%%%%%
\begin{equation}
I_{crust} \approx \frac{8\pi}{3}\int_{R_c}^{R}\ dr\ ({\cal E}+P)\  e^{-\phi(r)}\Big[1-\frac{2m(r)}{r}\Big]^{-1}\frac{\Bar{\omega}}{\Omega}\ r^4.
\label{eq:moic}
\end{equation}

%%%%%%%%%%%%%%%%%%%%%%%%%%%%%%%%%%%%%%%%%%%%%%%%%%%%%%%%%%%%%%%%%%%%%%%%%%%%%%%%%%%%%%%%%%
\subsection{Relative pasta layer thickness and mass}
\label{relativepasta}
It is shown in Refs. \cite{Lattimer_2007, Zdunik_2017} that the relative mass and thickness of the crust are correlated with radius, mass, and crust-core transition density, which depends on the choice of EoS. In the same line, Newton {\it et al.} \cite{Newton_2021} derived the relative thickness and mass of a single layer of pasta structure as,
\begin{equation}
\label{eq:rr}
    \frac{\Delta R_p}{\Delta R_c} \approx \frac{\mu_c-\mu_p}{\mu_c-\mu_0},
\end{equation}
\begin{equation}
\label{eq:pp}
    \frac{\Delta M_p}{\Delta M_c} \approx 1-\frac{P_p}{P_c}.
\end{equation}
Here, $\mu_c$, $\mu_p$, and $\mu_0$ are the baryon chemical potential at crust-core (CC) transition, the location at which the pasta structure starts and at the surface of the star. $P_p$ and $P_c$ are the pressure at the bottom of the pasta layer and at the CC transition. Furthermore,  it can be approximated that the moment of inertia ($I$) of the crust is directly proportional to its mass to the first order approximation \cite{Lorenz_1993}. Hence, it follows that
\begin{equation}
    \frac{\Delta M_p}{\Delta M_c} \approx  \frac{\Delta I_p}{\Delta I_c}. 
\end{equation}
%%%%%%%%%%%%%%%%%%%%%%%%%%%%%%%%%%%%%%%%%%%%%%%%%%%%%%%%%%%%%%%%%%%%%%%%%%%%%%%%%%%%%%%%%%

\subsection{Shear modulus and velocity}
\label{shearmodulus}
The shear modulus ($\mu$) of a BCC Coulomb lattice in a uniform electronic background (using the low-temperature limit) and  including electron screening effects as per  the Monte Carlo simulation \cite{Chugnov_2010} can be written as \cite{tews_2017, Sotani_2013},
%%%%%%%%%%%%%%%%%%%%%%%%
\begin{equation}
\label{eq:shearmodulus}
    \mu=0.1194\left(1-0.010Z^{2/3}\right)\frac{\rho_i\left(Ze\right)^2}{a}.
\end{equation}
%%%%%%%%%%%%%%%%%%%%%%%
Here, $\rho_i$ is the density of nuclei, $Ze$ the charge and $a=R_{WS}$. Eq. (\ref{eq:shearmodulus}) is applicable for the case of spherical nuclei, whereas, near the crust-core boundary, there is a possibility of stable pasta structures. Although the exact elastic nature of these "exotic structures" is still unknown, one expects a decrease in the rigidity of the crust. To model this behaviour, i.e., between the density region $\rho_{ph} \le \rho_b \le \rho_c$, where $\rho_{ph}$ us the density at which nonspherical shapes appear and $\rho_c$ represents the crust core transition density, respectively, a function which joins these regions smoothly is used and is written as \cite{Sotani_2012, Gearheart_2011, Passamonti_2016}
%%%%%%%%%%%%%
\begin{equation}
\label{eq:mubar}
    \Bar{\mu}=c_1\left(\rho_b-\rho_c\right)\left(\rho-c_2\right),
\end{equation}
%%%%%%%%%%%%%
where $c_1$ and $c_2$ are the constants determined from  the boundary condition that $\Bar{\mu}$ should connect with Eq. (\ref{eq:shearmodulus}) smoothly at $\rho_b=\rho_{ph}$ and become zero smoothly at crust-core boundary. The latter condition arises from the fact that shear speed becomes zero at the crust-core boundary. the shear speed the can be defined as \cite{tews_2017},
%%%%%%%%%%%%%%%
\begin{equation}
\label{eq:shearspeed}
    V_s=\sqrt{\frac{\mu}{\rho_d}},
\end{equation}
%%%%%%%%%%%%%%
with $\rho_d$ being the dynamical mass density. If the effects of neutron superfluidity are ignored, the dynamical mass density is equivalent to the total mass density, denoted by $\rho_d=\rho_m$ according to \cite{Steiner_2009}. Using the plane wave analysis of the crustal shear perturbation equation, the frequency of the fundamental torsional oscillation mode can be approximated and expressed as \cite{Samuelsson_2006, Gearheart_2011}
%%%%%%%%%%%%%%%%
\begin{equation}
\label{eq:freq}
    \omega_0^2 \approx \frac{e^{2\nu} V_s^2 (l-1)(l+2)}{2RR_c},
\end{equation}
%%%%%%%%%%%%%%
where $e^{2\nu}= 1-2M/R$, $R$ is the radius of the star, $R_c$ is the radius of the crust and  $l$  is the angular `quantum' number.
%%%%%%%%%%%%%%%%%%%%%%%%%%%%%%%%%%%%%

%%%%%%%%%%%%%%%%%%%%%%%%%%%%%%%%%%%%%%%%%%%%%%%%%%%%%%%%%%%%%%%%%%%%%%%%%%%%%%%%%%%%%%%%%%
\section{Magnetised nuclear matter}
Neutron stars are generally observed as pulsars and magnetars which have strong magnetic field. To simulate such environmental conditions, this work constructs a magnetised EoS of  nuclear matter by employing the E-RMF formalism. The energy spectrum of the proton, which gets modified due to the Landau level, is written as \cite{Broderick_2000, Strickland_2012}
%%%%%%%%%%%%%%%%
\begin{equation}
    E_p=\sqrt{k_z^2+\ols{M}_{n,\sigma_z}^{p^2}}+W-R/2,
\end{equation}
%%%%%%%%%%%%%%%
and for charged leptons (electron and muon) as
%%%%%%%%%%%%%%%%
\begin{equation}
    E_{e,\mu}=\sqrt{k_z^2+\ols{M}_{n,\sigma_z}^{{e,\mu}^2}},
\end{equation}
%%%%%%%%%%%%%%%
where 
%%%%%%%%%%%%%%
\begin{align}
        \ols{M}_{n,\sigma_z}^{(p)^2}=M_{(p)}^{*^2}+2\Big(n+\frac{1}{2}-\frac{1}{2}\frac{q}{|q|}\sigma_z \Big)|q|B.\\
        \ols{M}_{n,\sigma_z}^{(e,\mu)^2}=M_{(e,\mu)}^{2}+2\Big(n+\frac{1}{2}-\frac{1}{2}\frac{q}{|q|}\sigma_z \Big)|q|B.
\end{align}
%%%%%%%%%%%%%
Here, $\sigma_z$ denotes the spin aligned with the magnetic field $B$, $n$ refers to the principal quantum number, and $k_z$ represents the momentum aligned with the direction of the magnetic field. The effective mass for the proton is denoted by $M^*$. The neutron spectrum bears resemblance to that of the Dirac particle, and can be expressed as:
%%%%%%%%%%%%%%%%
\begin{equation}
    E_n=\sqrt{k^2+M_n^{*^2}}+W+R/2.
\end{equation}
%%%%%%%%%%%%%%%
The number and energy density at zero temperature and in the presence of a magnetic field is given by \cite{Broderick_2000}
%%%%%%%%%%%%%%%%
\begin{equation}
\label{eq:density}
    \rho_{i=e,\mu,p}=\frac{|q|B}{2 \pi^2} \sum_{\sigma_z} \sum_{n=0}^{n_{max}}k_{f,n,\sigma_z}^{i},
\end{equation}
%%%%%%%%%%%%%%%
\begin{eqnarray}
\label{eq:energy}
    E_{i=e,\mu,p}&=&\frac{|q|B}{4 \pi^2} \sum_{\sigma_z} \sum_{n=0}^{n_{max}} \Big[E_f^{i}k_{f,n,\sigma_z}^{i}     + \ols{M}_{n,\sigma_z}^{{i^2}} \ln \Big(\Big|\frac{E_f^i+k_{f,n,\sigma_z}^i}{\ols{M}_{n,\sigma_z}^{{i}}}\Big|\Big)\Big],
\end{eqnarray}
%%%%%%%%%%%%%%%
respectively. In above equations, $k_{f,n,\sigma_z}^i$ is defined by
%%%%%%%%%%%%%%%%
\begin{equation}
\label{eq:fermi_momentum}
    k_{f,n,\sigma_z}^{i^2}=E_f^{i^2}-\ols{M}_{n,\sigma_z}^{{i^2}},
\end{equation}
%%%%%%%%%%%%%%%
where the Fermi energies are fixed by the respective chemical potentials given by
%%%%%%%%%%%%%
\begin{align}
    E_f^{l=e,\mu}=\mu_{\mu,e},\\
    E_f^{b=p,n}=\mu_b-W \pm R/2.
\end{align}
%%%%%%%%%%%%
In Eq. (\ref{eq:density}) and (\ref{eq:energy}), the $n_{\rm max}$ is the integer for which the Fermi momentum remains positive in Eq. (\ref{eq:fermi_momentum}) and is written as
%%%%%%%%%%%%%
\begin{align}
    &n_{max}=\Bigg[\frac{E_f^{i^2}-M^{*^2}}{2|q|B}\Bigg], \; {\rm proton}\\ \nonumber
    &n_{max}=\Bigg[\frac{E_f^{i^2}-M^2}{2|q|B}\Bigg], \; \; {\rm electron} \; \& \; {\rm muon} \, .
\end{align}
%%%%%%%%%%%%
Here $[x]$ represents the greatest integer less than or equal to $x$. The scalar density for the protons is further determined as 
%%%%%%%%%%%%%%%%
\begin{equation}
    \rho_p^s=\frac{|q|BM^*}{2\pi^2}\sum_{\sigma_z} \sum_{n=0}^{n_{max}}\ln \Big(\Big|\frac{E_f^i+k_{f,n,\sigma_z}^i}{\ols{M}_{n,\sigma_z}^{{i}}}\Big|\Big).
\end{equation}
%%%%%%%%%%%%%%
The number, scalar, and energy density for the neutrons are similar to the field-free case and can be found in \cite{Kumar_2018, Patra_2002} and references therein. The total energy density is the sum of matter-energy density and the contribution from the electromagnetic field, $\frac{B^2}{8 \pi}$. Finally, the pressure can be written, keeping the thermodynamic consistency as
%%%%%%%%%%%%%%%%
\begin{equation}
\label{eq:press}
P= \sum_{i=n,p}\mu_i\rho_i-E \, .   
\end{equation}
It is often convenient to express the strength of the magnetic field  in terms of the critical magnetic field of an electron ($B_c$) as
\begin{equation}
    B^*=\frac{B}{B_c}, 
\end{equation}
where $B_c \sim 4.414 \cross 10^{13}$ G. A magnetic field  is  strongly quantizing if only the lowest Landau level is filled i.e. $\nu=\Big(n+\frac{1}{2}-\frac{1}{2}\frac{q}{|q|}\sigma_z \Big)=0$ \cite{Haenel_2007}.
%%%%%%%%%%%%%%%

\clearpage
\addcontentsline{toc}{section}{Bibliography}
\printbibliography

%% file: Chapter_3/CHAP3.tex
\chapter{\label{chap3} Hot infinite  nuclear matter}

\section{Symmetric nuclear matter}
Investigating the nuclear matter properties as a function of temperature is  one of the vital problems in nuclear physics due to its importance in HIC reactions, supernovae and proto-neutron star. The nuclear matter exhibits typical first order liquid-gas phase transition (LGPT) similar to a water molecule  while increasing the temperature of the system.  Mahi \textit{et al.} \cite{fermilabgroup1} and  Finn \textit{et al.} \cite{fermilabgroup2} suggested that the LGPT  of nuclear matter can be analysed by studying the variation of yield with mass and projectile energy in a multi fragment Proton-Xenon/Krypton reaction and found that the yield follows a power-law dependence \cite{fermilabgroup1,fermilabgroup2}. In these experiments, researchers studied inclusive production of nuclear fragments with atomic numbers $3 \le Z \le 14$, which were produced by protons in the incident energy range of 80-350 GeV colliding with krypton and Xenon targets. The primary objective of these investigations was to gain insights into the phenomenon of Liquid-Gas criticality in nuclear matter. This led to several experiments which explored the above mentioned critical nuclear properties in the subsequent years \cite{experiment1,experiment2}. The critical behaviour of nuclear matter is a crucial feature in HIC reactions and therefore numerous theoretical phase-transition predictions have been attempted by several authors over the last two decades. The basis of  these studies is thermodynamics at equilibrium and phase diagram of nuclear matter \cite{theoreticalcalulation2,theoreticalcalulation3}.

In the experiments to understand the LGPT  in nuclear matter \cite{84, 94, 02, 03, 08, ctnew}, the critical temperature $T_c$ is hardly constrained. There is large uncertainty in the value of $T_c$ among these experiments. Moreover, the model dependence in these experimental calculations arises inevitably. Therefore, the theoretical and experimental calculation of $T_c$ and in general LGPT  can not be compared one on one. Instead, to overcome the uncertainty, this work look for the dependence of $T_c$ on bulk matter properties such as incompressibility at zero temperature. The correlation of critical parameter among themselves can also be utilised to constrain the related quantities and consequently deducing $T_c$.  

To understand this, E-RMF formalism is employed and newly developed E-RMF  forces such as  G3 \cite{g3}, FSUGarnet \cite{Chen_2015} and IOPB-I \cite{Kumar_2018}  are used in the finite temperature limit.  These forces have comparable bulk matter properties at zero temperature although having different numbers of adjustable parameters and their values. Different values of couplings in these newly developed E-RMF sets motivate us to look for the contribution of different terms such as scalar-scalar, scalar-vector terms etc. on the critical values and EoS of SNM at finite temperature. It should be mentioned here that the bulk matter properties are not unique to a particular force as a different combination of these adjustable parameters can give the same bulk matter properties at zero temperature. These facts are used here to study the behaviour of these forces in the finite temperature limit and consequently analysing the LGPT qualitatively.

There is a surprising similarity in behaviour near the critical point among systems that are otherwise quite different in nature.  These systems can be partitioned based on their criticality and placed in some ``universal classes''.  The systems here liquid-gas system, which belongs to one universal class should have comparable values of critical exponents and compressibility factor (the deviation from ideal gas) keeping with the scaling hypothesis and renormalization \cite{scalinglaw}.  Therefore, a complete statistical study including critical exponents of EoSs derived from different force parameter is necessary to provide a complete qualitative and quantitative understanding of phase transition properties as they all are based on mean-field approximations.

%\subsection{\label{rd} Results and Discussions}
\subsection{Force parameters}
In the present study,  the critical behaviour of the LGPT in nuclear matter is investigated using three different E-RMF sets, namely  IOPB-I \cite{Kumar_2018}, G3 \cite{g3} and FSUGarnet \cite{Chen_2015}. The NL3 \cite{Lalazissis_1997} force, known for its success in finite system calculations, is used  for comparison with the obtained results. The NL3 force has been noted to accurately describe nuclear properties, such as quadrupole deformation and charge radius, not only for nuclei located away from the valley of stability, but also for those on the $\beta$-stability line \cite{abdul}.  FSUGarnet parameter set and its latest iteration, the IOPB-I, are relatively new and have been recognized for their ability to accurately reproduce the neutron-skin thickness, as well as other important bulk matter properties, to a satisfactory degree.  These both sets have positive $k_3$  and negative $k_4$ corresponding to cubic  and  quartic  terms  arising from the self couplings of the $\sigma$ meson. It is relevant to mention  that a large negative value of $k_4$ leads to the  divergence  of   solution  in  the  lighter  mass  region  of  the  periodic table. The  scalar and vector cross-couplings $\eta_1$ and $\eta_2$ in these sets are zero yet they have  the value of vector self-coupling $\zeta_0$ within the acceptable limit \cite{Arumugam_2004}. The positive $k_3$, $k_4$ and small $\zeta_0$ guarantee the agreement with Dirac-Brueckner-Hartree-Fock (DBHF) theory \cite{eft}.\\
The G3 parameter set contains all the mesons and coupling constants in Eq. \eqref{rmftlagrangian} and therefore describe properties like skin thickness and two-neutron separation energy exceptionally well. The main feature of G3 is that it include the $\delta$ mesons. These are an important degree of freedom for the infinite nuclear matter calculations. Furthermore, due to the finite   $\eta_1$ and $\eta_2$, the G3 set has $\zeta_0$ $\approx$ 1 and positive $k_3$ and $k_4$. All these coupling constants have their specific role in EoS and they, therefore, describe the characteristics of a force.

\begin{figure}[!tbh]
	\centering
		\includegraphics[width=0.7\linewidth]{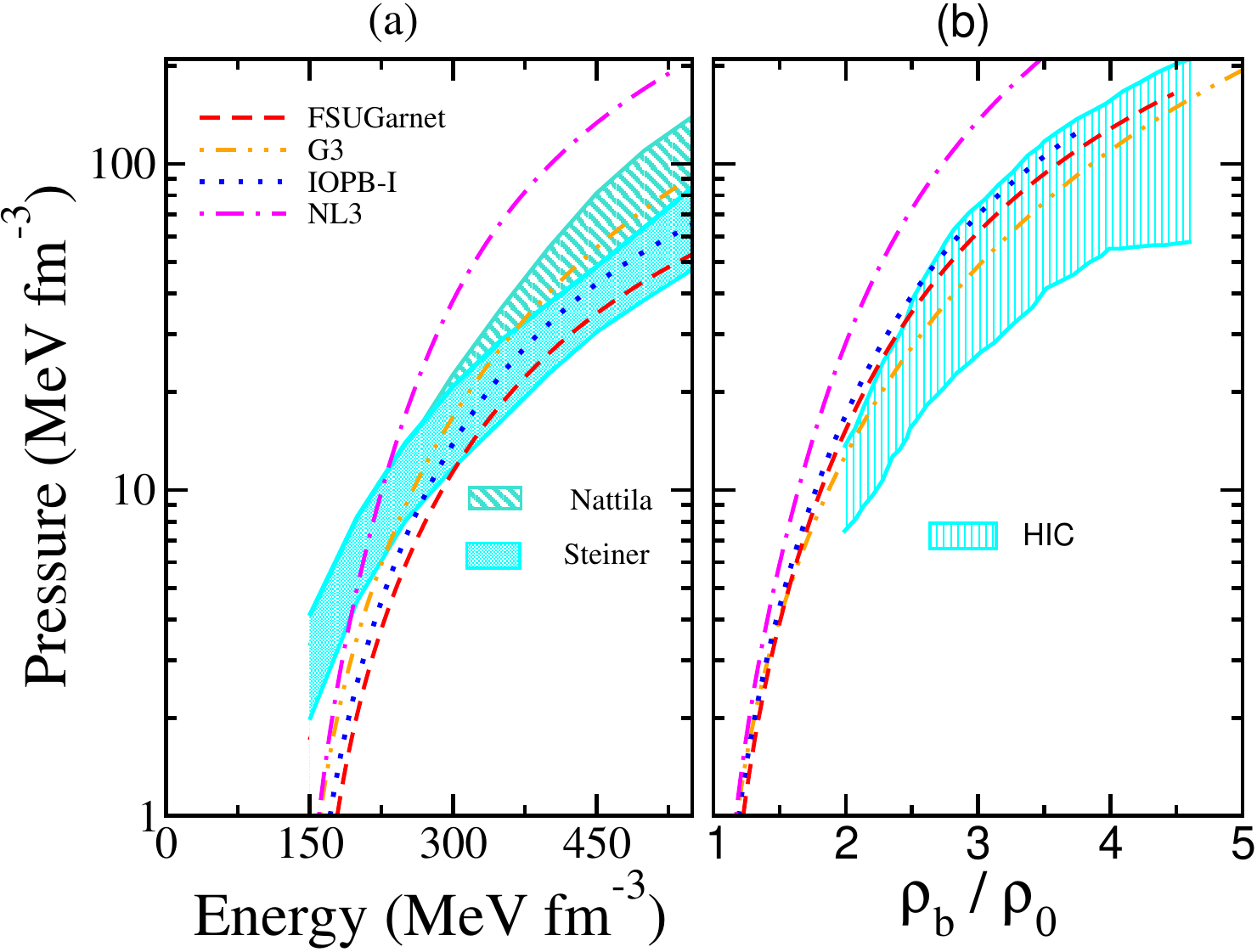}
	\caption{(a) The pressure-energy density relationship for four different parameter sets, FSUGarnet, G3, IOPB-I and  NL3, at $T=0$ is shown. The lower shaded region indicates the constraint from \cite{steiner} while  the upper shaded region represents the constraints on EoS for dense matter at zero temperature \cite{nattila}.\\
(b) The pressure-baryon density relationship is shown for the same four parameter sets along with experimental data obtained from the analysis of the EoS after incorporating pressure from both density and isospin asymmetry dependence \cite{hic}.}
	\label{eossnm}
\end{figure}
Any effective nuclear force is such that it can explain both finite and infinite nuclear matter.   However, it is difficult using the properties of nuclei as constraints. Consequently, one prefers to extrapolate the systematics of observables such as binding energy, saturation density, compressibility, effective mass etc.  The E-RMF parameter sets considered above have a very narrow range of binding energy, saturation density and effective nucleon mass. They also have compressibility in the  range as determined from
isoscalar giant monopole resonance (ISGMR) \cite{Colo_2014, Col_2004} result. Currently, this value is 240 $\pm$ 20 MeV \cite{Col_2004}. When compared with the experimentally derived  EoS's from various sources at zero temperature depicted in Fig. \ref{eossnm}, All the parameter sets are in reasonable agreement except the NL3 force due to its slightly larger incompressibility.  This disagreement can be neglected in this work as the domain of density useful for the LGPT is less than 0.15 $fm^{-3}$. Moreover, NL3 set is known to work exceptionally well in this density range. Therefore, it becomes essential to investigate the implication of these forces in the finite temperature limit realizing their comparable bulk matter properties at zero temperature.   Since the mean-field theory is thermodynamically consistent, a true  ``universal''  force should not only describe nuclear matter at zero temperature but it should adequately explain the finite temperature properties both qualitatively and quantitatively. The aim here is to understand the behaviour of these forces in the context of the LGPT as they have different scalar and vector self and cross-couplings.

%%%%%%%%%%%%%%%%%%%%%%%%%%%%%%%%%%%%%%%%%%%%%%%%%%%%%%%%%%%%%%%%%%%%%%%%%%%%%%%%%%%%%%%%%%%%%%%%%%%%%%
\subsection{Liquid-gas phase transition}
\begin{figure}
	\centering
		\includegraphics[width=0.8\linewidth]{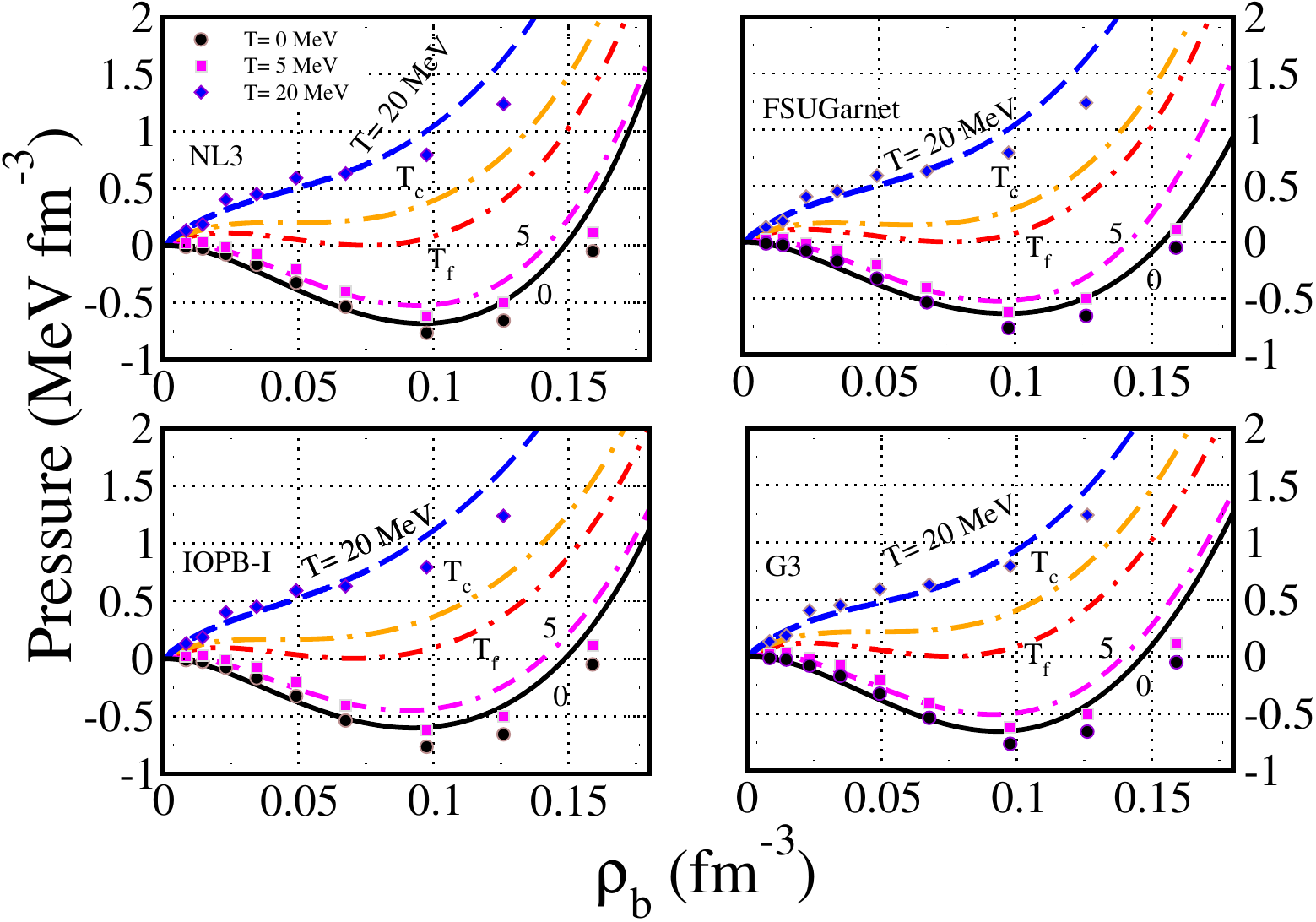} 
	\caption{ The variation of pressure with baryon density as a function of temperature. The symbols represent the microscopic calculation by \cite{fp}.}
	\label{pvd}
\end{figure}
Now let us discuss the implication of these force parameters at finite temperature explicitly near nuclear matter saturation density. The equation of state is shown in Fig. \ref{pvd} for the IOPB-I,  FSUGarnet, G3, and NL3 sets at temperature T= 0, 5, T$_f$, T$_c$, 20 MeV in the pressure vs. baryon density phase diagram. Here T$_f$ and T$_c$ stands for flash and critical temperature, respectively. It is evident that P-$\rho_b$ isotherms have a quite distinct pocket at a lower temperature. These isotherms show typical short-range van-der Waals like property as the nature of both van-der Waals and Nuclear force is the same. As one move towards higher temperature, the compressibility of nuclear matter decreases and therefore at a particular temperature, we will have a situation where pressure is no more negative i.e. P($\rho$, T$_f$)=0 and consequently $dP/d\rho = 0.0$. The highest temperature at which a self-bound system can exist in hydrostatic equilibrium, represented by $P=0$, is referred to as the flash temperature. The density at this temperature is known as the flash density. Once the temperature surpasses the flash temperature, the nuclear matter becomes unbound and starts to expand. Further increase in temperature results in the occurrence of an inflation point which is expressed as \cite{pwang}
\begin{equation}
\frac{\partial p}{\partial \rho}\bigg|_{T_c}=\frac{\partial^2 p}{\partial \rho^2}\bigg|_{T_c}=0.
\end{equation} 
This temperature is called critical temperature $T_c$ after which, the pressure is a monotonically increasing function of density. The pressure and density at the inflation point  are referred to as the critical pressure, denoted as $P_c$, and the critical density, denoted as $\rho_c$.
At the critical density, the second derivative of free energy ($F=E-TS$) as a function of baryon density is called critical incompressibility. This is given by
\begin{equation}
   K_c=9\rho_b^2 \frac{\partial^2}{\partial \rho_b^2} \frac{F}{\rho_b}\bigg |_{\rho_c}.
\end{equation}
This behaviour is similar to the first-order phase transition  of one-component water system where due to 
the latent heat of vaporization, the temperature does not increase unless the complete liquid is not 
vaporized. Table \ref{criticaldata} compiles all these critical values for all the sets along with some other theoretical and experimental predictions. \\ 
There is a large uncertainty among the several experimental values of critical temperature and it is hardly constrained \cite{84, 94, 02, 03, 08, ctnew}. This uncertainty mainly arises because all these experiments are performed for fragmentation reaction on finite nuclei and then extrapolated to evaluate the $T_c$ for the infinite matter. These experimental calculations are model dependent. Besides, various effects like finite size,  small time scale $\approx 10^{-(22-23)}$s in multi fragment reaction which makes it hard to study thermodynamic equilibrium, Coulomb interaction, isospin, angular momentum etc., also add to this uncertainty. Therefore, it is not very wise to theoretically address the quantitative nature of the nuclear matter phase transition using any mean-field calculation. Although one can realise based on recent experiments \cite{03,08, ctnew} that critical temperature should be greater than 15 MeV. 
	
\begin{sidewaystable}\renewcommand{\arraystretch}{.85}
		\caption{The  critical temperature $T_c$ (in $MeV$), critical pressure $P_c$ 
			(in $MeV fm$$^{-3}$) and critical density $\rho_c$ (in $fm$$^{-3}$) with the respective binding energy  
			(in $MeV$) at the critical point  along with the flash temperature $T_f$ and flash density $\rho_f$  for symmetric nuclear
			matter. The corresponding incompressibility, effective mass at critical density and compressibility factor $C_f$ and critical incompressibility $K_c$ is also shown.        The            experimental and other theoretical estimations of $T_c$  are also shown for the reference.} 
     	    \resizebox{\textwidth}{!} {\begin{tabular}{ c  c c c c  c  c  c  c  c c c c c}
			\hline
			\hline
			& $T_c$  & $P_c$ & $\rho_c$   & $\mathcal{E}/\rho_b$ at $T_c$ &T$_f$&$\rho_f$&$K_\infty$& $M^*_c/M$&$C_f$ & $K_c$   \\ 
			& $MeV$ & $MeV/fm$$^3$ & $fm$$^{-3}$  & $MeV$& $MeV$& $fm$$^{-3}$&$MeV$ &&& $MeV$  \\ 
			[0.5ex] 
			\hline
			FSUGarnet \cite{Chen_2015}  & 13.80 & 0.171 & 0.043 & -9.61&11.3&0.071&229.50&0.850&0.282&-68.35 \\  
			IOPB-I \cite{Kumar_2018} & 13.75 & 0.167 & 0.042 & -8.80&11.2&0.071&222.65&0.864&0.277&-69.95 \\ 
			G3  \cite{Kumar_2017} & 15.30 &  0.218 & 0.049 & -6.68&12.1&0.075&243.96&0.879&0.292&-82.91 \\ 
			NL3  \cite{Lalazissis_1997} & 14.60 & 0.202 & 0.046 & -8.38&11.8&0.070&271.38&0.846&0.276&-77.85 \\
			G1 \cite{Frun_1997}& 14.30&0.187&0.046&-8.24&11.5&0.075&215.00&0.877&0.285&-72.03\\
			G2 \cite{Frun_1997}  &14.30&0.184&0.043&-8.04&11.8&0.080&215.00&0.879&0.299&-78.77\\
			NL1 \cite{Lourenco} & 13.74 & 0.164& 0.041 &-9.71&11.2&0.070&211.70&0.872&0.290&--71.909\\
			NL2 \cite{Lourenco} & 18.63 & 0.361 & 0.056 &-3.47&14.3&0.085&399.20&0.861&0.345&-111.03\\
			NL-SH \cite{Lourenco} & 15.96 & 0.264 & 0.052& -6.70&12.7&0.080&355.36&0.846&0.315&-90.097\\
			
			FP (Microscopic) \cite{fp}& 17.5    $\pm$ 1.00 & - &- &-&-&-&240.00&-&-&-\\ 
			Exp. 1 (1984) \cite{84}& 12.0 $\pm$ 0.20 &- &- &- &-&-&-&-&-&- \\ 
			Exp. 2 (1994) \cite{94}& 13.1 $\pm$ 0.60 &- &- &- &-&-&-&-&-&- \\  
			Exp. 3 (2002) \cite{02}& 16.6 $\pm$ 0.86 &- &- &- &-&-&-&-&-&- \\  
			Exp. 4 (2003) \cite{03}& 20.0 $\pm$ 3.00 &- &- &- &-&-&-&-&-&- \\  
			Exp. 5 (2008) \cite{08}& 19.5 $\pm$ 1.20 &- &- &- &-&-&-&-&-&-\\
			& 16.5 $\pm$ 1.00 &- &- &  &-&-&-&-&- \\  
			Exp. 6 (2013) \cite{ctnew} & 17.9 $\pm$ 0.40 & 0.31 $\pm 0.07$ & 0.06 $\pm$ 0.01 &-&-&-&-&-&0.288&*\\[1ex]
			\hline
			\hline
		\end{tabular}} 
		\label{criticaldata}
	\end{sidewaystable}
Theoretically, there are two sets of RMF parameters. One, those have incompressibility in the accepted range as predicted by ISGMR result i.e. 240 $\pm$ 20 MeV. These parameters, which include the FSUGarnet, IOPB-I and G3 forces,  predict the critical temperature less than 15 MeV. One exception being G3 which estimates $T_c$ = 15.2 MeV. These sets are extensively used for Neutron star calculation where the nuclear matter is at high density. The others set, although estimate large $T_c$ but have large incompressibility and therefore are not used in nuclear astrophysical applications.  The calculations with the set FSUGarnet, IOPB-I, G3 along with the NL3 force are in reasonable agreement with the first set of forces.

The critical parameters obtained using mentioned E-RMF sets in Table \ref{criticaldata} shows some striking behaviour in terms of their correlations which otherwise are very difficult to establish analytically. The pearson correlation coefficient between two variables $x$ and $y$ can be calculated as,

\begin{equation}
 R_{xy}=  \frac{cov(x,y)}{\sigma_x\sigma_y}.
\end{equation}
Here,  $cov$ denotes the covariance, and $\sigma$ represents the standard deviation. Among the variables studied, $T_c$ and $P_c$ exhibit the most robust correlation, with a Pearson's coefficient of 0.991 and a p-value of 0.009. The p-value serves as the significance level (typically, a p-value of less than 0.05 is considered statistically significant for a 95\% confidence limit) indicating the strength of the correlation coefficient \cite{jawlik2016statistics}. This is in consistency with  classical van-der Waals (VDW) gas property; $T_c/P_c$ =8b, where b arises due to the repulsive interaction \cite{wanderwaal}. A similar trend is observed in $T_c-T_f$, $T_c-K_c$, $P_c-K_c$ and $\rho_c-K_c$ with p-value less then 0.05. These correlations are consistent with the standard analytical equation relating the $K_c$ and the ratio $P_c/\rho_c$ given by
\begin{equation}
    K_c + 18\frac{P_c }{\rho_c} = 0,
\end{equation}
Therefore, these correlation suggests the strong relationship of critical parameter among themselves. They become important as constraining $T_c$ is difficult  and a direct comparison with experimental data is not suitable. If however, extrapolating the finite nuclei experiments to infinite matter was possible accurately and critical temperature in these experiments could be measured precisely, one can easily get other finite temperature properties based on established correlations.\\
\begin{figure}
	\centering
		\includegraphics[width=.7\textwidth]{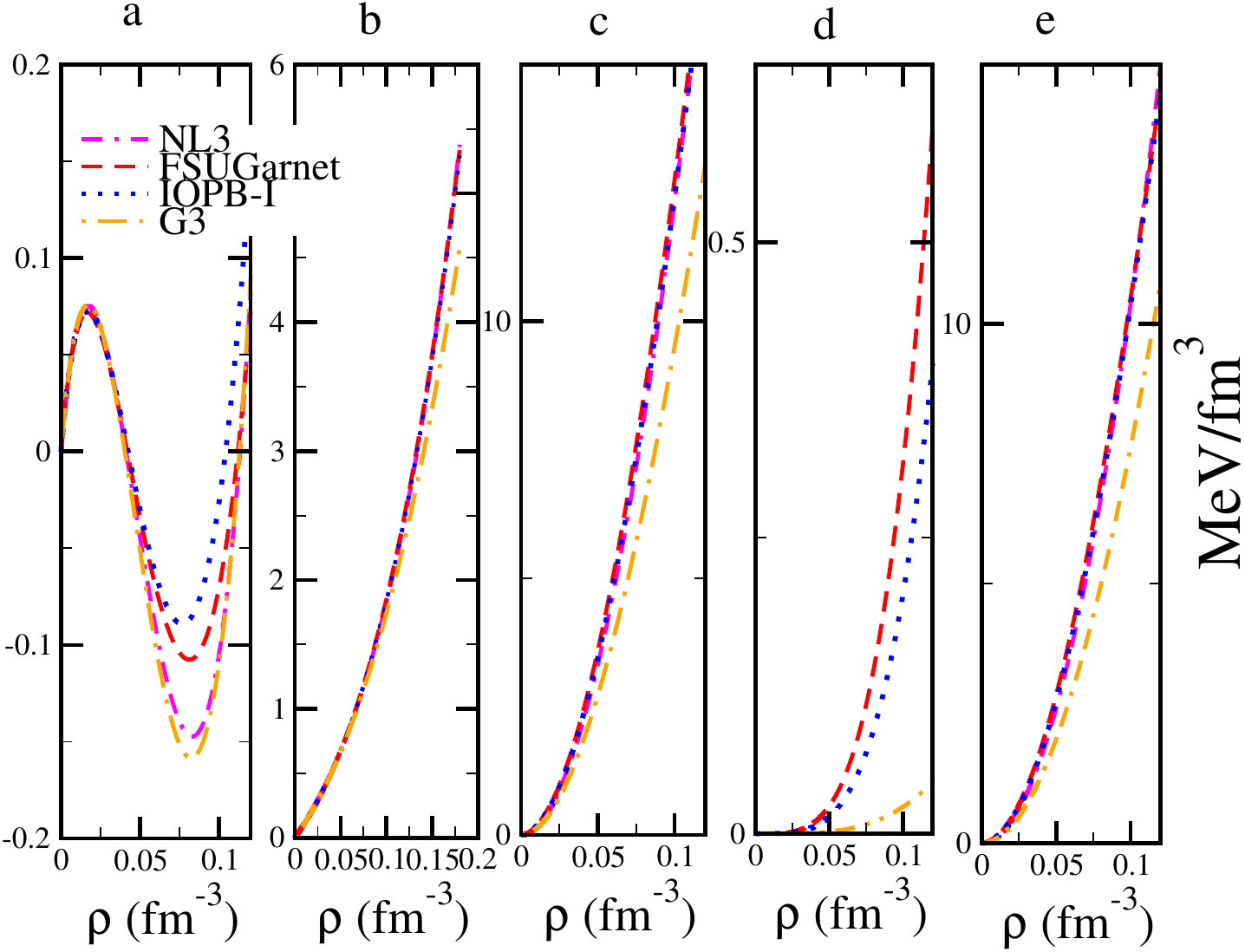} 
	\caption{The contribution arising from the different meson interaction on total pressure at T=10 MeV is shown for the different parameter sets. (a) Total Pressure, contribution due to (b)  kinetic part (c) scalar-self interaction (d) vector-self interaction and (e) scalar-vector cross interaction.}
	\label{comparison}
\end{figure}
On the other hand, weak correlation of critical parameters with that of saturation properties at zero temperature is also interesting. Naively, one could argue that the binding energy at saturation should be related to critical temperature. Instead, there is no such correlation between the binding energy $e_0$ and $T_c$ or $T_f$.  Similarly these models also do not satisfy the empirical relations \cite{rhocempirical} between saturation density  $\rho_0$ and $\rho_c$ or $\rho_f$. It is observed that the $T_c$ has a spread of 1.55 MeV while the saturation energy and saturation density lie within 0.27 MeV and 0.005 $fm^{-3}$, respectively. It indicates that the properties at saturation have no dictation over the position of critical points. Therefore, any  relation, relating properties at saturation and finite temperature  can not be generalised.
It should also be noted that each of the above force i.e. FSUGarnet, IOPB-I, G3 and NL3 have different adjustable parameters. One can not make any prediction directly by these adjustable parameters. In other words, none of the scalar and vector couplings (self and cross) ($k_3, k_4, \zeta_0$) individually has a direct correlation with the critical parameters. However, except the NL3 force, a greater value of $\zeta_0$ gives smaller $\rho_c$ and force with greater incompressibility estimates the larger $T_c$ .  A close investigation of the contribution arising from scalar and vector channel suggests very strongly that these forces only differ in contribution from vector self-coupling term and therefore, $\zeta_0$ becomes an important quantity in finite temperature EoS. This is shown in Fig.\ \ref{comparison} where contribution from different interaction in total pressure is shown at T=10 MeV. The only major difference is  in graph (d) which represents the vector-self interaction term in the pressure. The effect of small value of $\zeta_0$ in G3 and absence of it in NL3 is visible. The NL3 consequently have a large negative $k_4$ which is a possible reason for it to not show the common characteristic of other three forces. A soft contribution of vector-self interaction in total pressure therefore seems to be the reason for larger $T_c$ in G3 set. 
The compressibility factor  $C_f$, which signify the deviation from ideal gas is given by \cite{Lourenco}
\begin{equation}
\label{compressibilityfactor}
C_f=\frac{P_c V_c}{T_c}=\frac{P_c}{\rho_c T_c},
\end{equation}
can be calculated for the respective force parameters. The universal class corresponding to the liquid-gas system have the compressibility factor $C_f \approx$ 0.292  \cite{cfuniversal} following the law of corresponding states. Table \ref{criticaldata} also shows the value of $C_f$ for each force parameter. All the forces except NL2 and NL-SH,  which have high incompressibility have $C_f$ close to 0.292.  This further validates the importance of an EoS well within the acceptable incompressibility range \cite{Lourenco}. \\

The  parameter sets G3, IOPB-I and  FSUGarnet  are known to work remarkably at zero temperature high-density regime.This work also aims to find their validity at the low dense finite temperature limit.  For this,  EoS for each force in Fig. \ref{pvd} is also compared with the two-nucleon and three-nucleon interactions called $\nu_{14}$ and TNI \cite{fp} at T= 0, 5 and 20 MeV. This microscopic variational calculation is known to work better at low density but is not quite suitable for infinite nuclear matter \cite{variationalfail}. 
Whereas, here the primary focus is on the low-density behaviour of these forces at finite temperature. The EoS for each of the force at the mentioned temperature traces the microscopic calculation at density $<$ 0.06 MeV but then deviate at higher density.  This deviation gets better at a higher temperature and the parameter set G3 is nearest to the microscopic variational calculation at T=20 MeV. This is simply because the fitting procedure of any force does not take into account the critical temperature of the liquid-gas phase transition. The critical temperature, therefore, can be an important tool in making a force universally acceptable at both low and high-density regime at zero and finite temperature. It can be used along with the other bulk matter properties at zero temperature as a constraint for EoS. The parameter set G3 can be a good candidate for that. \\
\begin{figure}
	\centering
		\includegraphics[scale=0.4]{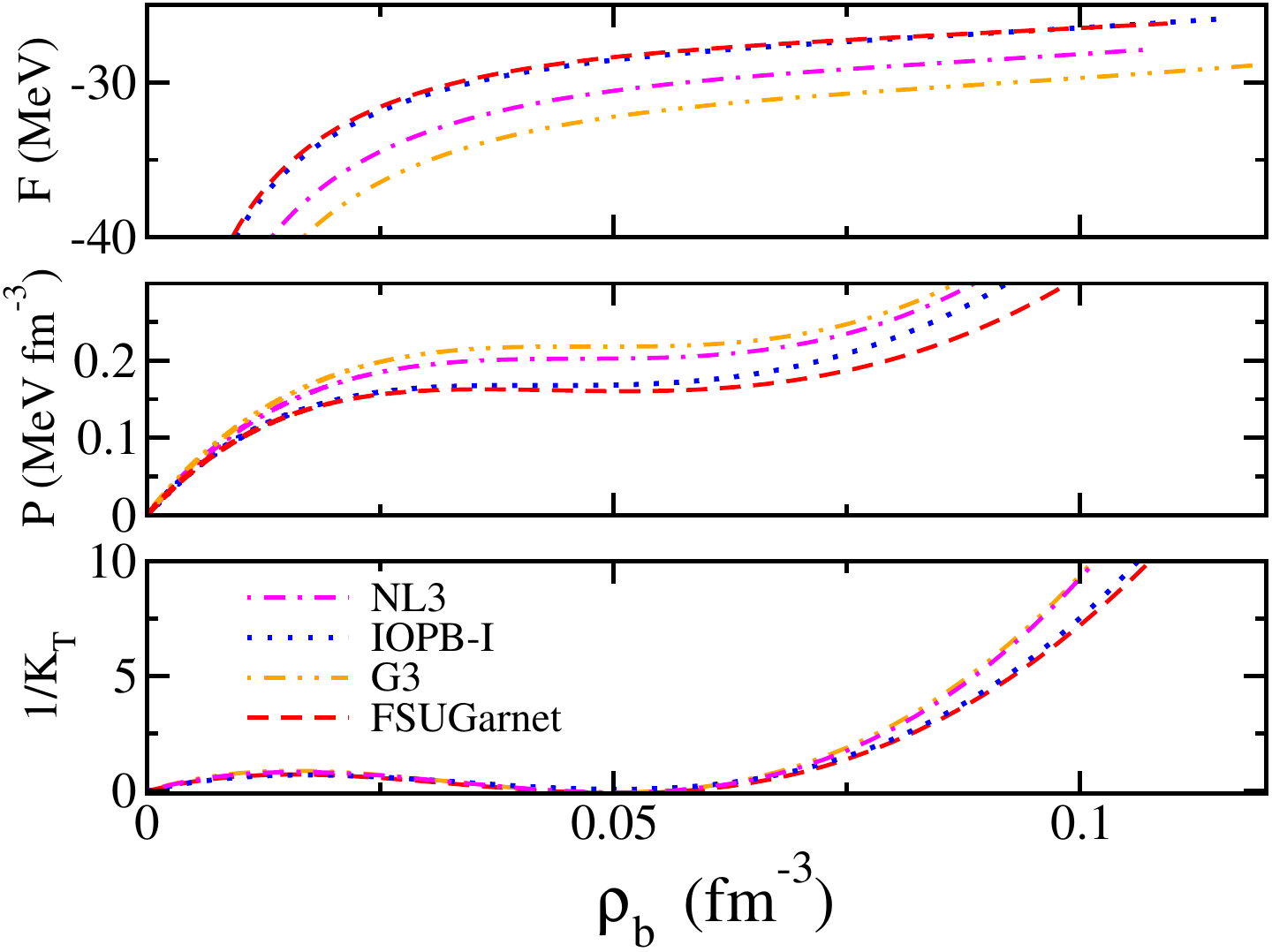}
	\caption{Variation of free energy, pressure and rigidity (1/K$_T$) with density for critical isotherm.}
	\label{rigidity}
\end{figure}
The temperature also impacts the incompressibility at saturation($K_\infty$) or rigidity ($K_T^{-1} = \rho_b \partial P/ \partial \rho_b$) of nuclear matter. It is defined as the curvature of the 
EoS at saturation. This curvature decreases with an increase in temperature . At 
large temperature, the nucleons are more prone to be free due to extra thermal energy and therefore 
incompressibility decreases when the temperature increases. 
Fig.\ \ref{rigidity} shows the variation of Free energy, pressure and rigidity (1/K$_T$) with density for critical isotherm for different parameter sets. For critical isotherm, the free energy is a smooth monotonically increasing function with G3   showing the least slope. Below the critical isotherm, the free energy has a well-defined pocket which disappears at T=T$_c$. The respective rigidity (1/K$_T$) is shown for each parameter set. It first increases and then starts decreasing and at the critical density, becomes zero. At critical density, one observes the inflation point in P-$\rho$ isotherm, which makes rigidity equals zero at the critical point.\\

\begin{figure}
	\centering
		\includegraphics[scale=0.4]{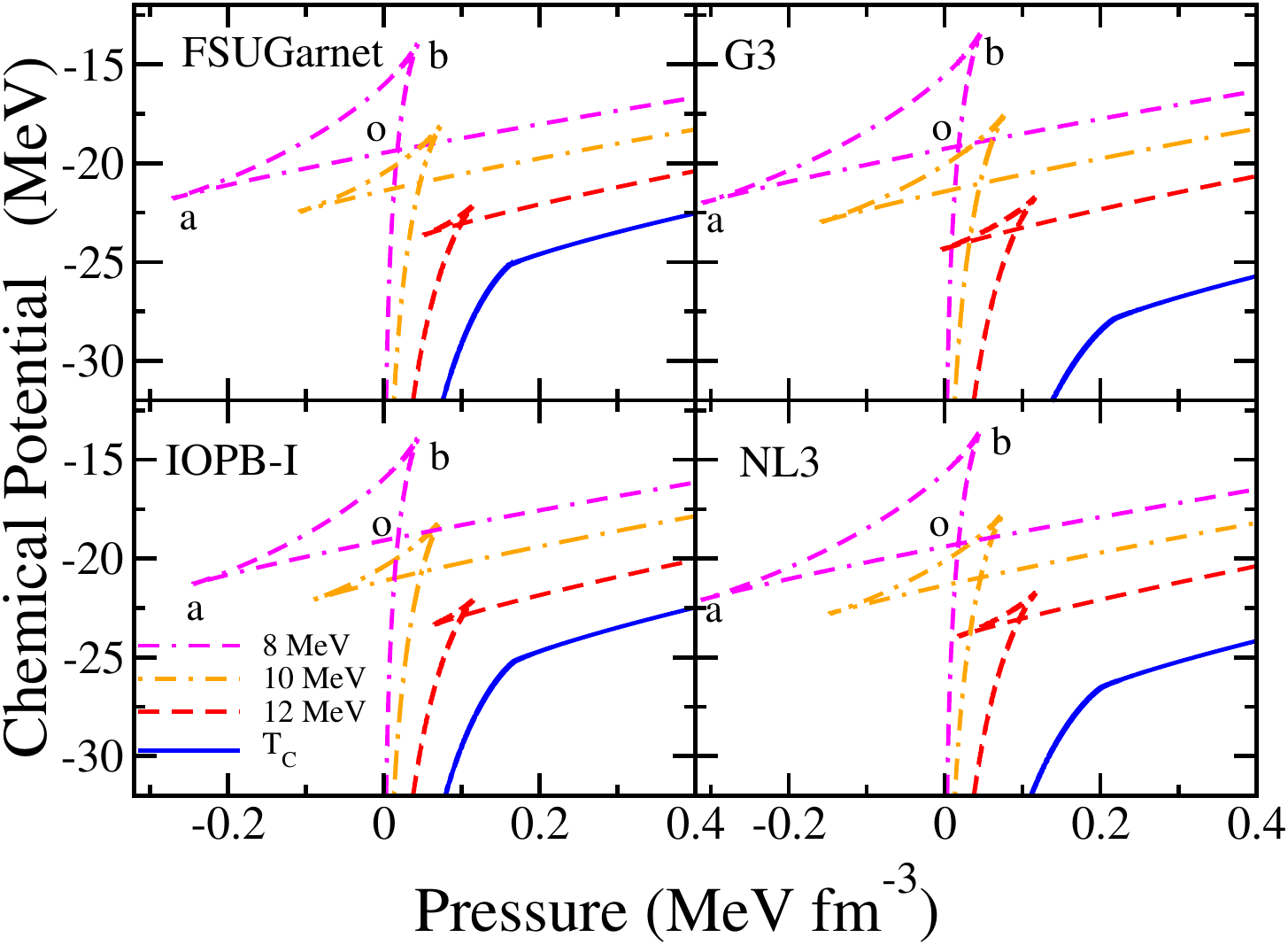}
	\caption{Variation of chemical potential as a function of pressure for various values of temperatures.}
	\label{chemicalpotentialpressure}
\end{figure}
The order of the transition is depicted in Fig. \ref{chemicalpotentialpressure} where   the  isothermal dependence of molar Gibbs potential which for a pure substance (here one component neutron-proton symmetric system) is nothing but the chemical potential  on pressure and temperature is shown. The chemical potential plays an important role in determining the values of Fermi function which consequently determine the baryon density. For each parameter set, as one moves from zero pressure to point `b', chemical potential increases sharply. The potential then decreases till point `a' and pressure becomes negative. The path `ba' marks the instability region. After point `a', the potential increases monotonically with a different slope as compared to the path 0 to `b'. The area of path `obao' is largest for the set G3 and NL3 for each temperature. Because the system wants to attain minimum potential and free energy, a real system will not follow the path `obao' and will take the path with the least potential. It will move from zero pressure to point `o' and then will again increase. The path `obao' keeps decreasing with an increase in temperature and vanishes as one reaches the critical temperature. At critical isotherm, the slope changes discontinuously validating the argument of nuclear matter undergoing a first-order LGPT at low density.  The set G3 and NL3 estimate the largest instability area for a given isotherm, suggesting the dominance of strong attractive $\Phi$ field over repulsive vector $W$ field at low density for these sets, which play important role in determining the instability boundary. This is directly the consequence of the value of $\zeta_0, k_3 ,k_4$ in these forces.
The low value of $\zeta_0$ in G3 and NL3 sets account for the less repulsive force as compared to the high value of $\zeta_0$ in FSUGarnet and IOPB-I forces. Therefore the instability area is inversely proportional to the value of $\zeta_0$ in finite temperature limit. 

\begin{figure}
	\begin{center}
			\includegraphics[scale=0.4]{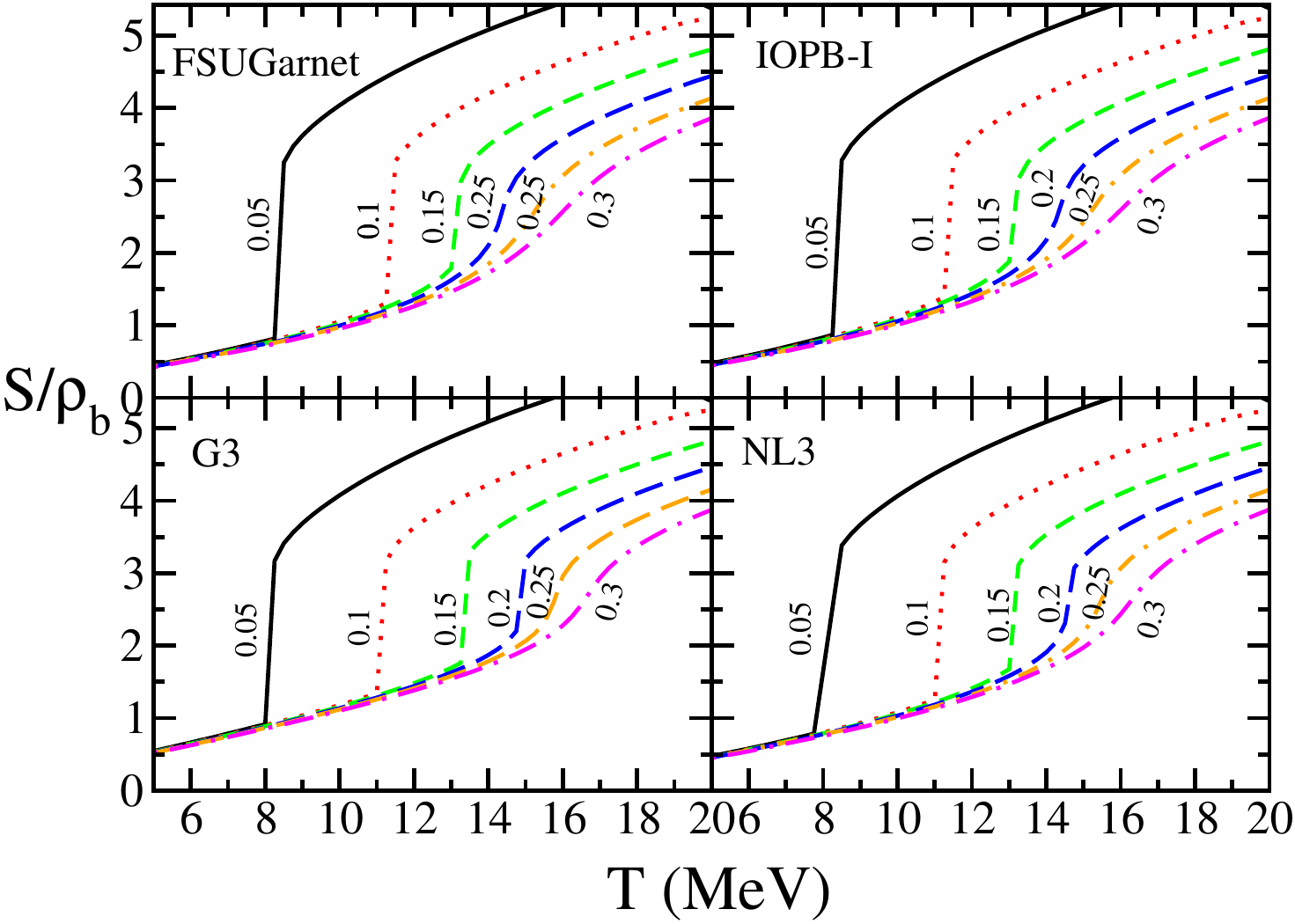}
		\caption{The entropy at constant pressure for 0.05, 0.1, 0.15, 0.2, 0.25 and  0.3 MeV fm$^{-3}$ as a function of temperature for various force parameters.}
		\label{isobarentropy}
	\end{center}
\end{figure}

The isobaric entropy as a function of temperature is shown in Fig.\ \ref{isobarentropy} for the IOPB-I,  FSUGarnet, G3, and NL3 sets. The values of pressure are 0.05, 0.1, 0.15, 0.2, 0.25 and  0.3 MeV fm$^{-3}$ from left to right.  There is a sharp discontinuity at the transition temperature marking the presence of two phases in the system. As one goes beyond the critical temperature (refer Table \ref{criticaldata}), the curve becomes smooth and continuous showing that system now only exists in the gaseous  phase. One can see a sharp difference in isobar for 0.2 MeV fm$^{-3}$ for all the sets. For IOPB-I and FSUGarnet, this pressure is above $P_c$ showing continuous curve and for G3 and NL3, this is very close to $P_c$  thus still showing a discontinuity. Furthermore, the relative separation between isobars decreases as we increase the pressure.\par
The isobaric entropy depends on the chemical potential as is shown in Eq.\ \eqref{entropyeq} and consequently dictated by the respective EoS. Again the vector self-coupling $\zeta_0$ plays an important role. It can, therefore, be apprehended that, if a force parameter is designed to have the $\zeta_0$ $\approx$ 1.0  keeping the incompressibility within the limit, it can address the qualitative as well as quantitative nature of phase transition. This is visible in these parameter sets. They only differ much in the value of $\zeta_0$ with almost the same bulk matter properties. The force G3 stands out due to its low $\zeta_0$ and presence of  cross-couplings ($\eta_1, \eta_2$) of scalar $\sigma$ and vector $\omega$ meson.  

\subsection{\label{StabilityAnalysis}Stability analysis}
\begin{figure}
	
	\centering
		\includegraphics[scale=0.4]{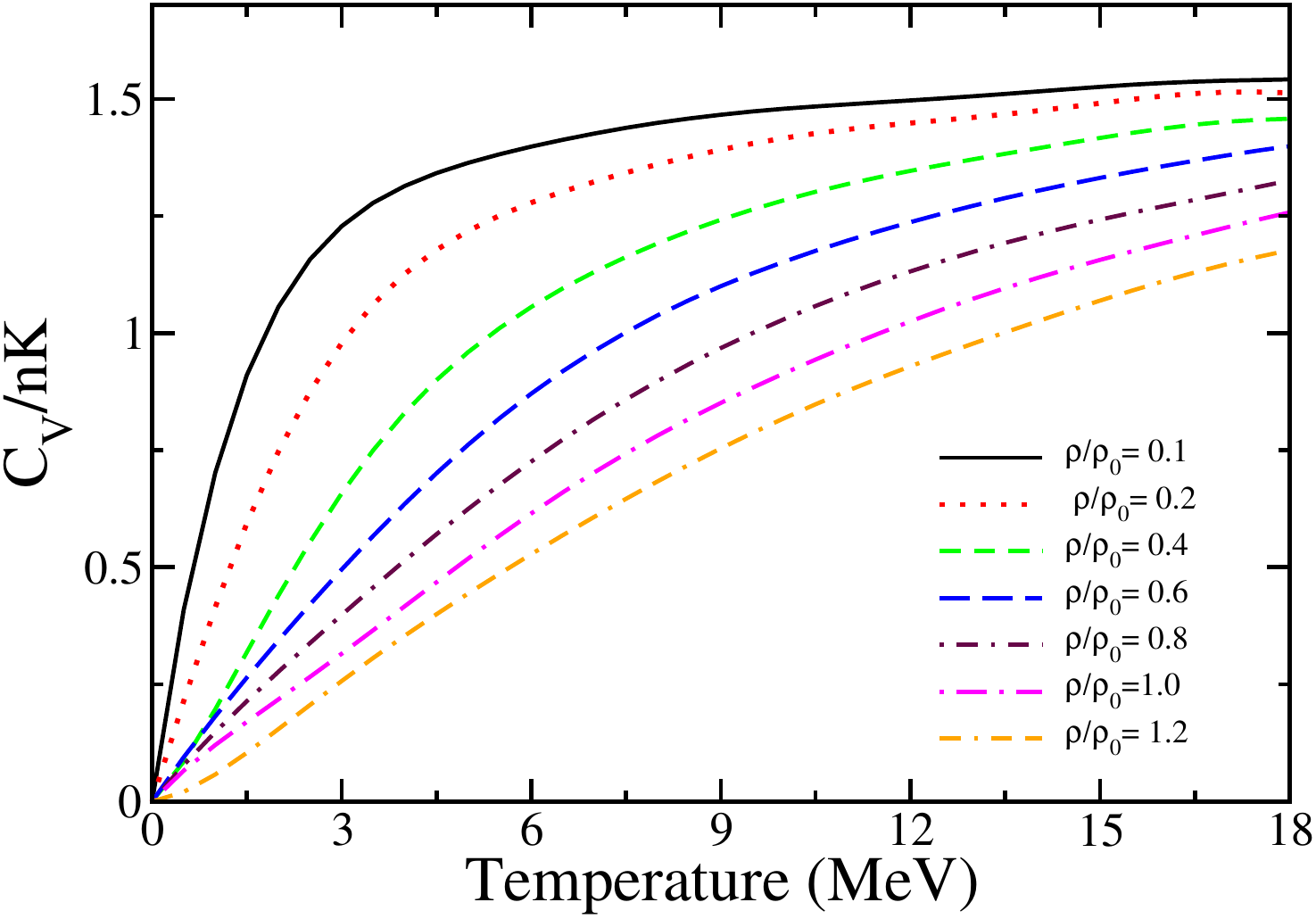}
	\caption{Specific heat at constant volume as a function of temperature for the set G3. }
	\label{specificheat}
\end{figure}

\begin{figure}
	\centering
		\includegraphics[scale=0.4]{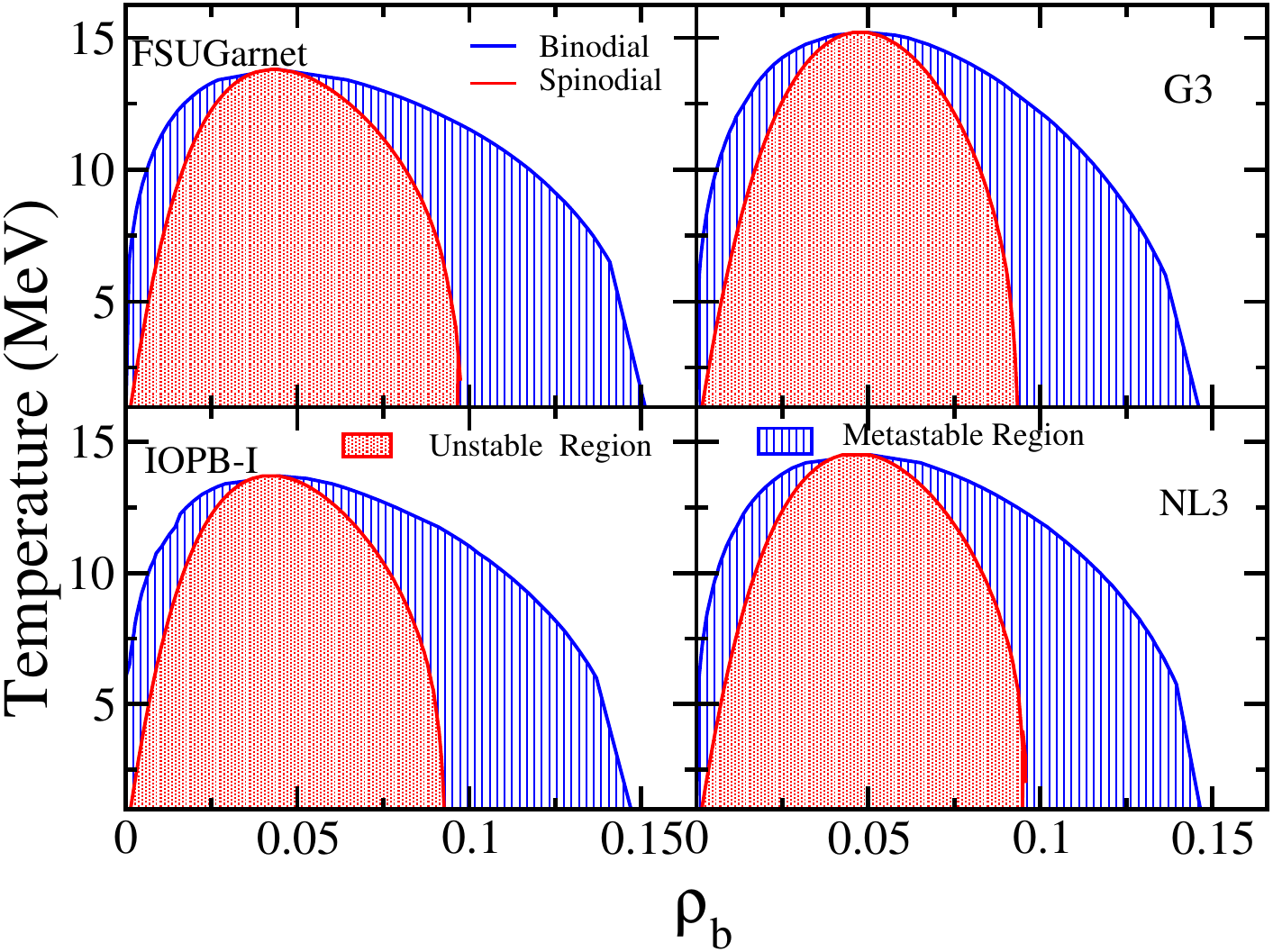}
	\caption{The binodal and spinodal curve for IOPB-I, FSUGarnet, G3, and NL3 sets.}
	\label{binodal}
\end{figure}

\begin{figure}
	\centering
		\includegraphics[scale=0.4]{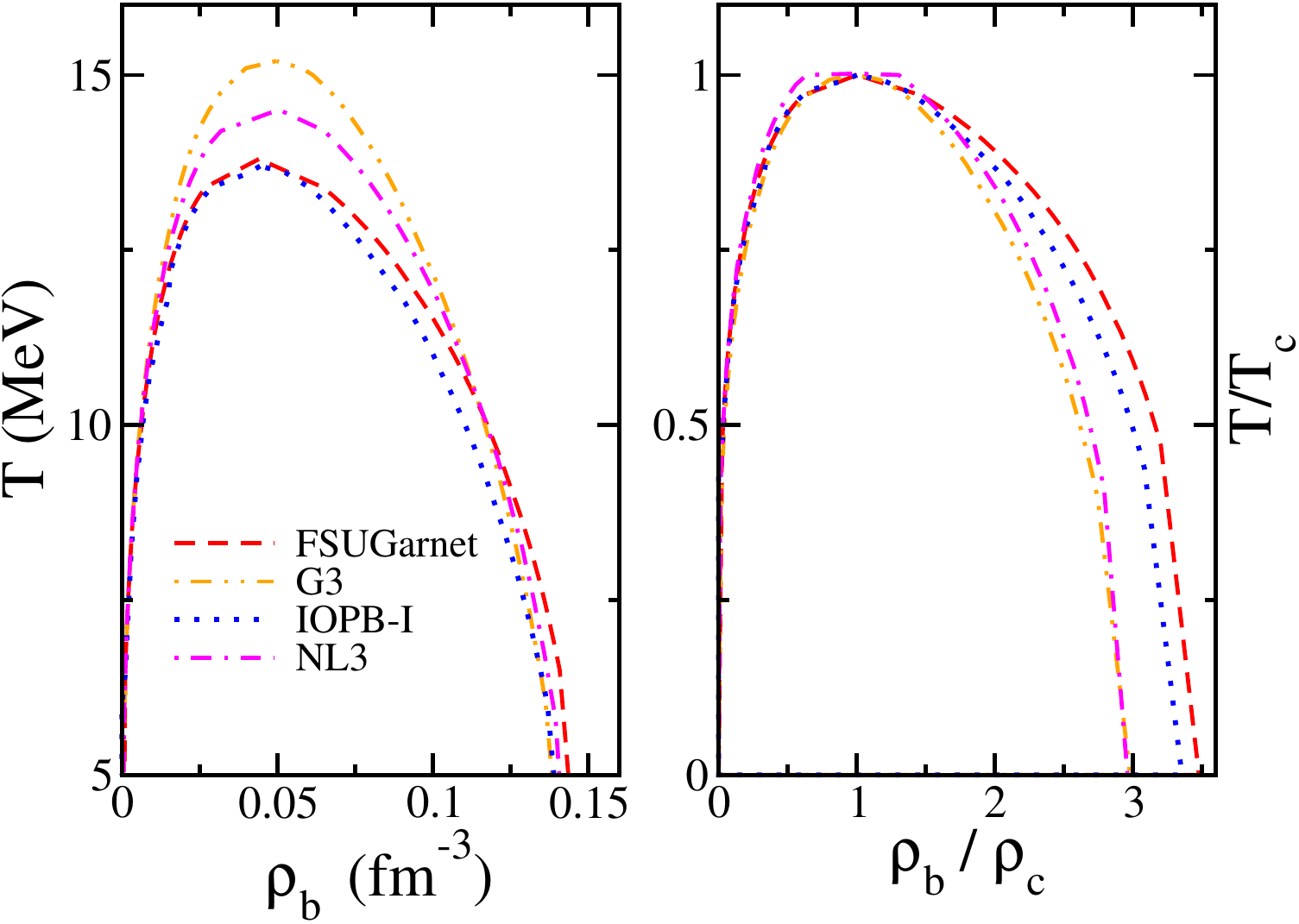}
	\caption{The binodal curve for IOPB-I,  FSUGarnet, G3, and NL3 sets  along with dimensionless density and temperature.}
	\label{binodal1}
\end{figure}

Let us consider the symmetric nuclear matter interacting only through strong interaction as Coulomb interaction is zero hypothetically due to absence of any source current \cite{zerocoloumb}. It, therefore, reduces to a one-component system and Gibbs phase rule then allows only one degree of freedom i.e. only one pressure for each temperature. The phase transition of nuclear matter liquid into gaseous phase start occurring when the stability conditions given by Eqs. \eqref{stability_con1} and \eqref{stability_con2}  are violated. 
This implies that a system should follow the inequalities
%\begin{eqnarray}
\begin{align}
C_V >0 \; or\; \Big[\frac{\partial S}{\partial T}\Big]_\rho >0,\\
\label{ineq}
K_T>0\; or  \Big[\frac{\partial P}{\partial \rho}\Big]_T >0,
\end{align}
%\end{eqnarray}
which corresponds to the dynamical and mechanical stability, to prevent phase change. If anyone of these is violated, a system with more than one phase is energetically favourable. 
In these calculations, the dynamical stability condition is never violated. This is shown in Figure  \ref{specificheat} where the specific heat at constant volume of SNM for the set G3 is plotted against the temperature for various values of densities. All other sets show similar behaviour. It is clear that in the limit $T \rightarrow 0$, the specific heat vanishes as $dS/dT \rightarrow 0$. At higher temperature, the specific heat asymptotically approaches
the noninteracting limit 3/2 for low densities. For a large number of particles, the specific heat has a linear dependence with temperature which extends to large values of temperature.     
In Fig. \ref{pvd}, one can observe that for each isotherm, there are three values of densities for each small positive pressure for $T< T_c $. At very low density, the thermal and Fermi degeneracy pressure makes the total pressure positive. As density increases slightly, the attractive force arising due to attractive scalar $\Phi$ field, increases, which try to reduce the volume of the system and pressure goes from positive to negative. This corresponds to the part where $dP/d\rho < 0$. It violates the mechanical stability inequality in Eq.\ \eqref{ineq} and system transform into a two-phase system. Instability is only a sufficient condition for phase separation to take place, therefore, it may even occur in the one phase-stable system. This is the case of the metastable state \cite{metastablestate}. Once the two-phase system is favoured over the one-phase system, the solution of Eqs. \eqref{chmeicalpotentialeqi} and \eqref{pressureeqi} gives us the coexistence densities i.e. $\rho_g$ for gaseous phase and $\rho_l$ for liquid phase for each isotherm. This density pair for each temperature and pressure then determine the coexistence phase boundary or binodal. Also, the locus of points where the second derivative of free energy is zero for each isotherm marks the boundary of spinodal. The point where the binodal and spinodal curve meet defines the critical parameters given in Table \ref{criticaldata}. This is shown in Fig.\ \ref{binodal} where the binodal or coexistence boundary and spinodal or instability boundary is plotted for IOPB-I,  FSUGarnet, G3, and NL3 sets. The middle red shaded region marks the region of instability and outer blue region signify the metastable region. The outer blue solid line is the coexistence boundary which gives us the ($\rho_g, \rho_l$) couple.
The liquid at saturation density is in equilibrium with the zero density gaseous phase at zero temperature. As one increases the temperature, the liquid coexistence density decreases and gaseous coexistence density increases until the critical temperature is encountered. Beyond critical temperature, the liquid-gas coexistence vanishes and the system only remain in the gaseous phase. At small temperatures i.e. below 5-6 MeV, there is no solution to Eqs. \eqref{chmeicalpotentialeqi} and \eqref{pressureeqi} and system only remain in one stable liquid phase. For more insight and closer comparison among all the force parameters, the binodal are plotted in Fig.\ \ref{binodal1}. Here, critical density and temperature are used for scaling in the right panel. As discussed, the least value of $\zeta_0$ gives us the maximum binodal area for G3. The NL3 shows a similar behaviour. On the other hand, IOPB-I and  FSUGarnet  with comparable $\zeta_0$ show similar behaviour if we look at the right panel of the Fig.\ \ref{binodal1}. The comparatively large instability region in G3 is the consequence of soft scalar-self coupling as also evident in graph (c) of Fig.\ \ref{comparison}.\\

%%%%%%%%%%%%%%%%%%%%%%%%%%%%%%%%%%%%%%%%%%%%%%%%%%%%%%%%%%%%%%%%%%%%%%%%%%%%%%%%%%%%%%%%%%%%%%
\subsection{Latent heat}
\begin{figure}
	\centering
		\includegraphics[scale=0.4]{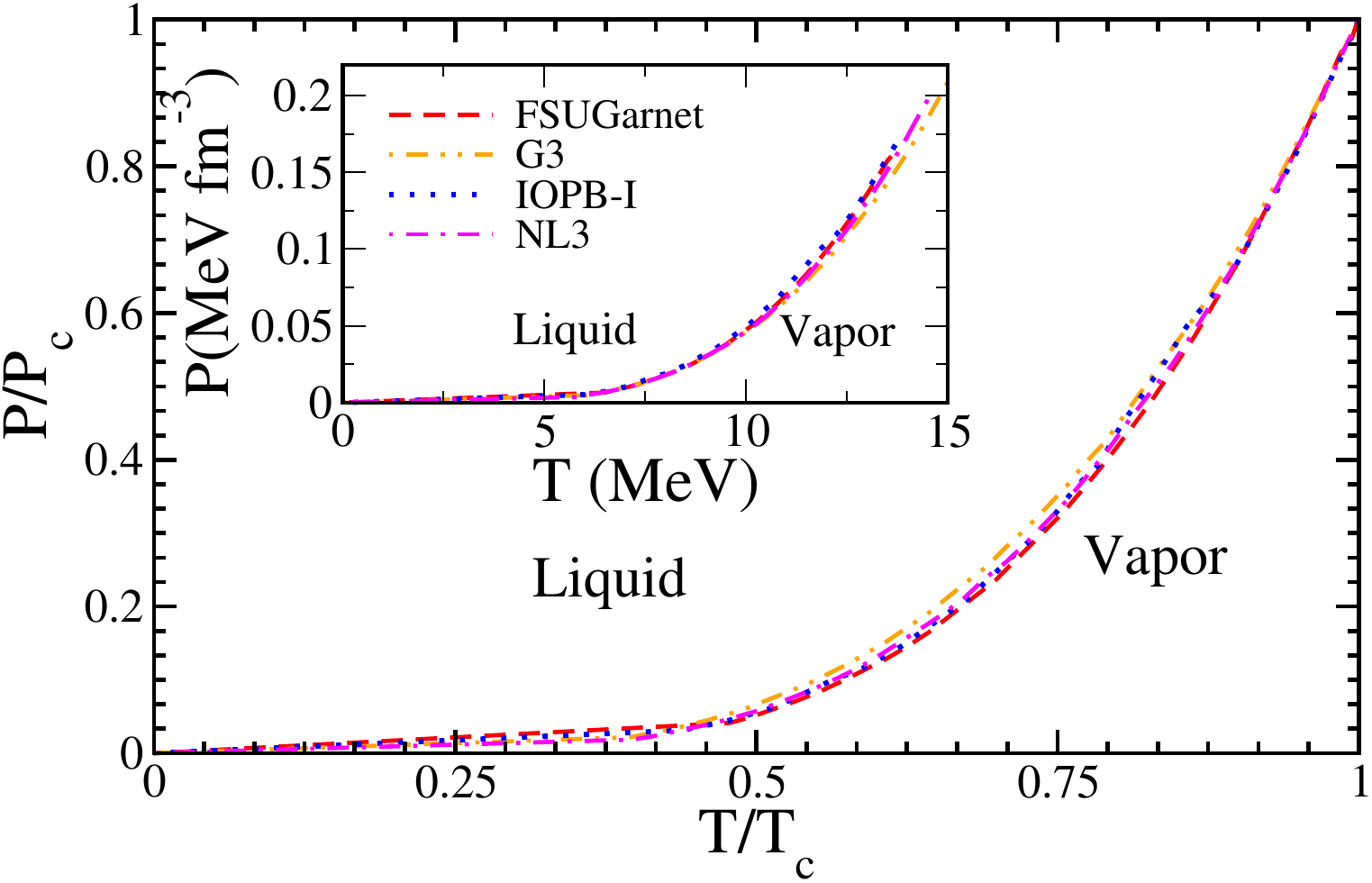}
	\caption{Variation of vapour pressure with temperature for IOPB-I,  FSUGarnet, G3, and NL3 sets.}
	\label{ptcurve}
\end{figure}

\begin{figure}
	\begin{center}
			\includegraphics[scale=0.4]{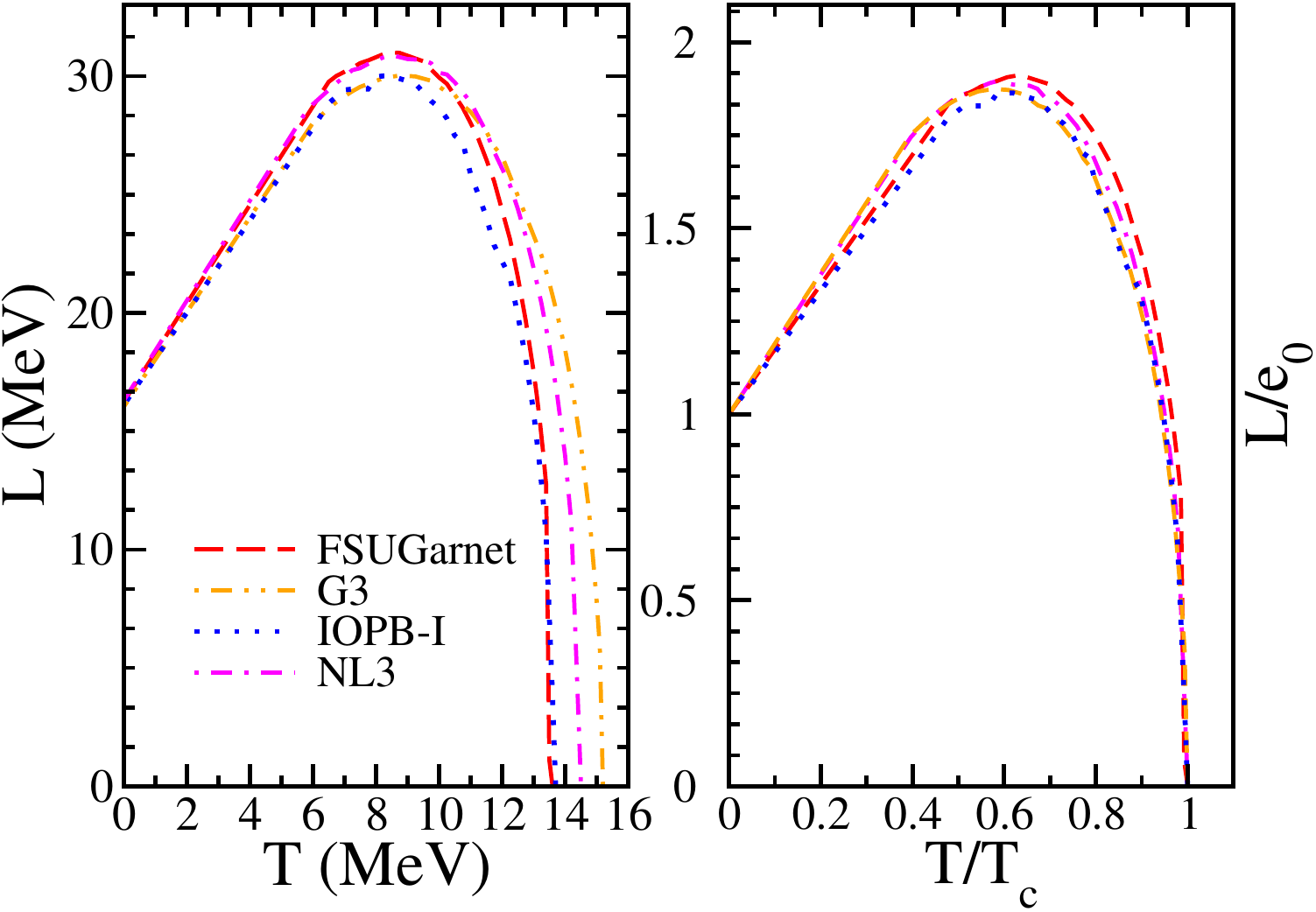}
		\caption{Latent heat of vaporization for symmetric nuclear matter $L=T(s_g-s_l)$ for the IOPB-I,  FSUGarnet, G3, and NL3 sets.}
		\label{lheat}
	\end{center}
\end{figure}

One of the unique characteristics of the first-order phase transition is that the first-order derivatives of Gibbs free energy are discontinuous. It gives rise to the release of heat at a constant temperature which is known as the latent heat of vaporization. The latent heat can be determined using the Clausius Clapeyron equation given in Eq.\ \eqref{classiousclapron}. This equation requires the derivative of pressure with temperature along the binodal or coexistence curve.  This is called vapour pressure. Fig.\ \ref{ptcurve} shows the variation of vapour pressure along with the temperature for the NL3, FSUGarnet, IOPB-I, and G3 sets. The outer graph is in the reduced dimensionless form where the respective critical parameter is used to making them dimensionless. The inner graph uses absolute values of the nuclear matter parameters. There is not much deviation in different force parameter sets. At higher temperatures, the curve has a linear dependence which can be expressed as $\pi = a \tau$, where $\pi$ and $\tau$ are reduced pressure and temperature, respectively. This characteristic helps to determine the behaviour of the curve in the neighbourhood of critical parameters. At lower temperature, pressure rises very slowly and there is a sudden increase after 0.5$T_c$ for all forces. Below this temperature, system only remains in one stable liquid phase and therefore system encounter very less pressure. At higher temperature, the thermal excitation and existence of a two-phase system drive the pressure towards higher values.\\
Fig.\ \ref{lheat} shows the variation of latent heat of vaporization with temperature for different parameter sets using the relation $L=T(s_g-s_l)$ by virtue of Clausius Clapeyron equation. This method is numerically more stable as compared to the calculation of the derivative of the vapour pressure shown in Fig.\ \ref{ptcurve}. The values of latent heat are checked with both the methods and they estimate a similar trend. The figure shows the characteristic behaviour of latent heat with temperature. In the zero-temperature limit, we can write the Clausius Clapeyron equation as \cite{0templheateq}
\begin{equation}
\small
\lim_ {T \to 0 } L(T) = \lim_{T \to 0} T\Big(\frac{1}{\rho_g}-\frac{1}{\rho_l}\Big) \frac{dP_{vapour}}{dT}= -\mu_g = -\mu_l= e_0,
\end{equation}
because the liquid-gas coexistence density approaches zero for vanishing temperature. Therefore, in this limit, latent heat is simply the heat required to remove a nucleon from the saturated liquid with energy density e$_0$. As one increases the temperature, the latent heat rises linearly up to a maximum value which is close to $\approx$ 30 MeV. This linearity at low temperature indicates the absence of interactions for low-density gas. This trend is similar to all the forces considered and there is very less deviation among different parameters in both left panel where absolute values are plotted and in the right panel where reduced parameters are depicted to minimise the mean-field dependence i.e. effect of different vector and scalar self and cross-couplings of the various models. The latent heat reaches the peak value in between 8-10 MeV or (0.5-0.7)T$_c$ in all cases and then fall sharply to zero at the critical temperature. The maximum $L/e_0$ in right panel has a very narrow range of 1.8-1.9. The Latent heat, therefore, is a thermally correlated parameter. Furthermore, the maximum latent heat L$_H$  is the least correlated parameter at finite temperature. A similar trend is seen if we use flash temperature as the scaling parameter. Therefore, flash temperature along with the critical temperature is suitable to constrain the EoS for finite temperature. At low temperature, the latent heat has linear relationship with temperature and  as the system reaches near the critical temperature, all the force parameters give the same slope in the right panel of the graph which is a significant validation of the well defined thermodynamical theories \cite{0templheateq}.

\subsection{\label{scaling}Scaling laws}
The critical points related to the phase transition are of special interest as they are used to subsequent characterisation.     
The behaviour of a given system in the neighbourhood of these points forms a fundamental problem in view of the phase transition phenomenon. Near the critical point, various physical quantities encounter the singularity. It becomes essential to express these singularities in terms of power laws.
These power laws are characterised by critical exponents and in turn, determine the qualitative nature of a given system.  The exponent should also follow well known scaling laws which make only two exponents independent at a given time. The nature of exponents is such that they do not differ much from one system to other in one universality due to their dependence on very less number of parameters. The E-RMF calculations are needed to numerically satisfy these laws to check their independence from the underlying interactions. Therefore, these are used here to   qualitatively check the consistency of our analysis.
The exponents $\beta$, $\Lambda$, $\delta$, $\alpha$ and $\tau$ are defined as \cite{jbeliot}
\begin{eqnarray}
\rho_l-\rho_g &\sim & \Big(\frac{T_c-T}{T_c}\Big)^\beta, \\
K_T &\sim &\Big| \frac{T-T_c}{T_c}\Big|^{-\Lambda},\\
|P-P_c|_{T_c}& \sim &\Big|\frac{\rho-\rho_c}{\rho_c}\Big|^\delta ,\\
C_V & \sim &(T_c-T)^{-\alpha},
\end{eqnarray}
and finally, from Fisher's analysis \cite{fisher},
\begin{eqnarray}
C_f= \frac{\zeta(\tau)}{\zeta(\tau-1)}.
\end{eqnarray}
A simple power law fitting was done in the form $y=aX^b$ and corresponding values of these exponents along with the mean field reaults and other liquid-gas systems are shown in Table \ref{criticalexponent}. All the critical exponents are close to their respective values derived from mean field results.
\begin{table}
	\centering
	\caption{Calculated critical exponent for  IOPB-I,  FSUGarnet, G3, and NL3 parameter sets. }
	\begin{tabular*}{\linewidth}{c @{\extracolsep{\fill}} ccccccc}
		\hline
		\hline
		Critical   & NL3 & IOPB-I & FSUGarnet & G3 & Mean  & Liquid-Gas   \\ Exponent&&&&&Field&System \cite{gasliquidsystem}\\
		\hline
		$\beta$ &0.37 &0.40&0.39  &0.40  &0.5  &0.32-0.35  \\ 
		$\Lambda$ &0.97 &1.12  &1.02&1.01  & 1 &1.2-1.3  \\ 
		$\delta$ & 3.65 &3.62 &3.50  &3.38 & 3 &4.6-5.0  \\ 
		$\alpha$ & 0.00 &0.00 &0.00 &0.00& 0 &0.0-0.2  \\ 
		$\tau$ &2.209 & 2.210 &2.167 &2.224&- &2.1-2.25  \\ 
		\hline  
		\hline  
	\end{tabular*}
	
	\label{criticalexponent}
\end{table}
The consistency check of the critical exponents can be done using standard scaling laws or thermodynamic inequalities \cite{scalinglaw}. These are given by Rushbrooke, Griffiths and Widom inequalities and can be respectively written as \cite{jbeliot} 
\begin{eqnarray}
\alpha + 2 \beta + \gamma \ge 2,\\
\alpha +  \beta(1+ \delta) \ge 2,\\
\beta(\delta-1) - \gamma \le 0,
\end{eqnarray}
and from Fisher's droplet model as ,
\begin{eqnarray}
\frac{\beta}{\gamma}- \frac{\tau-2}{3-\tau} = 0.
\end{eqnarray}

Table \ref{scalinglaw} compiles all the calculated results of these scaling laws
for each parameter set. The critical exponents derived from each parameter satisfy these scaling laws up to a good approximation. We can, therefore, strongly conclude that E-RMF parameter sets with incompressibility comparable to accepted experimental value are very consistent in the finite temperature limit satisfying the statistical inequalities. They can consequently be used in the calculation of hot EoS, which is of primary importance in processes like binary neutron star merger, where temperature plays a very important role.   
\begin{table}[ht]
	\centering
	\caption{Result of scaling laws for IOPB-I,  FSUGarnet, G3, and NL3 parameter sets.}
	\begin{tabular*}{\linewidth}{c @{\extracolsep{\fill}}  ccccc}
		\hline
		\hline
		Force Parameter & Rushbrooke  & Griffiths & Widom & Fisher \\
		\hline    
		
		NL3&1.72&1.74  &0.02& 0.13 \\
		IOPB-I&1.92  & 1.84 &-0.07&0.09  \\
		FSUGarnet& 1.79  & 1.75 &-0.04&0.10   \\
		G3&1.80  &1.75  & -0.05&0.10 \\
		\hline     
		\hline     
	\end{tabular*}
	
	\label{scalinglaw}
\end{table}

%%%%%%%%%%%%%%%%%%%%%%%%%%%%%%%%%%%%%%%%%%%%%%%%%%%%%%%%%%%%%%%%%%%%%%%%%%%%%%%%%%%%%%%%%%%%%%%%%%%%%%%%%%%%%%%%%%%%%%%
\section{Asymmetric nuclear matter}
%\subsection{ Introduction} 
%Core-collapse supernova is nature's one of the brightest optical display where million-year life of a giant star ($M > 8M\odot$) is put to an end violently and abruptly within fractions of a second \cite{couch2017mechanism, ccsn}. The exact mechanism of collapse explosion is still not well understood even after several decades of thorough investigations. In recent years, such explosion has been studied using several ab-initio core-collapse simulations where the hydrodynamics equations are solved numerically \cite{Muller2010, ccsnsimulation}.  These simulations estimate that the explosion energy of $\approx$ 10 $^{51}$ erg is attained within the time scale of $\ge$ 1s  \cite{Sawada_2019}.  The temperature of the matter rises to 20 MeV and the density of the bounce can vary up to two times the nuclear saturation density.  The short time scale of collapse does not allow the matter to reach $\beta$ equilibrium and calculations are usually done at a fixed asymmetry $\alpha =\frac{\rho_n-\rho_p}{\rho_n+\rho_p} \approx$ 0.4 \cite{Alam2017,Nishizaki1994}.  

Investigation of  the EoS for isospin-asymmetric nuclear matter (ANM) is relevant in various areas of nuclear physics ranging from finite nuclei to infinite matter.  Not only the understanding of its ground state is important, but  its behaviour at finite temperature is equally significant. The finite temperature behaviour of ANM is relevant in context to astrophysical events such as gamma-ray bursts, neutron star mergers, proto-neutron stars and early universe \cite{Liu_2010}.  Furthermore, the composition of matter inside the neutron star is highly asymmetric and  impact its transport and cooling process which are governed by the  direct URCA process \cite{brown2018}. In view of above, a systematic understanding of ANM at finite temperature is highly desirable.

The central motivation of this study is to perform a detailed analysis of the EoS for dilute and hot homogeneous asymmetric nuclear matter within E-RMF formalism. The aim is to understand the nuclear matter properties like symmetry energy $F_{sym}$, slope parameter ($L_{sym}$), skewness parameter ($Q_{sym}$) and curvature parameter ($K_{sym}$) as a function of temperature.  These are significant properties of ANM and  are often used to constrain the EoS around saturation density.  Several finite temperature effects such as thermal effects on various state variable, isothermal and isentropic incompressibility are addressed. The results presented in this section focus on differentiating between the realistic inhomogeneous phase in a supernova and the ideal homogeneous phase, specifically below subnuclear density. The discussion of the phase transition properties in asymmetric nuclear matter (ANM) is then compared to symmetric matter, considering their dependence on the incompressibility $K$ and slope parameter $(L_{sym})$. With this study, the aim is to verify the trends available in various studies \cite{Alam2017, SHARMA2020121974, Avancini2006, Avancini2004} where effect of symmetry energy and its derivative is discussed on the instability of ANM. Establishing these trends is of primary importance as they serve as the bridge between various nuclear matter properties which are not measured directly from the experiments. In symmetric nuclear matter the trends are seen among the properties at critical temperature  \cite{vishal2020}. The properties at ground state do not necessarily dictate the critical properties of phase transition. However, for ANM, the symmetry energy and its slope parameter decide the energetic  and therefore impact the  instabilities occurring in the system. 

%\subsection{ Results and Discussions }

%%%%%%%%%%%%%%%%%%%%%%%%%%%%%%%%%%%%%%%%%%%%%%%%%%%%%%%%%%%%%%%%%%%%%%%%%%%%%%%%%%%%%%%%%%%%%%%%%%%%%%%%%%%%%%%%%%%
\subsection{\label{model}  Model properties}
\begin{figure}
	\centering
		\includegraphics[scale=.4]{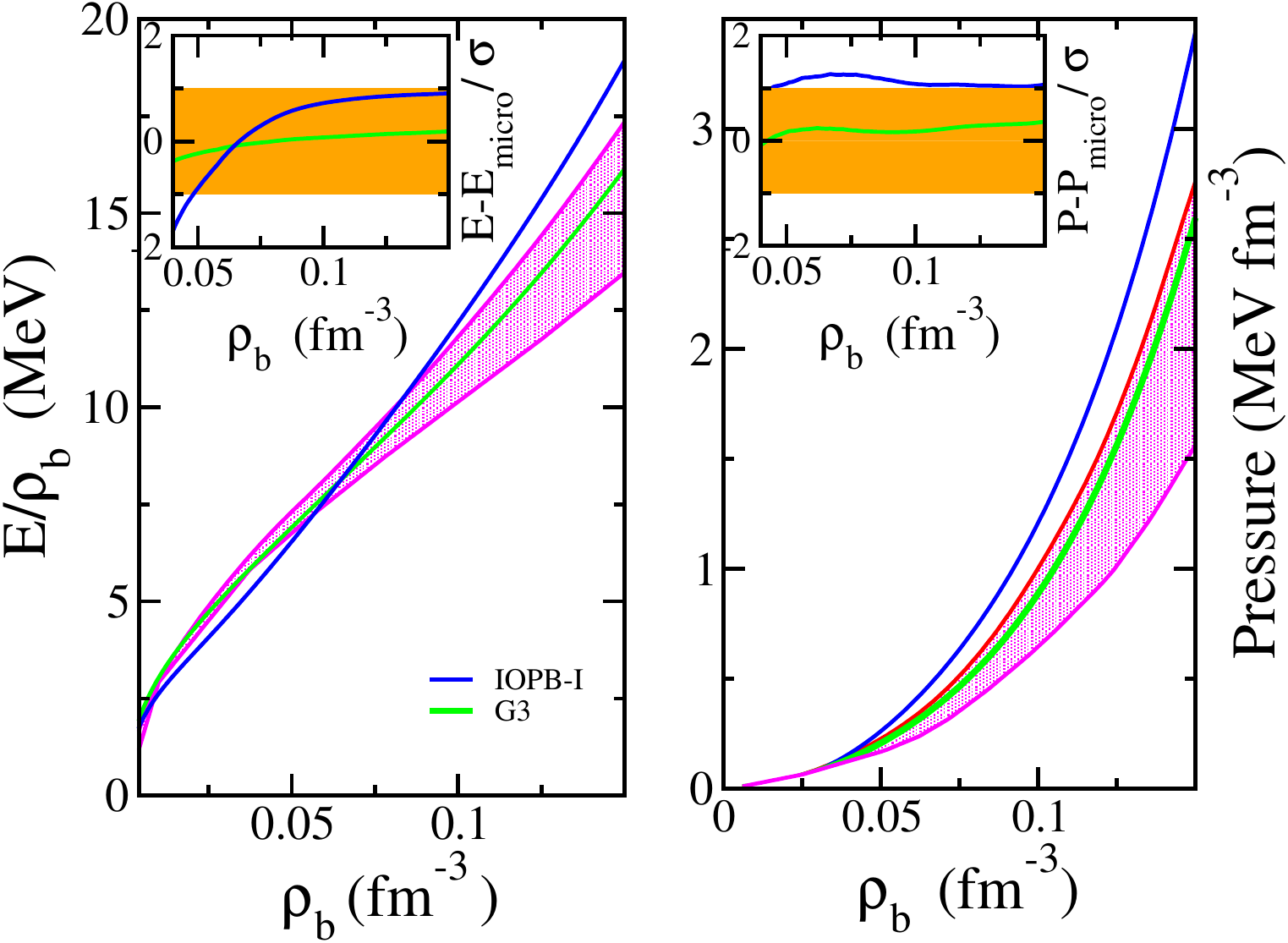}
	\caption{ EoS of nuclear matter below saturation density for pure neutron matter at T=0 MeV. The shaded magenta region corresponds to the microscopic chiral EFT (NN + 3N) \cite{Hebeler2013} calculations. Inner graph  represents the difference between the neutron matter energy and pressure and the average energy and pressure with 1$\sigma$ calculation uncertainty area.}
	\label{eospnm}
\end{figure}
To investigate the ANM in the finite temperature limit, two  E-RMF parameteres, namely, IOPB-I \cite{iopb} and G3 \cite{g3} are considered which have been discussed earlier. Fig. \ref{eospnm} compares the neutron matter binding energy and pressure with the microscopic calculations based on chiral effective field theory (EFT) with realistic two and three-nucleon interactions \cite{Hebeler2013}.  Inner graph  represents the difference between the neutron matter energy and pressure and the average energy and pressure normalised to the  uncertainty of the microscopic calculations($\sigma$=$\delta P$) . The uncertainties are represented by orange band, and they indicate that the points that lie inside this band are within the $1\sigma$ error limits. The G3 set satisfy nicely the microscopic constraints whereas IOPB-I also fall within the $1\sigma$ error limit  below saturation density. Both parameter also satisfy the constraint form collective flow data in heavy-ion collisions and  Kaon experiment along with the GW170817 gravitational wave constraints \cite{iopb}. These features along with the agreement of bulk matter properties with empirical data  are the motivation to  study the E-RMF sets with and without the $\delta$ meson. $\delta$ meson couplings are a necessary feature in the dense asymmetric nuclear matter. This work also intends to investigate the effects of  $\delta$ meson  in the dilute ANM in the finite temperature limit.

%\begin{table}
%    \centering
%        \caption{Bulk matter properties of nuclear matter for the IOPB-I and G3 parameter and their corresponding empirical values.}
%    \begin{tabular*}{\linewidth}{c @{\extracolsep{\fill}}  ccccc}
%    \hline
%    \hline
%    & IOPB-I & G3 & Empirical Value &  \\ \hline
%$\rho_0 (fm^{-3})$ & 0.149       &0.148    &    0.148/0.185  \cite{bethe}     &  \\
%$E_0$ (MeV)  & -16.10  &-16.02    &       -15.0/-17.0   \cite{bethe}       &  \\
%M*/M         &0.593    &    0.699&       0.55/0.6 \cite{marketin2007}    &\\
%$J$ (MeV)  &33.30        & 31.84   &        30.0/33.70    \cite{DANIELEWICZ20141}     &  \\
%$L$ (MeV)  &63.58        & 49.31   &      35.0/70.0      \cite{DANIELEWICZ20141}     &  \\
%$K_{sym}$ (MeV)  & -37.09       &-106.07    &   -174.0/-31.0  \cite{zimmerman2020measuring}            &  \\
%$Q _{sym}$ (MeV)  &     862.70   &  915.47  &     -494/-10 \cite{cai2017constraints}            &  \\
%$K$ (MeV)  & 222.65        &243.96    &   220/260     \cite{GARG201855}         &  \\
%    \hline     
%    \hline     
%    \end{tabular*}
%\label{bulkproperties}
%\end{table}

\subsection{\label{ftp} Finite temperature properties}
\begin{figure}[h]
	\centering
		\includegraphics[scale=0.5]{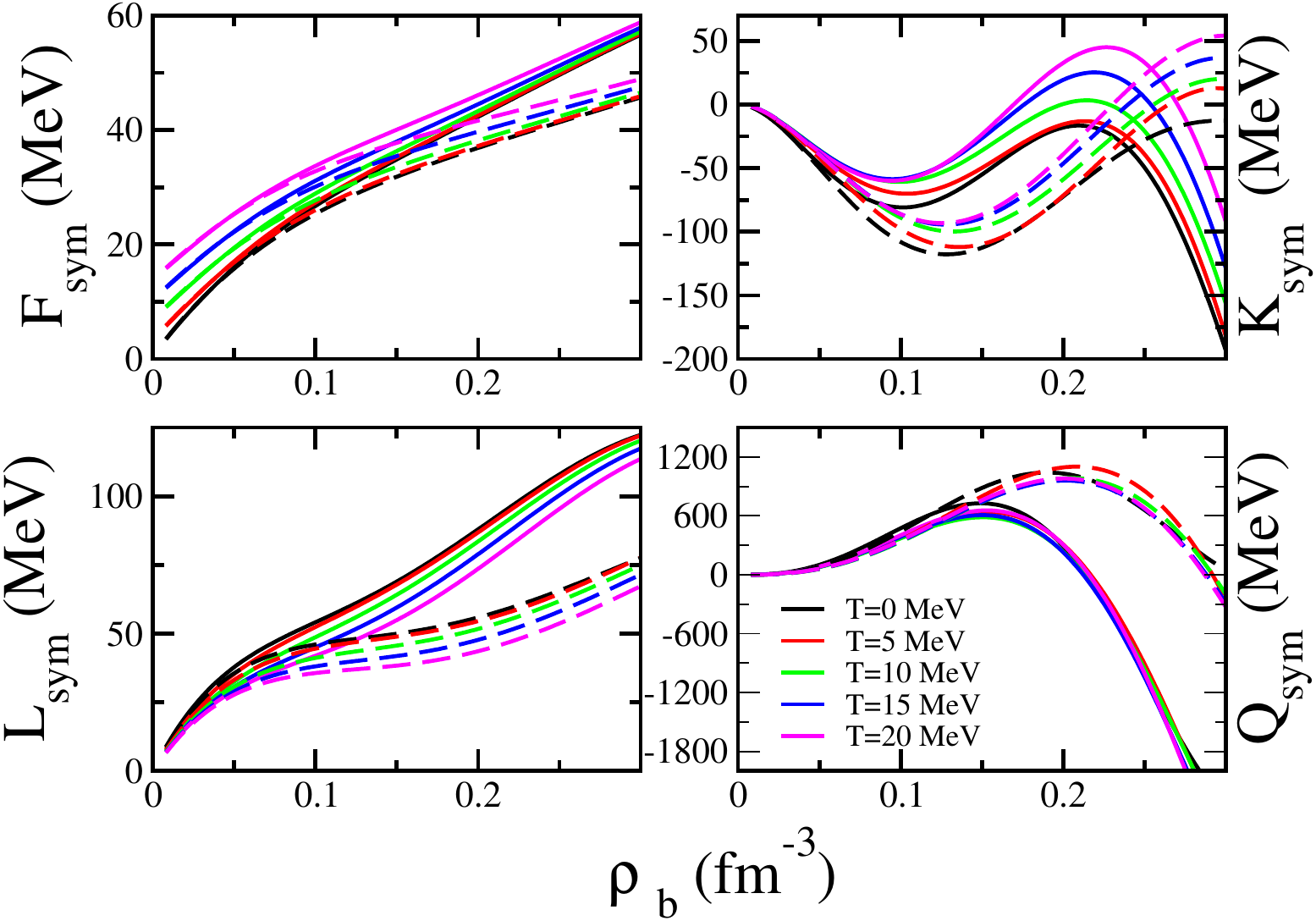}
	\caption{Variation of free symmetry energy $F_{sym.}$, slope parameter $L$, curvature $K_{sym}$, and isovector skewness parameter $Q_{sym}$  with density at various temperature for IOPB-I (solid lines) and G3 (dashed lines) sets. }
	\label{symenergy}
\end{figure}

\begin{figure*}
  
  \centering
\subfloat[]{%
  \includegraphics[height=8cm,width=.49\linewidth]{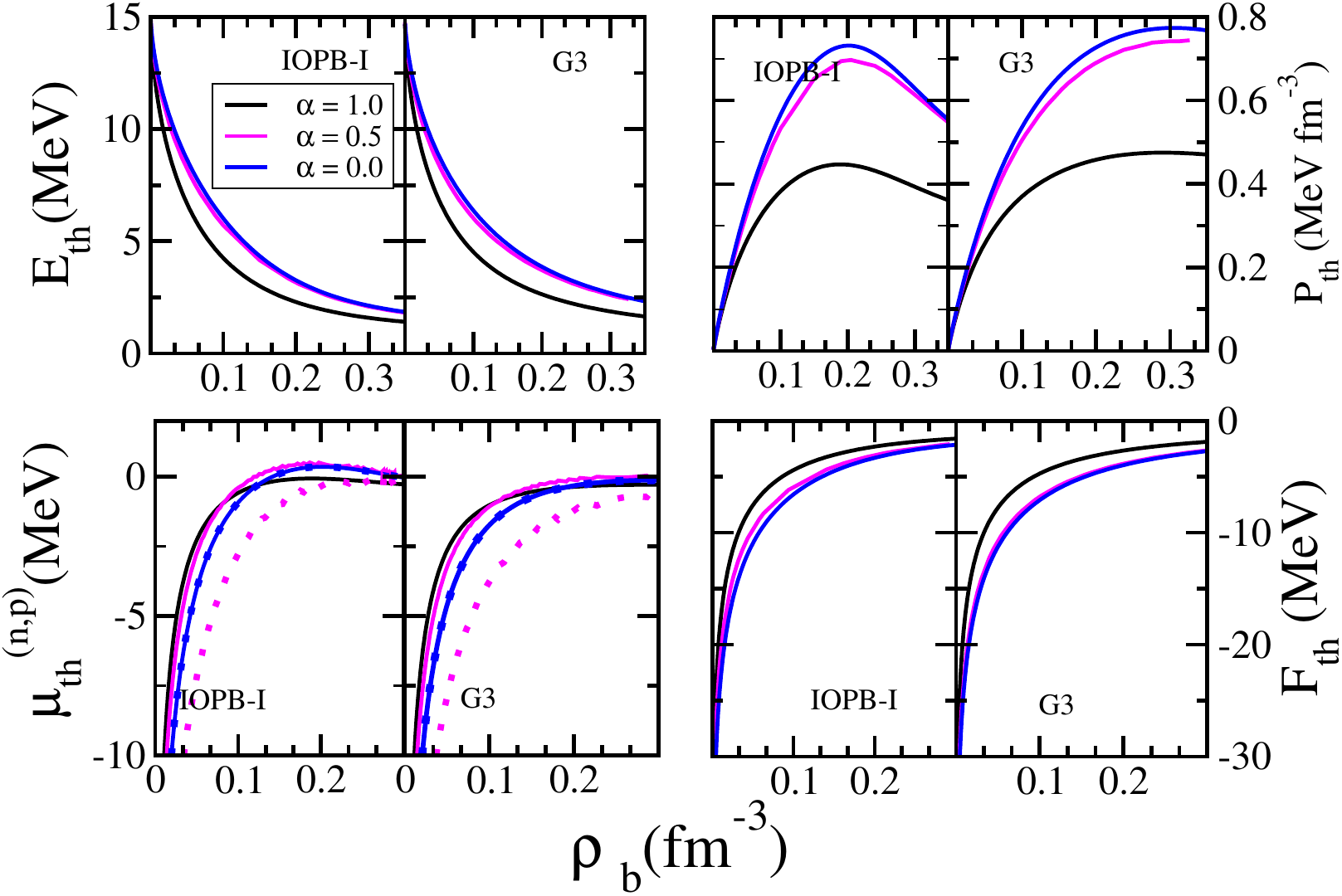}%
}\hfill
\subfloat[]{%
  \includegraphics[height=8cm,width=.49\linewidth]{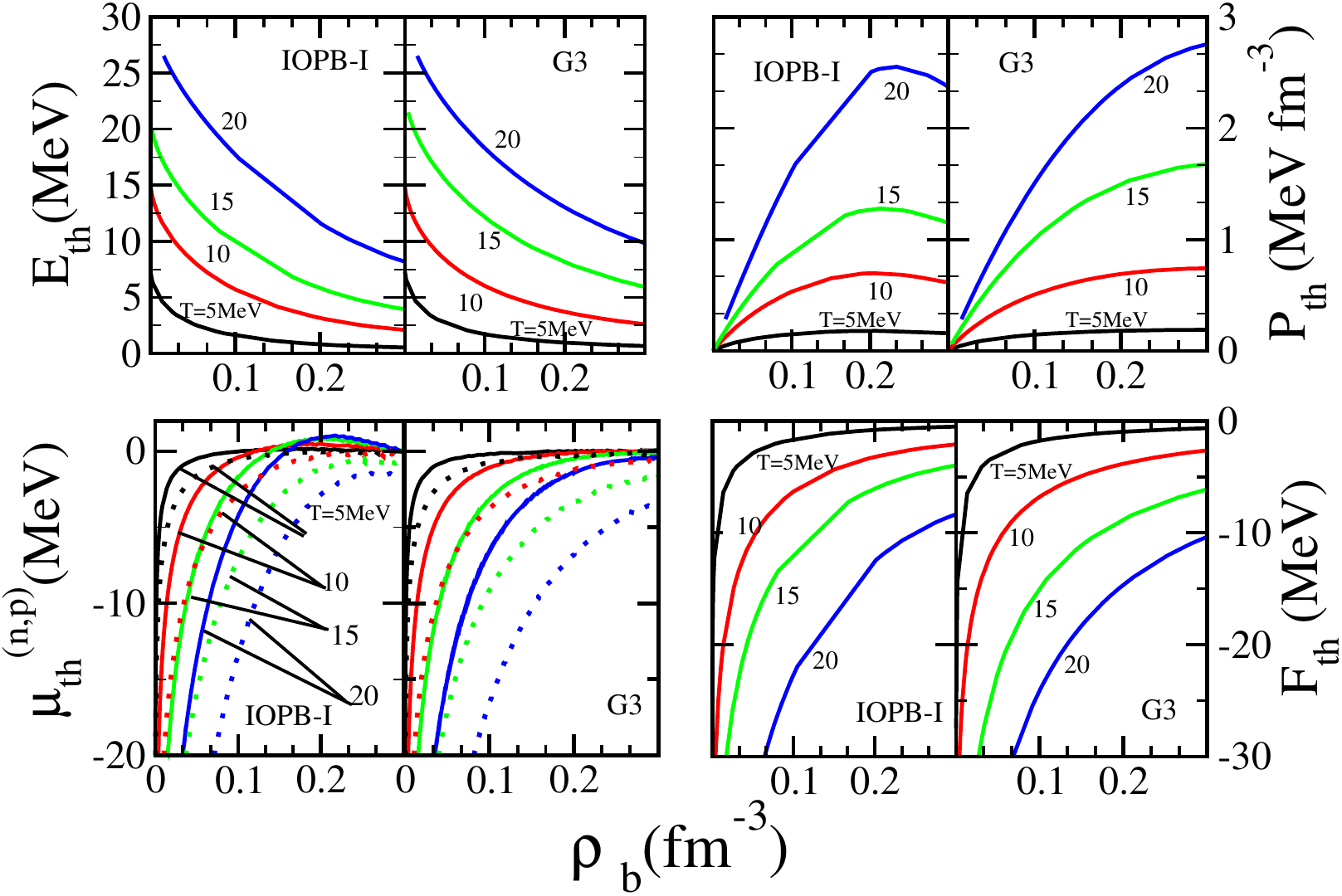}%
}
\caption{(a) Thermal energy, pressure, chemical potential  and free energy density  for fixed T=10 MeV for various $\alpha= 0, 0.5, 1$ (b) same as in (a) but for various T  = 5, 10, 15, 20 MeV at fixed $\alpha=0.5$. The solid lines are for neutron chemical potential and the same colour dotted line represent the proton chemical potential.}
\label{thermaleffect}  
\end{figure*}

The nuclear symmetry energy (NSE) is one of a crucial properties of asymmetric nuclear matter governing several areas of nuclear matter calculations like reaction dynamics, phase stability and cooling in the neutron star, etc. Temperature dependence of NSE is one of the inputs in the calculations of dynamical evolution of neutron star and isoscaling analyses of heavy-ion induced  reactions. NSE is not a directly measurable quantity in experiments and is extracted from the observables related to it. Despite numerous theoretical and experimental efforts, it is still not a very precise parameter even for cold nuclear matter. In Fig. \ref{symenergy},  the variation of Free nuclear symmetry energy (FNSE)  (more relevant quantity in finite temperature case, see Eq.\ \eqref{freesymenergy} ), the slope parameter $L_{sym}$, the isovector incompressibility $K_{sym}$, and the isovector skewness $Q_{sym}$ with the density up to 2 times the saturation density is shown.  The Free NSE is scaled towards higher magnitude due to decrease in entropy preserving its characteristic shape at a higher temperature for both IOPB-I and G3 sets.  At higher density, the temperature range considered here does not affect FSNE much. The slope parameter which has a direct correlation with neutron skin thickness, electric dipole polarizabilities, etc. also follows the similar trend for both the sets. NSE estimated from these sets are also consistent with the HIC Sn + Sn and IAS data \cite{tsang2009}. The low values of  $F_{sym}$ and $L_{sym}$ are the result of cross-couplings of $\rho$ meson with $\omega$ meson in IOPB-I set and coupling of $\sigma$ meson with $\omega$ meson in G3 set which predicts even lower $F_{sym}$ due to the presence of isovector scalar $\delta$ meson. The $\delta $ meson has the positive effect on binding energy and  helps to estimate the $F_{sym}$, $L_{sym}$ and $ K_{sym}$ within the permissible limit  \cite{singh2014}. The sinusoidal  variation of $K_{sym}$  with density is also shown.   $K_{sym}$  is constrained recently by combining the data from PSR J0030+0451 and GW170817 estimating the $K_{sym}$ = $102^{+71}_{-72}$ MeV within $1\sigma$ error \cite{zimmerman2020measuring} . The IOPB-I and G3 both fall within this constraint. The Variation of $Q_{sym}$ with density is almost independent of temperature for the IOPB-I set and a small variation is observed for G3 set. The $Q_{sym}$ is the least constrained property in any experiment and several models predict it with a large variation \cite{kumar2020warm}.

To study the finite temperature effect, one can isolate the thermal part of a given function according to equations \eqref{thermalpart}. The subtraction scheme applies to only those variables which depend on the kinetic energy density \cite{Constantinou2015}. Fig. \ref{thermaleffect} shows the thermal effect on various state variables at a fixed temperature and $\alpha$ for IOPB-I and G3 parameter set. The common observation is that i) at the fixed temperature, the thermal energy decreases with density. The difference due to asymmetry disappears at high densities and thermal effects become weak and thermal energy gets density independent with asymptotically tending to zero, ii) At very low density, the thermal energy and pressure have linear T dependence as for a free Boltzmann gas (non-degenerate limit). This linearity is changed when matter becomes increasingly degenerate, iii) Temperature effects are more prominent in thermal pressure as compared to thermal energy and iv) The thermal chemical potential becomes saturated after saturation density.  In references \cite{Constantinou2014, Constantinou2015},  the thermal effects are found to be dominated by the behaviour of effective mass. These calculations were done for the Skyrme and APR forces where the effective mass has different origin when compared to the Dirac mass of relativistic forces \cite{nonreleffmass}. The Dirac mass in the relativistic formalism arises from the self-energy of the nucleon in the Dirac equation whereas the effective mass in the non-relativistic formalism arises from the momentum dependence of single particle potential \cite{chen2007}. However, both of these masses impact the thermal contribution on the state variables in somewhat  similar way.

\begin{figure}[h]
	\centering
\includegraphics[scale=0.5]{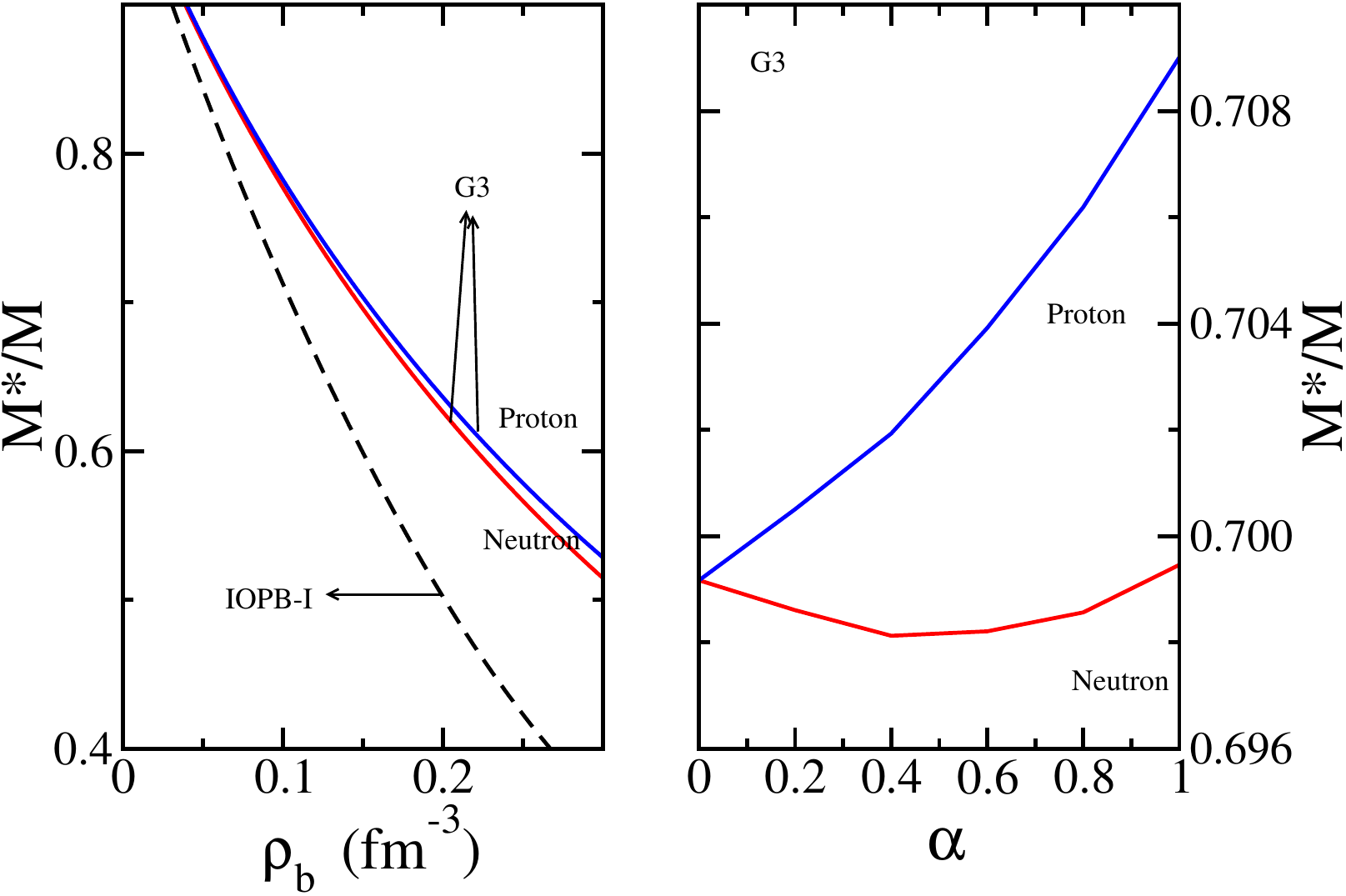}
	\caption{ The effective mass for IOPB-I and G3 set. The left panel shows the mass splitting for PNM in  G3 set (solid lines) as compared to the IOPB-I set (dashed line). In the right panel, the effective mass at saturation density is shown for the G3 set.}
	\label{effectivemass}
	
\end{figure}

In Fig. \ref{effectivemass}, the density dependence of Dirac effective mass for PNM is shown in the left panel. The effective mass for G3 decreases at a relatively slower pace   as compared to IOPB-I set. Due to the presence of $\delta$ meson, neutron and proton mass gets split which is not the case for IOPB-I set due to the absence of $\delta$ meson. In the right panel, the effective mass at saturation density is plotted for different values of $\alpha$ for G3 parameter set. This $\delta$ meson mechanism on effective mass is an in important phenomenon in studying the drip line nuclei which are of astrophysical interests \cite{Horowitz2001} and is analysed in experiments such as PREX \cite{prex}. The effective mass is the input for the computation of energy, pressure and chemical potential which is determined self consistently.  The behaviour of effective mass  therefore clearly dictates the thermal pressure and thermal energy. The G3 set with larger effective mass estimates greater thermal contribution on state variables as compared to the IOPB-I set with smaller effective mass. This is consistent with the Fermi-liquid theory  and non-relativistic calculations \cite{Constantinou2014}.  For IOPB-I set, thermal pressure increases and then decreases beyond the saturation density but for G3, it gets saturated at higher density.  Furthermore, the quantitative difference in thermal energy and pressure between IOPB-I and G3 set is due to the difference in the self-coupling of isoscalar-scalar $\sigma$ meson which is responsible for 3N interaction that plays an important role in determining the thermal pressure and energy. These behaviour are analogous to chiral 2N and 3N interaction, although with larger thermal contribution as compared to the many-body self-consistent Green’s function method \cite{Carbone2019}. The decrease in thermal pressure  after reaching to its maximum is the combining effect of incompressibility of EoS at zero temperature and how rapid is the finite temperature pressure.
    
Understanding the thermal effects on relevant state variables holds significant importance in large-scale simulations, such as those involving supernovae, neutron star crusts, and neutron star binary systems. In these simulations, EoS at any given temperature is estimated by incorporating the thermal contribution into an arbitrary cold EoS to account for heating \cite{hshen, Janka_1993}. However, this methodology, based on the ideal gas law, deviates when applied at higher densities and temperatures, especially if non-nucleonic components are considered \cite{raduta2022equations}. The thermal component in these calculations is often parameterized using Eq. \eqref{thermalpart} with a constant thermal index ($\Gamma$), neglecting the impact of degeneracy on the thermal pressure. At high densities and finite temperatures, a portion of the available energy is used to lift degeneracy rather than solely contributing to additional thermal support  \cite{Raithel_2019}. Consequently, this leads to a net reduction in the thermal pressure at high densities, as depicted in Fig. \ref{thermaleffect}, compared to the prediction for an ideal fluid. However, by properly incorporating the ideal fluid component (along with a contribution from relativistic particles with density-dependent thermal effects), it is possible to obtain the EoS at finite temperatures with reasonable accuracy \cite{Raithel_2019}. This approach reduces the computational cost of these simulations, as it eliminates the need for a full self-consistent calculation at every temperature. nevertheless, since the thermal effects depend on the nuclear model, one should perform self-consistent calculations for a more robust and consistent outcome from large scale simulations.

The chemical potential at fixed T=10 MeV has interesting behaviour. $\mu_n$ at $\alpha=1$ is crossed over by  $\mu_n$ at $\alpha<1$ with increasing  density while that is not the case for $\mu_p$. Moreover, the crossings of $\mu_n$ occur at a higher density at larger temperature. Chemical potential becomes saturated at higher density because of the increasing degeneracy at higher density. The different nature of chemical potential is again the consequence of effective mass along with the self and cross-coupling of $\sigma$ meson. The thermal free energy tends to zero with increasing density like the chemical potential.
Comparing IOPB-I and G3 sets for thermal properties, it is seen that the parameter set G3 has additional $\delta$ meson coupling whose contribution increases with density. This contribution directly impacts the effective mass  which in turn decides the behaviour of various variables studied above. The $\delta$ meson along with the $\sigma$ meson therefore has the direct contribution in thermal properties of EoS.  

\begin{figure}
	\centering
		\includegraphics[scale=0.5]{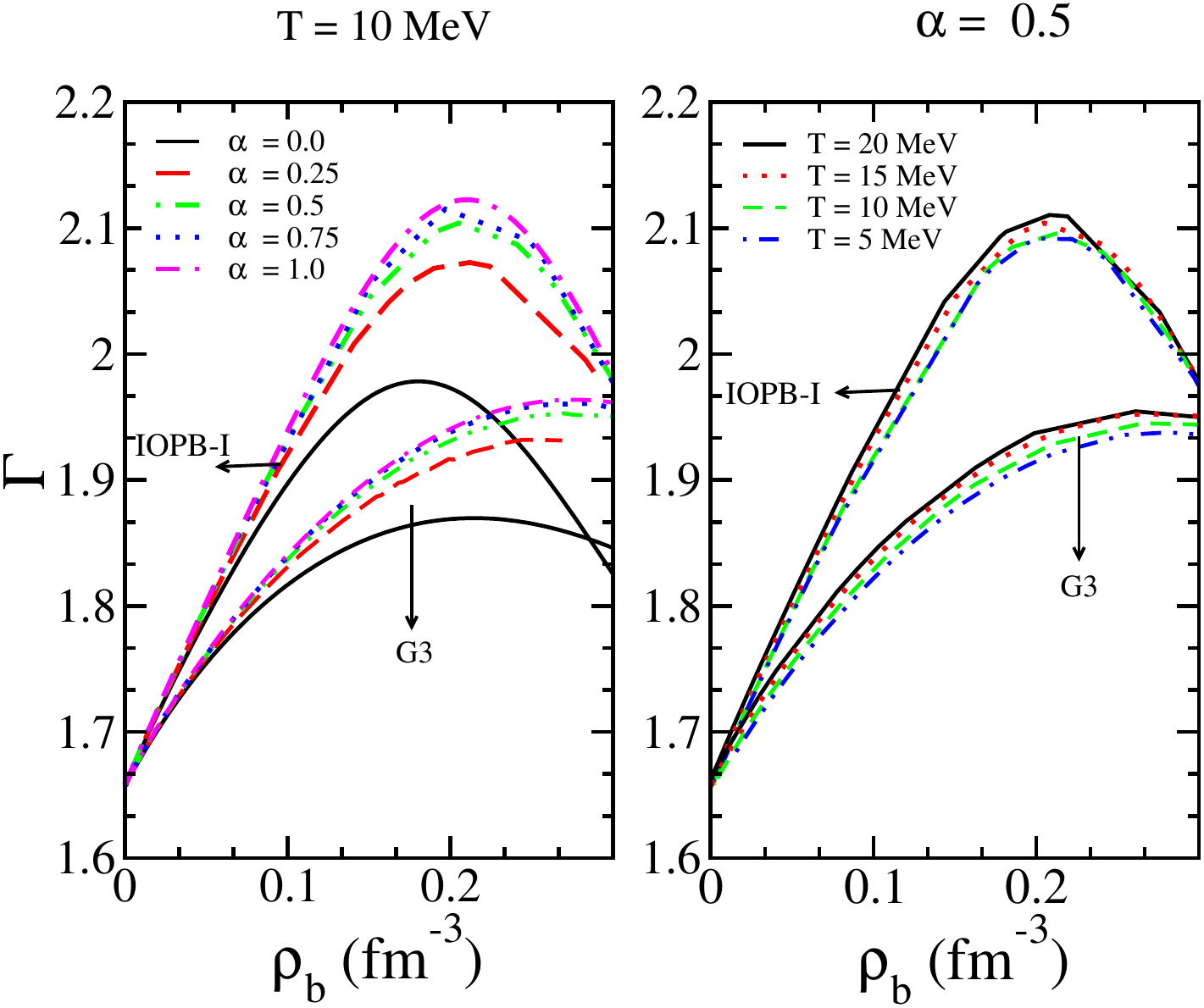}
	\caption{Thermal index for IOPB-I (solid lines) and G3 set (dash line) for fixed temperature in the left panel and fixed $\alpha$ =0.5 in the right panel. }
	\label{thermalindex}
\end{figure}

Fig. \ref{thermalindex} shows the variation of the thermal index ($\Gamma$) with density for  IOPB-I and G3 set at fixed temperature and asymmetry. By comparing  the case of fixed temperature and $\alpha$ with those of thermal energy and pressure (Fig. \ref{thermaleffect}) it is certain that $\Gamma$ depends mainly on i) the stiffness of pressure ii) behaviour of effective mass with respect to density and iii) $\alpha$.  For $\rho \rightarrow 0$, $\Gamma$ approaches the  non-relativistic ideal gas index $\frac{5}{3}$. $\Gamma$ is very sensitive to the asymmetry at a fixed temperature which is opposite to the non-relativistic calculations where the peak of $\Gamma$ is insensitive to asymmetry \cite{Constantinou2015}. Furthermore, it is  immune to temperature change for fixed $\alpha$. For IOPB-I set, the maximum $\Gamma$ is 2.1 for PNM while 1.97 for SNM. The G3 set reports these values to be 1.96 and 1.87, respectively. The G3 set with larger effective mass estimates the lower pressure and therefore larger thermal index as compared to the IOPB-I set with lower effective mass. These results of $\Gamma$ from the newly developed E-RMF sets around saturation density is in the same range are  in the same range as for other EoS used in CCSN simulations \cite{yasin2020}.   However, it is to be noted that, in the astrophysical simulation like binary star and proto-neutron star, $\Gamma$ is taken as a constant while here it varies with the density.  The behaviour of $\Gamma$ is in agreement  with EFT theory \cite{Carbone2019}. 

\begin{figure}
  \centering
\subfloat[]{%
  \includegraphics[height=8cm,width=.49\linewidth]{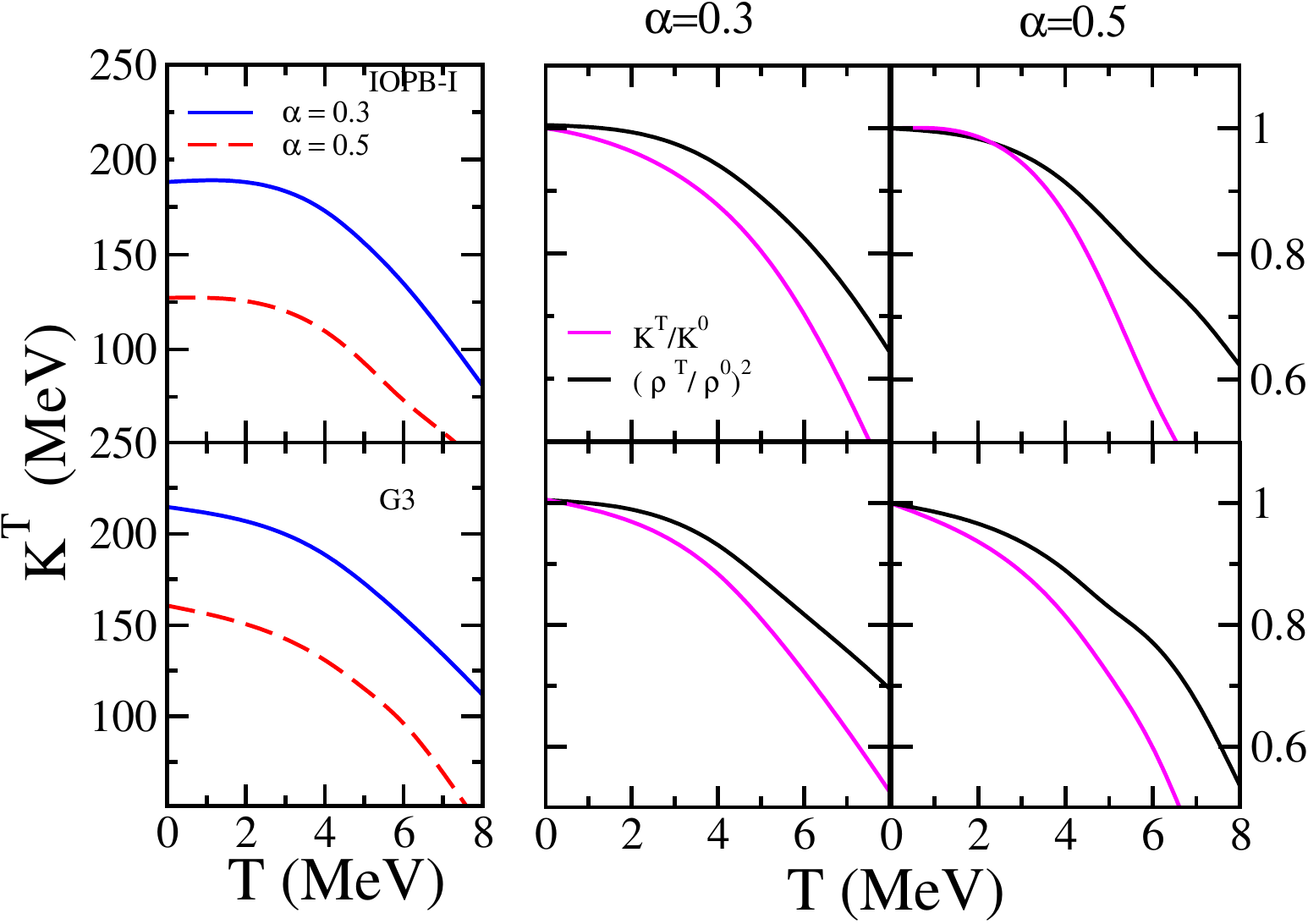}%
}\hfill
\subfloat[]{%
  \includegraphics[height=8cm,width=.49\linewidth]{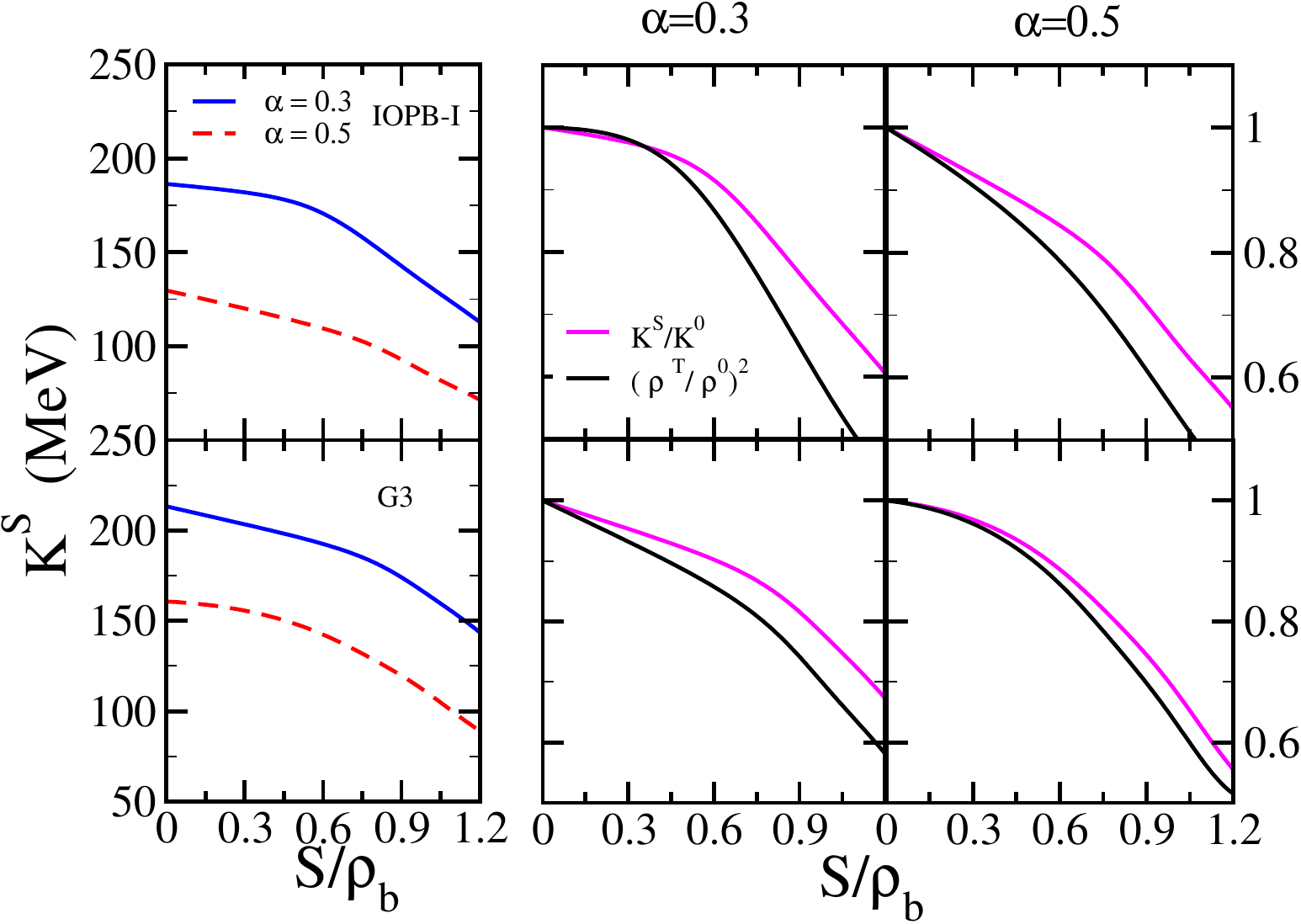}%
} 
\caption{(a) Isothermal and (b) Isentropic incompressibility at $\alpha$ = 0.3 and 0.5 for IOPB-I and G3 set at saturation density.}
  \label{incompressibility}
\end{figure}

Nuclear matter incompressibility along the isothermal ($K^T$) and isentropic ($K^S$) paths is shown in Fig. \ref{incompressibility} for IOPB-I and G3 set for two values of $\alpha$ i.e. 0.3 and 0.5. These values are taken due to their relevance in a core-collapse supernovae. At finite temperature, the incompressibility can be defined within two channels. One being the isothermal and other isentropic incompressibility defined according to Eqs. \eqref{isothermal} and \ref{isentropic}, respectively. The isentropic incompressibility is more relevant quantity in context to supernova explosion as the time scale of collapse is less than 1 second and the process is adiabatic instead of isothermal. It prompts us to use energy instead of free energy ( see Eq.  \eqref{isentropic}). The incompressibility (both isothermal and isentropic) decreases  quadratically with temperature with G3 having higher magnitude at each temperature and entropy. It also decreases with increasing asymmetry.  The temperature dependence of $K^{T,S}/K^0$ and $(\rho^{T,S}/\rho^0)^2$ is shown in context to their relation with respective incompressibility ( see Eqs.\ \eqref{isothermal} and \eqref{isentropic}.) Their behaviour remains almost similar irrespective of change in asymmetry. These results satisfy the calculations carried out using microscopic approaches \cite{modarres1998, BOMBACI19939} thereby suggesting that these newly developed parameters not only  describe finite nuclei and cold nuclear matter but can also be used in studying the phenomenon at finite temperature such as proto-neutron star and supernova explosion.  

\subsection{  Liquid-gas phase transition}
The asymmetric nuclear matter (ANM) is a two-component system with two conserved charges Q (B, $I_3$). In a two-component system, although the total charge remains conserved, their ratio can be different in different phases. The constraint on T, Q, and $\rho$ which determine the energetic of the system, forces vapour pressure and chemical potential to change during the phase transition. Apart from mechanical instability, the diffusive instability (fluctuations on the charge concentration) appears and is more relevant to describe the ANM \cite{Muller1995}.  In the literature, it has been argued that asymmetric nuclear matter exhibits only one dominant type of instability, which is primarily of isoscalar nature, rather than two types (mechanical and chemical). This implies that the instability is  governed by density fluctuations, which can be interpreted as a  liquid-gas separation \cite{Margueron_2003, DUCOIN2007407, Alam_2016}. The present work however explore the ANM using the seperate mechanical and spinodal instabilities in line with H. M\"uller \textit{et al.} \cite{Muller1995}. The phase transition in the ANM is therefore described by the following three regions:
\begin{enumerate}
    \item Isothermal Spinodal (ITS): describe the mechanical instability given by $\frac{\partial P}{\partial \rho_b}$. It defines the critical temperature in the symmetric matter.
    \item Diffusive Spinodal (DS):  describe the chemical instability. It essentially means that energy is required to add extra protons in the system at a fixed temperature and pressure. The  critical isobar $P_c$ is estimated by finding a inflation point  $\frac{\partial \mu_p}{\partial \alpha}\Big|_{P_c}$= 0. The corresponding T=$T_c$ and $\rho = \rho_c$ are called the critical temperature and density, respectively. 
    \item Coexistence Curve (CE): Set of points where Eq. \eqref{instabilitycondition} along with the Gibbs conditions are satisfied. This curve may contain the critical points. Unlike symmetric matter case, here CE or binodal is 2 dimensional.
\end{enumerate}

\begin{figure}
	\centering
		\includegraphics[scale=0.5]{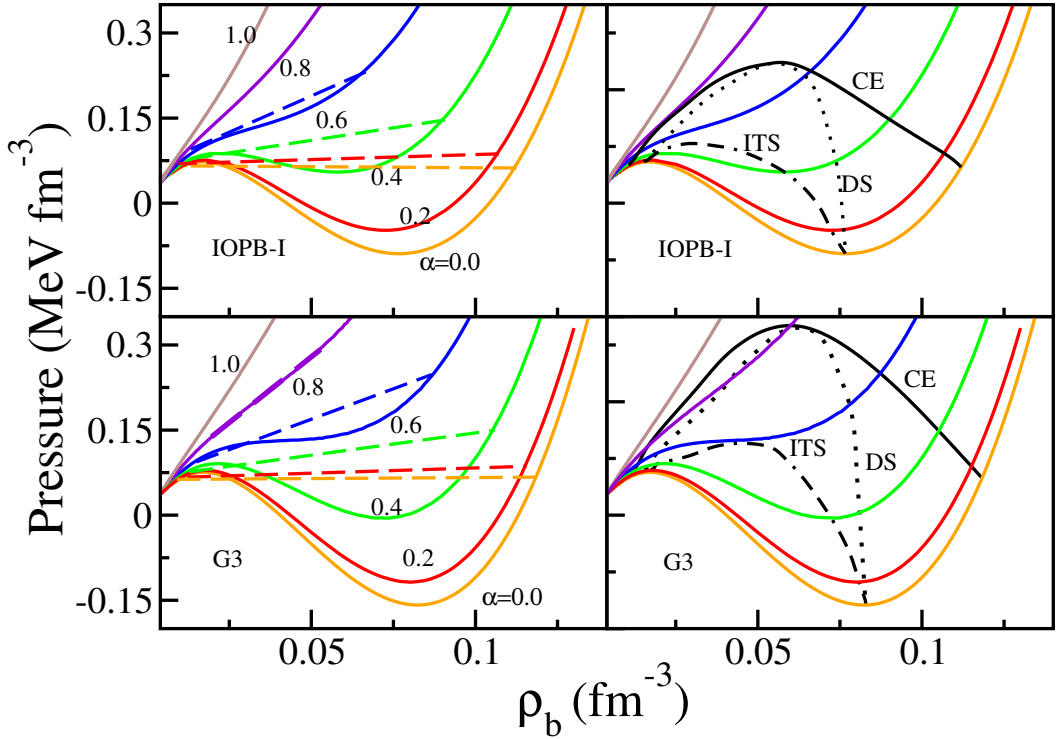}
	\caption{EoS of nuclear matter at various $\alpha$ at T= 10 MeV along with the ITS, DS and CE curves for IOPB-I and G3 sets.}
	\label{eos}
\end{figure}
The complexity of phase transition in the asymmetric nuclear matter is shown in Fig. \ref{eos}. As one move from symmetric to ANM, a new behaviour distinct to the two-component system is allowed. Asymmetry is held constant during the phase transition which forces the system to change its chemical potential and consequently the pressure (shown by dashed line in left panel). Due to charge fluctuation during this phase transition, the diffusive instability appears and plays more important role than mechanical instability in describing the phase transformation. The right panel shows all  three curves i.e. ITS, DS and CE and it is visible that diffusive instability has larger area as compared to mechanical instability.
\begin{figure*}
  \centering
\subfloat[]{%
  \includegraphics[height=7cm,width=.49\linewidth]{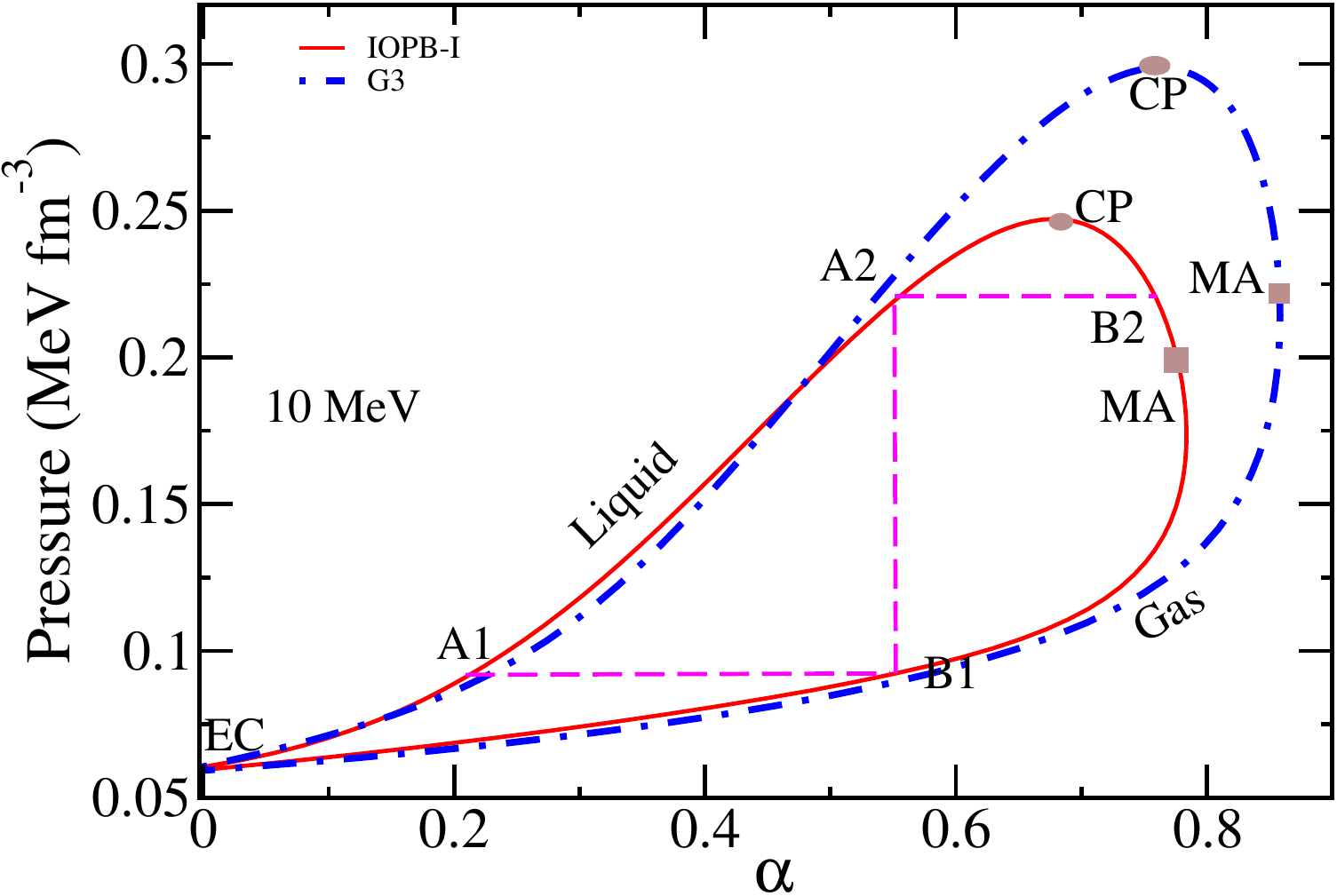}%
}\hfill
\subfloat[]{%
  \includegraphics[height=7cm,width=.49\linewidth]{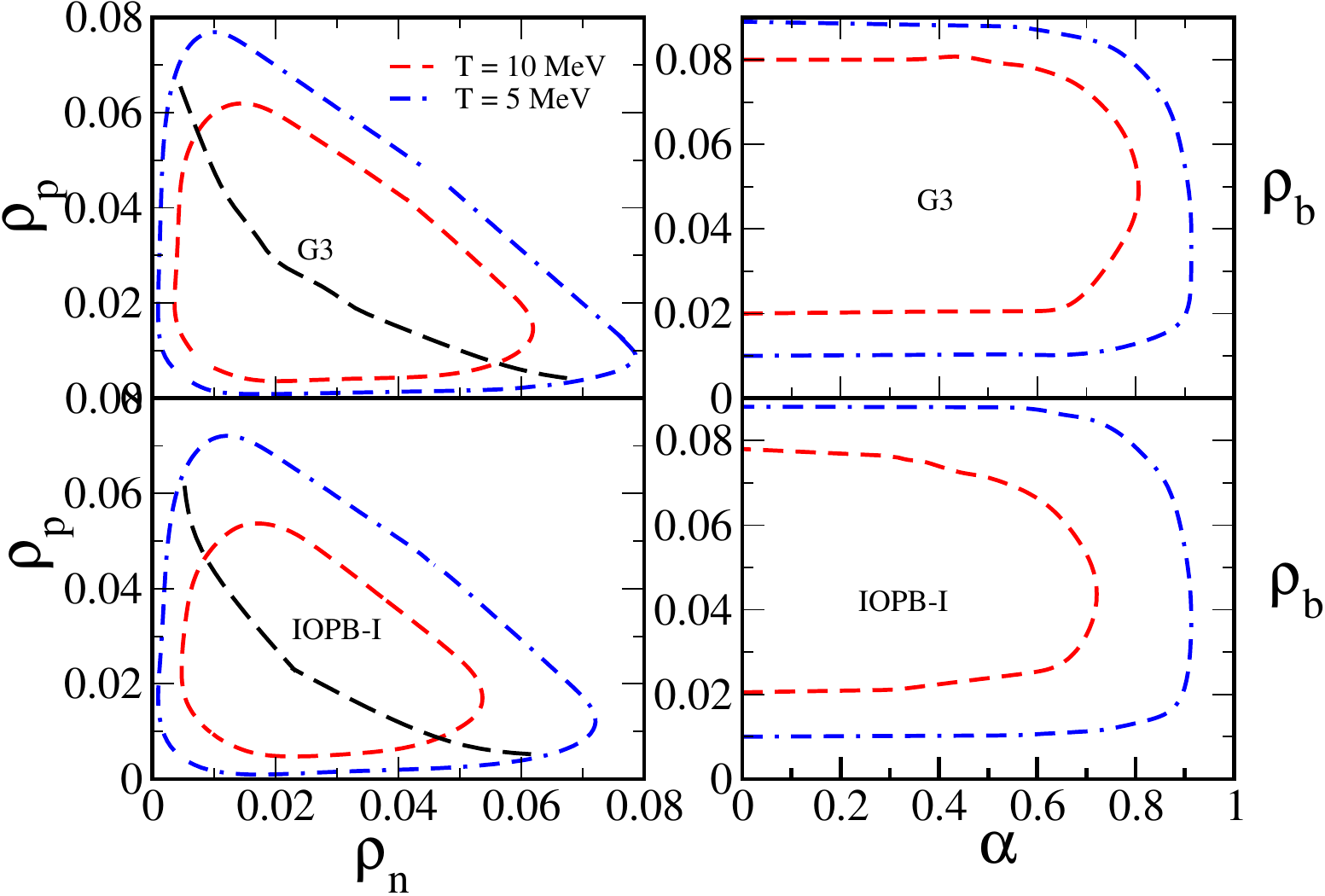}%
}  
 \caption{The binodal surface on the P-$\alpha$ plane is shown on the left (a). The spinodal boundary in $\rho_n$-$\rho_p$ plane and $\alpha$-$\rho_b$ plane is shown on the right (b). The black dashed line on the spinodal shows the critical points.}
  
\label{binodalspinodal}
\end{figure*}

Binodal as per the Gibbs condition given in Eqs.  \eqref{chmeicalpotentialeqi} and \eqref{pressureeqi} at T=10 MeV is plotted in Fig.  \ref{binodalspinodal} by geometrical construction where a rectangle is drawn on the chemical potential isobars of neutrons and protons \cite{lattimer1978neutron}. It is characterised by the point of equal construction (EC), the point of maximal asymmetry (MA) and the critical point which determine the edge of instability area. In the phase coexistence region, the proton fraction of two-phase changes (a unique feature of two-component system) and the phase with higher asymmetry exhibits a lower density or vice-versa. At the critical temperature of symmetric matter, all the three points (EC, MA, CP) coincide and the surface becomes a point. The vertical dashed magenta line indicates that during the phase transition, $\alpha$ remain constant and both phases follow different paths i.e. liquid follow the path A1-A2 while gas phase evolves from B1 to B2 during the isothermal compression. Finally, the system leaves the instability at A2. This condition is called stable condensation. On the other hand, when the system is prepared with $\alpha> \alpha_c$ ($\alpha$ at CP), it operates in the gaseous phase only and this unique phenomenon is called retrograde condensation. The spinodal according to Eq.\ \eqref{instabilitycondition} is plotted on the right side of Fig. \ref{binodalspinodal} on both $\rho_n$-$\rho_p$, $\alpha$-$\rho_b$ planes.  Fig.\ \ref{eos} and \ref{binodalspinodal} provide a complete description of phase transition in asymmetric nuclear matter.

In symmetric nuclear matter, the compressibility is the deciding factor for critical parameters of phase transition whereas, the phase transition in the ANM is characterised by symmetry energy. This can be verified from Eq.\ \eqref{fsym} where the contribution of iso-spin asymmetry is reflected from the free symmetry energy $F_{sym}$ and its slope $L_{sym}$. $K_{sym}$ and $Q_{sym}$ are the higher order derivatives of FNSE in the Taylor series which are still not well constrained. Two E-RMF sets IOPB-I and G3 are used to account for the various EoS properties on the phase transition in the ANM. The detailed analysis of phase transition on the symmetric matter using the IOPB-I and G3 set is discussed in \cite{vishal2020}. For ANM, the asymmetry in density is introduced by $\rho$ meson and is dictated by cross coupling $\Lambda_\omega(R^2W^2)$. The G3 and IOPB-I set has $\Lambda_\omega$= 0.038 and 0.024, respectively. The corresponding values of $J$ and  $L$ at T=0 MeV are given in Table \ref{bulkproperties} whereas their finite temperature dependence is shown in Fig. \ref{symenergy}. The G3 set has an additional mass asymmetry introduced by $\delta $ meson. The $\delta$ meson allows one to vary the  $L_{sym}$ without altering the symmetry energy $F_{sym}$. At a given temperature, the G3 set has larger coexistence area and large values of CP and MA as compared to the IOPB-I set due to  the $\delta$ meson. A large coexistence area favours highly asymmetric gas in coexistence with less asymmetric dense fluid. This has a direct consequence for the core-crust transition and crust structure of neutron star. Opposite to SNM, where $\zeta_0$ plays the determining role, the value of $\Lambda_{\omega}$ decides the ANM which in turn affects the $L_{sym}$. A greater $\Lambda_\omega$ usually gives smaller $L_{sym}$ and vice-versa. 

\begin{figure}[h]
	\centering
		\includegraphics[scale=0.5]{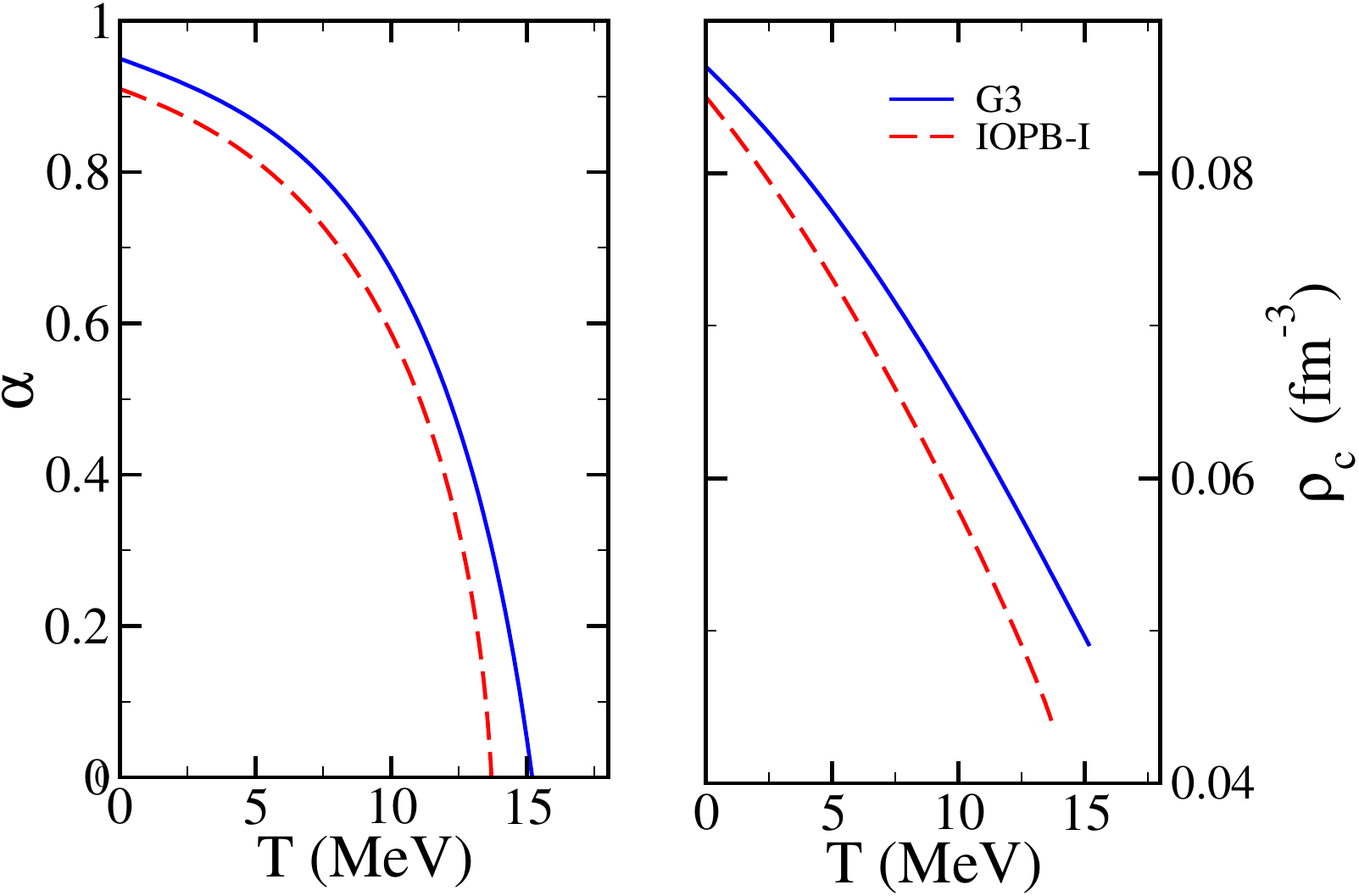}
	\caption{ $\alpha$ as a function of temperature and corresponding $\rho_c$ for the IOPB-I and G3 set.}
	\label{criticalvalue}
\end{figure}

The spinodal for the G3 set also has a larger area at any given temperature as compared to the IOPB-I set. One can observe the major variation among two sets in the coexistence densities in the $\alpha-\rho_b$ plane.  This means that the densities where different structure in non-homogeneous phase occur will be different. This property is again determined by the $L_{sym}$. The G3 set with smaller $L_{sym}$ estimates the larger $\alpha$ and $\rho_c$ at any given temperature. This is shown in Fig. \ref{criticalvalue}
 where the dependence of $\alpha$ and $\rho_c$ is shown on temperature. The $\alpha-T$ plots signify the temperature at which the diffusive instability disappears (also called critical temperature). This critical temperature is not similar to symmetric matter where mechanical instability decides the phase transition but is determined according to $\frac{\partial
  \mu_p}{\partial \alpha}|_{P,T} = 0$ and  $\frac{\partial^2
  \mu_p}{\partial \alpha^2}|_{P,T} = 0$. In the E-RMF sets with constant couplings, the inflation point for proton and neutron coincides having synchronous behaviour. This might not be the case with density-dependent coupling sets \cite{Fedoseew2015, pring}. $\alpha$ decreases smoothly at low temperatures but after $T>0.5T|_{\alpha=0}$, there is a steep fall in the $\alpha$. The G3 set estimates larger $\alpha$ at a particular T due to its smaller value of $L_{sym}$ and greater value of $\Lambda_\omega$. This same trend is observed in the $\rho_c$. These trends are similar to the references \cite{Alam2017, SHARMA2020121974}, where any one coupling in a parameter set was varied keeping other fixed to obtain different $L_{sym}$. The agreement of those trends while comparing two different parameter sets with almost same symmetry energy indicated that the correlation between different  properties of phase transition still holds as in case of SNM \cite{louren2017} and these can be exploited to constrain the EoS which do not take critical temperature into the account \cite{vishal2020,yang2019}.

%%%%%%%%%%%%%%%%%%%%%%%%%%%%%%%%%%%%%%%%%%%%%%%%%%%%%%%%%%%%%%%%%%%%%%%%%%%%%%%%%%%%%%%%%%
\subsection{ \label{electron} Effect of electrons}
 In a physical system, the electrons are present so that Coulomb energy do not diverge. They  are included in EoS as a free non-interacting relativistic Fermi gas described by \cite{Avancini2006}

\begin{figure}
	\centering
		\includegraphics[scale=0.5]{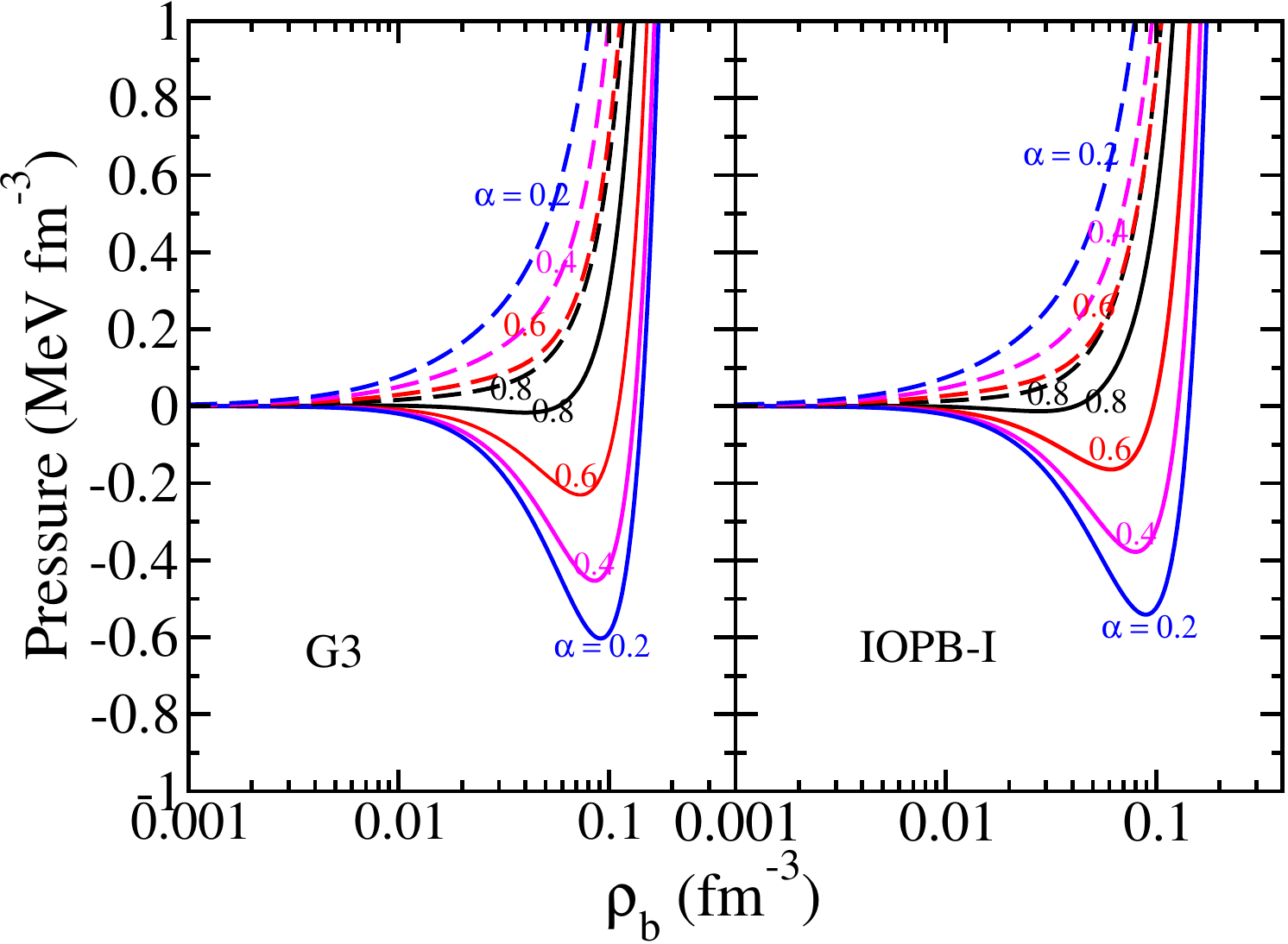}
	\caption{The EoS with and without the electrons. Solid line represents the nuclear matter without electrons and dashed line represents with electrons.}
	\label{eoselectron}
\end{figure}
\begin{equation}
L_e =  \bar{\psi_e}[i\gamma_\mu \partial^\mu - m_e]\psi_e,
\end{equation}
where $L_e$ is the Lagrangian, $m_e$ is the mass of electron. Since the electrons only compensate the  proton charge, we have $\rho_p$=$\rho_e$=$\frac{1}{\pi^2}\int k^2 dk (n_{ke}-\bar{n}_{ke})$. Here, $n_{ke}$ and $\bar{n}_{ke}$ are the Fermi integral for electrons and positrons. Figure \ref{eoselectron} shows the effect of electrons on the EoS for IOPB-I and G3 parameter sets at T=0 MeV. The effect of electrons is dominant for the matter with less asymmetry as the electron density becomes high to compensate for the larger proton density. Electrons are taken as non-interacting particles and therefore the underlying nature of a parameter set is unaltered.   Electrons have high Fermi energy and therefore, makes the system devoid of the instability. Both the IOPB-I and G3 sets have no spinodal when electrons are included for T= 5 MeV. No spinodal means that stellar matter at $\beta$ equilibrium will be uniform at temperature above 5 MeV \cite{Avancini2008}. This is  consistent with the various calculations of neutron star core-crust transition. 

To further understand the implication of electrons in the EoS, the adiabatic index is studied. In the processes like supernovae explosions and neutron stars, the compression and rarefactions modes of vibration are adiabatic or isentropic instead of isothermal \cite{Constantinou2015} . The adiabatic index is related to the stiffness of EoS and is given by 

\begin{equation}
    \Gamma_s=\frac{\rho_b}{P}\frac{\partial P}{\partial \rho_b} \Big|_s.
\end{equation}
\begin{figure}
	\centering
		\includegraphics[scale=0.28]{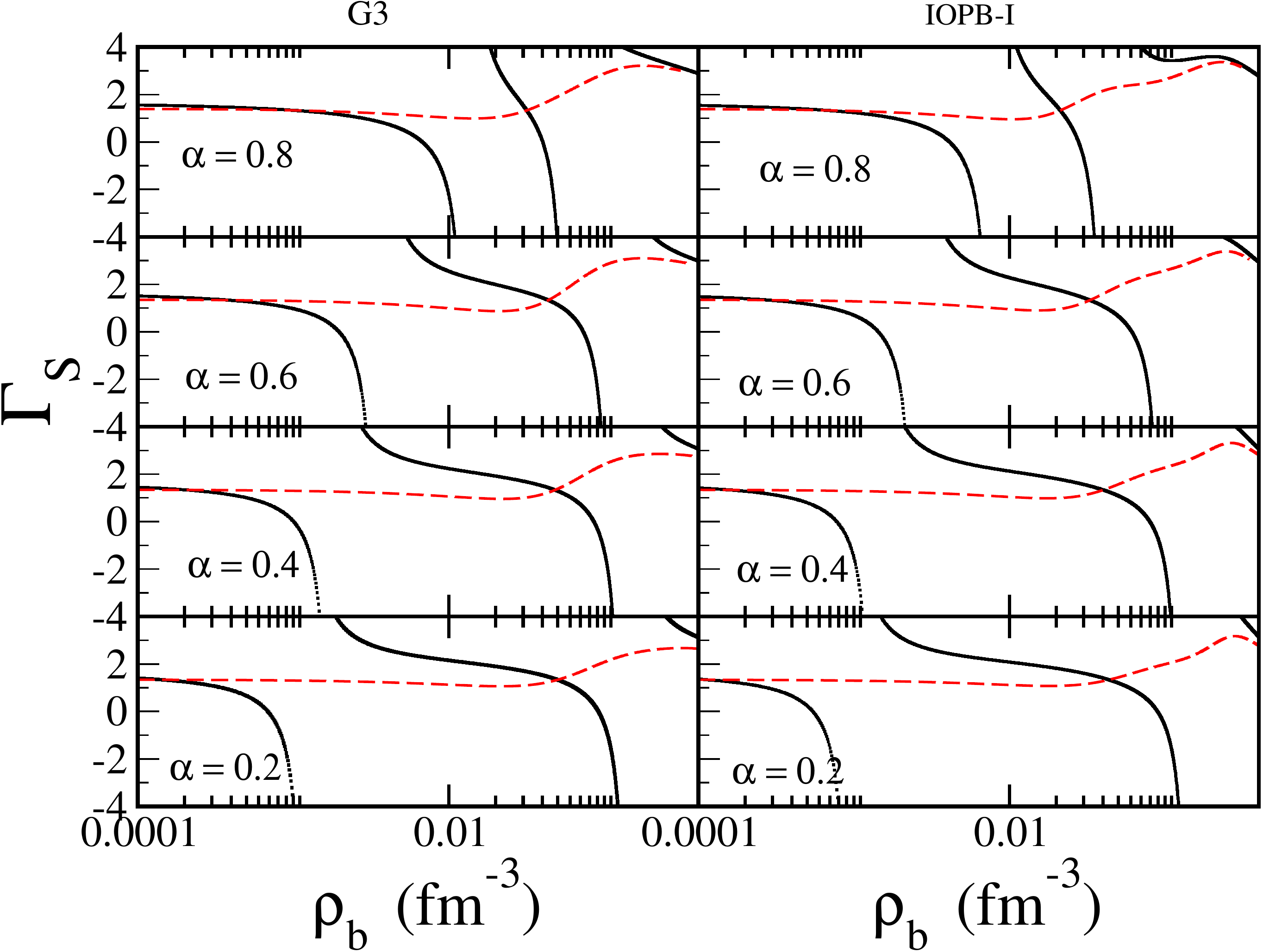}
	\caption{Adiabatic index $\Gamma_{s=0}$  for the G3 and IOPB-I set with various asymmetry. Solid black line include only nucleons while red line indicates the presence of electrons.}
	\label{adiabaticindex}
\end{figure}
$\Gamma_{s=0}$ for the two models employed here is shown in 
Fig. \ref{adiabaticindex}. The solid black curve represents the nucleon only while the red dashed curve includes the contribution from electrons. $\Gamma_{s=0}$ corresponding to nucleons goes negative in some density regions showing the mechanical instability. For low and high densities, it varies asymptotically. The inclusion of electrons restores the mechanical instability and value of $\Gamma_{s=0}$ increases gradually around subsaturtaion density and become asymptotically constant at low and high densities. These observations can be understood quantitatively by examining the baryon and electron pressure as shown in Fig. \ref{eoselectron}.  For $\rho \rightarrow$ 0, $\Gamma_{s=0}$ tends to the 4/3 which is due to the relativistic electrons and  is an important requirement for the stability of supernova simulation. As asymmetry rises, this value goes to 5/3 for pure neutron matter. Although the underlying properties for $\Gamma_s$ are same for both the models, but the position of instability and highest value of $\Gamma_s$ in case of matter with electrons, is essentially determined by the pressure due to baryon.  The presence of  electron in the system also impacts the  speed of sound $C_s^2=\frac{\partial p}{ \partial E}$. Addition of electrons do not yield the nonphysical region in low density as is seen in the  nuclear matter system without leptons. At higher density, the electrons impart significant impacts on the more symmetric matter making it smoother as compared to the ANM. 

\section{ Summary }

In summary, the present chapter investigates various finite temperature properties of isospin symmetric and asymmetric nuclear matter over a wide range of density and pressure. The E-RMF formalism employing the  FSUGarnet, IOPB-I and G3 forces, along with one of the most used NL3 parameter set, are used in the finite temperature limit realizing their narrow range of bulk matter properties at zero temperature.  The estimated critical parameters agree with other RMF forces available in literature, which have incompressibility in the range 240 $\pm$ 20 MeV.

The critical parameters, i.e. $T_c$, $P_c$ and $\rho_c$, are clearly model dependent. However, there is a direct correlation among parameters such as $\rho_c-\rho_f$, $T_c-T_f$, $P_c-T_c$, etc. The critical temperature is not constrained in theoretical as well as in experiments. Consequently, there is large uncertainty in the value of critical temperature among experiments as well as theoretical calculations with the same bulk matter properties.   A little discrepancy in the experimental and theoretical values might be attributed to two main factors: (1) symmetric nuclear matter being an ideal system which is too difficult to simulate in experiments and (2) finite size effect and very short time scale of multi fragment reactions which make it difficult to study thermodynamic equilibrium. 

The vector-self coupling ($\zeta_0$) is a determining factor in the finite temperarture calculations using E-RMF formalism. The force with a lower value of $\zeta_0$ estimates a large coexistence area and large $T_c$. Similarly, $T_c$ is positively correlated with incompressibility at saturation except in NL3 set. A large negative value of $k_4$ makes the NL3 behave differently in some cases as compared to other forces. The parameter sets used in the present work estimate the compressibility factor (C$_f$) close to the universal value of liquid-gas systems. Whereas, the parameters having higher incompressibility deviate much from this value. Moreover,  there is also a need to look for higher-order vector interaction terms which might influence the EoS at finite temperature keeping other properties intact. The consistency of the calculations is checked using critical exponents and scaling laws and emphasize that E-RMF parameter sets with acceptable incompressibility are consistent and reliable at finite temperature limit. All the critical exponents are very close to the mean-field results and experimental liquid-gas systems.

Further, this chapter considers the dilute homogeneous nuclear matter at different values of temperature and isospin asymmetry because of their relevance in astrophysical simulations.   The temperature dependence of free nuclear symmetry energy ($F_{sym}$) and its higher-order derivatives are discussed. The  $F_{sym}$  increases with temperature at a given density due to a decrease in entropy density. The higher-order derivative of  $F_{sym}$ preserves the zero temperature behaviour with a slight change in magnitude which shows that one can use the zero-temperature value of these parameters to compare the relevant quantities at any given temperature.

It is further observed that the thermal effects in E-RMF formalism depend mainly on the density dependence of Dirac effective mass. The Dirac effective mass is calculated self consistently, which depends on the  $\sigma$ and $\delta$ mesons.   A larger Dirac effective mass corresponds to larger thermal effects on the state variables. A similar effect of effective mass on the thermal contribution is seen in non-relativistic formalisms, although both Dirac mass and effective mass in non-relativistic differ in their origin. The thermal effects are also sensitive to isospin asymmetry. The isospin asymmetry also impacts the peak of the isothermal thermal index ($\Gamma$)  at a fixed temperature. It is seen that the underlying nature of the thermal contribution to a state variable at a given isospin asymmetry remains the same with increasing temperature across the forces used in this study.   The change in magnitude of thermal contribution is principally attributed to their zero temperature variation, which results from their different nuclear matter observables such as incompressibility, symmetry energy and its higher order derivatives etc.

The phase transition is studied for the asymmetric nuclear matter considering a two-component system with two conserved charges, i.e. Baryon number and isospin. The G3 set, due to its low $L_{sym}$, estimates the higher value of maximal asymmetry and critical pressure. The presence of $\delta$ meson positively affects binding energy and therefore influences the boundary of spinodal. The critical density and asymmetry are also larger for the G3 set, which can be attributed to its lower $L_{sym}$. The Value of $L_{sym}$ is determined mainly by cross-coupling of $\rho$ and $\omega$ meson and $\delta$ meson. One can say that a larger value of $\Lambda_\omega$ estimates the larger instability in asymmetric nuclear matter. Critical asymmetry is a quadratic function of temperature and exhibits different behaviour in the low and high-temperature range. These trends are also consistent with other relativistic studies available in the literature. 

Finally, the effect of the electrons in EoS of nuclear matter and its instability is investigated. Electrons, due to their high Fermi energy, make the system devoid of instabilities. The adiabatic index ($\Gamma_{S=0}$) of matter with and without the inclusion of electrons is studied. The $\Gamma_S$ with electrons become asymptotically constant at low and high densities with a small variation near the saturation density. The density of this hump predominantly depends on the baryon pressure. The electron being a non-interacting particle, does not alter the underlying nature of the force parameter.  

%The present calculations can be extended to study the various astrophysical processes, such as the supernova explosion and neutron star crust, where the nuclear matter is dilute and at some finite temperature. The low-density matter results in the formation of clusters which subsequently impacts various cooling and transportation process. 
%Furthermore, the idea of the thermal effect on state variables will help in reducing the computational cost and time of the numerical calculations in large simulations such as impersonating supernovae explosions and neutron star binary collisions. Such analysis is essential to estimate an equation of state for a wide range of densities and will be carried out in future work. 

\clearpage
\addcontentsline{toc}{section}{Bibliography}

\printbibliography

%% file: Chapter_4/CHAP4.tex
\chapter{\label{chap4} Properties of hot finite nuclei and associated correlations with infinite nuclear matter}

\section{Introduction}
One astonishing universality in the laws of nature is the resemblance between the nuclear and the molecular force. Therefore one may arrive at the notion that a hot nucleus should undergo a liquid-gas phase transition (LGPT) like a classical liquid drop. The multifragmentation process in heavy-ion-induced reactions (HIR) is analogous to the evaporation of a water droplet under the influence of temperature. Therefore, investigating the temperature at which a nucleus undergoes the phase transition  is an important factor in these reactions  \cite{PhysRevLett.72.3321, tlimexpdata, WU1997385}. This aspect also plays a paramount role in describing the clusterized supernovae matter.

%There have been several qualitative attempts to study the limiting temperature of nuclei in terms of Coulomb instability, where the  EoS of infinite matter is taken from  various  theoretical frameworks such as Skyrme effective NN interaction \cite{PhysRevC.44.2505, PhysRevC.69.014602}, microscopic EoS such as Friedman and Pandharipande, finite temperature relativistic Dirac-Brueckner, chiral perturbation theory \cite{microscopic, m1, NICOTRA2005118}, EoS considering the degeneracy of the Fermi system \cite{PhysRevC.39.169} relativistic calculations using quantum hadrodynamics and Thomas-Fermi approach \cite{PhysRevC.47.2001, PhysRevC.49.3228, PhysRevC.55.R1641},  Gogny interactions \cite{PhysRevC.54.1137},  chiral symmetry model \cite{PhysRevC.59.3292}.  Some calculations  have been carried out by analyzing the plateau in caloric curve obtained from various experimental observations \cite{PhysRevC.65.034618}. These calculations  give a qualitative picture of $T_{l}$ and it is seen that $T_{l}$ is  model dependent and hence needs to be investigated for appropriate outcome. 

In the present work, E-RMF theory  is used to understand the properties of LGPT in nuclei and most importantly the temperature at which the nucleus undergoes multifragmentation and loses its entity. The E-RMF formulation calculates the volume energy of infinite nuclear matter on which the finite size corrections: surface, symmetry, Coulomb  are added to evaluate the properties of a realistic nucleus.  The idea behind using the E-RMF framework for the bulk volume energy part is that the nuclear drop is usually surrounded by a nucleon gas in complete thermodynamic equilibrium. To calculate the properties of such a system, one usually needs to solve the Gibbs conditions \cite{vishalasymmetric} where it is expected that the same equation of state (EoS) are used for  the gaseous as well as the liquid phase. 

The aim of present study is twofold: First, to investigate the properties of hot isolated nuclear drop by studying the variation of thermodynamic variables such as excitation energy, entropy, level density, fissility  etc. The results are compared  with available experimental or microscopic theoretical calculations \cite{BONCHE1984278, Quddus_2018}.  The second and  important part of this work is the qualitative analysis of the limiting temperature of a hot nucleus.
In HIRs, nuclei can be heated to their limiting temperature which provides an opportunity to investigate the collective motion of nucleons, and their highly chaotic and disordered behaviour at high excitation energy. The E-RMF parameter sets namely FSUGarnet, G3, IOPB-I, and most successful NL3 \cite{iopb} are used to calculate  the volume energy of a  nucleus. The temperature-dependent surface energy term depends on the  $T_c$ which is calculated for these individual E-RMF parameter sets. In the analysis of critical properties of infinite nuclear matter using these E-RMF sets in Chapter \ref{chap3} , it was found that the  $T_c$ is not a well-constrained quantity and the majority of E-RMF sets that satisfy the relevant observational and experimental constrains on EoS underestimates it.  Since the experimental value of  $T_c$ is calculated by extrapolating the data from multifragmentation reaction data on finite nuclei, it is interesting to see the variation of $T_{l}$ of finite nuclei using different E-RMF forces.  To further generalize the relationship between various saturation properties of infinite nuclear matter, its critical properties, and the limiting properties of a hot nucleus, fifteen E-RMF parameter sets are used that lie within the allowed incompressibility range and satisfy other constrains  \cite{duttra}. An effort is made to establish correlations among these properties.

\section{\label{excitation}Excitation energy, level density and fissility}

\begin{figure}
    \centering
    \includegraphics[scale=0.5]{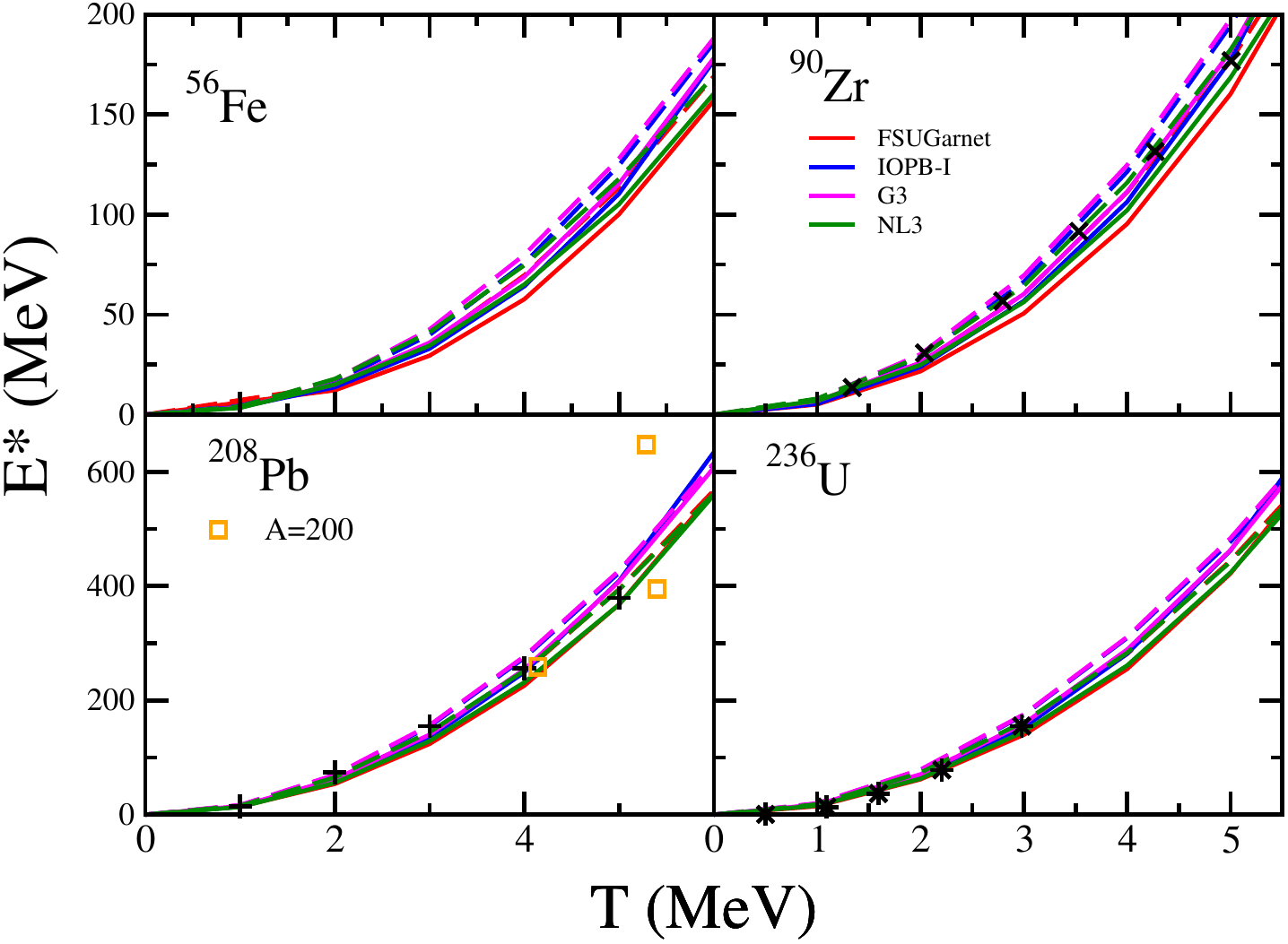}
    \caption{The excitation energy of $^{56}$Fe, $^{90}$Zr, $^{208}$Pb and $^{236}$U as a function of temperature for NL3, IOPB-I, FSUGarnet and G3 sets. The solid lines represents calculation from Eq.\ \eqref{s1} and dashed lines are from Eq.\ \eqref{s2}. The theoretical data in black cross is taken from \cite{carlson2016}, plus \cite{BONCHE1984278} from  and star from \cite{Quddus_2018}. The experimental values for A $\approx$ 200 are taken from \cite{a200}.}
    \label{ex}
\end{figure}

Let us begin with the discussion of the caloric curve which is the relation between excitation energy and temperature for the three isolated spherical nuclei i.e. $^{56}$Fe, $^{90}$Zr, $^{208}$ Pb and  $^{236}$U which is formed when thermally fissile $^{235}$U absorb a thermal neutron. In experiments, the temperature of the nucleus is not measured directly and it is calculated using  excitation energy which can be obtained  using resonance or energy of evaporated particles. Above mentioned   nuclei are most  studied nuclear systems and their microscopic calculations are available in literature.  
Fig. \ref{ex}  shows the caloric curve for these nuclei using the four E-RMF sets FSUGarnet, IOPB-I, G3, and NL3.
The estimations of theoretical caloric curves from the E-RMF are in reasonable agreement with microscopic calculations \cite{carlson2016, BONCHE1984278, Quddus_2018}. The experimental value for mass A $\approx$ 200 extracted from \cite{a200} also align with our calculations for T $<5$ MeV. The deviation at higher temperature and excitation energy may be associated with the production of heavier particles in the multifragmentation process which may change the 
 energy of the system. The behaviour of different parameter sets is tightly constrained and the spread of curves becomes narrower as one moves from $^{56}$Fe to $^{208}$Pb. The effect of different parametrizations of surface energy from Eqs.\ \eqref{s1} and \eqref{s2} is also visible. Eq.\ \eqref{s2} derived from the semi-classical Seyler-Blanchard interaction estimates a steeper slope for caloric curve as compared to the Eq.\ \eqref{s1} based on thermodynamic equilibrium of sharp interface between liquid and gaseous phase. It is because the Eq.\ \eqref{s1} estimates relatively lower surface energy at any given temperature.

For a particular nucleus, the G3 set with the largest effective mass ($m^*/m$=0.699) estimates the steepest caloric curve while the FSUGarnet with the smallest ($m^*/m$=0.578) corresponds to the softest caloric curve. The effective mass in E-RMF  formalism is determined from the strength of scalar field because of  NN interaction. The G3 set due to small scalar self couplings $k_3, k_4$ and  scalar-vector cross couplings $\eta_1 , \eta_2$ estimates the softest scalar field while the FSUGarnet yields the stiffest scalar field. The scalar field consequently determine the mechanical properties of the system and therefore, the effective mass becomes a crucial saturation property at finite temperature. The effective mass which is obtained self consistently also determine the chemical potential and kinetic energy of nucleons which are essential inputs for the thermal properties calculations. Furthermore, the G3 set  estimates the softest repulsive contribution arising from the vector self coupling $\zeta_0$. The combined effect of scalar and vector field determine the critical temperature.  The parameter set G3 and FSUGarnet  estimate  the largest and smallest $T_c$ among these four sets (see Table \ref{criticalparamaters}). Therefore, in finite nuclei, the thermal contribution of energy essentially depends on the combined effect of effective mass, $T_c$ and the zero-temperature EoS. It may be noted  that the saturation properties are not unique and different combinations of mesons coupling can yield the similar nuclear matter properties. Therefore, it is  relevant to analyze the finite temperature properties of the nuclear matter in terms of saturation properties and not the coupling constants.

\begin{figure}
    \centering
    \includegraphics[scale=0.5]{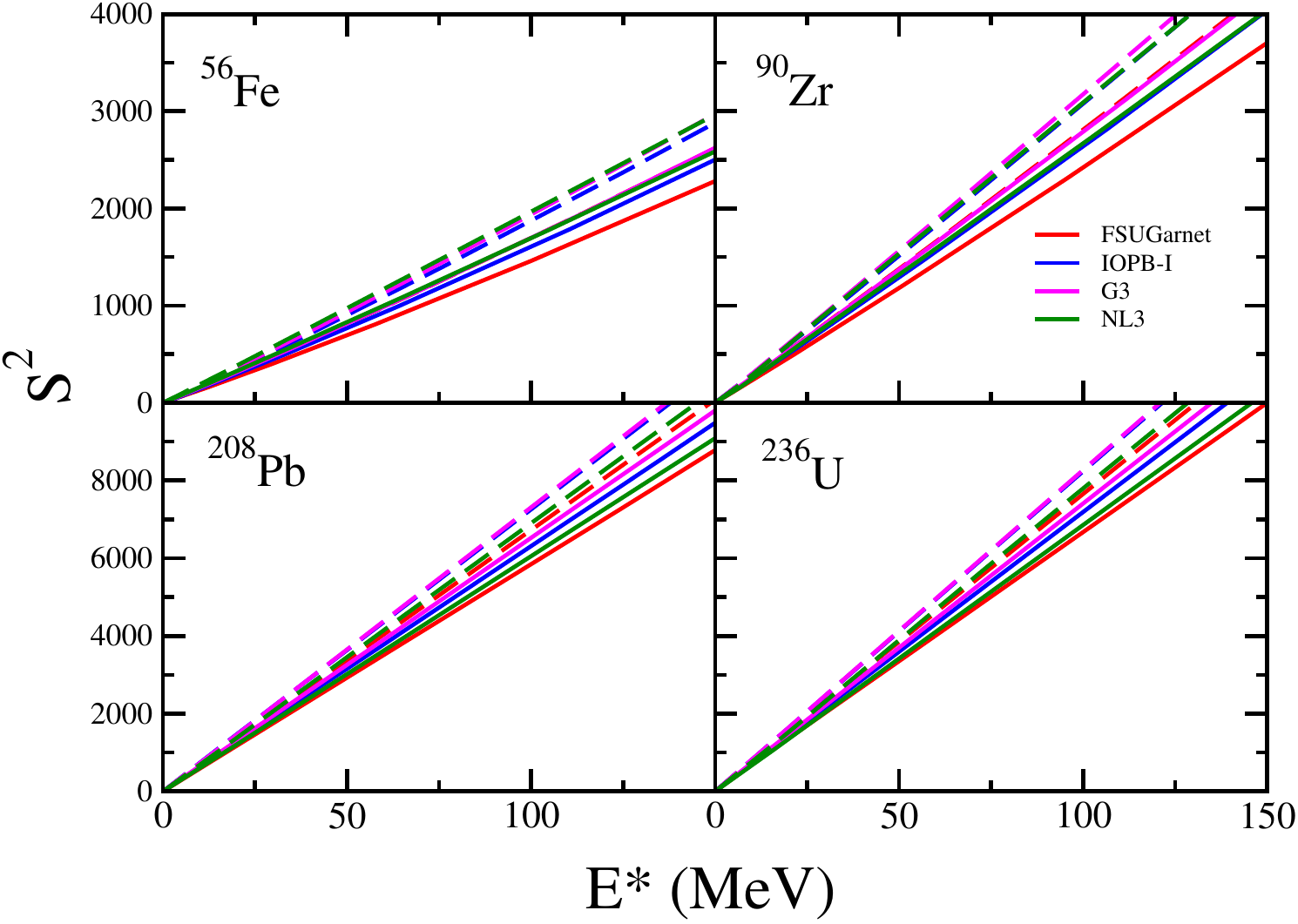}
    \caption{The Relation between square of entropy and excitation energy for the systems as in Fig \ref{ex}.}
    \label{entropy}
\end{figure}

In the Fermi gas model, the point of minimum entropy in the transition state nucleus corresponds to its minimum excitation energy (E$^*$) \cite{PhysRevLett.25.386}. Therefore,  the relation of the square of entropy and E$^*$ is shown in Fig. \ref{entropy} for the systems considered in Fig. \ref{ex}. The square of entropy increases monotonically with  the E$^*$ signifying a disordered and chaotic nucleus. The disorder increases with mass number implying a more violent multifragmentation process once the nucleus reaches its limiting temperature T$_l$.   
Eq.\ \eqref{s2} estimates higher entropy at a given E$^*$ as compared to  \ref{s1}. For a particular nucleus, the spread of different E-RMF sets increases with E$^*$. This effect can  be  attributed to the effective mass and $T_c$ of a particular E-RMF parameter. The present model have not considered the shell correction which deviates the straight-line behaviour of this curve at low temperature, where shell structure is still intact. These shells melt at around $E^*\approx40$ MeV or $T\approx1-2$ MeV \cite{PhysRevC.91.044620}.  The behaviour of $S^2$ is in agreement with results in  \cite{carlson2016, Quddus_2018}. 
 
\begin{table}
    \centering
        \caption{The level density parameters obtained using different expression of Eq.  \eqref{lebeldensityeq} for the NL3, IOPB-I, FSUGarnet and G3 parameter set.}
    \begin{tabular*}{\linewidth}{c @{\extracolsep{\fill}}  cccccc}
    \hline
    \hline
 Element             & Forces    &  \multicolumn{3}{l}{ $a$ ($MeV$$^{-1}$) Using Eq.\ \eqref{s2}} \\ \hline
                    &           & ${E^*}/{T^2}$    & $S^2/4E^*$   & $S/2T$ \\ 
\hline                  
\multirow{4}{*}{$^{56}$Fe} & NL3       & 4.695 & 4.931  & 4.789\\
                    & FSUGarnet & 4.582 & 4.323 & 4.357  \\
                    & IOPB-I    & 5.033 & 4.789 & 4.808  \\
                    & G3        & 5.149 & 4.942  & 4.963 \\ \hline
\multirow{4}{*}{$^{90}$Zn} & NL3       & 7.267  & 7.740 & 7.491 \\
                    & FSUGarnet & 7.102 & 7.185 & 7.072  \\
                    & IOPB-I    & 7.812  & 7.857 & 7.758  \\
                    & G3        & 7.872 & 8.065 & 7.930 \\
\hline                    
\multirow{4}{*}{$^{208}$Pb} & NL3       & 15.683 & 17.030 & 16.394  \\
                    & FSUGarnet & 15.752  & 16.725  & 16.233\\
                    & IOPB-I    & 16.998  & 18.040 & 17.531\\
                    & G3        & 17.126 & 18.191 & 17.683\\
\hline                    
\multirow{4}{*}{$^{236}$U}  & NL3       & 17.64   & 19.19   & 18.463 \\
                    & FSUGarnet & 17.761 & 18.946 & 18.353\\
                    & IOPB-I    & 19.196 & 20.400 & 19.818 \\
                    & G3        & 19.296 & 20.532 & 19.949 \\
\hline
\hline
    \end{tabular*}
\label{avalue}
\end{table}

The caloric curve gives us the opportunity to study the level density parameter ($a$) which plays a crucial role to understand the particle spectra and nuclear fission. Level density signifies the available excited state level at a given energy. In order to study the level density  Eq.\ \eqref{lebeldensityeq} is used, which is fitted on the value of $a$ with R-squared value $>$ 0.99. The level density parameters obtained using different expression of Eq.\ \eqref{lebeldensityeq} are listed in  Table \ref{avalue}. The level density calculated from all the three equations (Eqs. \eqref{lebeldensityeq}) are comparable. A larger effective mass and T$_c$ corresponds to the larger level density as in the case of G3.  These calculations are performed using   Eq.\ \eqref{s1}. On the other hand Eq.\ \eqref{s2} estimates lower magnitude of level density although the trend remains same. The value of level density lie  within the  empirical relations  $A/11.93$ from \cite{11.93} and A/14.75 from \cite{14.75}. Nuclear level density can also be studied in terms of temperature where one can take the relevant ratio in a straightforward manner i.e.,  $a=E^*/T^2$ at a particular temperature.  The G3 set with largest effective mass yield the largest temperature-dependent level density. The above analysis of thermal properties advocates the importance of effective mass over other saturation properties. 

\begin{figure}
    \centering
    \includegraphics[scale=0.5]{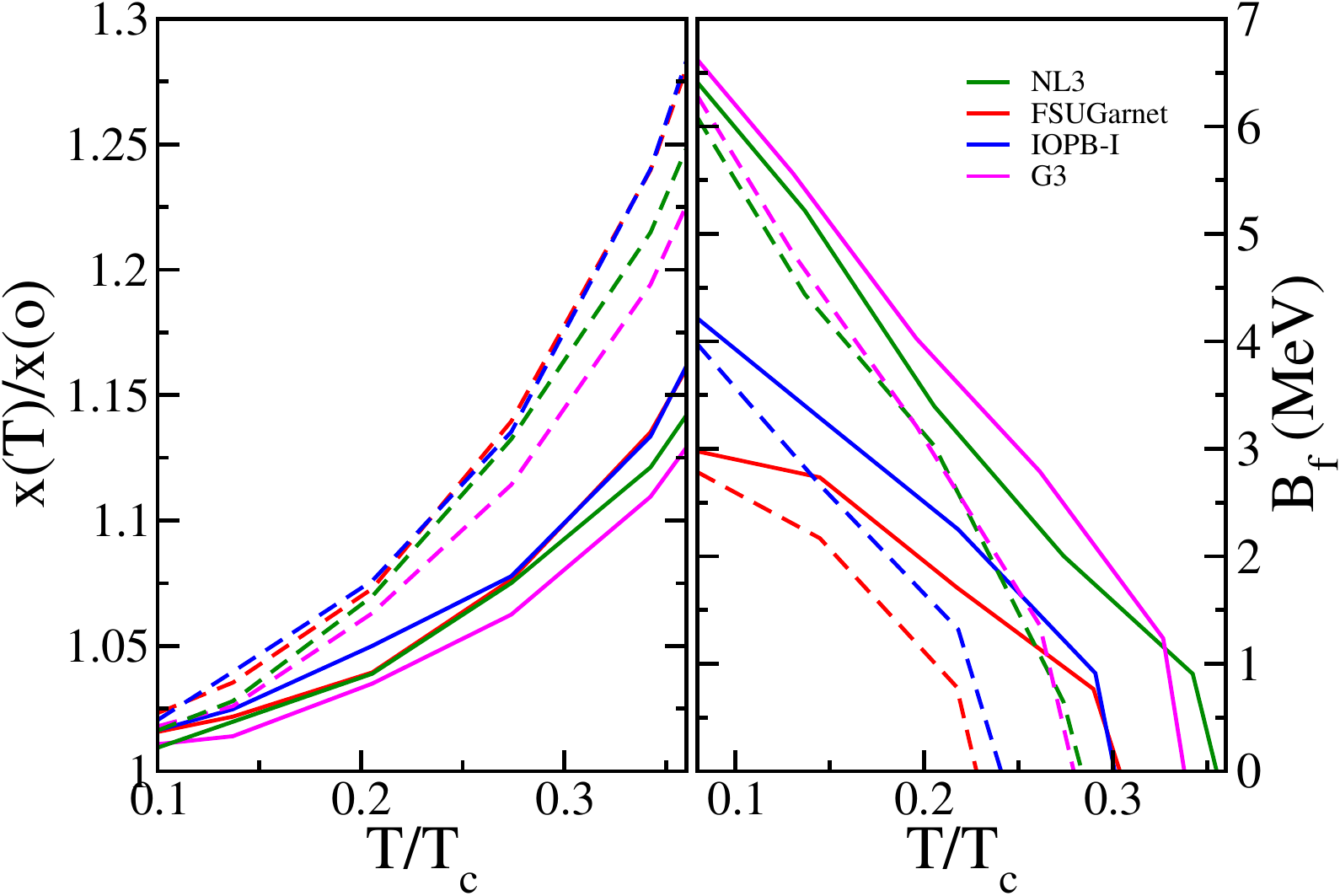}
    \caption{The fissility parameter $x(T)/x(T=0)$ as a function of $\frac{T}{T_c}$ for  $^{236}U$ using parameter sets  NL3, IOPB-I, FSUGarnet and G3 on the left panel. Right panel shows the Liquid-drop fission barrier for $^{236}U$. Solid and dashed lines have the same meaning as in Fig. \ref{ex}.}
    \label{barrierfissility}
\end{figure}

Fig. \ref{barrierfissility} shows the variation of fissility as a function of  $T/T_c$ using Eq.\ \eqref{fissilityeq} with different forces and both the parametrizations of temperature dependence of surface energy i.e. Eq.\ \eqref{s1} and \eqref{s2}.  Fissility characterizes the stability of a charged nuclear drop against fission. In general, when Coulomb free energy $\mathcal{F}_{col}$ becomes twice the surface free energy $\mathcal{F}_{surf}$, the spherical liquid drop become critical towards spheroidal deformation and split into two equal parts. 
%One thing to note here is that, similar to a classical liquid drop, on increasing the temperature, the nuclear liquid drop becomes more spherical \cite{Quddus_2018} i.e shell structure becomes trivial and deformations in the nucleus vanish. Therefore, a drop can not undergo spontaneous fission only by the temperature and one always needs external disturbance like a thermal neutron in case of $^{235}$U. Although, at a certain maximum temperature $T_{l}$, the nucleus will undergo multifragmentation process.
The fissility for  $^{236}$U increases exponentially with temperature suggesting that the surface energy decreases much faster on increasing the temperature. Eq.\ \ref{s2} has steeper slope than Eq \ref{s1} which is again the result of lower surface energy in case of Eq.\ \eqref{s1}. The fission barrier decreases with temperature and almost vanishes for $T/T_c$=0.4 for all the forces. G3 parameter set estimates the largest barrier and FSUGarnet the lowest which may be due to their effective mass. The effective mass controls the mechanical properties and consequently determine the equilibrium density of the nuclear liquid drop. One may notice  in Fig. \eqref{barrierfissility} the dominant effect of $T_c$  as these quantities do not include the volume term (see Eq. 
\eqref{fissilityeq}). The FSUGarnet and IOPB-I show the similar trend with almost similar $T_c$. G3 parameter set estimates the softest fissility and largest fission barrier followed by the NL3 set as their value of $T_c$ are 15.3 and 13.75, respectively. The vanishing points of liquid-drop fission barrier are aligned with their respective value of $T_c$ (see Table \ref{criticalparamaters}).

\section{\label{limitingtemperature}Limiting temperature}
Determination of the temperature at which a hot nucleus drop will undergo multifragmentation by loosing its entity, is one of a challenging problem in nuclear physics. Experimentally it is difficult to estimate  $T_{l}$  and other related properties such as specific heat for a particular nucleus as there are large number of nucleons involved. Although, theoretically one can study these properties by applying appropriate constrains. In that context, this section considers a simplistic approach to determine the $T_{l}$ of a nucleus. Employing our assumption stated in  \textbf{Chapter \eqref{methodology}},  and solve  Eqs. \eqref{coexcondition}. These Equations will not have any solution for a given T,  $\rho_v$ and $\rho_l$  for temperature greater than $T_{l}$ signifying that the nucleus can no longer exist. 
\begin{figure}
    \centering
    \includegraphics[scale=0.5]{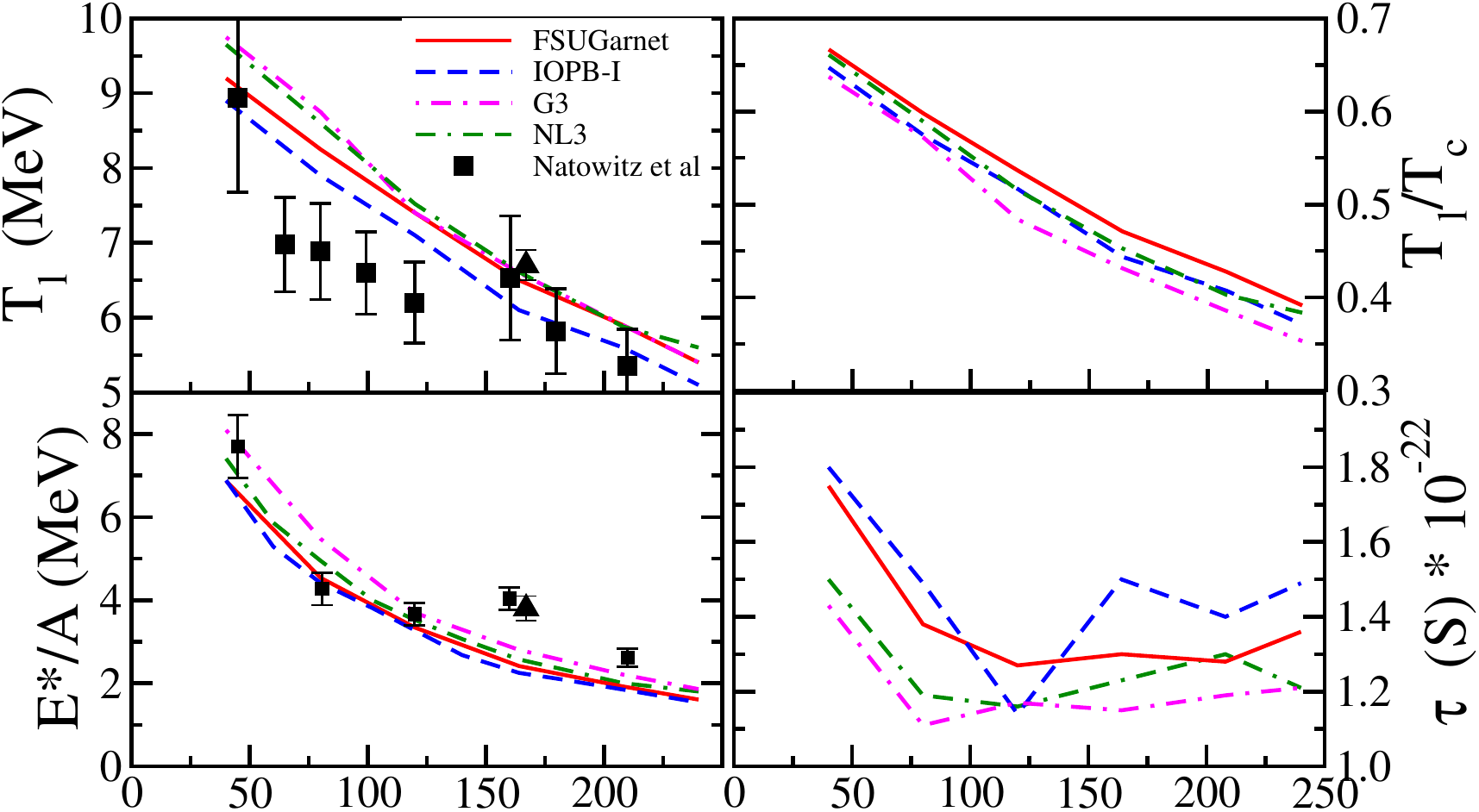}
    \caption{The limiting temperature T$_l$, the ratio of $\frac{T_l}{T_c}$, limiting excitation energy per nucleon and life time of nuclear liquid drop at the limiting temperature as a function of mass number A for the nuclei on $\beta$-stability line. The temperature dependent expression used here is Eq. \eqref{s1}. Experimental points in solid square are taken from \cite{natowitz} for T$_{l}$ which are calculated using double isotope yield ratio and thermal bremsstrahlung measurements and from \cite{PhysRevC.65.034618} for excitation energy. The points represented in upper triangle are taken from  the fisher droplet model derived from \cite{fisher}. }
    \label{tlim}
\end{figure}
\begin{figure}
    \centering
    \includegraphics[scale=0.4]{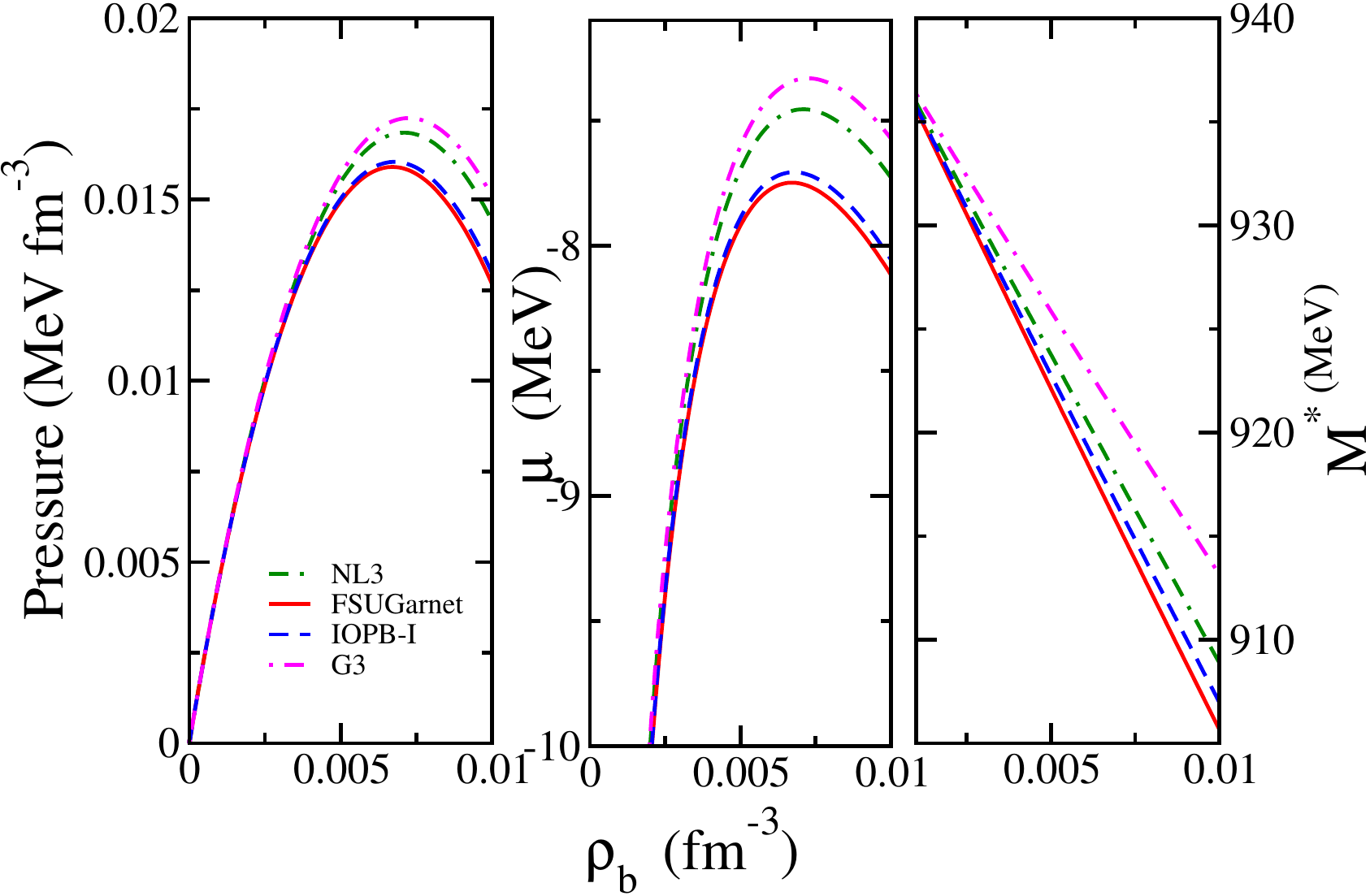}
    \caption{EoS, Chemical potential and effective mass at low density at T=5 MeV for the NL3, IOPB-I, FSUGarnet and G3 parameter sets.}
    \label{forceprop}
\end{figure}
Fig. \ref{tlim} shows  the variation of limiting temperature $T_{l}$, $T_{l}/T_c$, limiting excitation energy ($E^*(T_{l})/A$) and the life time ($\tau$)  of nucleus at limiting temperature as a function of mass number for the nuclei along $\beta$ stability line where the atomic number can be written as \cite{a200} 
\begin{equation}
    Z=0.5 A -0.3\cross10^{-2}A^{\frac{5}{3}}.
\end{equation}
The value of  $T_{l}$ decreases exponentially with increasing mass number as the Coulomb energy rises due to larger Z. At lower Z,  $T_{l}$ decreases at faster pace because the Coulomb component dominates the surface and symmetry energy of liquid drop. At a higher mass number, the situation becomes a little different. There is competition between Coulomb, surface, and symmetry terms. On moving from low to higher mass number along the $\beta$ stability line, the $Z/A$ ratio decreases. The decrease in the $Z/A$ ratio weakens the A dependence causing  $T_{l}$ to increase. On the other hand, the symmetry and surface energy increase with the increase in mass number which tries to bring down the $T_{l}$. For comparison, points determined from phenomenological analysis \cite{natowitz, PhysRevC.65.034618, fisher} for the T$_{l}$ and $E^*(T_{l})$ are also shown. The results from E-RMF forces are within reasonable agreement. 

The value of $T_{l}$ for a particular nucleus and a particular EoS depends on the $T_{c}$ of infinite nuclear matter  and the low density ( $\rho_0<0.01$) variation of EoS which determine the properties of surrounding gaseous phase.  Since the finite-size corrections are employed externally, they are the same for every EoS. To understand the effect of EoS, Fig. \ref{forceprop} shows
the EoS, chemical potential ($\mu$), and effective mass ($M^*$) calculated using
the FSUGarnet, IOPB-I, G3, and NL3 parameter sets for the density range significant for nuclear vapor surrounding the hot nucleus. Chemical potential is a function of temperature-dependent effective mass which consequently determines the chemical equilibrium between nuclear gas and nuclear drop. The IOPB-I and FSUGarnet have similar ground state saturation properties and they have similar behaviour at T=5 MeV.  The incompressibility of the NL3 and G3 sets are 271.38 and 243.96 MeV, respectively, but their behaviour is opposite in the low-density regime. G3 set estimates the maximum value of pressure, and effective mass at any given density. This is the reason G3 set have larger value of T$_c$ than the NL3 set. This trend in Fig. \ref{forceprop} for different EoS,  validates the variation of   $T_{l}$ in Fig. \ref{tlim}, where the magnitude of T$_l$ explicitly depends upon the low density EoS. In other words, to understand the effect of EoS on the $T_{l}$ one has to take into account the $T_{c}$ and low density behaviour of EoS instead of incompressibility at saturation.

Additionally, the ratio $T_{l}/T_{c}$ represents the finite size effect, which denotes how $T_l$ behaves in comparison to $T_c$. When there are no finite size effects, $T_l$ will be equal to $T_c$. It reduces up to  0.3$T_{c}$ for heavy nuclei. Furthermore, there is still model dependence in the $T_l/T_c$ . The larger effective mass yields smaller $T_l/T_c$ which is  clear from the fact the FSUGarnet and G3 estimate the largest and smallest $T_l/T_c$.  Limiting excitation energy per nucleon is calculated at $T_{l}$ and  calculations from E-RMF forces agree with the phenomenological calculation \cite{PhysRevC.65.034618}. These calculations are also performed using Eq.\ \eqref{s1} as there were no significant difference between the values of $T_{l}$ calculated from Eqs.\ \eqref{s1} and \eqref{s2}. However, Eq.\ \eqref{s2}  estimates the  larger excitation energy for a given nucleus as compared to Eq.\ \eqref{s1}. Eqs.\ \eqref{s1} and \eqref{s2} are frequently used in various calculations such as statistical equilibrium analysis and supernovae matter. In that context,  these equations correctly estimate the finite nucleus observables with slight difference in magnitude.  Eq.\ \eqref{s1} has a slight edge as it is  consistent with the surface energy estimated from thermal Hartree-Fock approximation \cite{SAUER1976221}. The present calculations show better agreement with  experimental and theoretical values when using Eq.\ \eqref{s1} as well. However, the judicious use of these can be made depending on the problem such as supernova where the thermal energy plays a very important role.

 \begin{figure}
  \centering
\subfloat[]{%
  \label{fixz}
  \includegraphics[height=6cm,width=.49\linewidth]{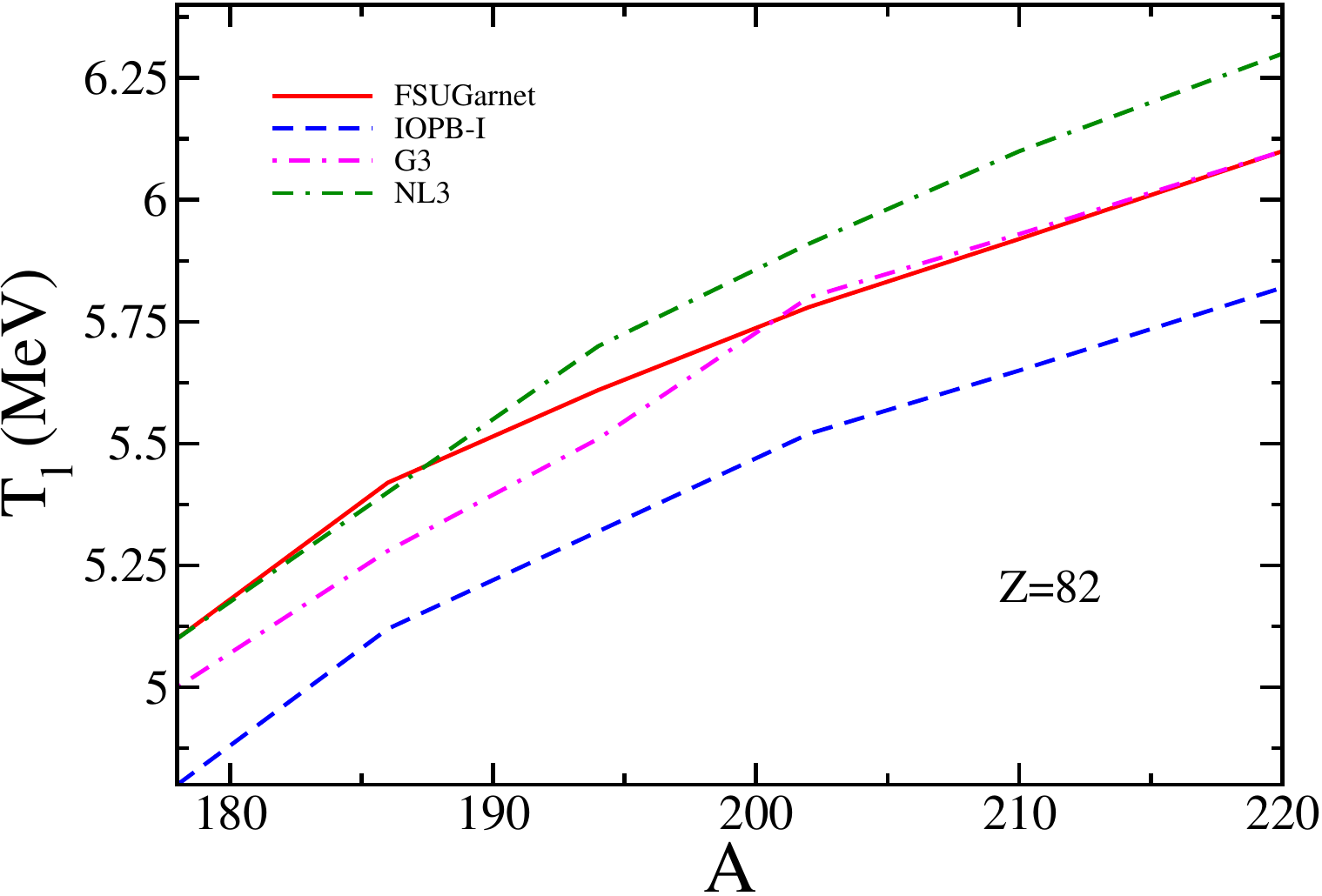}%
}\hfill
\subfloat[]{%
  \label{fixn}
  \includegraphics[height=6cm,width=.49\linewidth]{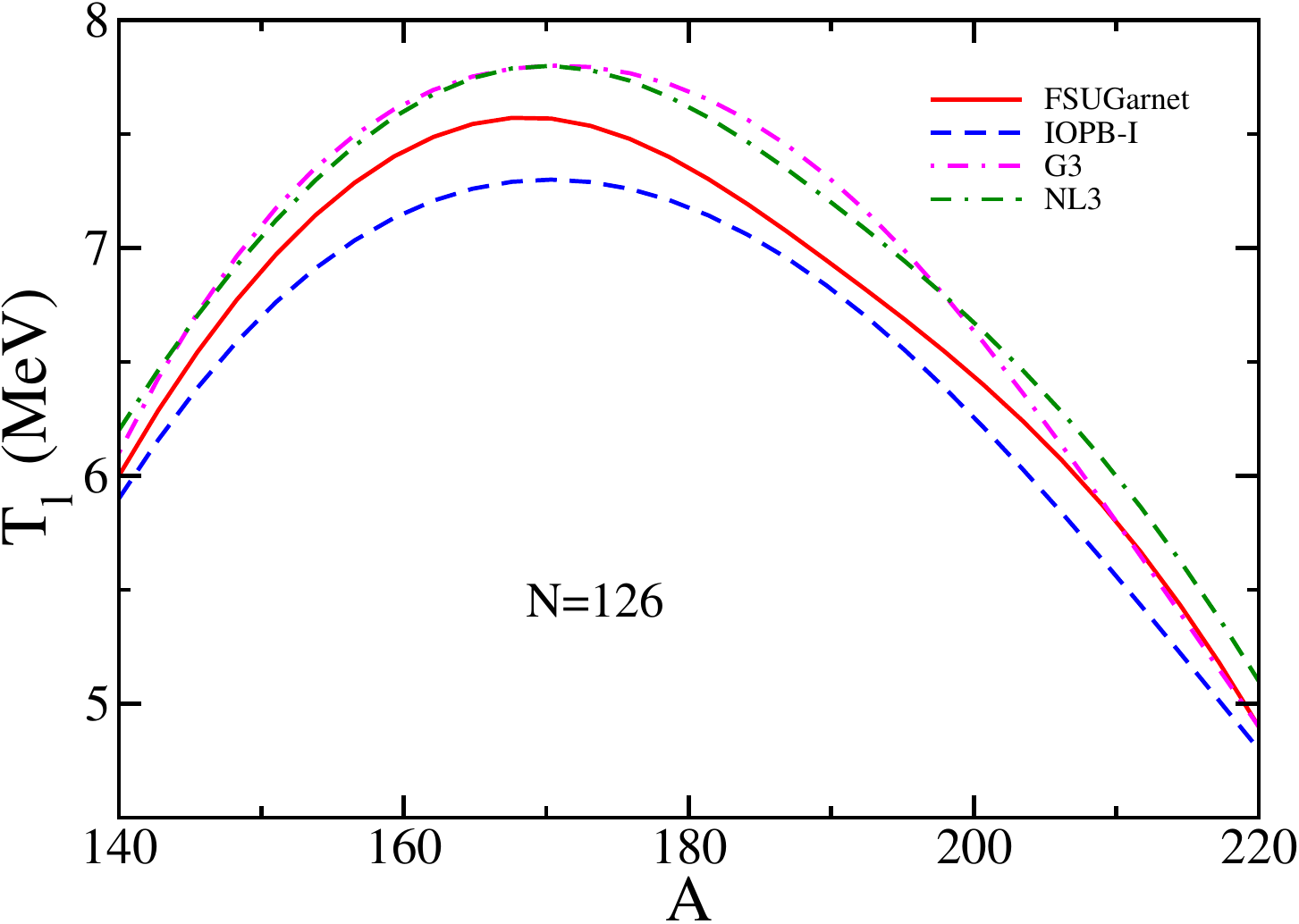}%
} 
\caption{(a) Limiting temperature for fix atomic number Z=82 as a function of mass number calculated from the expression \eqref{s2} (b) Same as in (a) but for fix neutron number N=126. }
\end{figure}

To further understand the behaviour of  $T_{l}$, the lifetime of hot nucleus ($\tau$)  using Eq.\ \eqref{timeeq} is calculated. As the temperature dependence of neutron-capture cross-section is not considered, these values will slightly underestimate the lifetime but the trend will remain the same. The radius R which is the input for Eq.\ \eqref{timeeq}   is determined after solving the coexistence Eqs.\ \eqref{coexcondition} for a particular nucleus. It is seen that the nuclear gas surrounding the nuclear liquid plays a significant role in  determining the T$_l$. In terms of lifetime, a larger pressure and smaller density  corresponds to a less stable liquid drop and therefore, lower lifetime. The IOPB-I set that estimates the lower T$_l$ for a given nucleus yields the higher lifetime.  It is apparent that the lifetime $\tau$ is of the order of $10^{-22}$S at $T_{l}$ for all the nuclei on the $\beta$ stability line. Nuclei at the lower mass range are slightly more stable than heavy nuclei. This time scale is just enough for a nucleus to allow thermalization. This also states the fact that at $T_{l}$ the nucleus is highly unstable and will undergo violent multifragmentation which has the time scale of $10^{-22}$s \cite{KARNAUKHOV200691, lifetimevalue}.

In Fig. \ref{fixz}, the variation of $T_{l}$ is shown for a fixed atomic number Z=82 and  Fig. \ref{fixn} demonstrates the behaviour for a fixed neutron number N=126. For a fixed atomic number, the $T_{l}$ rises $\approx$ 1.5 MeV when one move from A=178 to A=220 or from  $Z/A$ = 0.46 to 0.37. The increase in $T_{l}$ with a decrease in $Z/A$ ratio is because Coulomb free energy reduces as the radius of nuclear liquid drop increases as a function of charge number. The surface energy then dominates over the Coulomb energy which helps in preserving the surface of the drop at a much higher temperature. This trend is confirmed with the non-relativistic Hartree-Fock calculation where the solution becomes unstable after a certain temperature \cite{BONCHE1985265}. When one keeps the neutron number fixed, there is an interesting binodal type trend in the values of $T_{l}$ with increasing mass number. $T_{l}$ increases with increasing Z and reaches its maximum at A $\approx$ 170. It then decreases at a faster rate on further increasing the value of Z. This effect is the result of competition between Coulomb and surface energy at lower and higher mass region. This shape of the graph then signifies that a more stable isotope will survive the larger temperature $\approx 7$  $MeV$. In Figs. \ref{fixz} and \ref{fixn}, the trends of EoS are similar to the ones obtained at low density regime.

\section{\label{correlation}Correlations}
In the analysis of a hot nucleus and its limiting temperature, it was seen that the critical temperature $T_c$ of infinite nuclear matter affects the observables through Eqs.\ \eqref{s1} and \eqref{s2}. They also depend on the properties of EoS such as effective mass and  low density behaviour of a particular EoS. The $T_c$ which is basically an inflation point on critical isotherm, is one of the most uncertain parameter in nuclear matter studies. The value of $T_c$ is an important factor in calculation of finite nuclei as well as supernovae matter and neutron star crust \cite{PhysRevC.79.035804}. Hence it becomes important to relate the $T_c$ of a particular EoS to its saturation properties. In the previous chapter, it was observed that the critical temperature $T_c$ is not a  well constrained quantity.  It requires a comprehensive statistical analysis of nuclear  properties at critical points and saturation properties of cold nuclear matter as their analytical relationship is difficult to establish.  For this, the present work considers fifteen E-RMF  parameter sets  satisfying relevant constrains \cite{duttra, iopb, vishalasymmetric, Quddus_2018, bka} on EoS and first of all calculate the properties at critical point of LGPTin infinite matter.

%%%%%%%%%%%%%%%%%%%%%%%%%%%%%%%%
\begin{sidewaystable}\renewcommand{\arraystretch}{1.05}
\tabcolsep 0.2cm \caption{ The zero temperature incompressibility K, binding energy $e_0$, saturation density $\rho_0$, effective mass M$^*$ and  critical temperature $T_c$, pressure $P_c$, density $\rho_c$ along with flash temperature $T_f$, density $\rho_f$, incompressibility $C_f$ and effective mass at $T_c$ for infinite symmetric nuclear matter using the several forces.} \label{criticalparamaters}
%\begin{ruledtabular}
\hspace{1.5cm}
%\begin{adjustbox}{angle=90}
\begin{tabular}{ ccccccccccccc}
 \hline
\hline
  Parameter & K      & $e_0$     &  $\rho_0$    &  $m^*/m$    & $T_c$    & $P_c$    & $\rho_c$     & $T_f$     & $\rho_f$     & C$_f$ & $m^*_c/m$   \\
  & MeV     & MeV     &  fm$^{-3}$    &      & MeV    & MeV fm$^{-3}$    & fm$^{-3}$     & MeV     & fm$^{-3}$     &   & \\ 
  
\hline
  
 G2   \cite{g2}        & 215.00 & -16.10 & 0.153 & 0.664 & 14.30 & 0.181 & 0.0432 & 11.80 & 0.080 & 0.293 & 0.879 \\
IOPB-I   \cite{iopb}   & 222.65 & -16.10 & 0.149 & 0.593 & 13.75 & 0.167 & 0.0424 & 11.20 & 0.071 & 0.286 & 0.864 \\
Big Apple \cite{bigapple}  & 227.00 & -16.34 & 0.155 & 0.608 & 14.20 & 0.186 & 0.0441 & 11.45 & 0.073 & 0.297 & 0.876 \\
BKA22   \cite{bka}   & 227.00 & -16.10 & 0.148 & 0.610 & 13.90 & 0.178 & 0.0442 & 11.33 & 0.072 & 0.290 & 0.855 \\
BKA24  \cite{bka}    & 228.00 & -16.10 & 0.148 & 0.600 & 13.85 & 0.177 & 0.0450 & 11.31 & 0.073 & 0.284 & 0.845 \\
FSUGarnet  \cite{iopb} & 229.50 & -16.23 & 0.153 & 0.578 & 13.80 & 0.171 & 0.0430 & 11.30 & 0.071 & 0.288 & 0.850 \\
FSUGold  \cite{fsugold} & 230.00 & -16.28 & 0.148 & 0.600 & 14.80 & 0.205 & 0.0460 & 11.90 & 0.074 & 0.301 & 0.844 \\
IUFSU   \cite{iufsu}   & 231.31 & -16.40 & 0.155 & 0.610 & 14.49 & 0.196 & 0.0457 & 11.73 & 0.074 & 0.296 & 0.862 \\
FSUGold2 \cite{fsugold2}  & 238.00 & -16.28 & 0.151 & 0.593 & 14.20 & 0.187 & 0.0450 & 11.51 & 0.073 & 0.293 & 0.855 \\
BKA20   \cite{bka}   & 240.00 & -16.10 & 0.146 & 0.640 & 15.00 & 0.209 & 0.0458 & 11.91 & 0.073 & 0.304 & 0.868 \\
G3    \cite{iopb}      & 243.96 & -16.02 & 0.148 & 0.699 & 15.30 & 0.218 & 0.0490 & 12.10 & 0.075 & 0.291 & 0.879 \\
NL3* \cite{nl3*}      & 258.27 & -16.31 & 0.150 & 0.590 & 14.60 & 0.202 & 0.0466 & 11.70 & 0.075 & 0.297 & 0.861 \\
Z27v1   \cite{z27v}    & 271.00 & -16.24 & 0.148 & 0.800 & 18.03 & 0.304 & 0.0515 & 13.70 & 0.077 & 0.327 & 0.914 \\
NL3  \cite{iopb}       & 271.38 & -16.29 & 0.148 & 0.595 & 14.60 & 0.202 & 0.0460 & 11.80 & 0.070 & 0.301 & 0.846 \\
TM1   \cite{tm1}      & 281.10 & -16.26 & 0.145 & 0.630 & 15.60 & 0.236 & 0.0486 & 12.09 & 0.076 & 0.311 & 0.862 \\
\hline

Exp/Emp & 240  \cite{duttra}   & -16 \cite{vishalasymmetric} & 0.166 \cite{vishalasymmetric}  & 0.63  \cite{FURNSTAHL1998607}  & 17.9 \cite{tcexp} & 0.31\cite{tcexp} & 0.06 \cite{tcexp} & -  & -& 0.288 \cite{vishalsymmetric} & -\\
 & $\pm$ 20 & $\pm$ 1  & $\pm$ 0.019  & $\pm$ 0.05  & $\pm$ 0.40  & $\pm$ 0.07 &  $\pm$ 0.01 & -  & - & -& -\\
\hline
\hline
    \end{tabular}
%\end{adjustbox}
\\
%\end{ruledtabular}
\end{sidewaystable}
%%%%%%%%%%%%%%%%%%%%%%%%%%%%%%%%%
Table \ref{criticalparamaters}  presents the saturation properties of cold nuclear matter i.e. incompressibility (K), binding energy ($e_0$), saturation density ($\rho_0$), effective mass (M$^*$) and  critical temperature ($T_c$), pressure ($P_c$), density ($\rho_c$) along with flash temperature ($T_f$), density ($\rho_f$), incompressibility factor ($C_f$) and effective mass at ($T_c$) for infinite symmetric nuclear matter using different force parameters. For further details on these quantities please see Ref. \cite{vishalsymmetric}. For the correlation analysis, this work considers variety of forces with different meson couplings, which include up to the quartic order scalar and vector terms  in order to have a generalized analysis of E-RMF forces.  The E-RMF sets satisfying the allowed incompressibility range and other observational constrains underestimate the critical values of temperature, density and pressure when compared to experimental data \cite{tcexp}.

\begin{figure}
    \centering
    \includegraphics[scale=0.45]{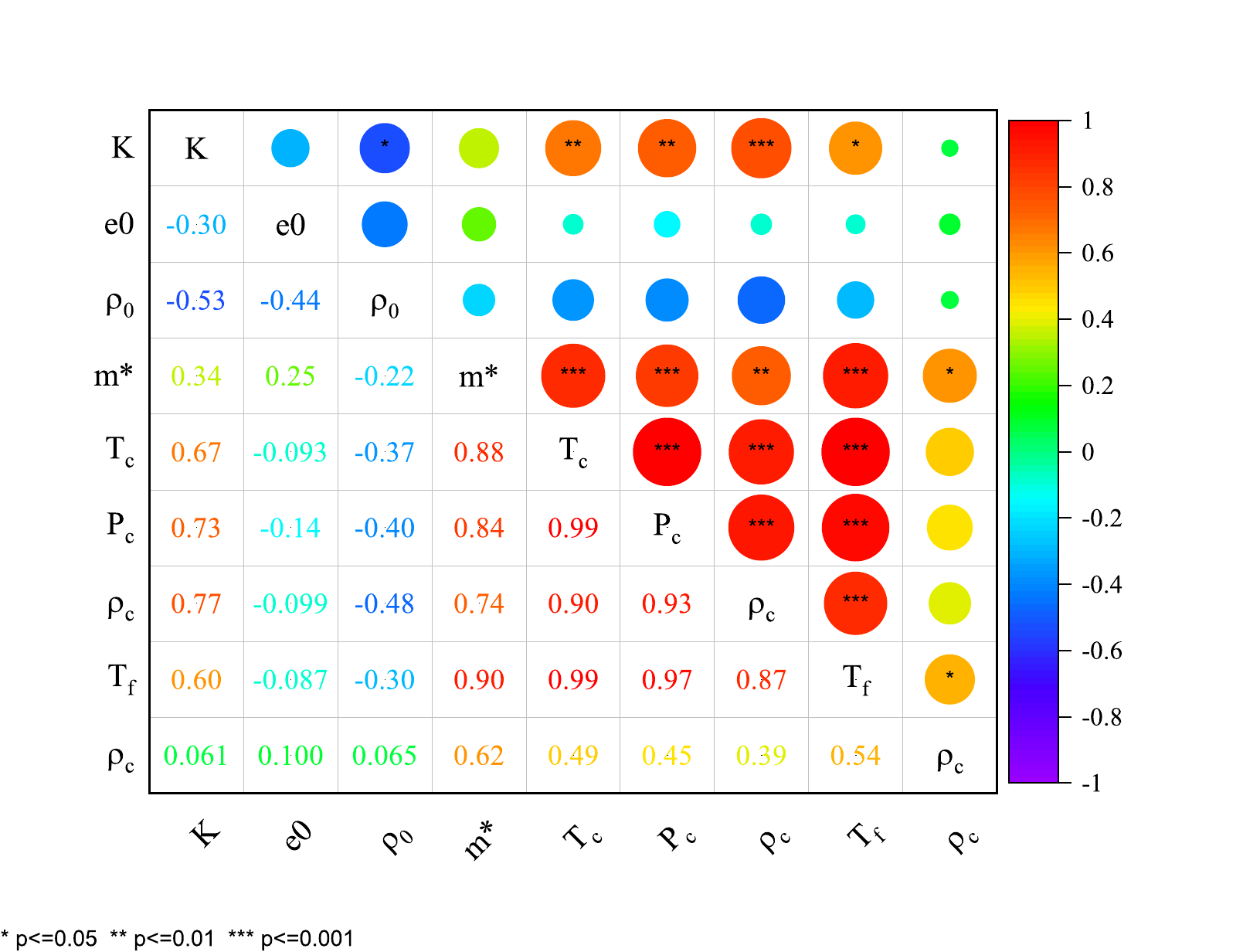}
    \caption{The Pearson correlation matrix for the critical parameter for infinite symmetric nuclear matter and some cold nuclear matter properties. The number of stars in a circle represents the p-value given at the bottom. The strength of correlation is colour mapped.}
    \label{tctp}
\end{figure}

Next, the Pearson Correlation matrix \cite{pearson} for variables calculated in Table \ref{criticalparamaters} is calculated and the results are  shown in Fig. \ref{tctp}. The colour coded correlation  matrix also shows the statistical significance in form of p-value \cite{pearson} for different confidence interval i.e. 95\%, 99\% and 99.9\%. The binding energy ($e_0$) and saturation density ($\rho_0$) of cold infinite nuclear matter have very weak strength of correlation with the critical properties at finite temperature. This is against the natural intuition that binding energy of infinite matter should impact  the $T_c$.

The incompressibility on the other hand shows positive correlation with critical properties i.e. $T_c$, $p_c$, $\rho_c$ and $T_f$. Although this correlation does not exceed the value of 0.77. Therefore, it  can be concluded that the saturation properties of cold nuclear matter do not significantly impact the value of critical parameter individually. The reason for this can be the fact that saturation properties are calculated at saturation density $\rho_0\approx$ 0.16 fm$^{-3}$, whereas, the nuclear matter convert from liquid to gaseous phase at $\approx$ 0.25-0.3 $\rho_0$. The behaviour of EoS in this density region is not always as per the properties at saturation, as noticed  in Fig. \ref{forceprop}. One exception is the effective mass which shows a strong positive correlation with critical properties. This is in line with our analysis in the previous chapter  of  infinite nuclear matter that finite temperature properties in E-RMF formalism are  governed by the effective mass. This behaviour is consistent with the non-relativistic formalisms as well, although the definition of effective mass is different in both the cases \cite{vishalasymmetric}.  

From Table \ref{criticalparamaters}, we see that the parameter sets G3 and Z27v1 have relatively high effective mass and a high value of $T_c$. A high positive correlation between $m^*$ and $T_c$ in Fig. \ref{tctp} suggests the same. Therefore, one way to construct a model at par with experimental findings is to exploit this property of effective mass. This fact was also considered in \cite{PhysRevC.94.045207}. However, the prescribed range of effective mass 0.58 $\le$ m$^*$/m $\le $ 0.68 in agreement with spin-orbit splitting experiments \cite{FURNSTAHL1998607} should be kept in mind. The Z27v1 set does not satisfy this constrain and it was also not considered in \cite{duttra}, from where the constrains on EoS are taken for this study. Therefore, no standard RMF and E-RMF parameter sets, that satisfy all the available constrains can reproduce the experimental value of the critical parameter for infinite nuclear matter and hence needs more analysis especially on the low-density regime of EoS. Moreover, the effective mass dependence of thermal properties will also be useful in the microscopic calculations, where the concept of $T_c$ is not explicitly used for the surface energy calculation.

\begin{figure}
    \centering
    \includegraphics[scale=0.5]{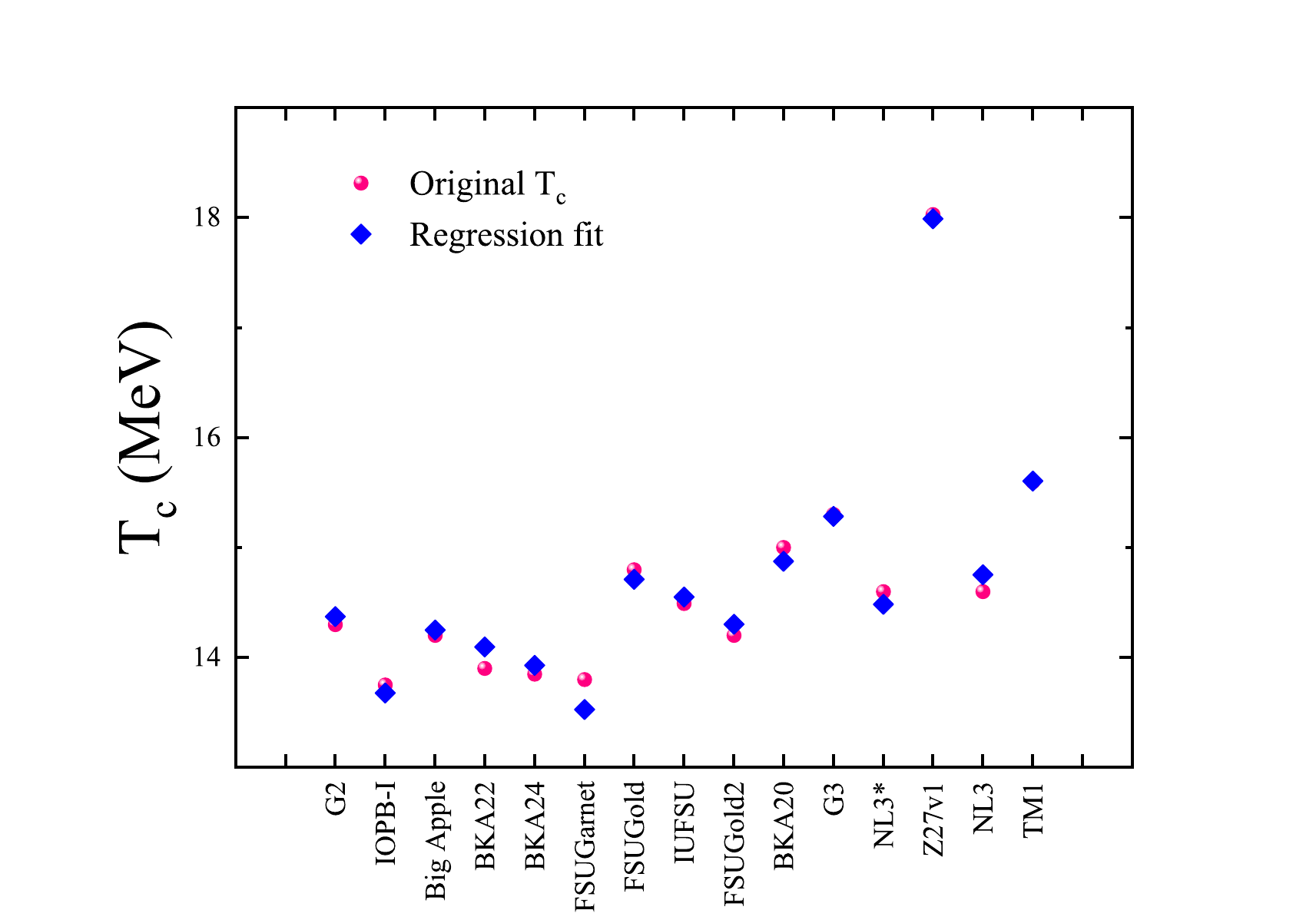}
    \caption{Actual value of $T_c$ from different forces and regression fit values calculated from Eq.\ \eqref{regfiteq}.}
    \label{regfit}
\end{figure}

The low correlation means that the variables are acting as independent parameters. This is also justified as the properties like K, $\rho_0$, $e_0$, and $m^*$ are the inherent characteristic properties of an EoS.   The critical temperature therefore can be understood as a  result of competition between various nuclear matter observables. To demonstrate this,  a very simple multiple linear regression (MLR) fit is considered of the following form.

\begin{equation}
\label{regfiteq}
    T_c= \beta_0+ \beta_1K+\beta_2e_0+\beta_3\rho_0^{(1/3)}+\beta_4m^*,
\end{equation}
where, all the variables are in MeV except $\rho_0$ which is in MeV$^3$ and coefficients have relevant dimensions with $\beta_{0,1,2,3,4}$=-11.5033, 0.00201, -4.32248, -0.52433, 0.01795. These coefficients are statistically significant as well for 95 \% confidence interval.  In Fig. \ref{regfit},  the result of Eq.\ \eqref{regfiteq} against the actual $T_c$ from Table \ref{criticalparamaters} is shown. The regression equation estimate the  $T_c$ excellently with R-square=0.987. The fitted regression equation suggests that the binding energy and saturation density has opposite variation with  $T_c$.  The regression equation \ref{regfiteq} is better than the empirical relations suggested in \cite{RIOS201058} based on Lattimer–Swesty and Natowitz predictions. This is because the larger degrees of freedom are considered in this equation. However, this will yield a strange value of $T_c$ when all the saturation properties tend to zero. This equation gives an useful insight in the form of free coefficient $\beta_0$ which  suggests that there is a missing link between our current understanding of critical temperature and its relationship with the saturation properties. The $\beta_0$ becomes inevitable as the equation then gives a bad fitting. 

Unlike saturation and critical properties, the critical parameters are strongly correlated with each other, except the flash density $\rho_f$. The flash density seems to be model independent with standard deviation $=$ 0.0025.  It is to note that these correlations are for the E-RMF sets considered in Table \ref{criticalparamaters} and are not universal. However, the selected parameter sets have a wide range of meson couplings and nuclear matter saturation properties. Moreover, the results are presented with the statistical significance of Pearson  correlation to make it as general as possible. 

\begin{table}[]
    \centering
        \caption{Limiting temperature $T_l$ ($MeV$), chemical potential $\mu$ ($MeV$), pressure $P$ ($MeV$ fm$^{-3}$), gas density $\rho_v$ ($fm^{-3}$), liquid density $\rho_l$ ($fm^{-3}$), radius $R$ ($fm$) and lifetime $\tau$ ($\tau\cross\exp{-22}$ Sec) of $^{208}$Pb nucleus for several forces.}
    \begin{tabular*}{\linewidth}{c @{\extracolsep{\fill}}  ccccccccc}
 \hline
 \hline
  Parameter  & $T_l$   & $\mu$      & P & $\rho_v$    & $\rho_l$  & R & $\tau$ \\
\hline
 G2            & 5.4  & -8.55  & 0.0162   & 0.0075  & 0.147 & 6.964 & 1.49 \\
IOPB-I        & 5.88 & -9.61   & 0.0200     & 0.0084  & 0.143 & 7.028 & 1.29 \\
Big Apple     & 5.37 & -8.36 & 0.0191   & 0.0075 & 0.148 & 6.948& 1.47\\
BKA22         & 5.46 & -8.65  & 0.0197   & 0.0076  & 0.142 & 7.045 & 1.42\\
BKA24         & 5.51 & -8.73  & 0.0196   & 0.0075  & 0.142 & 7.045 & 1.40\\
FSUGarnet     & 5.9  & -9.48   & 0.0240    & 0.0082  & 0.148 & 6.948 & 1.28 \\
FSUGold       & 5.92 & -9.21   & 0.0239   & 0.0085  & 0.143 & 7.028 & 1.18\\
IUFSU         & 5.69 & -8.97  & 0.0224   & 0.0081  & 0.149 & 6.933 & 1.34\\
FSUGold2      & 5.59 & -8.88   & 0.0207   & 0.0078  & 0.145 & 6.996 & 1.38 \\
BKA20         & 5.85 & -9.03 & 0.0238   & 0.0085 & 0.140  & 7.078 & 1.18\\
G3            & 5.9  & -9.22   & 0.0245   & 0.0087  & 0.141 & 7.061 & 1.19\\
NL3*          & 5.74 & -9.08  & 0.0220    & 0.0082 & 0.144 & 7.012 & 1.30\\
Z27v1         & 6.95 & -10.49  & 0.0369   & 0.0110   & 0.14  & 7.078 & 0.80\\
NL3           & 5.88 & -9.17   & 0.0213   & 0.0084 & 0.144 & 7.012 & 1.21\\
TM1           & 5.85 & -8.63 & 0.0250    & 0.0086 & 0.138 & 7.112 & 1.09\\
\hline 
\hline
    
    \end{tabular*}
\label{tlimforces}
\end{table}

After establishing the relationship between critical properties and saturation properties of cold nuclear matter, let us extend these correlations to limiting properties. Table \ref{tlimforces} presents the values of $T_{l}$, chemical potential $\mu$ , pressure (P), gas density ($\rho_g$), liquid density ($\rho_l$), radius (R) and lifetime ($\tau$) of $^{208}$Pb nucleus for the forces considered in Table \ref{criticalparamaters}. To establish the relation of different properties in Table \ref{tlimforces}, the correlation  matrix is calculated for limiting properties of $^{208}$Pb nucleus, critical properties of infinite nuclear matter $T_c$ and saturation properties of cold nuclear matter. 

\begin{figure}
    \centering
    \includegraphics[scale=0.5]{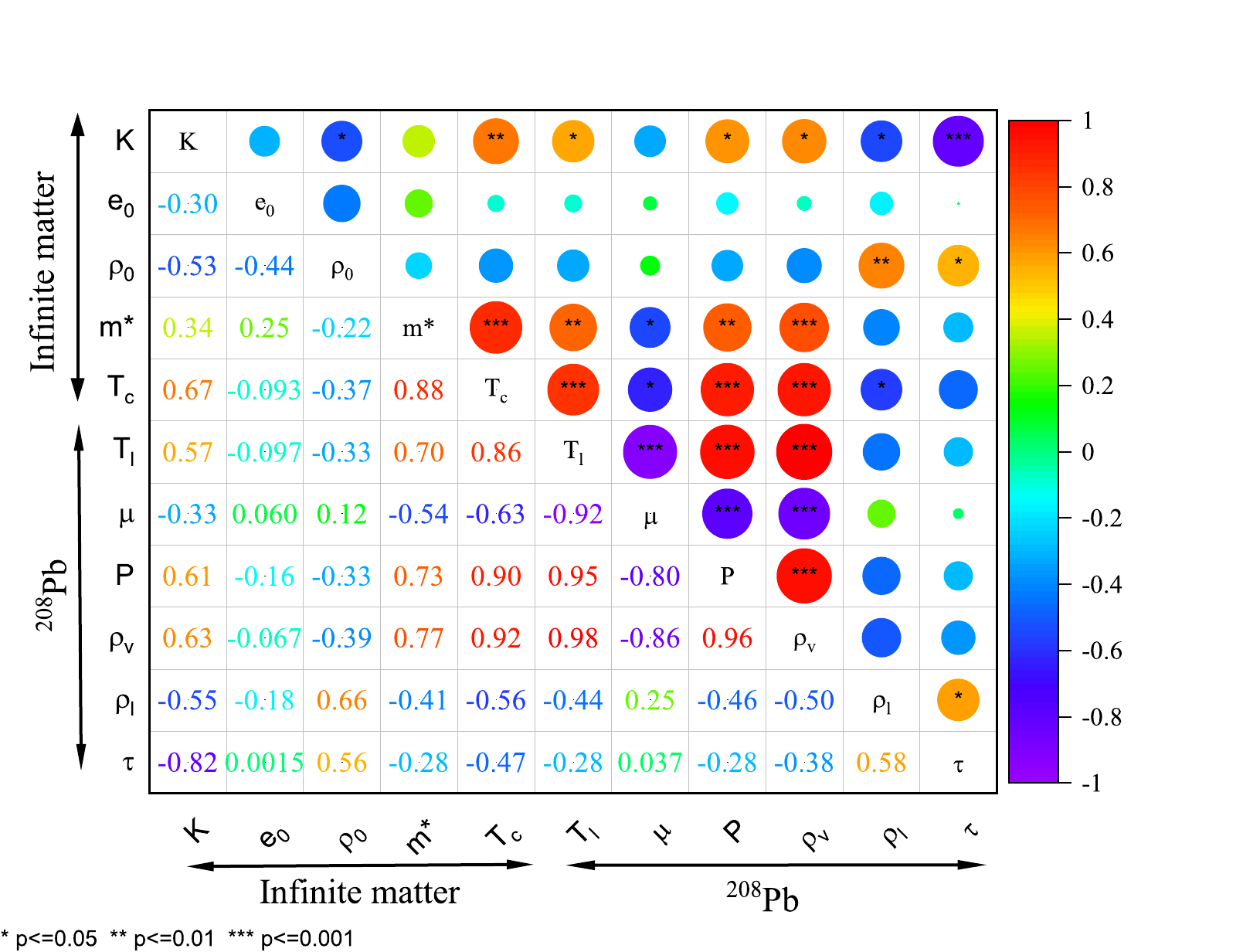}
    \caption{ The Pearson correlation matrix for the critical parameter for infinite symmetric nuclear matter, some cold nuclear matter properties and limiting properties for the $^{208}$Pb. }
    \label{tctl}
\end{figure}

Once again, the binding energy and saturation density of cold nuclear matter is weakly correlated with the limiting properties. The incompressibility  shows a weak correlation with the limiting properties which is in agreement with the analysis of  the low density behaviour of EoS. However, it is  correlated negatively with the lifetime of nucleus. This is justified as the stiff EoS corresponds to the larger pressure, which in turn make the nucleus less stable surrounded in a nucleon gas. 
The effective mass is strongly correlated with the limiting properties. A strong correlation between T$_c$ and effective mass then suggest that the limiting properties of a nucleus essentially depend on the T$_c$ and M$^*$ of the model applied. This statement has a far reaching implication as the majority of the calculations employing statistical thermodynamics as well as compress liquid-drop model (CLDM) in astrophysical applications heavily depend on the value of T$_c$ for surface energy. Also in the microscopic calculations where the surface energy is determined using the derivative of mean-fields, effective mass plays the determining role. On the other hand, the limiting properties for $^{208}$Pb i.e. limiting temperature (MeV), chemical potential (MeV), pressure (MeV fm$^{-3}$), gas density ($fm^{-3}$), liquid density ($fm^{-3}$) and radius ($fm$) are tightly correlated. A higher  $T_{l}$  means that the chemical potential will be smaller and the equilibrium pressure and gas density will be larger.

\section{Summary}
 
In summary, the E-RMF framework is used to analyze the thermal properties of hot nuclei. The free energy of a nucleus is estimated by using temperature and density-dependent parameters of the liquid-drop model. The surface free energy is parametrized  using two approaches based on the sharp interface of the liquid-gaseous phase and the semi-classical Seyler-Blanchard interaction. The later parametrization estimates relatively stiff behaviour of excitation energy, entropy, and fissility parameter. The estimations of these properties are in reasonable agreement with the available theoretical microscopic calculations and experimental observations. 

It has been observed that the thermal properties of the finite nuclear system are influenced strongly by the effective mass and critical temperature ($T_c$)  of the E-RMF parameter sets employed. A larger effective mass corresponds to the higher excitation energy, level density, limiting temperature, and limiting excitation energy. The limiting temperature also depends on the behaviour of  EoS at subsaturation densities which helps to calculate the properties of surrounding nuclear gas in equilibrium with the hot nucleus. A stiff EoS at subsaturation density corresponds to the larger limiting temperature. The temperature-dependent liquid-drop fission barrier is also influenced by the $T_c$. A larger $T_c$ estimates a larger temperature where the barrier vanishes.

Finally  a detailed correlation matrix analysis to account for the large deviations in the value of critical parameters among various E-RMF sets have been performed.  The effective mass shows a strong positive correlation with the critical parameters namely ($T_c$,  $\rho_c$, $P_c$) and limiting temperature of $^{208}Pb$ nucleus, which is consistent with the analytical analysis. 
The binding energy and saturation density act as independent parameters which prompts us to establish a simple multiple linear regression (MLR) between the $T_c$ and saturation properties of cold nuclear matter. Our MLR equation fits the original  $T_c$ and gives useful relationship between saturation properties and critical temperature.

%The present calculations can be extended to various astrophysical problems. A similar situation is encountered in supernovae explosion and neutron star crust, where the nuclei are surrounded in a nuclear and relativistic electron gas. The model dependence can also be studied within statistical multifragmentation calculations. Furthermore, a comprehensive analysis is required to address the anomaly in the magnitude of the critical temperature of nuclear matter by employing the low-density correction in the EoS. 

\clearpage
\addcontentsline{toc}{section}{Bibliography}
\printbibliography

%% file: Chapter_5/CHAP5.tex
\chapter{\label{chap5} Crustal properties of cold catalyzed non-accreting neutron star}
\section{Introduction}
%%%%

Determining the structure of the neutron star from the surface to interiors in a unified way is one of the main problems in neutron star physics. Apart from a small region of the outer crust, the structure of the neutron star is mainly dependent on the equation of state (EoS). A substantial amount of research has been carried out in the last two decades to constrain the EoS based on many experimental and theoretical observations \cite{Fattoyev_2012, Abbott_2018, Tuhin_2018, Miller_2019, Raaijmakers_2019, Greif_2020}. The GW170817 event \cite{Abbott_2017, Abbott_2018} provides an upper limit on the tidal deformability while the massive pulsars such as PSR J0740+6620 \cite{Cromartie_2019} ,   PSR J0348+0432 \cite{Antoniadis_2013} and PSR J1614–2230 \cite{Demorest_2010} estimate that the maximum neutron star mass should be greater than $2~M_\odot$. There are just a few EoSs which have been used to calculate the neutron star structure in the entire density range within a unified approach and satisfy the relevant constraints \cite{Dutra_2014}. The unified treatment of the neutron star is essential as various properties such as crust-core transition density, pressure, the crustal moment of inertia, etc., are very sensitive to the choice of EoS \cite{Dutra_2021}. These properties and the structure of the crust, which essentially depends on the subsaturation behaviour of EoS, have a significant impact on the transport and thermodynamical properties of the neutron star. This work aims to provide a unified treatment of the structure of the neutron star within the effective relativistic mean-field (E-RMF) approach using the cold catalyzed matter approximation (CCM). The CCM means that the star is in thermal and $\beta-$ equilibrium, valid for any non-accreting neutron star \cite{Miller_2019}. 

In this chapter, the calculations start from the surface of the star with a density greater than $10^{-10}$ fm$^{-3}$ where all the atoms are completely ionized, and electrons form a degenerate Fermi gas. Below this density, the electrons are still bounded to the nuclei, and one can use generalized Thomas-Fermi (TF) theory to calculate the properties of this thin layer \cite{Feynman_1949, Haenel_2007}.  The only input in the calculation of outer crust is the atomic mass evaluations. The recently measured atomic mass evaluation (AME) 2020 mass table \cite{Huang_2021} is used in this work, which is available up to isospin asymmetry of 0.3. Mass evaluations are not possible for more neutron-rich nuclei in the laboratory, so the need to use a mass model arises. For this,  the nuclear mass model calculated from the Hartree-Fock-Bogoliubov (HFB) \cite{Samyn_2002} method using the accurately calibrated Brussels-Montreal \cite{Eya_2017} energy-density functionals, such as, BSk14, BSk24, and BSk26 \cite{hfb14, hfb2426} are used. The HFB approach is a highly precise formalism used in various calculations concerning nuclear masses for the highly neutron-rich nuclei. The onset of neutron drip marks the beginning of the inner crust, which has an intricate structure making it a challenging problem.

To estimate the inner crust structure, the compressible liquid drop model (CLDM) \cite{Mackie_1977, Newton_2021} is  used. The CLDM is recently applied in the work of Refs. \cite{Carreau_2019, Carreau_2020, carreau2020modeling} where the energy-density functional is taken in the form of meta-modeling, a technique developed to mimic the original relativistic or non-relativistic functional using the isoscalar and the isovector energy of the EoS \cite{Margueron_2018} and for the Bayesian inference of neutron star crust properties \cite{Newton_2021}.  The meta-modeling reduces the computational difficulties when studying the statistical properties such as Bayesian inference to constrain the EoS.  Although this formalism reasonably imitates the EoS at low density but deviates at extremely low and high density, thereby estimating different neutron star results as the original EoS. This work uses the technique developed by Carreau {\it {\it et al.}} \cite{Carreau_2019, code} and modifies it to use the exact E-RMF formalism for the calculation of bulk and finite-size contribution of the cluster. This will preserve the underlying properties of a parameter that may otherwise be lost in the meta-modeling.

The aim of this chapter is twofold: First, to develop three unified EoS, namely FSUGarnet-U, IOPB-I-U, and G3-U  with available core EoSs, such as FSUGarnet \cite{Chen_2014}, IOPB-I \cite{Kumar_2018}, and G3 \cite{Kumar_2017}. EoS from the outer crust to the liquid core is constructed using the experimental mass from the AME2020 data \cite{Huang_2021}, mass table of HFB-26 \cite{hfb2426}, available mass excess of neutron-rich nuclei \cite{wolf2013, welker2017, Beck_2021} and the E-RMF sets FSUGarnet \cite{Chen_2015}, IOPB-I \cite{Kumar_2018}, and G3 \cite{Kumar_2017}. Only spherical geometry is assumed for the estimation of inner crust structure in this chapter. Second, it studies the neutron star properties such as the $M-R$ relation, the moment of inertia and the influence of the crust on the moment of inertia in the form of fractional moment of inertia (FMI) which plays an important role to understand the pulsar glitch behaviour \cite{Basu_2018, Eya_2017}. Pulsar glitches are the sudden jump in the spin frequency usually attributed to the depth of their interior superﬂuid from the surface. Therefore, these glitches are related to the crust thickness and act as the laboratory to test the validity of nuclear models.  

%\section{\label{results} Results and Discussions}
%%%%
%%%%%%%%%%%%%%%%%%%%%%%%%%%%%%%%%%%%%%%%%%%%%%%%%%%%%%%%%%%%%%%%%%%%%%%%
\section{Outer crust}
%%%%
In the outer crust of the cold nonaccreting neutron star, the neutron-rich nuclei are embedded in a BCC lattice arrangement, ensuring that the cell's Coulomb energy is minimized. These nuclei are stable against the $\beta-$decay by surrounding uniform relativistic electron gas. To calculate the composition of the outer crust of a neutron star,  the Gibbs free energy in Eq. (\ref{eq:gibbsminimization}) is minimized at fixed pressure where the atomic mass table serves as an input. I use the most recent AME2020 data \cite{Huang_2021} along with the recently measured mass excess of $^{77-79}$ Cu taken from \cite{welker2017}, $^{82}$Zn from \cite{wolf2013} and $^{151-157}$Yb \cite{Beck_2021}  for the known masses and extrapolate them using the microscopic HFB calculation namely HFB-24, HFB-26 \cite{hfb2426}, and HFB-14 \cite{hfb14},  which are based on BSk functional  characterized by unconventional Skyrme forces along with the most recent FRDM(2012) \cite{MOLLER20161} mass table.
%%%%
\begin{figure}
    \centering
    \includegraphics[scale=0.7]{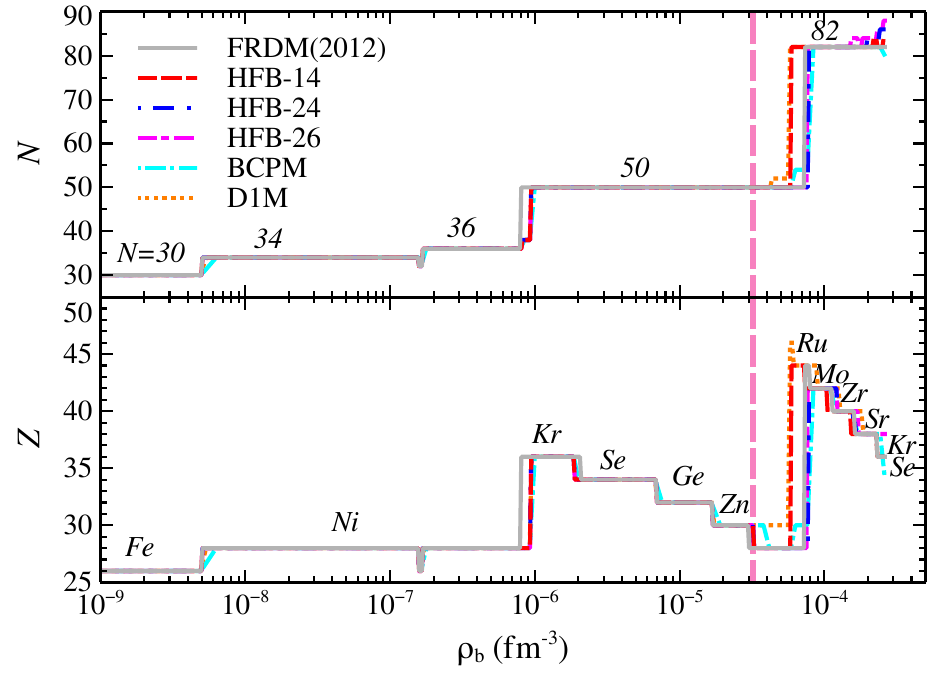}
    \caption{The neutron number ($N$) and proton ($Z$) and  in the outer crust as a function of density. The experimental data are taken from AME2020 when available \cite{Huang_2021}. The unknown mass are taken from microscopic calculations HFB-14 \cite{hfb14}, HFB-24 , HFB-26 \cite{hfb2426} along with the FRDM(2012) mass table \cite{MOLLER20161}. A comparison with BCPM  \cite{BKS_2015} and D1M \cite{sym13091613} is also shown. In addition the experimental mass of $^{82}$Zn \cite{wolf2013}, $^{77-79}$Cu \cite{welker2017} and $^{151-157}$Yb \cite{Beck_2021} are also considered. Vertical dashed line represents the boundary where prediction from experimental masses ends. }
    \label{fig:zndistribution}
\end{figure}
%%%%
%%%%
\begin{figure}
    \centering
    \includegraphics[scale=.8]{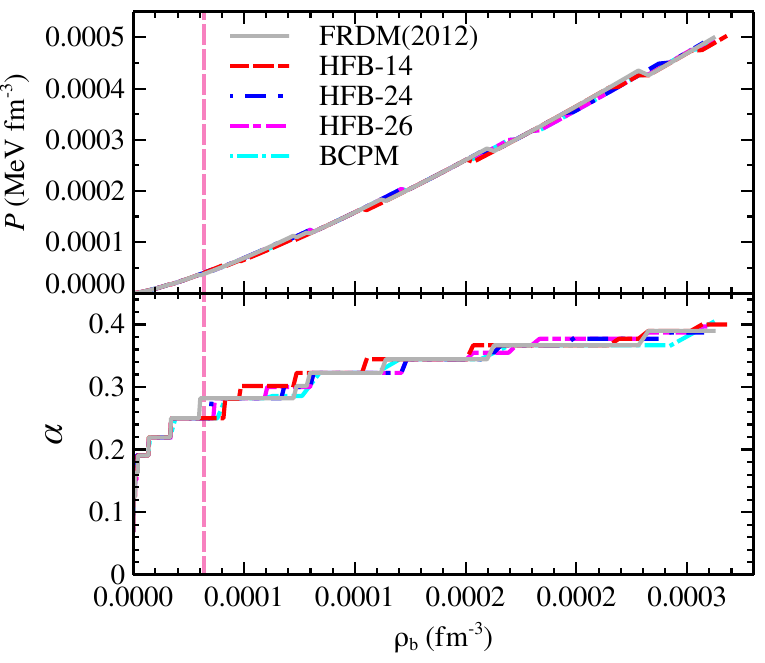}
    \caption{EoS of outer crust is shown for different mass models in the upper panel. The lower panel shows the global asymmetry in the outer crust as a function of density. Vertical dashed line represents the boundary where prediction from experimental masses ends. }
    \label{fig:eosoc}
\end{figure}
%%%%

The composition of outer crust  is shown in Fig. \ref{fig:zndistribution} for the various mass models. In addition to the HFB computed mass excess, it also shows the result from most recent FRDM(2012) \cite{MOLLER20161}, BCPM \cite{BKS_2015} and D1M \cite{sym13091613} Gogny interaction for a comparative analysis. The outermost layer is occupied by the $^{56}$Fe nucleus accompanied by the layer of $^{28}$Ni nucleus in the intermediate densities. The persistent existence of  magic shell nuclei is also visible in $Z=28$ and $N=50$, $82$ plateau due to their enhanced binding energies. The layer of $N=50$ starts at density $\approx 10^{-6}$ fm$^{-3}$ and is characterized by the staircase structure signifying the decrease in atomic number due to the electron capture process. It leads to the appearance of more and more neutron-rich nuclei once we move deeper into the crust. The composition of the outer crust is determined solely from the experimental mass table up to the density $3.2\times 10^{-5}$ fm$^{-3}$ for the HFB-26, which is marked by the dashed vertical line in Fig. \ref{fig:zndistribution}. The composition is model-independent until this density which is clear from the fact that all the curves overlap each other. It may be noted that the value of this density is slightly lower than the value determined from the AME2016 data.

As we move deeper into the outer crust, the need to apply a mass model to calculate the mass excess of extremely neutron-rich nuclei arises as these values are difficult to obtain in a laboratory setup. However, various advanced radioactive beam facilities are working toward measuring the properties of these neutron-rich nuclei in order to have a better understanding of the unconventional regime \cite{Dilling_2006, Beck_2021}. The highly precise HFB calculations and those obtained from the FRDM(2012), BCPM, and D1M predict the appearance of the $N=82$ layer at high density (near the transition to the inner crust), which is also marked by the staircaselike structure.  However, the model dependency is clearly visible in this case. The HFB calculations using HFB-14, HFB-24, and HFB-26 are close to the calculation of highly successful FRDM. For comparison of different models, Table \ref{tab:twonuc} shows the last two layers of the outer crust, where the last element corresponds to the layer just before the transition into the inner crust. In the entire outer crust, one can see a strong effect of closed proton and neutron shells on the composition, except for the outermost layer of $^{56}$Fe nucleus. The existence of nuclei with $Z=28$ and $N=50$ is the consequence of experimental fact whereas, $N=82$ can be treated as the artifact of extrapolation via the microscopic mass table used. In addition to these, there appears a thin layer of $^{121}$Y at the density 0.0001596 fm$^{-3}$ using the HFB-24 mass model. The existence of an odd mass or charge number in the outer crust is not considered in the  calculations of BPS \cite{BPS_1971} and signifies a possible ferromagnetic phase transition in a neutron star. Although one needs a more precise evaluation of the mass of odd-nuclei as it can alter the composition \cite{Pearson_2018} of the outer crust.
%%%%%%%%%%%%%
\begin{table}

\centering
\caption{The last two layers of nucleus in the outer crust predicted from the different models.}
\label{tab:twonuc}
\scalebox{0.92}{
\begin{tabular}{llllllll}
\hline
\hline
Model & Element & $Z$  & $N$  &
\begin{tabular}[c]{@{}l@{}}$ \hfill \rho_{max}$\\ (fm$^{-3}$)\end{tabular}  &
\begin{tabular}[c]{@{}l@{}} \hspace{0.5cm} $P$\\(MeV fm$^{-3}$) \end{tabular} &
\begin{tabular}[c]{@{}l@{}} \hspace{0.5cm} ${\cal E}$\\(MeV fm$^{-3}$) \end{tabular} & \hspace{0.2cm} $\alpha$
\\\hline
\multirow{2}{*}{HFB-14}    &   $^{122}$Sr      & 38 & 84 & 2.2799E-04 & 4.2566E-04 & 0.2137 & 0.377 \\
                           &   $^{120}$Kr      & 36 & 84 & 2.6712E-04 & 5.0108E-04 & 0.2505 & 0.400 \\
\hline                           
\multirow{2}{*}{HFB-24}    &    $^{122}$Sr     & 38 & 84 & 2.3720E-04 & 4.4874E-04 & 0.2224 & 0.377 \\
                           &    $^{124}$Sr     & 38 & 86 & 2.5675E-04 & 4.8804E-04 & 0.2407 & 0.387 \\
\hline                           
\multirow{2}{*}{HFB-26}    &    $^{122}$Sr     & 38 & 84 & 2.2799E-04 & 4.2566E-04 & 0.2137 & 0.377 \\
                           &    $^{126}$Sr     & 38 & 88 & 2.6188E-04 & 4.9052E-04 & 0.2456 & 0.397 \\
\hline                           
\multirow{2}{*}{FRDM}      &$^{120}$Sr         & 38 & 82 & 2.2799E-04 & 4.3515E-04 & 0.2137 & 0.367 \\
                           &$^{118}$Kr         & 36 & 82 & 2.6188E-04 & 4.9909E-04 & 0.2456 & 0.390 \\
\hline                           
\multirow{2}{*}{BCPM}      &$^{120}$Sr         & 38 & 82 & 2.4265E-04 & 4.7276E-04 & 0.2275 & 0.367 \\
                           &$^{114}$Se         & 34 & 80 & 2.6155E-04 & 4.8422E-04 & 0.2453 & 0.404 \\
\hline                           
\multirow{2}{*}{D1M}      & $^{122}$Zr        & 40 & 82 & 1.7990E-04 & 3.3165E-04 &       0.1685 & 0.344 \\
                           &$^{120}$Sr         & 38 & 82 & 2.4420E-04 & 4.7680E-04 &  0.2289      & 0.367 \\
\hline  \hline
\end{tabular}%
}
\end{table}
%%%%
%%%%%%%%%%%%%%outercrust HFB-26%%%%%%%%%%%%%%%%%
\begin{table}
\centering
\caption{The composition and EoS of outer crust. The experimental atomic mass evaluations are taken from AME2020 \cite{Huang_2021} when available. The unknown mass are taken from microscopic calculations HFB-26 \cite{hfb2426} .  In addition the experimental mass of $^{82}$Zn \cite{wolf2013}, $^{77-79}$Cu \cite{welker2017}  and $^{151-157}$Yb \cite{Beck_2021} are also considered. The upper part is obtained from the experimental data and the lower part from the HFB-26 results.}
\label{tab:oceosdata}
\renewcommand{\tabcolsep}{0.3cm}
\renewcommand{\arraystretch}{1.0}
\begin{tabular}{lllll}
\hline
\hline
\begin{tabular}[c]{@{}l@{}} \hspace{0.2cm} $ \rho_b$ \\(fm$^{-3}$) \end{tabular} &
\begin{tabular}[c]{@{}l@{}} \hspace{0.5cm} $P$\\(MeV fm$^{-3}$) \end{tabular}  &
\begin{tabular}[c]{@{}l@{}} \hspace{0.5cm} ${\cal E}$\\(MeV fm$^{-3}$) \end{tabular} & $Z$ & $N$ \\
\hline
1.0000E-09 & 2.9973E-11 & 9.3046E-07 & 26 & 30 \\
4.9730E-09 & 3.4018E-10 & 4.6275E-06 & 26 & 30 \\
5.0724E-09 & 3.3533E-10 & 4.7201E-06 & 28 & 34 \\
1.5597E-07 & 4.0911E-08 & 1.4522E-04 & 28 & 34 \\
1.5909E-07 & 4.1697E-08 & 1.4812E-04 & 26 & 32 \\
1.6552E-07 & 4.3999E-08 & 1.5411E-04 & 26 & 32 \\
1.6883E-07 & 4.3634E-08 & 1.5719E-04 & 28 & 36 \\
8.0697E-07 & 3.5983E-07 & 7.5177E-04 & 28 & 36 \\
8.2311E-07 & 3.5457E-07 & 7.6682E-04 & 28 & 38 \\
9.2696E-07 & 4.1587E-07 & 8.6361E-04 & 28 & 38 \\
9.4550E-07 & 4.1607E-07 & 8.8089E-04 & 36 & 50 \\
1.8538E-06 & 1.0258E-06 & 1.7278E-03 & 36 & 50 \\
1.8909E-06 & 1.0090E-06 & 1.7623E-03 & 34 & 50 \\
6.8498E-06 & 5.6411E-06 & 6.3900E-03 & 34 & 50 \\
6.9868E-06 & 5.5275E-06 & 6.5179E-03 & 32 & 50 \\
1.6699E-05 & 1.7692E-05 & 1.5592E-02 & 32 & 50 \\
1.7033E-05 & 1.7260E-05 & 1.5904E-02 & 30 & 50 \\
3.2099E-05 & 4.0208E-05 & 2.9994E-02 & 30 & 50 \\
\\
3.2741E-05 & 3.9028E-05 & 3.0595E-02 & 28 & 50 \\
7.5214E-05 & 1.1838E-04 & 7.0370E-02 & 28 & 50 \\
7.6718E-05 & 1.1094E-04 & 7.1779E-02 & 42 & 82 \\
1.2098E-04 & 2.0367E-04 & 1.1328E-01 & 42 & 82 \\
1.2340E-04 & 2.0062E-04 & 1.1554E-01 & 40 & 82 \\
1.5042E-04 & 2.6126E-04 & 1.4090E-01 & 40 & 82 \\
1.5343E-04 & 2.6250E-04 & 1.4372E-01 & 40 & 84 \\
1.6940E-04 & 2.9956E-04 & 1.5871E-01 & 40 & 84 \\
1.7278E-04 & 3.0065E-04 & 1.6189E-01 & 38 & 82 \\
1.7624E-04 & 3.0869E-04 & 1.6513E-01 & 38 & 82 \\
1.7977E-04 & 3.1695E-04 & 1.6844E-01 & 38 & 82 \\
1.8336E-04 & 3.1834E-04 & 1.7182E-01 & 38 & 84 \\
2.2799E-04 & 4.2566E-04 & 2.1372E-01 & 38 & 84 \\
2.3255E-04 & 4.2767E-04 & 2.1801E-01 & 38 & 86 \\
2.5171E-04 & 4.7532E-04 & 2.3601E-01 & 38 & 86 \\
2.5675E-04 & 4.7774E-04 & 2.4074E-01 & 38 & 88 \\
2.6188E-04 & 4.9052E-04 & 2.4557E-01 & 38 & 88 \\
\hline \hline
\end{tabular}%
%}
\end{table}
%%%%%%%%%%%%%%%%%%%%%%%%%%%%%%%%%%%%%
Fig. \ref{fig:eosoc} shows the equation of state and the variation of global isospin asymmetry in the outer crust and tabulated data for HFB-26 in Table \ref{tab:oceosdata}.  The outer crust is marked by the discontinuous transition in the density at some pressure values, indicating a change of equilibrium nucleus. The pressure and chemical potential remain constant during the transition from one nucleus to another resulting in the finite shift in baryon density of the system. However, it is shown in Ref.  \cite{Jog_1982} that the transition between one layer to another layer takes place through a thin layer of the mixed state of two species with a pressure interval of $\approx$ 10$^{-4} P$. It should be noted here that the pressure of the outer crust is mainly determined from the relativistic electron gas as suggested in Eq. (\ref{ocpress}). The HFB calculations estimate similar EoS for the outer crust except at the points where the transition in the nucleus layers takes place. One can see that the majority of the outer crust is determined from the nuclear mass models, which are used to calculate the mass excess of neutron-rich nuclei. The inner layers of heavy nuclei account for the maximum mass of the outer crust. It is also seen that the asymmetry increases monotonically with density, although relatively at a slower pace at high density in the outer crust, reaching $\approx 0.4$ at the transition from outer to the inner crust. The relative difference among different HFB mass models is also visible, attributed to their different symmetry energy. The symmetry energy plays a prominent role in determining the outer and inner crust structure and will be discussed in the next section.

%%%%%%%%%%%%inner crust%%%%%%%%%%%%%
\section{Inner crust}
With the increase in density or the distance from the star's surface, neutron chemical potential increases monotonically. When the chemical potential exceeds the rest mass of the neutron, the neutron starts dripping out of nuclei making the onset of the inner crust. Since no such system can be produced in terrestrial laboratories as neutrons evaporate, the inner crust inevitably becomes model dependent. This study uses the E-RMF model to calculate the properties of the inner crust using three recently developed parameter sets, namely IOPB-I \cite{Kumar_2018}, FSUGarnet \cite{Chen_2014},  and G3 \cite{Kumar_2017}. The bulk properties of these three E-RMF forces are provided in Table \ref{bulkproperties} along with the theoretical or experimental constraints. 
%%%%
\begin{table}
\centering
\caption{Bulk matter properties such as saturation density ($\rho_{\rm sat}$), binding energy ($E_0$), effective mass ($M^*$), symmetry energy ($J$), slope parameter ($L$), second ($K_{sym}$) and third ($Q_{sym}$) order derivative of symmetry energy , incompressibility ($K$) of nuclear matter for the NL3, FSUGarnet, IOPB-I and G3 parameter and their corresponding empirical values.}

%\renewcommand{\arraystretch}{2}
%\scalebox{0.85}{
\begin{tabular*}{\linewidth}{c @{\extracolsep{\fill}}ccccccc}
\hline
\hline
&NL3 & IOPB-I & G3 & FSUGarnet&Empirical Value &  \\ \hline
$\rho_{\rm sat}$ (fm$^{-3})$&0.148 & 0.149       &0.148 &0.153    &    0.148/0.185  \cite{bethe} &  \\
$E_0$ (MeV) & -16.29 & -16.10  &-16.02   &-16.23 &   -15.0/-17.0   \cite{bethe} && \\
$M^*/M$  & 0.595&0.593  &  0.699 &0.578 &  0.55/0.6 \cite{marketin2007}&\\
$J$ (MeV)& 37.43 &33.30  & 31.84  &30.95 &30.0/33.70 \cite{DANIELEWICZ20141}&  \\
$L$ (MeV) & 118.65  &63.58  & 49.31  &51.04 &35.0/70.0  \cite{DANIELEWICZ20141}&  \\
$K_{sym}$ (MeV)&101.34 & -37.09&-106.07&59.36& -174.0/31.0 \cite{zimmerman2020measuring} &  \\
$Q _{sym}$ (MeV) &177.90& 862.70 &915.47 &130.93 & -494/-10 \cite{cai2017constraints} &  \\
$K$ (MeV)  &271.38 & 222.65 &243.96 &229.5 &  220/260 \cite{GARG201855} &  \\
\hline     
\hline     
\end{tabular*}
\label{bulkproperties}
\end{table}
%%%%

\begin{figure}
    \centering
    \includegraphics[scale=0.8]{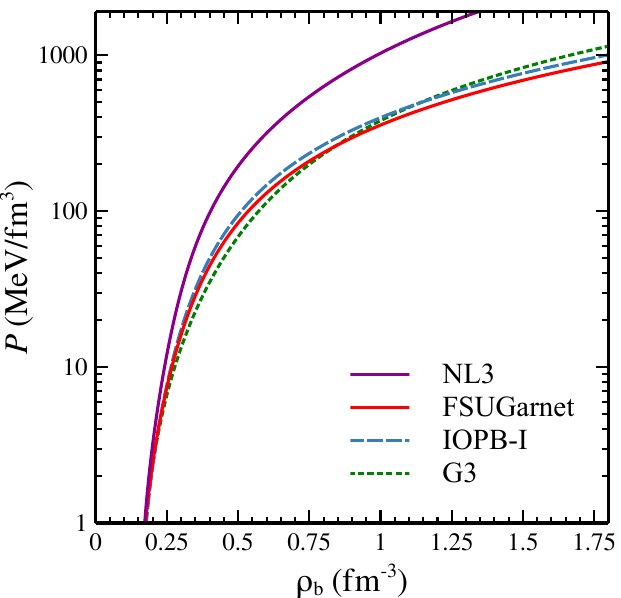}
    \caption{EOSs of the Nuclear matter for NL3 set with other three considered sets.}
    \label{fig:nm_eos}
\end{figure}

For a comparison, Fig. \ref{fig:nm_eos} shows  the EoS of the nuclear matter for three considered E-RMF parameter sets along with one RMF parameter set NL3 \cite{Lalazissis_1997}. It is observed that the NL3 is the stiffest EoS compared to the other three E-RMF sets. Hence, the predicted NM properties such as incompressibility, symmetry energy and its slope parameter etc. for NL3 case is quite larger as compared to other three as shown in Table \ref{bulkproperties}. Also the predicted properties does not satisfy the empirical/experimental data.
On the other hand, E-RMF  parameters satisfy various constraints on EoS and are used in this work for the complete description of the neutron star. The structure and properties of the inner crust are calculated using the famous CLDM, assuming the existence of spherical clusters surrounded by the gas of dripped neutrons throughout the inner crust. The bulk energy of the cluster in Eq. (\ref{ecluster}) and neutron gas is calculated using the E-RMF parameter sets FSUGarnet, IOPB-I, and G3, ensuring numerical and physical consistency. 

\begin{figure}
    \centering
    \includegraphics[scale=0.6]{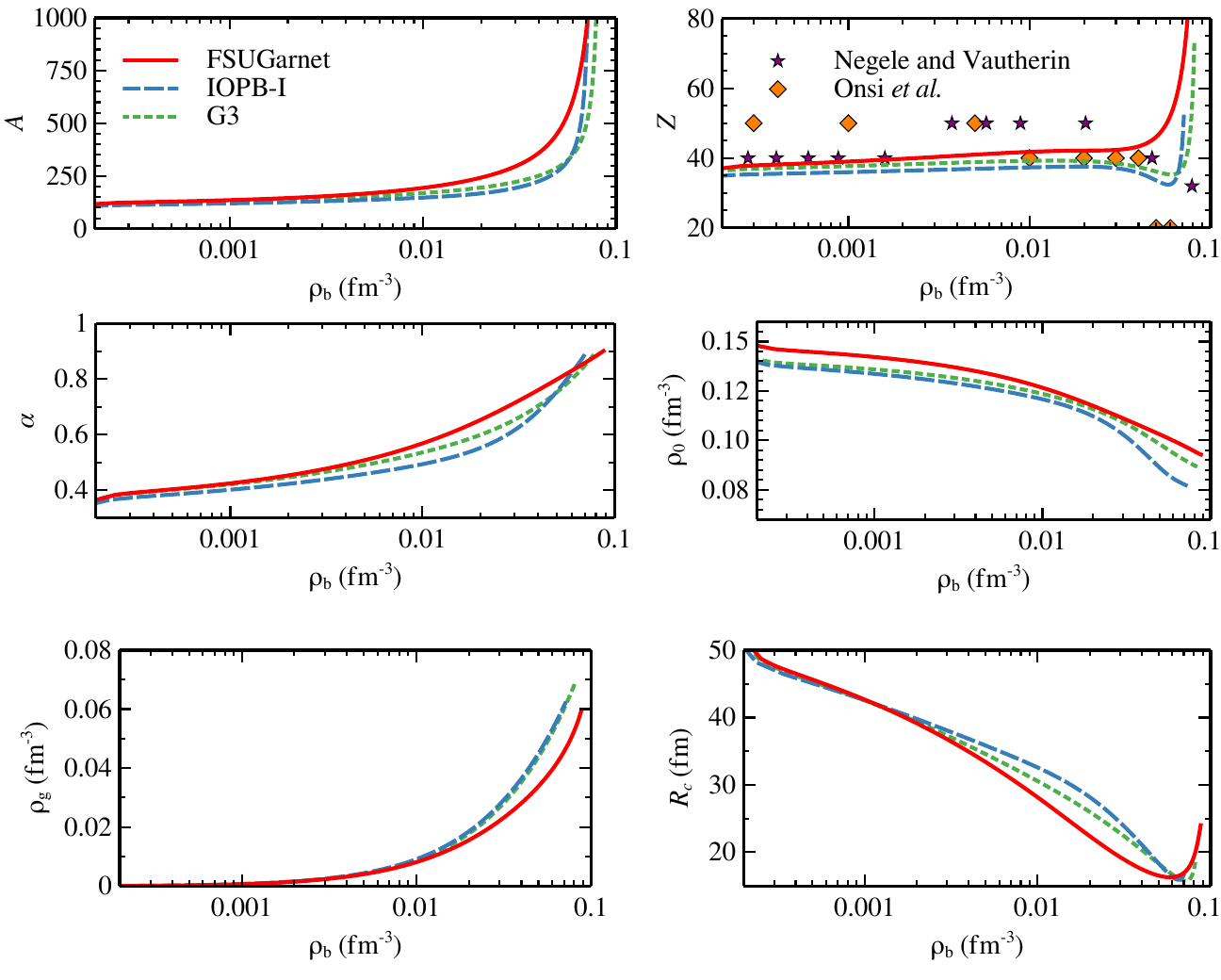}
    \caption{The variation of mass number $A$, proton number $Z$, asymmetry $\alpha$, average cluster density $\rho_0$, the neutron gas density $\rho_g$ and the radius of cell with the baryon density $\rho_b$ in the inner crust of neutron star with FSUGarnet, IOPB-I, and G3 E-RMF parameter set. The quantum calculation by Negele and Vautherin \cite{NEGELE1973298} and Onsi {\it {\it et al.}} \cite{Onsi_2008} are also shown.}
    \label{fig:icprop}
\end{figure}

The most important aspect in the calculation of inner crust structure is the parametrization of the surface and curvature energy of the cluster. The curvature energy helps to understand the surface energy of the cluster better and is an integral part of the modified liquid-drop formula \cite{Pomorski_2003}. Since we do not have the significant knowledge of surface energy of very neutron-rich nuclei from the laboratory experiments, we resort to the fitting of semiempirical formula such as given in Eq. (\ref{ecluster}).  In order to fit the surface and curvature energy of CLDM  with the experimental mass, a parameter space $\boldsymbol{S}= \{\sigma_0, b_s, \sigma_{0,c}, \beta,\alpha,p\}$ is defined which is fitted to the experimental mass obtained from AME2020 table \cite{Huang_2021}. The goodness of reproduction of experimental binding energy is measured by the penalty function $ \chi^2(\boldsymbol{S})$ as \cite{Dobaczewski_2014}
%%%% 
\begin{equation}
    \chi^2(\boldsymbol{S})=\frac{1}{N}\sum_{i=1}^{N}\Big(\frac{(\mathcal{O}_i(s)-\mathcal{O}^{exp}_i)^2}{\Delta \mathcal{O}_i^2} \Big),
\end{equation}
%%%%
where $N$ is the degree of freedom, $\mathcal{O}_i(s)$ stands for the calculated energy of cluster, $\mathcal{O}^{exp}_i$ for the experimental binding energy and $\Delta \mathcal{O}_i$ for adopted systematic theoretical error of 0.1 MeV \cite{carreau2020modeling}. The value of $p$, which takes care of isospin asymmetry dependence of surface energy, is taken to be 3. This is a favorable choice in  various calculations of surface energy \cite{Lattimer_1991, Avancini_2009}, and $\alpha$ is taken to be 5.5 as prescribed in \cite{Newton_2012}. The parameter space $\boldsymbol{S}$ then reduces to four variables whose values for different E-RMF parameter sets used in this study are given in Table \ref{tab:surfaceparameter}.
%%%%
\begin{table}
\centering
\caption{The fitted value of surface and curvature energy parameters for the FSUGarnet, IOPB-I, and G3 force parameter sets. The value of $\alpha$ and $p$ is taken to be 5.5 and 3,  respectively. Experimental binding energy is taken from AME2020 table \cite{Huang_2021}. }
\label{tab:surfaceparameter}
\scalebox{1.1}{
\begin{tabular}{lllll}
\hline
\hline
Parameter & 
\begin{tabular}[c]{@{}l@{}} \hspace{0.2cm} $\sigma_0$ \\(MeV fm$^{-2}$) \end{tabular}& \hspace{0.2cm} $b_s$ &
\begin{tabular}[c]{@{}l@{}} \hspace{0.2cm} $\sigma_{0,c}$ \\(MeV fm$^{-1}$) \end{tabular} & \hspace{0.2cm} $\beta$ \\
\hline
FSUGarnet &   1.13975            & 29.39987      &   0.07819  &  0.44021\\ \hline
IOPB-I    &   0.97594            & 16.35460      &   0.09064  &  0.81485\\ \hline
G3        &   0.88424            & 26.58373      &   0.09921  &  0.93635\\ 
\hline
\hline
\end{tabular}%
}
\end{table}
%%%%
The importance of fitting individual parameter set for the experimental mass excess instead of taking the same value for all the parameter sets is clear from the Table \ref{tab:surfaceparameter}, where one can see a substantial difference in fitted parameters of surface and curvature energy. The neutron star's inner crust and crustal properties are susceptible to the surface and curvature energy, making this step essential for the CLDM calculation. 

After fixing the surface parameters, let us now calculate the composition of the neutron star inner crust, which is shown in Fig. \ref{fig:icprop} as a function of baryon density for the FSUGarnet, G3, and IOPB-I parameter sets. The number of nucleons $A$ inside the cluster increase  monotonically with increasing density. One can see a steep rise in the number of nucleons when approaching the crust-core transition density, thereby indicating that the matter is transiting to a homogeneous phase of nucleons and leptons.  The variation of charge number is also shown in Fig. \ref{fig:icprop}. It is observed that the $Z\approx$ 40 dominates over the majority of the inner crust. This feature is analogous with the quantum calculation carried by Negele and Vautherin \cite{NEGELE1973298} which predicts the dominance of $Z=40$ at lower densities and $Z=50$ at higher densities along with the calculations by Onsi {\it {\it et al.}} \cite{Onsi_2008}. The distinctive feature of these works is the existence of strong proton quantum-shell effects in the nuclear cluster with $Z=40$ and $50$ in the inner crust of the neutron star. One may note that the $Z=40$ is not a magic number in ordinary nuclei but corresponds to a filled proton subshell.  Recent calculation by BCPM \cite{BKS_2015} and D1M \cite{sym13091613} also indicated the same feature of inner crust. 
%%%%
%%%%

\begin{figure}
    \centering
    \includegraphics[scale=0.8]{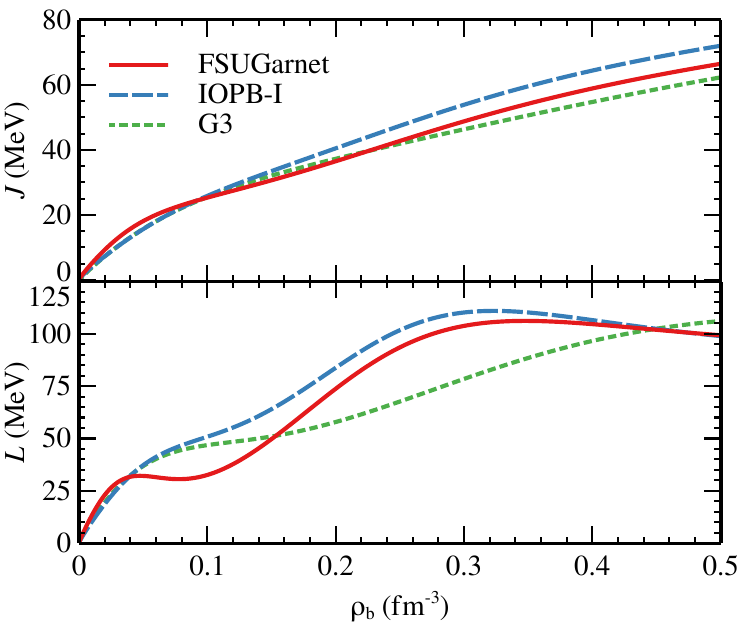}
    \caption{The density dependent symmetry energy ($J$) and slope parameter ($L$) for different E-RMF parametrizations. }
    \label{fig:symmenergy}
\end{figure}

Distribution of mass and charge number in inner crust within CLDM formalism primarily depends on two parameters; a) the isovector surface parameter $p$ in Eq. (\ref{eq:sigma}) which is responsible for the isospin dependence of surface energy, and b) the density-dependent symmetry energy or slope parameter of the EoS used to calculate the bulk energy of cluster. It is observed that the surface parameter $p=3$ correctly estimates the inner crust properties such as crust-core transition density in agreement with the dynamical \cite{Boquera_2019} or thermodynamical \cite{Bao_2020} formalisms and is used in various works such as Refs. \cite{Avancini_2009, Avancini_2008}. In the same context, inner crust calculation are performed   with $p=3$. %We have checked that the larger value of $p$ estimates a higher nucleon and charge number.%
Furthermore, it is an artifact of the literature that nuclear symmetry energy plays a vital role in the structural properties of a neutron star, such as radii, the moment of inertia, crust-core transition density, etc. \cite{fattoyev_2013}. Additionally, it was observed in Ref. \cite{Pearson_2018} that the symmetry energy correlates with the EoS of the inner crust for the Brussels–Montreal functionals. Recently,  Dutra {\it {\it et al.}} \cite{dutra2021neutron}  suggested that the mass and thickness of the crust are more sensitive to the symmetry energy compared to other saturation properties. Taking motivation from these facts and to ascertain the effect of symmetry energy ($J$) and slope parameter ($L$)  on the equilibrium distribution of inner crust, these quantities are shown in Fig. \ref{fig:symmenergy} for the FSUGarnet, IOPB-I, and G3 parameter sets. All these sets follow the constraints from the experimental flow data \cite{Kumar_2018, Tsang_2012}. The behaviour of $J$ and $L$ of parameter sets used is different for different density regions.  At sub-saturation densities ($<0.1$ fm$^{-3}$), which is relevant for the inner crust, the FSUGarnet shows the maximum symmetry energy followed by IOPB-I and G3. The behaviour of the slope parameter (L), however, changes and for the most part, in the subsaturation region, FSUGarnet estimates the smallest slope parameter as compared to the IOPB-I and G3 parameter set. This slope parameter behaviour  suggests that the higher symmetry energy or lower slope parameter of an EoS in the sub-saturation density region corresponds to the larger nucleon and charge number of clusters inside neutron star crust. This fact is also verified in Ref.   \cite{Oyamatsu_2007} which used macroscopic nuclear models to study the inner crust of the neutron star.
%%%%
%%%%
%%%%

\begin{figure}
    \centering
    \includegraphics[scale=0.7]{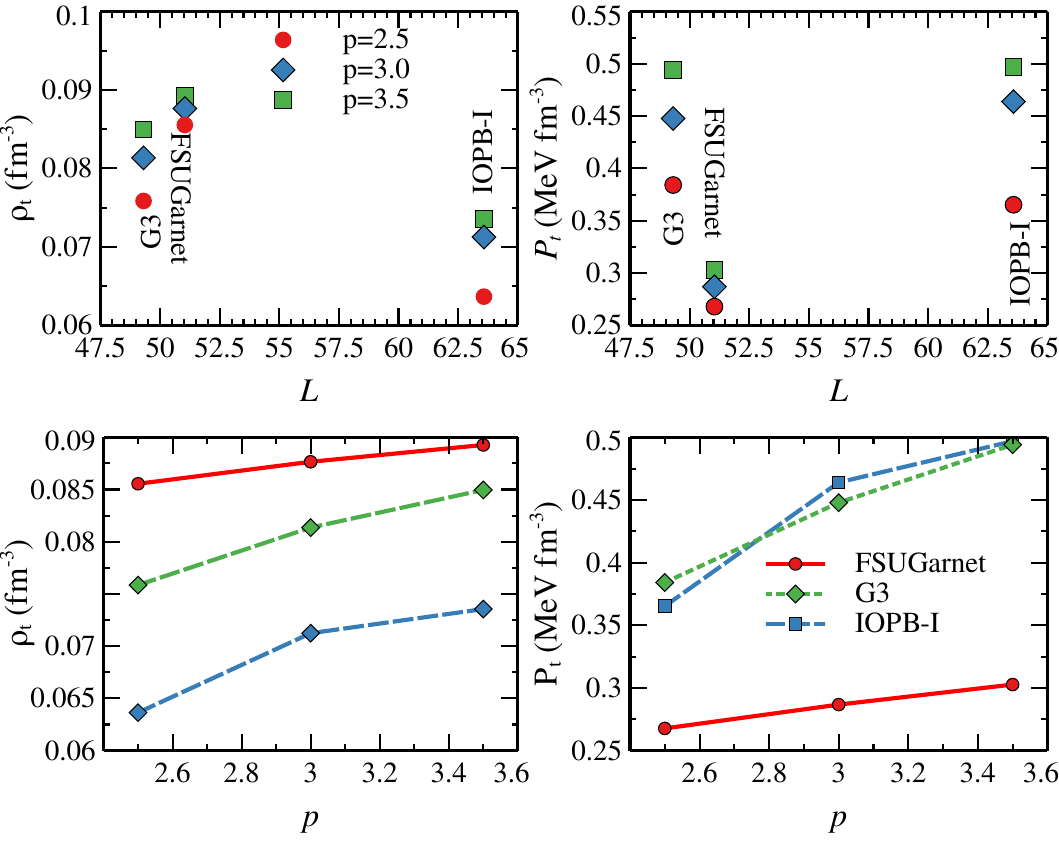}
    \caption{Crust-core transition density and pressure as a function of slope parameter $L$ and $p$ (Eq. (\ref{eq:sigma})) for the FSUGarnet, IOPB-I, and G3 parameter sets.}
    \label{fig:cctransition}
\end{figure}
%%%%

With increasing density or distance from the star's surface, the spherical cluster becomes more and more asymmetric and dilute. The asymmetry $\left(\alpha=\frac{\rho_n-\rho_p}{\rho_n+\rho_p}\right)$ reaches $\approx 0.9$ when reaching the crust boundary, and the density of cluster ($\rho_0$) becomes comparable to the density of neutron gas ($\rho_g$) surrounding these clusters. It should be mentioned that the terms associated with iso-vector meson coupling affect the asymmetry of the system. But in accordance to the mathematical conventions, the terms with high powers of iso-vector mesons are less effective, so, the linear term decides the asymmetry factor considerably. The  asymmetry at crust boundary are 0.896, 0.900, 0.902, \& 0.894 for NL3, FSUGarnet, IOPB-I and G3 sets, respectively. However, the FSUGarnet shows the largest asymmetry and density of cluster as one starts moving toward the core from the outer crust of neutron star, while IOPB-I the least owing to the behaviour of their symmetry energy. Finally, the radius of the WS cell decreases with density while the cluster keeps growing in size. This leads the cluster to get closer and closer to form a large cluster and ultimately convert to homogeneous matter when reaching the crust-core boundary. The slope parameter has an inverse effect on the density of neutron gas and WS cell radius. A larger $L$ corresponds to the smaller neutron gas density and radius of the cluster. 

Fig. \ref{fig:cctransition}  shows the crust-core transition from the crust side using Eq. (\ref{eq:cctransition}). As discussed, the EoS of the inner crust is sensitive to the choice of surface parameters $p$ and slope parameter $L$. To examine this, the relationship between the transition density $\rho_t$ and pressure $P_t$ is displayed with respect to varying values of $L$ and $p$. The G3 parameter set predicts a larger transition density as compared to the IOPB-I set owing to its smaller $L$, while FSUGarnet does not follow the trend. In general practice, the $\rho_t$ and $P_t$ are anti-correlated to the saturation value of $L$ for a given EoS. However, one can notice in Fig. \ref{fig:symmenergy} that the behaviour of $L$ is different for below and above saturation density. Therefore, if we consider the behaviour of $L$ in the subsaturation density region, the trends in the crust-core transition density could be understood more precisely. The FSUGarnet set with the least $L$ estimates the larges transition density, and IOPB-I with maximum $L$ estimates the lowest crust-core transition density. The transition pressure follows the same trend, however, in the opposite way. The isovector surface parameter $p$ seems to act similarly to the symmetry energy. The transition pressure and density are positively correlated with the value of $p$. This fact suggests the importance of isospin-dependent surface tension in the CLDM calculation of inner crust. Furthermore, the correlation of transition density and pressure of crust-core transition is in harmony with the trends obtained from \cite{Lattimer_2007, Carreau_2019}. According to a recent study by Bao-An Li and Macon Magno \cite{Bao_2020}, the curvature parameter $K_{sym}$ appears to have a greater impact than the slope parameter $L$ in determining the density at which the transition between the crust and core of a neutron star occurs, using equations of state derived from meta-modeling. This work also finds a similar behaviour of $\rho_t$ while comparing the value of $K_{sym}$ from Table \ref{bulkproperties}. 
%%%%
\begin{figure}
    \centering
    \includegraphics[scale=0.8]{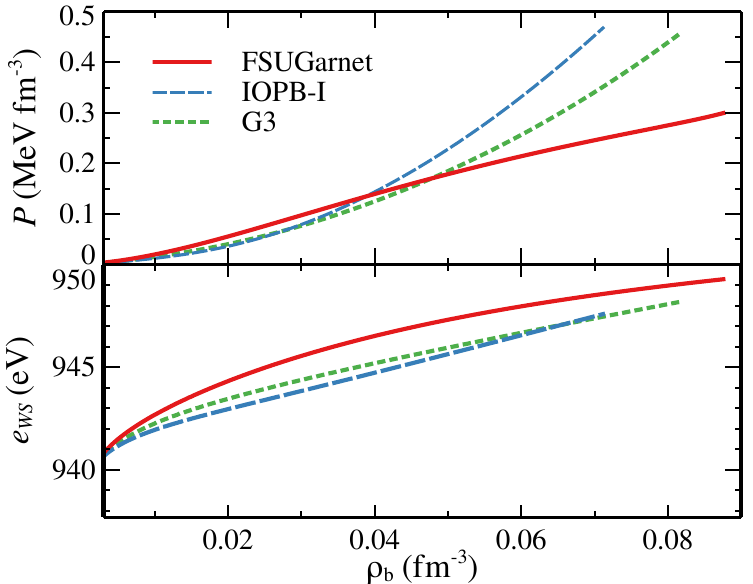}
    \caption{The EoS for the inner crust and equilibrium value of WS cell energy using the E-RMF parameter sets FSUGarnet, IOPB-I and G3. }
    \label{fig:eosic}
\end{figure}
%%%%

It is clear from the above discussion that the structure of the inner crust is susceptible to the behaviour of density-dependent symmetry energy and slope parameter in the sub-saturation density region. In the E-RMF framework, the symmetry energy is controlled mainly by the cross-coupling ($\Lambda_\omega$) of isoscalar-vector ($\omega$) and isovector-vector ($\rho$) mesons [see Eq. (\ref{rmftlagrangian})]. In addition, the parameter set  G3  takes the $\delta$ meson as the additional degree of freedom which helps to change the variation of $L$ and $J$ to reproduce the theoretical and observational constraints \cite{Singh_2014}. The $J$ and $L$ also play a crucial role in estimating the instability in the homogeneous nuclear matter \cite{Vishal_2021}. Therefore, $\Lambda_\omega$ becomes an essential parameter in the E-RMF forces that govern various aspects of the neutron star structure. 

Fig. \ref{fig:eosic} shows the EoS of the inner crust for the FSUGarnet, IOPB-I, and G3 E-RMF parameter sets along with the WS cell energy and the tabulated data in Table \ref{tab:iceosdata}. One may see that the inner crust is primarily  model-dependent, where the stiffness is related to the behaviour of symmetry energy or slope parameter. Higher symmetry energy at subsaturation densities corresponds to the larger $e_{WS}$, which is the case with FSUGarnet in Fig. \ref{fig:eosic}.  The behaviour of G3 and IOPB-I is similar, with IOPB-I estimating a comparatively stiffer EoS which is also in accordance with the behaviour of the symmetry energy. Therefore, it is apparent that the symmetry energy and its derivative predominantly decide the inner crust structure. However, one needs a detailed statistical study of various E-RMF parameter sets (e.g., Bayesian and correlation analysis) to comment on the ambiguities.
One may further note that, unlike in the outer crust, the pressure of the inner crust is mainly dependent on the neutron gas surrounding the clusters. Therefore, the parameters used must follow the necessary constraints on the pure neutron matter (PNM). It is seen that the FSUGarnet, IOPB-I, and G3 reasonably satisfy the results obtained using microscopic chiral EFT \cite{Vishal_2021}, making these parameters suitable for the calculation of inner crust EoS. 

%%%%
\begin{figure}
    \centering
    \includegraphics[scale=0.7]{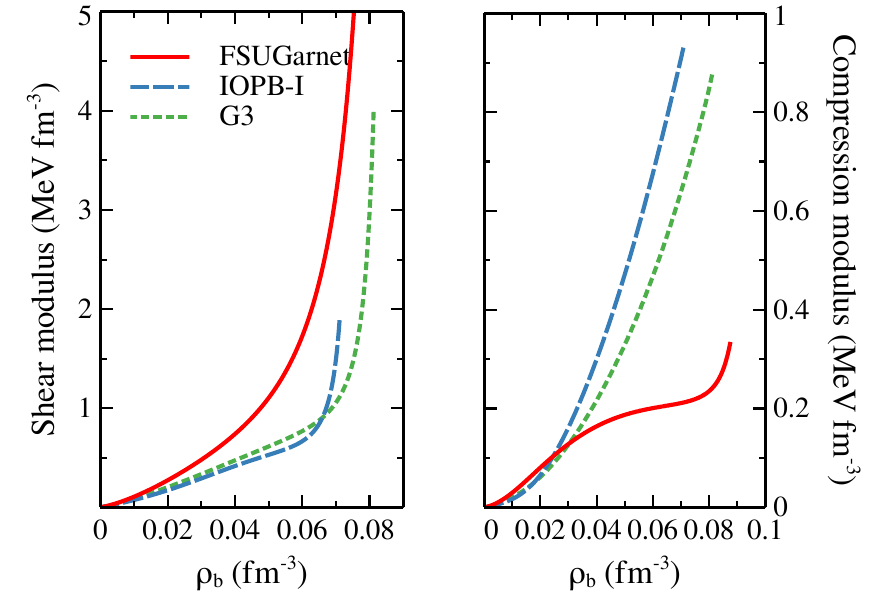}
    \caption{The effective shear and compression modulus for BCC lattice in the inner crust.}
    \label{fig:modulus}
\end{figure}
%%%%

It should be noted that this work is restricted to  spherically symmetric WS cell for the calculation of  inner crust of the neutron star. However, as one approaches the crust-core boundary,  there might be an energetic preference for nonspherical shapes (rod, slab, tube, bubble, etc.) commonly known as ``nuclear pasta'' \cite{Ravenhall_1983, Avancini_2008, Martin_2015, Maruyama_1998}. These structures influence various properties of neutron star crust such as crustal oscillation modes, crust cooling, crust shattering, magnetic field evolution, etc. \cite{Newton_2021}. Nevertheless, it is  seen that the existence of pasta structure is sensitive to the approximations made and  minute energy differences exist between spherical and nonspherical cell shapes. Therefore, nuclear pasta structures have a weak impact on the EoS  \cite{Pearson_2018} and the WS cell composition \cite{Pearson_2020}  and hence they do not affect the global properties of neutron stars, such as the mass-radius profile. The quantitative analysis of pasta structure are discussed in the next chapter.
The EoS of the crust is found to have a significant influence on the fundamental seismic shear mode, which manifests as a quasiperiodic oscillation in the giant flares discharged by neutron stars with strong magnetic fields. \cite{Steiner_2009, Sotani_2012}. In that context, it is  assumed that the neutron star crust is made up of an isotropic BCC poly-crystal whose elastic properties are a function of two elastic moduli: shear ($\mu$) and compression modulus ($K$). These are written as \cite{Haensel_2008}
%%%%
\begin{equation}
    \begin{aligned}
    &K=\rho_b\frac{\partial P}{\partial \rho_b}=\Gamma P,\\
    &\mu=0.1194\frac{\rho_b (Ze)^2}{R_{cell}},
    \end{aligned}
\end{equation}
%%%%
where $\Gamma$ is the adiabatic index. The variation of shear and compression modulus as a function of baryon density is shown in Fig. \ref{fig:modulus}. The shear modulus depends on the distribution of $Z$ and the size of the cell, which is a smoothly increasing function of average baryon density as shown in Fig. \ref{fig:icprop}. As a result, the shear modulus increases continuously on moving toward the core. The FSUGarnet and IOPB-I show the maximum and minimum values of $\mu$. A higher value of $\mu$ means that the fundamental shear mode will have a higher frequency. The compression modulus also increases with density and has an opposite trend as compared to the shear modulus. 
%%%%
\begin{figure}
    \centering
    \includegraphics[scale=0.9]{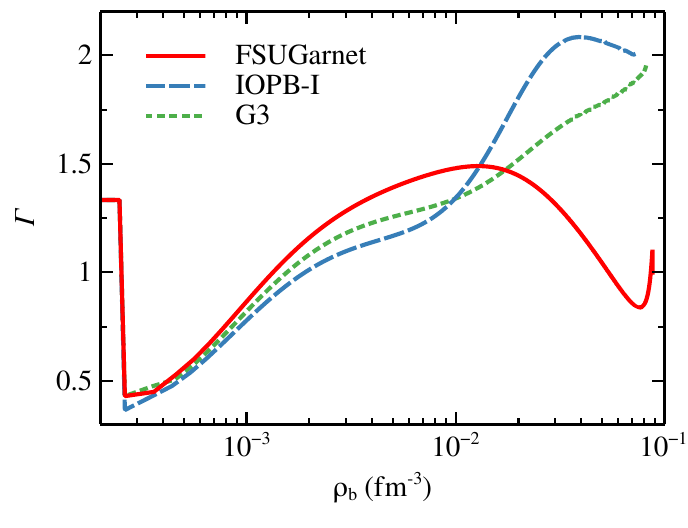}
    \caption{Adiabatic index of the inner crust calculated from the FSUGarnet, IOPB-I, and G3 E-RMF forces.}
    \label{fig:adiabatic}
\end{figure}
%%%%

Finally, the adiabatic index, which determines the response of the crust toward the compression and decompression, is plotted in Fig. \ref{fig:adiabatic} from the outer layer of outer crust till the transition of inner crust to the core. As the pressure in the outer crust is prominently determined from the ultrarelativistic electron gas,  the $\Gamma$ becomes equal to 4/3. The onset of the inner crust is marked by dripped neutrons which soften the EoS. This results in a decrease in the value of $\Gamma$ considerably. As the density in the crust increases, the neutron gas density increases resulting in more and more pressure of neutron gas. As a consequence, the $\Gamma$ increases and reaches up to $\approx$ 2 on reaching the crust-core transition. The FSUGarnet shows a relatively lower value of $\Gamma$ at CC point, which can be explained based on the behaviours of its compression modulus in Fig. \ref{fig:modulus}. The results are in agreement with the microscopic calculation using  three-body forces\cite{BKS_2015}.
%%%%%%%%%%%%%%%%%%%%%%%%%%%%%%%%%%%%%
%%%%%%%%%%%%inner crust%%%%%%%%%%%%%
\begin{sidewaystable}\renewcommand{\arraystretch}{.85}
\centering
\caption{Composition and EoS of inner crust with the IOPB-I, FSUGarnet, and G3 E-RMF parameter sets. The table includes the values of pressure ($P$), energy (${E}$), mass ($A$) and charge ($Z$) of the cluster and the radius ($R_c$) of cell.}
\label{tab:iceosdata}
\resizebox{\textwidth}{!}{%
\begin{tabular}{l|lllll|lllll|lllll}
\hline
\hline
&\multicolumn{5}{c|}{IOPB-I} & \multicolumn{5}{c|}{FSUGarnet} & \multicolumn{5}{c}{G3} \\ 
\hline
\cmidrule(lr){2-6}\cmidrule(lr){7-11}\cmidrule(lr){12-16}
%%-------------------------------------------------------------------------%%
\begin{tabular}[c]{@{}l@{}} \hspace{0.2cm} $\rho_b$ \\(fm$^{-3}$) \end{tabular}&
\begin{tabular}[c]{@{}l@{}} \hspace{0.5cm} $P$ \\(MeV fm$^{-3}$) \end{tabular} &
\begin{tabular}[c]{@{}l@{}} \hspace{0.5cm} ${ E}$ \\(MeV fm$^{-3}$) \end{tabular}&
\hspace{0.2cm} $A$ &
\hspace{0.2cm} $Z$ & 
\begin{tabular}[c]{@{}l@{}} \hfill $R_c$ \\(fm) \end{tabular}&
%%-------------------------------------------------------------------------%%
\begin{tabular}[c]{@{}l@{}} \hspace{0.5cm} $P$ \\(MeV fm$^{-3}$) \end{tabular} &
\begin{tabular}[c]{@{}l@{}} \hspace{0.5cm} ${E}$ \\(MeV fm$^{-3}$) \end{tabular}&
\hspace{0.2cm} $A$ &
\hspace{0.2cm} $Z$ & 
\begin{tabular}[c]{@{}l@{}} \hfill $R_c$ \\(fm) \end{tabular}&
%%-------------------------------------------------------------------------%%
\begin{tabular}[c]{@{}l@{}} \hspace{0.5cm} $P$ \\(MeV fm$^{-3}$) \end{tabular} &
\begin{tabular}[c]{@{}l@{}} \hspace{0.5cm} ${ E}$ \\(MeV fm$^{-3}$) \end{tabular}&
\hspace{0.2cm} $A$ &
\hspace{0.2cm} $Z$ & 
\begin{tabular}[c]{@{}l@{}} \hfill $R_c$ \\(fm) \end{tabular}
\\
\hline
0.0003 & 0.000512 & 0.281414 & 112.7301 & 35.3635 & 47.0669 & 0.000529 & 0.281392 & 124.2469 & 37.9120 & 47.7178 & 0.000529 & 0.281396 & 121.0933 & 36.9770 & 47.3051 \\
0.0023 & 0.002196 & 2.162921 & 127.4424 & 36.4676 & 39.2188 & 0.002666 & 2.163182 & 148.1375 & 39.9832 & 38.3637 & 0.002421 & 2.163018 & 139.7759 & 38.3019 & 38.6883 \\
0.0043 & 0.004378 & 4.046355 & 134.4834 & 36.8530 & 36.5161 & 0.006011 & 4.047617 & 162.3130 & 40.8027 & 34.4444 & 0.005088 & 4.046883 & 149.7114 & 38.7594 & 35.4098 \\
0.0063 & 0.006857 & 5.930916 & 139.6914 & 37.0827 & 34.8056 & 0.010280 & 5.933861 & 174.1880 & 41.2951 & 31.7416 & 0.008233 & 5.932151 & 157.3900 & 39.0083 & 33.2786 \\
0.0083 & 0.009637 & 7.816308 & 144.0866 & 37.2410 & 33.5129 & 0.015351 & 7.821608 & 184.9821 & 41.6146 & 29.6417 & 0.011770 & 7.818483 & 163.9397 & 39.1544 & 31.6794 \\
0.0103 & 0.012790 & 9.702446 & 148.1121 & 37.3580 & 32.4303 & 0.021143 & 9.710712 & 195.1915 & 41.8254 & 27.9171 & 0.015679 & 9.705767 & 169.8407 & 39.2376 & 30.3825 \\
0.0123 & 0.016404 & 11.589212 & 152.0036 & 37.4464 & 31.4627 & 0.027576 & 11.600992 & 205.0751 & 41.9625 & 26.4566 & 0.019966 & 11.593800 & 175.3510 & 39.2765 & 29.2779 \\
0.0143 & 0.020568 & 13.476612 & 155.9021 & 37.5109 & 30.5624 & 0.034569 & 13.492286 & 214.7894 & 42.0488 & 25.1970 & 0.024647 & 13.482594 & 180.6262 & 39.2813 & 28.3051 \\
0.0163 & 0.025362 & 15.364671 & 159.8925 & 37.5534 & 29.7039 & 0.042041 & 15.384688 & 224.4444 & 42.1009 & 24.0977 & 0.029740 & 15.372041 & 185.7704 & 39.2576 & 27.4281 \\
0.0183 & 0.030855 & 17.253335 & 164.0333 & 37.5730 & 28.8741 & 0.049912 & 17.277980 & 234.1273 & 42.1316 & 23.1307 & 0.035264 & 17.262138 & 190.8534 & 39.2087 & 26.6240 \\
0.0203 & 0.037103 & 19.142629 & 168.3629 & 37.5681 & 28.0663 & 0.058107 & 19.172182 & 243.9109 & 42.1518 & 22.2749 & 0.041231 & 19.152846 & 195.9276 & 39.1367 & 25.8779 \\
0.0223 & 0.044151 & 21.032522 & 172.9082 & 37.5368 & 27.2774 & 0.066553 & 21.067174 & 253.8634 & 42.1702 & 21.5146 & 0.047654 & 21.044189 & 201.0313 & 39.0433 & 25.1792 \\
0.0243 & 0.052030 & 22.923195 & 177.6925 & 37.4765 & 26.5066 & 0.075186 & 22.962877 & 264.0506 & 42.1949 & 20.8366 & 0.054541 & 22.936130 & 206.1971 & 38.9293 & 24.5210 \\
0.0263 & 0.060762 & 24.814495 & 182.7338 & 37.3848 & 25.7538 & 0.083950 & 24.859370 & 274.5427 & 42.2329 & 20.2305 & 0.061897 & 24.828700 & 211.4524 & 38.7959 & 23.8976 \\
0.0283 & 0.070358 & 26.706495 & 188.0512 & 37.2593 & 25.0194 & 0.092795 & 26.756579 & 285.4087 & 42.2906 & 19.6879 & 0.069724 & 26.721787 & 216.8215 & 38.6439 & 23.3053 \\
0.0303 & 0.080822 & 28.599194 & 193.6626 & 37.0978 & 24.3043 & 0.101674 & 28.654369 & 296.7274 & 42.3745 & 19.2015 & 0.078025 & 28.615431 & 222.3302 & 38.4740 & 22.7407 \\
0.0323 & 0.092152 & 30.492652 & 199.5922 & 36.8984 & 23.6087 & 0.110554 & 30.552771 & 308.5798 & 42.4907 & 18.7654 & 0.086799 & 30.509675 & 228.0027 & 38.2873 & 22.2015 \\
0.0343 & 0.104339 & 32.386812 & 205.8659 & 36.6596 & 22.9332 & 0.119398 & 32.451663 & 321.0590 & 42.6453 & 18.3746 & 0.096044 & 32.404474 & 233.8671 & 38.0848 & 21.6857 \\
0.0363 & 0.117373 & 34.281724 & 212.5165 & 36.3809 & 22.2781 & 0.128181 & 34.351065 & 334.2689 & 42.8449 & 18.0250 & 0.105759 & 34.299832 & 239.9538 & 37.8677 & 21.1916 \\
0.0383 & 0.131239 & 36.177311 & 219.5887 & 36.0623 & 21.6434 & 0.136879 & 36.250965 & 348.3255 & 43.0963 & 17.7132 & 0.115943 & 36.195666 & 246.2988 & 37.6376 & 20.7177 \\
0.0403 & 0.145923 & 38.073709 & 227.1356 & 35.7053 & 21.0292 & 0.145474 & 38.151365 & 363.3607 & 43.4070 & 17.4362 & 0.126592 & 38.092189 & 252.9423 & 37.3964 & 20.2631 \\
0.0423 & 0.161410 & 39.970881 & 235.2303 & 35.3131 & 20.4360 & 0.153951 & 40.052167 & 379.5305 & 43.7854 & 17.1918 & 0.137707 & 39.989108 & 259.9348 & 37.1462 & 19.8267 \\
0.0443 & 0.177688 & 41.868732 & 243.9690 & 34.8913 & 19.8641 & 0.162296 & 41.953359 & 397.0108 & 44.2410 & 16.9782 & 0.149287 & 41.886623 & 267.3362 & 36.8900 & 19.4077 \\
0.0463 & 0.194741 & 43.767368 & 253.4835 & 34.4481 & 19.3141 & 0.170499 & 43.854858 & 416.0135 & 44.7846 & 16.7940 & 0.161332 & 43.784718 & 275.2192 & 36.6310 & 19.0056 \\
0.0483 & 0.212560 & 45.666732 & 263.9520 & 33.9958 & 18.7874 & 0.178552 & 45.756858 & 436.7864 & 45.4290 & 16.6382 & 0.173844 & 45.683268 & 283.6740 & 36.3737 & 18.6201 \\
0.0503 & 0.231136 & 47.566890 & 275.6233 & 33.5507 & 18.2858 & 0.186450 & 47.659061 & 459.6274 & 46.1891 & 16.5101 & 0.186825 & 47.582352 & 292.8160 & 36.1233 & 18.2510 \\
0.0523 & 0.250460 & 49.467791 & 288.8484 & 33.1352 & 17.8120 & 0.194192 & 49.561660 & 484.8955 & 47.0827 & 16.4096 & 0.200280 & 49.482021 & 302.7876 & 35.8868 & 17.8984 \\
0.0543 & 0.270527 & 51.369389 & 304.1277 & 32.7789 & 17.3698 & 0.201772 & 51.464453 & 513.0301 & 48.1317 & 16.3370 & 0.214213 & 51.382168 & 313.7734 & 35.6728 & 17.5629 \\
0.0563 & 0.291330 & 53.271847 & 322.1954 & 32.5213 & 16.9645 & 0.209193 & 53.367562 & 544.5713 & 49.3622 & 16.2927 & 0.228631 & 53.282866 & 326.0136 & 35.4924 & 17.2451 \\
0.0583 & 0.312864 & 55.174940 & 344.1495 & 32.4164 & 16.6034 & 0.216456 & 55.270952 & 580.1935 & 50.8072 & 16.2781 & 0.243539 & 55.184066 & 339.8258 & 35.3602 & 16.9466 \\
0.0603 & 0.335122 & 57.078838 & 371.7034 & 32.5429 & 16.2972 & 0.223564 & 57.174654 & 620.7521 & 52.5076 & 16.2947 & 0.258946 & 57.085810 & 355.6361 & 35.2952 & 16.6694 \\
0.0623 & 0.358096 & 58.983547 & 407.6620 & 33.0209 & 16.0623 & 0.230523 & 59.078454 & 667.3394 & 54.5159 & 16.3450 & 0.274861 & 58.988009 & 374.0316 & 35.3235 & 16.4164 \\
0.0643 & 0.381778 & 60.888911 & 456.9247 & 34.0513 & 15.9241 & 0.237340 & 60.982554 & 721.3793 & 56.8996 & 16.4321 & 0.291295 & 60.890753 & 395.8363 & 35.4807 & 16.1918 \\
0.0663 & 0.406154 & 62.795116 & 528.8127 & 36.0048 & 15.9275 & 0.244023 & 62.886854 & 784.7565 & 59.7475 & 16.5605 & 0.308257 & 62.794055 & 422.2481 & 35.8176 & 16.0017 \\
0.0683 & 0.431207 & 64.701956 & 643.1830 & 39.6507 & 16.1563 & 0.250586 & 64.791356 & 860.0130 & 63.1784 & 16.7356 & 0.325756 & 64.697854 & 455.0602 & 36.4091 & 15.8547 \\
0.0703 & 0.456922 & 66.609668 & 848.7924 & 46.8294 & 16.7832 & 0.257041 & 66.696056 & 950.6534 & 67.3544 & 16.9651 & 0.343803 & 66.602142 & 497.0797 & 37.3702 & 15.7640 \\
0.0723 &  &  &  &  &  & 0.263407 & 68.600956 & 1061.6230 & 72.5019 & 17.2589 & 0.362407 & 68.506998 & 552.9294 & 38.8868 & 15.7498 \\
0.0743 &  &  &  &  &  & 0.269708 & 70.505948 & 1200.0577 & 78.9449 & 17.6307 & 0.381575 & 70.412409 & 630.7147 & 41.2793 & 15.8444 \\
0.0763 &  &  &  &  &  & 0.275974 & 72.411149 & 1376.5995 & 87.1585 & 18.0986 & 0.401315 & 72.318190 & 745.8048 & 45.1431 & 16.1026 \\
0.0783 &  &  &  &  &  & 0.282245 & 74.316548 & 1607.4969 & 97.8569 & 18.6875 & 0.421639 & 74.224643 & 930.1836 & 51.6896 & 16.6213 \\
0.0803 &  &  &  &  &  & 0.288574 & 76.222148 & 1918.1697 & 112.1331 & 19.4311 & 0.442582 & 76.131600 & 1256.6567 & 63.5913 & 17.5768 \\
0.0823 &  &  &  &  &  & 0.295044 & 78.127848 & 2348.6311 & 131.6563 & 20.3718 &  &  &  &  &  \\
0.0843 &  &  &  &  &  & 0.301780 & 80.033651 & 2959.3448 & 158.8194 & 21.5538 &  &  &  &  &  \\
0.0863 &  &  &  &  &  & 0.309003 & 81.939748 & 3824.6786 & 196.1983 & 22.9867 &  &  &  &  &  \\
0.0883 &  &  &  &  &  & 0.316890 & 84.005883 & 4964.3837 & 242.5463 & 24.6032 &  &  &  &  & \\
\hline
\hline
\end{tabular}%
}
\end{sidewaystable}
%%%%
\section{Neutron star unified EoS, $M-R$ relation}
\label{res:NS_EOS}
\begin{figure}
    \centering
    \includegraphics[scale=0.9]{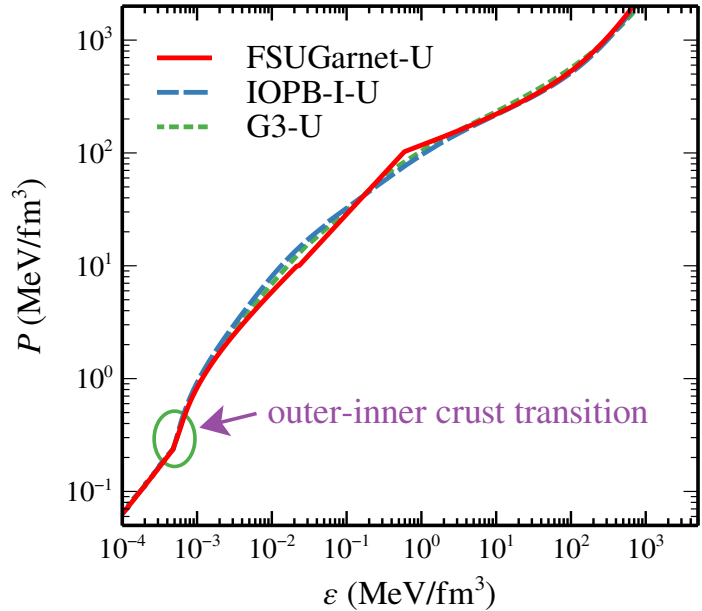}
    \caption{The unified EoSs for FSUGarnet-U, IOPB-I-U, and G3-U sets. The green line represents the outer-inner crust transition.}
    \label{fig:unifiedeos}
\end{figure}

The core EoS of the neutron star is calculated with E-RMF formalism for FSUGarnet, IOPB-I, and G3 parameter sets. For the crust part, both outer and inner crust EoS are used. The unified EoS is then constructed by matching the crust-core density and pressure, and is shown in Fig. \ref{fig:unifiedeos} for FSUGarnet, IOPB-I, and G3 sets.  The unified EoSs are named as FSUGarnet-U, IOPB-I-U, and G3-U, respectively and one can find from the GitHub link\footnote{{\bf \url{https://github.com/hcdas/Unified_eos}}}. The green circle represents the outer-inner crust transition. The crust-core transition is different for different forces because it is model-dependent. With these EoSs, one can now calculate the neutron star's mass, radius, and moment of inertia. 
%%%%
%%%%

\begin{figure}
\centering
\includegraphics[scale=0.9]{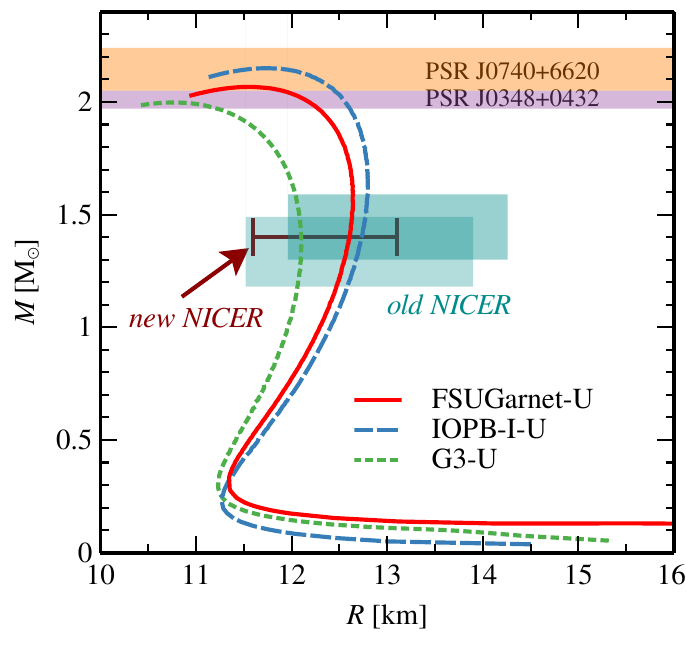}
\caption{The $M-R$ relations for three unified EoSs such as FSUGarnet-U, IOPB-I-U, and G3-U. The horizontal bars represent the PSR J0740+6620 \cite{Cromartie_2019} (light orange) and PSR J0348+0432 \cite{Antoniadis_2013} (light violet). The old NICER data are also shown with two boxes from two different analysis \cite{Miller_2019,Riley_2019}. The double-headed red line represents the radius constraints by the Miller {\it {\it {\it et al.}}} \cite{Miller_2021} for 1.4 $M_\odot$ neutron star termed as new NICER data.}
\label{fig:mr}
\end{figure}

Let us now calculate the mass and radius of the neutron star using Eqs. (\ref{eq:pr} and \ref{eq:mr}) for a fixed central density. The $M-R$ profile is calculated for the whole star which is depicted in Fig. \ref{fig:mr} for considered sets. The maximum mass of the all the sets satisfy $\sim 2\ M_\odot$ limit. The maximum mass constraints from different massive pulsars such as PSR J0348+0432 ($M = 2.01\pm{0.04} \ M_\odot$) \cite{Antoniadis_2013} and PSR J0740+6620 ($M = 2.14_{-0.09}^{+0.10} \ M_\odot$) \cite{Cromartie_2019} are shown. The radius constraints given by Miller {\it {\it {\it et al.}}} \cite{Miller_2019} and Riley {\it {\it {\it et al.}}} \cite{Riley_2019} are shown with two dark cyan boxes termed as {\it old NICER}. The {\it new NICER} data is also shown from the study of PSR J0030+0451 with X-ray Multi-Mirror Newton for canonical star with $R_{1.4} = 12.35 \pm 0.75$ km \cite{Miller_2021}. From the figure it is clear that all the considered EoSs satisfy all constraints; such as maximum mass by two different pulsars and canonical radius by both NICER data.
%%%%%
%%%%
\begin{figure}
    \centering
    \includegraphics[scale=0.7]{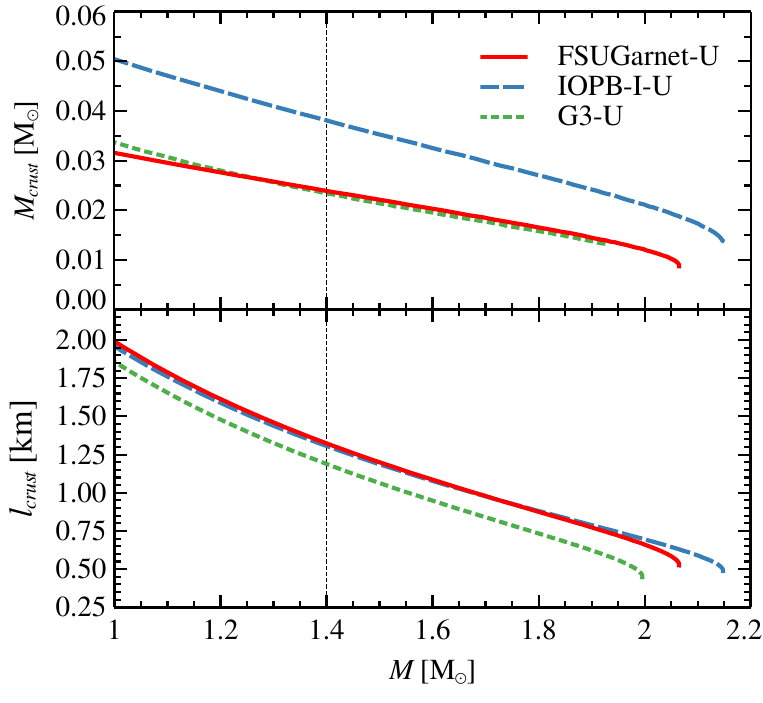}
    \caption{{\it Upper:} The mass of the crust as a function of mass for three unified EoSs. {\it Lower:} The length of the crust as a function of mass. The black dotted line represents the canonical neutron star mass.}
    \label{fig:crust}
\end{figure}
%%%%
The mass and thickness of the crust for three unified EoSs are calculated using the formula $M_{crust} = M-M_{core}$, and $l_{crust} = R-R_{core}$, respectively.  The $M_{core} (R_{core})$ is the mass (radius) of the neutron star core. The variation of mass and thickness of the crust is plotted in Fig. \ref{fig:crust} for three EoSs. It is seen that the crust is thicker for low mass neutron star, and it drops continuously with increasing neutron star mass. Similar results are obtained  for the crust mass as well. The  mass and thickness of the crust for all considered EoSs are given in Table \ref{tab:NS_observables}.
%%%%%%%%%%%%%%%%%%%%%%%%%%%%%%%%%%%%%%%%%%
\section{Moment of inertia of the neutron star}
\label{res:MOI}
\begin{figure}
    \centering
    \includegraphics[scale=0.7]{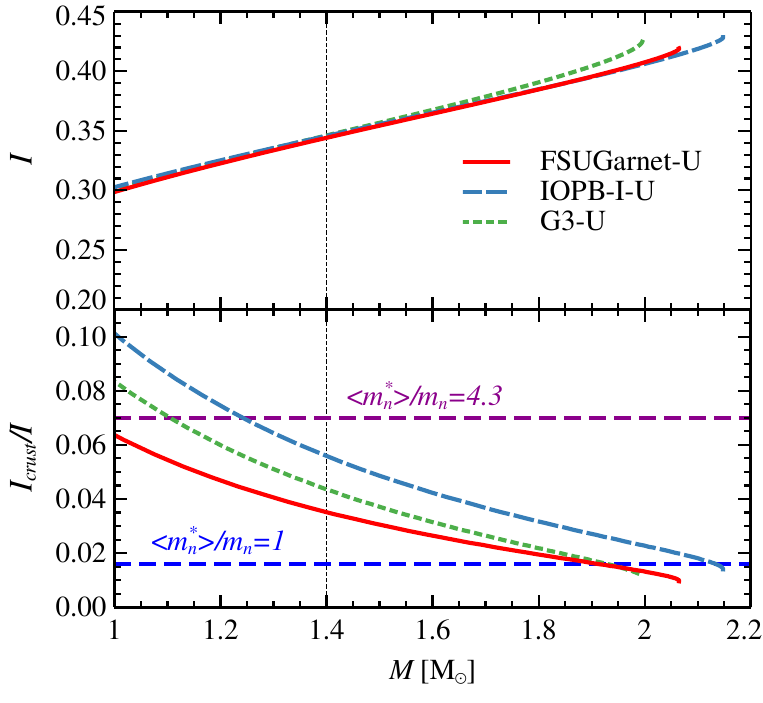}
    \caption{{\it Upper:} The normalized moment of inertia as a function of mass for three unified EoSs. {\it Lower:} The fractional moment of inertia as a function of mass. The dashed dark magenta and dark blue lines represent the Vela pulsar data (see text for details).}
    \label{fig:mom}
\end{figure}

The moment of inertia (MI) of the neutron star is calculated for a uniformly rotating case (slow rotation) as described in Eq. \ref{eq:moi}. The total normalized MI of the neutron star is shown in the upper panel of Fig. \ref{fig:mom} for three unified EoSs. The $I$ increases with the masses of the neutron star as it depends on the mass of the star. The $I$ for considered sets is almost  same up to $1.6 \ M_\odot$ and then slightly diverges. This is because the core part of EoS is  model-dependent. Some theoretical predictions believe that the relation between $I$ and $M$ is universal \cite{Lattimer_2005, LATTIMER_2016, Landry_2018}. It means that one can predict the nature of $I$ from the observed mass of the star. 

The crustal MI of the neutron star is calculated using Eq. (\ref{eq:moic}) from the crust-core transition radius $R_c$ to the surface of the star $R$. The fractional moment of inertia ($I_{crust}/I$) is depicted in the lower panel of Fig. \ref{fig:mom}. It is seen that for a massive  neutron star, the lesser moment of inertia is stored in the crust. In this case, the maximum mass, FMI for the canonical star, FMI$_{1.4}$ predicted by IOPB-I-U EoS is $2.149\ M_\odot$ and $\approx 0.057$, respectively. For FSUGarnet-U and G3-U cases, the masses and FMI$_{1.4}$ are ($2.065 \ M_\odot$, 0.044) and ($1.996 \ M_\odot$, 0.036), respectively as given in Table \ref{tab:NS_observables}. The blue and violet dashed lines represent the minimum value needed to justify the Vela glitch with \cite{Andersson_2012} and without \cite{Link_1999} crustal entrainment. The details on the crustal entrainment are discussed in the following section. It is evident that the crustal moment of inertia is sensitive to the crust's mass and radius, which subsequently depends on the crust-core transition density and the pressure. Therefore, accurate estimation of these properties is an essential and unified treatment of EoS become pivotal.

%%%%
%%%%
\begin{table*}
\caption{The neutron star properties such as maximum mass ($M_{\rm max}$), maximum radius ($R_{max}$), canonical radius ($R_{1.4}$), normalized maximum MI ($I_{max}$), normalized canonical MI ($I_{1.4}$), maximum FMI (FMI$_{max}$), canonical FMI (FMI$_{1.4}$), mass of the crust ($M_{\rm crust}$), and length of the crust ($l_{crust}$) for FSUGarnet, IOPB-I, and G3 EoSs.}
\label{tab:NS_observables}
\begin{tabular}{llllllllll}
\hline \hline
EoSs &
  \begin{tabular}[c]{@{}l@{}}$M_{\rm max}$\\ ($M_\odot$)\end{tabular} &
  \begin{tabular}[c]{@{}l@{}}$R_{\rm max}$\\ (km)\end{tabular} &
  \begin{tabular}[c]{@{}l@{}}$R_{1.4}$\\   (km)\end{tabular} &
  $I_{\rm max}$ &
  $I_{1.4}$ &
  FMI$_{\rm max}$ &
  FMI$_{1.4}$ &
  \begin{tabular}[c]{@{}l@{}} $M_{\rm crust}$\\ ($M_\odot$)\end{tabular} &
  \begin{tabular}[c]{@{}l@{}}$l_{\rm crust}$\\(km)\end{tabular}  \\ \hline
 IOPB-I-U & 2.148 & 11.947 & 13.301 & 0.429 & 0.346 & 0.014 & 0.057 & 0.013  & 0.490 \\ \hline
 FSUGarnet-U& 2.065 & 11.775 & 13.170  & 0.419  & 0.344 & 0.010 & 0.044 &0.009 &0.528  \\ \hline
 G3-U& 1.996  &10.942 & 12.598 & 0.426  & 0.346 & 0.011 & 0.036  & 0.010 & 0.451  \\ \hline \hline
\end{tabular}
\end{table*}
%%%%
\section{Pulsar glitch }
\label{res:pulsar_glitch}

\begin{figure}[ht]
    \centering
    \includegraphics[scale=0.6]{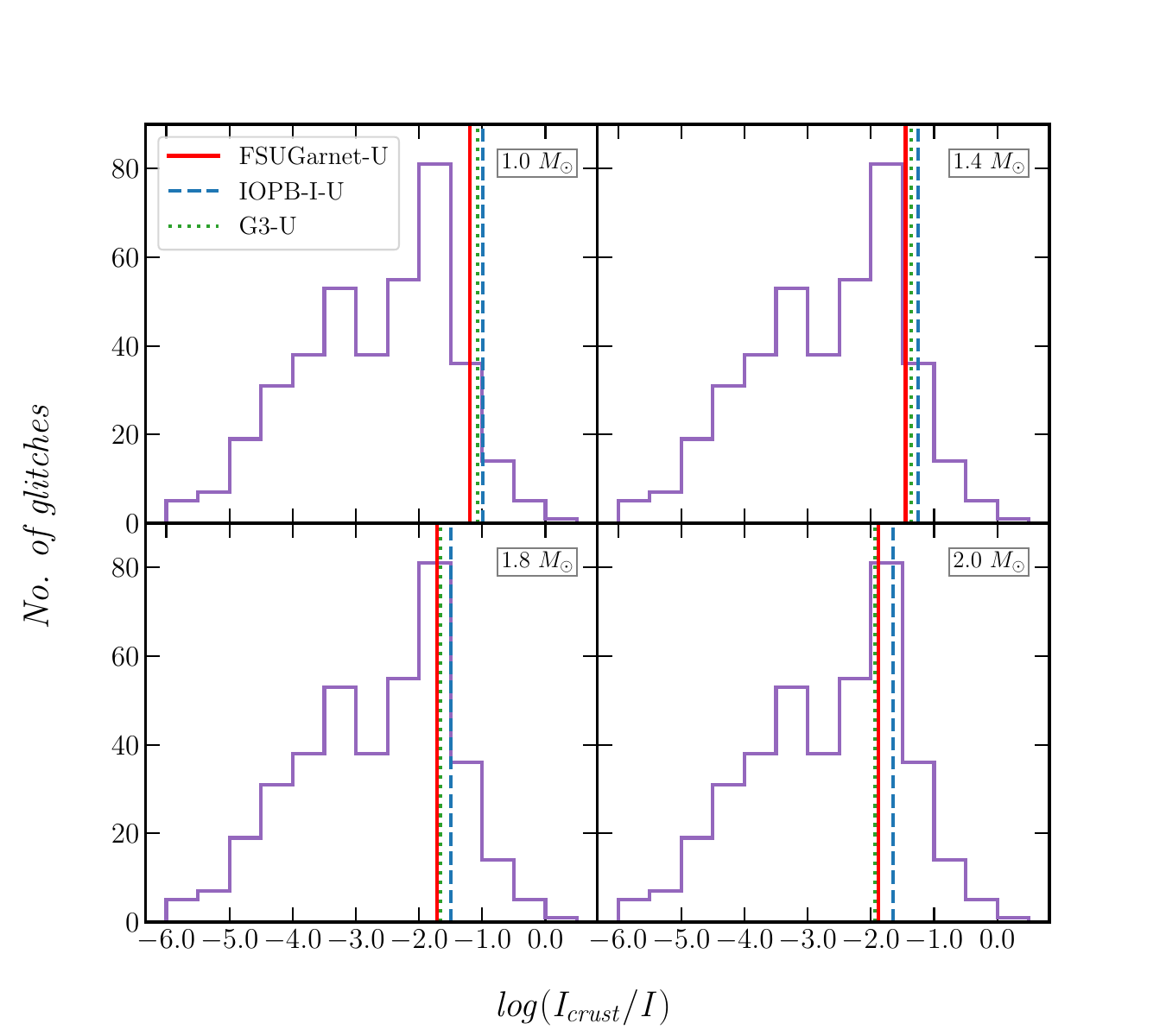}
    \caption{Distribution of $I_{crust}/I$ calculated using 581 glitches \cite{Espionza_2011}. The vertical lines are the FMI for FSUGarnet-U, IOPB-I-U, and G3-U EoSs.}
    \label{fig:glitch}
\end{figure}

Pulsars are rotating neutron stars and they emit regular pulses of radiation that usually last from milliseconds to seconds. These objects have incredibly powerful magnetic fields that funnel particles into two streams along the magnetic poles, producing intense beams of light. Using a technique called pulsar timing, scientists can measure the rotation of the star's crust by observing the pulsating emission in the radio frequency range \cite{Basu_2018}. This involves estimating the rotational speed and glitch activity by calculating the arrival time of the pulses.

The glitches are produced due to the sudden spin-ups in the radio pulsars. This is because the angular momentum transfers from the superfluid component of the stellar interior to the solid crust. Therefore, there is a change of MI from the superfluid to the rest of the star. The fractional crustal moment of inertia (FMI) is the ratio of the total MI to the crustal MI ($I_{crust}/I$), and it is related to the characteristic pulsar glitches properties \cite{Basu_2018, Eya_2017},
%%%%
\begin{equation}
    \frac{I_{crust}}{I}=2\tau_c\frac{1}{t_i}\Big(\frac{\Delta \nu}{\nu}\Big)_i,
\label{eq:glitch}
\end{equation}
%%%%
where $\tau_c$ is the characteristic age of the pulsar, $t_i$ is the time elapsed before the $i$th glitch since the preceding glitch and $\Big(\frac{\Delta \nu}{\nu}\Big)_i$ is fractional frequency jump. From the above relation, one can compare the theoretical FMI with the observational results. 

The strong coupling between the neutron superfluid and the solid crust inside a neutron star is attributed to nondissipative entrainment effects \cite{Chamel_2013, G_gercino_lu_2014}. These effects impose a limit on the amount of angular momentum that can be transferred during a glitch event. The significance of the entrainment coupling is dependent on the neutron effective mass $m^*_n$ present in the inner crust, which is directly proportional to the ratio of unbound neutrons to those that are not entrained, as explained in reference \cite{carreau2020modeling}. By taking these entrainment effects into account, equation (\ref{eq:glitch}) can be expressed as:
%%%%
\begin{eqnarray}
 \frac{I_{crust}}{I}=2\tau_c\frac{\left<m_n^*\right>}{m_n}\frac{1}{t_i}\Big(\frac{\Delta \nu}{\nu}\Big)_i,
\label{eq:glitch_ent}
\end{eqnarray}
%%%%
where $\left<m_n^*\right>$ is the average effective mass of neutrons in the inner crust. The ratio of the $\left<m_n^*\right>/m_n$ has value 4.3 \cite{Andersson_2012} and the ratio becomes one ($\left<m_n^*\right>=m_n$) where no crustal entrainment are considered \cite{Link_1999}. 
%%%%

%%%%

Fig. \ref{fig:glitch} shows the FMI estimated from the observed 581\footnote{{\bf \url{http://www.jb.man.ac.uk/pulsar/glitches.html}}} glitches catalogue \cite{Espionza_2011}. With addition to this, the theoretical FMI are calculated using Eqs. (\ref{eq:moic} and \ref{eq:moi}) for three unified EoSs with different masses of the star. The FMIs for theoretical calculations are well consistent with peak in case for $1.8\ M_\odot$ and $2.0\ M_\odot$ masses. 
%%%%
\section{Summary}

%%%%
In summary, this chapter provides the unified treatment of EoS of the neutron star, namely FSUGarnet-U, IOPB-I-U, and G3-U. These EoSs successfully replicate the data obtained from various sources such as NICER, pulsars, and glitch events. Hence, these  unified EoSs may be used for future exploration of more neutron star properties such as transport, cooling, inspiral etc. It consider different physics for various layers beginning from the outer crust to the inner core within the E-RMF framework. The outer crust is treated within the well-known variational BPS formalism, while the structure of the inner crust is calculated using the compressible liquid drop model. It uses the most recent atomic mass evaluation AME2020 and the highly precise microscopic HFB mass models along with the experimental mass of available neutron-rich nuclei to find the equilibrium composition of the outer crust. The EoS and composition of outer crust calculated from different mass models are compared, which points towards the persistent existence of $Z=28$ and $N= 50$ and $82$ nuclei. The majority of mass models predict the presence of even mass nuclei in the outer crust except for the HFB-14, which indicate a thin layer of $^{121}$Y at high pressure suggesting a possible ferromagnetic behaviour of neutron star. 

The inner crust is treated with the CLDM formalism using the E-RMF framework to calculate the bulk and finite-size energies of the cluster. The composition of the inner crust using the CLDM is in harmony with the available microscopic predictions. The  mass, asymmetry, and gas density increase monotonically with  baryon density or star's depth while the cluster becomes dilute. It is seen that the equilibrium configuration of the inner crust is strictly model-dependent and depends mainly on symmetry energy and slope parameter  in the subsaturation density regime, and the surface energy parametrization. A higher value of symmetry energy or lower slope parameter results in the larger  mass and charge of the cluster. The crust-core transition density ($\rho_t$) and pressure ($P_t$) are calculated from the crust side and it is seen  that these values are sensitive to the isovector surface parameter $p$ and slope parameter $L$. The values of $\rho_t$ for the FSUGarnet-U, IOPB-I-U, and G3-U are found to be 0.08755, 0.07114, and 0.08125 fm$^{-3}$ whereas, the $P_t$ is calculated as 0.46793, 0.31415 and 0.45284 MeV fm$^{-3}$, respectively.    The neutron star properties such as mass, radius, and the moment of inertia are calculated with three unified EoSs viz. FSUGarnet-U, IOPB-I-U, and G3-U. The masses predicted by the three EoSs are well consistent with the different massive pulsars data. The predicted canonical radii are well within  the old and NICER constraints limits. The crustal mass and thickness are also calculated with three unified EoSs. It is  observed that the crust is thicker for low mass neutron star, and it drops continuously with increasing neutron star mass.

The moment of inertia is calculated for a slowly rotating neutron star. The MI increases with increasing the star's mass, and it is almost unchanged around $1.6\ M_\odot$, then it diverges. From the theoretical predictions, it is believed that there exist some Universal relations between MI and mass of the neutron star. In  future, it is expected that more pulsars detection (glitch events) and binary neutron star merger events may put  tight constraints on the MI.

To illustrate the glitch event in pulsar due to sudden spin-up in the radio frequency, the Fractional moment of inertia  for three EoSs is calculated. It is  observed that the more massive a neutron star is, the less MI stores in its crust. The constraint on the FMI is ensured by putting Vela pulsars data with and without entrainment of the crust. Finally, the FMI from the theoretical and observed data (approximately for 581 glitches) are compared. The theoretical prediction is well consistent with the highest peak for canonical to maximum mass star. This implies that the maximum number of glitches observed so far are well compatible with our theoretical results. 

This work restricts itself to the spherically symmetric Wigner-Seitz cell as nonspherical structures do not affect the EoS significantly. However, the existence of nonspherical structures  close to the crust-core interface have various observational consequences. Therefore, to access the impact of pasta structures, one shall perform a comprehensive analysis of neutron star crust including nonspherical shapes. The pasta structures are discussed in the next chapter.
\

\clearpage
\addcontentsline{toc}{section}{Bibliography}
\printbibliography

%% file: Chapter_6/CHAP6.tex
\chapter{\label{chap6}  Nuclear pasta in the inner crust of neutron star and its implications}
\section{Introduction}
%%%%%%%%%%%%%%%%%%%%%%
In the previous chapter, three unified EoSs were constructed using the E-RMF framework employing the CLDM formalism. We considered only spherical symmetric shapes in the inner crust to estimate various crustal properties of the neutron star. Since nonspherical configurations influence the microscopic properties of the neutron star, it is essential to have a unified treatment of EoS (same EoS from surface to the core) considering all the possible pasta structures. Therefore, to comprehensively understand the impact of pasta structure, this chapter extends the previous calculations for the case of nonspherical shapes. It consider 13 well-known parameter sets, namely, BKA24 \cite{Aggarwal_2010}, FSU2 \cite{Chen_2014}, FSUGarnet \cite{Chen_2015}, G1 \cite{Furnstahl_1997}, G2 \cite{Furnstahl_1997}, G3 \cite{Kumar_NPA_2017}, GL97 \cite{NKGb_1997}, IUFSU \cite{Fattoyev_2010}, IUFSU$^*$ \cite{Fattoyev_2010}, IOPB-I \cite{Kumar_2018}, SINPA \cite{Mondal_2016}, SINPB \cite{Mondal_2016} and TM1 \cite{Sugahara_1994}. Using these parameter sets, this chapter aims to construct the neutron star model by evaluating the unified EoS considering the existence of nonspherical shapes in the inner crust. In view of the recent Bayesian inference of crust properties, the mass and thickness of the pasta structures are calculated and  their dependency on the model used is investigated. The related properties such as the shear modulus of the crust and the frequency of fundamental torsional oscillation mode in context to the soft gamma repeaters (SGRs) are also investigated. Finally, the global properties of the neutron star from the unified EoSs such as mass-radius ($M-R$) profile, total crust mass ($M_{\rm crust}$), and thickness ($l_{\rm crust}$), moment of inertia ($I$), fractional moment of inertia ($I_{\rm crust}/I$), etc. are calculated.

\section{Model properties}
%%%%%%%%%%%%%%%%%%%%%%%%%%%%%%%%
In this work,  the finite-size effects such as surface, curvature, Coulomb, etc. are calculated using the CLDM formalism. This method has been widely used to calculate the structure of the crust and other crustal properties such as pairing, thermal, entrainment properties, etc. \cite{BONCHE1981496, CHAMEL2006263, CHAMEL2005109}. For the equation of state (EoS), it uses thirteen effective relativistic mean-field parameter sets to investigate the influence of pasta structures on neutron star properties. We show the saturation properties of the parameter sets in Table \ref{tab:forceproperties} along with the available empirical/experimental values. The motivation of taking these parameter sets lies in the fact that these sets  are the only few among hundreds of relativistic parameters \cite{Dutra_2016}, that reasonably satisfy  the observational constraints from  different massive pulsars such as PSR J0348+0432 ($M = 2.01\pm{0.04} \ M_\odot$) \cite{Antoniadis_2013}, PSR J0740+6620 ($M = 2.14_{-0.09}^{+0.10} \ M_\odot$) \cite{Cromartie_2019} and the radius constraints given by Miller {\it et al.} \cite{Miller_2019}, Riley {\it et al.} \cite{Riley_2019} and PSR J0030+0451 with X-ray Multi-Mirror Newton for canonical star with $R_{1.4} = 12.35 \pm 0.75$ km \cite{Miller_2021}. In addition, these sets also reproduce the finite nuclear properties at par with the experimental values and abide by the relevant nuclear matter constraints on EoS such as flow and kaon data, isoscalar giant monopole resonance, etc. \cite{Dutra_2014}. These sets are differentiated from each other by a wide range of saturation properties and various mesons self and cross-couplings.
%%%%%%%%%%%%%%
\begin{sidewaystable}\renewcommand{\arraystretch}{1.05}
\centering
\caption{Saturation properties of nuclear matter such as saturation density ($\rho_{\rm sat}$), binding energy ($B/A$), effective mass ($M^*/M$), incompressibility ($K$), symmetry energy ($J$, $J^{0.05}$ ), slope parameter ($L$, $L^{0.05}$) at saturation density and at $\rho=0.05$ fm$^{-3}$, curvature  of symmetry energy ($K_{\rm sym}$)  of nuclear matter for 13 relativistic parameter sets.}
\label{tab:forceproperties}
\renewcommand{\arraystretch}{.8}
\begin{tabular}{lllllllllll}
\hline \hline
\begin{tabular}[c]{@{}l@{}} Parameter \\ sets \end{tabular}  &\hspace{0.1cm} $\rho_{\rm sat}$ &\hspace{0.1cm} $B/A$ & $M^*/M$ & \hspace{0.1cm}$K$ &\hspace{0.1cm} $J$ & \hspace{0.1cm}$L$ &\hspace{0.1cm} $K_{sym}$& \hspace{0.1cm} $J^{0.05}$ & \hspace{0.1cm} $L^{0.05}$&\hspace{0.1cm} $\Delta r_{np}^{^{208}Pb}$ \\ \hline
BKA24 \cite{Aggarwal_2010} & 0.147 & -15.95 & 0.600 & 227.06 & 34.19 & 84.80 & -14.95& 14.53 & 33.88 & 0.240\\ 
FSU2 \cite{Chen_2014}& 0.150 & -16.28 & 0.593 & 238.00 & 37.62 & 112.80 & -24.25& 13.16 & 35.72 &0.287 \\ 
FSUGarnet \cite{Chen_2015} & 0.153 & -16.23 & 0.578 & 229.50 & 30.95 & 51.04 & 59.36& 18.07 & 32.11 &0.162 \\ 
G1 \cite{Furnstahl_1997}& 0.153 & -16.14 & 0.634 & 215.00 & 38.50 & 123.19 & 96.87& 12.96 & 35.51 &0.281\\ 
G2 \cite{Furnstahl_1997}& 0.154 & -16.07 & 0.664 & 215.00 & 36.40 & 100.67 & -7.28& 13.3 & 34.81 &0.256 \\ 
G3 \cite{Kumar_NPA_2017}& 0.148 & -16.02 & 0.699 & 243.96 & 31.84 & 49.31 & -106.07& 15.66 & 36.78 &0.180 \\ 
GL97 \cite{NKGb_1997}& 0.153 & -16.30 & 0.780 & 240.00 & 32.50 & 89.40 & -6.37& 11.95 & 31.00 & -\\ 
IUFSU \cite{Fattoyev_2010}& 0.155 & -16.40 & 0.670 & 231.33 & 31.30 & 47.21 & 28.53& 17.80 & 33.85 &0.160 \\ 
IUFSU* \cite{Fattoyev_2010} & 0.150 & -16.10 & 0.589 & 236.00 & 29.85 & 51.508 & 7.87& 15.73 & 32.26 &0.164 \\ 
IOPB-I \cite{Kumar_2018}& 0.149 & -16.10 & 0.650 & 222.65 & 33.30 & 63.58 & -37.09 & 15.60 & 37.2 &0.221 \\ 
SINPA \cite{Mondal_2016}& 0.151 & -16.00 & 0.580 & 203.00 & 31.20 & 53.86 & -26.75& 17.02 & 33.59 &0.183 \\ 
SINPB \cite{Mondal_2016}& 0.150 & -16.04 & 0.634 & 206.00 & 33.95 & 71.55 & -50.57& 14.98 & 36.70 &0.241 \\ 
TM1 \cite{Sugahara_1994}& 0.145 & -16.30 & 0.634 & 281.00 & 36.94 & 111.00 & 34.00& 13.45 & 36.47 & 0.271\\ 
\hline
EMP/EXP  & 0.148/0.185    & -15.0/-17.0   & 0.55/0.6 &  220/260  &30.0/33.70 & 35.0/70.0   & -174.0/31.0 &-&-&0.212/0.354 \\
 &   \cite{bethe}  &    \cite{bethe} &  \cite{marketin2007}&   \cite{GARG201855} & \cite{DANIELEWICZ20141}&  \cite{DANIELEWICZ20141} &  \cite{zimmerman2020measuring} &&& \cite{Adhikari_2021}  \\
\hline
\hline
\end{tabular}%
\end{sidewaystable}
%%%%%%%%%%%%

Among the parameter sets, GL97 \cite{NKGb_1997} contains only the nonlinear self couplings ($k_3$ and $k_4$) of $\sigma$ mesons, which reduces the incompressibility at par with the accepted range \cite{DANIELEWICZ20141}. TM1 \cite{Sugahara_1994} set takes into account the self-coupling of $\omega$-meson ($\zeta_0$) to soften the EoS at higher density. Parameter sets FSU2 \cite{Chen_2014}, IUFSU \cite{Fattoyev_2010}, IUFSU$^*$ \cite{Fattoyev_2010}, SINPA \cite{Mondal_2016}, SINPB \cite{Mondal_2016} incorporate the cross-coupling ($\Lambda_\omega$) between $\rho-\omega$ meson which helps in better agreement with the skin thickness ($r_n-r_p$)  and the symmetry energy data \cite{Todd_2003, Todd_2005}. The parameter sets based on E-RMF such as G1 and G2 \cite{Furnstahl_1997} consider the cross-couplings  $\eta_1$, $\eta_2$ and $\eta_\rho$ while excluding $\Lambda_\omega$. These sets give a soft EoS consistent with the kaon and flow data \cite{Arumugam_2004}. In the line of E-RMF, recent forces FSUGarnet \cite{Chen_2015}, IOPB-I \cite{Kumar_2018} and G3 \cite{Kumar_NPA_2017} are designed for the calculation of finite nuclei and neutron star properties. G3 set contains all the couplings present in Eq. (\ref{rmftlagrangian}) and has an additional $\delta$ meson which is an important ingredient in the high-density regime \cite{Singh_2014}. All these forces are extensively used in the literature to estimate various nuclear matter properties ranging from nuclear reaction to nuclear structure and neutron star properties. 

%%%%%%%%%%%%%%%%%%%%%%%%%%%%%%%%%%%%%%%%%%%%%%%%%%
\section{Pasta phase within CLDM approximation}
\begin{figure}
    \centering
    \includegraphics[scale=0.5]{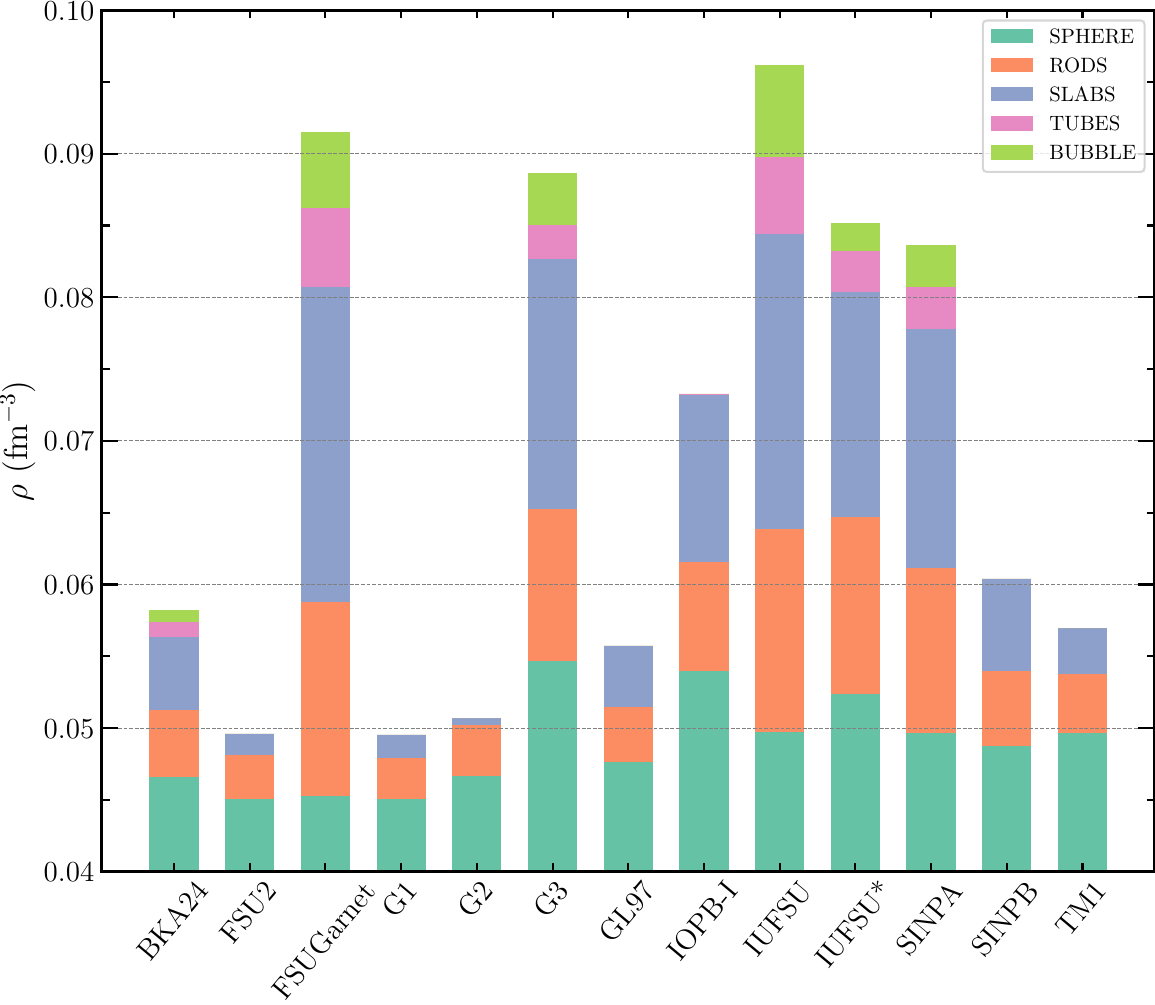}
    \caption{Comparison of the sequence of ground state pasta phase appearance for various functional.}
    \label{fig:past_bar}
\end{figure}

%%%%%%%%%%%%%%%%%%%%%%%%%%%%%%%%%%%%%%%%%%%%%%%%%%
Fig. \ref{fig:past_bar} presents the result of the calculations for the pasta phase in the inner crust of the neutron star using various relativistic parameters  using the CLDM approximation. Different colors represent the density regions where different pasta structures dominate. The edge in each bar represents the transition density of inhomogeneous crust to liquid homogeneous core. It is seen that the spherical geometry dominates for the majority of the inner crust extending up to $\rho \approx 0.05$ fm$^{-3}$ from the outer crust boundary, which is in agreement with various semi-classical and microscopic calculations \cite{Pearson_2018, Pearson_2020, BKS_2015}. There are two categories of parameter sets; one (FSU2, G1, G2, GL97, IOPB-I, SINPB, TM1) that estimates the pasta structure sequence as spheres $\rightarrow$ rods $\rightarrow$ slabs, and second (BKA24, FSUGarnet, G3, IUFSU, IUFSU$^*$, SINPA) that follow spheres $\rightarrow$ rods $\rightarrow$ slabs $\rightarrow$  tubes $\rightarrow$ bubbles. The parameter sets in the latter category are the ones that seem to give a higher density ($\rho_c$) at which the crust-core transition takes place. As one can see that the appearance of different pasta structures is sensitive to the applied model,  one needs to investigate the model dependence.
%%%%%%%%%%%%%%
%%%%%%%%%%%%

The sensitivity of pasta phase appearance can be attributed to two main factors: a) the parametrization of surface and curvature energy and b) the EoS for the bulk and neutron gas surrounding the clusters. Since pasta phase appearance takes place in the region where matter is highly neutron-rich, the correct parametrization of surface and curvature tension Eqs. (\ref{eq:sigma} and \ref{eq:ecurv}) becomes important. For this, as in previous chapter, the surface and curvature parameters in Eqs. (\ref{eq:sigma}) and (\ref{eq:ecurv}) i.e. parameter space $\boldsymbol{S}= \{\sigma_0, b_s, \sigma_{0,c}, \beta,\alpha,p\}$ is fitted with the experimental atomic mass evaluation of AME2020 \cite{Huang_2021} using a suitable penalty function \cite{Dobaczewski_2014, Parmar_2022, Carreau_2019}.  The surface energy plays a seminal role in determining the crustal properties of the neutron star, and therefore, fitting this parameter space for individual EoS is essential to appropriately estimate the surface energy rather than using the same value for all the models. Additionally, there exists a minor energy difference between various pasta structures \cite{Pearson_2018}, and hence, the finite size corrections in terms of surface and curvature term become crucial.   
The value of $p$, which takes care of the isospin asymmetry dependence of surface energy, is taken to be 3. This is a favourable choice in various calculations of surface energy \cite{Lattimer_1991, Avancini_2009, Carreau_2019, Avancini_2009}, A lower/higher value of the surface parameter $p$ results in a larger/smaller value of the surface tension. A smaller surface tension consequently predicts larger crust-core transition pressure and density (see Fig. 6 of \cite{Parmar_2022}). This further impacts the sequence of pasta configuration in Fig. \ref{fig:past_bar}. On varying the value of $p$ from $2.5$ to $3.5$, it is  observed that the number of pasta structures does not change, but the density at which they occur increases slightly for the higher value of $p$. The value of  $\alpha$ is taken as 5.5 \cite{Newton_2013} . Values of rest of the parameter space $\boldsymbol{S}$ is given in Table \ref{tab:surfaceparameter} for all the models considered in Table \ref{tab:forceproperties}. It is evident that the surface parameter $b_s$ has the largest deviation among $ \{\sigma_0, b_s, \sigma_{0,c}, \beta\}$. For minor departures from isospin symmetry, the $b_s$ value denotes the modification in the surface and curvature tensions. Furthermore, the choice of the simplified mass formula, Eq.\ \ref{eq:energy_sys}, is conceptually limited by the fact that mere knowledge of the nuclear mass is not sufficient to derive the surface and curvature contribution because of the partial compensation between nuclear bulk and the surface. This work does not consider the shell effects as the shell corrections to huge semiclassical objects as pasta structures are expected to be small \cite{Pearson_2022}.    
%%%%%%%%%%%%%
\begin{table}
\centering
\caption{The fitted value of surface and curvature energy parameters for various force parameters. The value of $\alpha$ and $p$ is taken to be 5.5 and 3, respectively. Experimental binding energy is taken from AME2020 table \cite{Huang_2021}. }
\label{tab:surfaceparameter}
\scalebox{1.1}{
\begin{tabular}{lllll}
\hline
\hline
Parameter & 
\begin{tabular}[c]{@{}l@{}} \hspace{0.2cm} $\sigma_0$ \\(MeV fm$^{-2}$) \end{tabular}& \hspace{0.2cm} $b_s$ &
\begin{tabular}[c]{@{}l@{}} \hspace{0.2cm} $\sigma_{0,c}$ \\(MeV fm$^{-1}$) \end{tabular} & \hspace{0.2cm} $\beta$ \\
\hline
BKA24 & 0.99339 & 14.3342 & 0.07965 & 0.7711 \\\hline
FSU2 & 0.96665 & 8.77776 & 0.09014 & 0.88746 \\\hline
FSUGarnet & 1.13964 & 29.3893 & 0.07844 & 0.44268 \\\hline
G1 & 0.93641 & 5.55101 & 0.09977 & 0.97866 \\\hline
G2 & 0.99538 & 8.81859 & 0.09672 & 0.85788 \\\hline
G3 & 0.88424 & 26.5837 & 0.09921 & 0.93635 \\\hline
GL97 & 0.73897 & 17.1523 & 0.12018 & 1.19306 \\\hline
IOPB-I & 0.97594 & 16.3546 & 0.09064 & 0.81485 \\\hline
IUFSU & 1.19953 & 30.2177 & 0.07691 & 0.31875 \\\hline
IUFSU* & 1.04205 & 34.2857 & 0.08197 & 0.62258 \\\hline
SINPA & 1.02767 & 24.5575 & 0.08667 & 0.69476 \\\hline
SINPB & 1.03574 & 15.2161 & 0.08332 & 0.70222 \\\hline
TM1 & 0.79998 & 7.35242 & 0.10278 & 1.14013\\\hline \hline
\end{tabular}%
}
\end{table}
%%%%%%%%%%%
%%%%%%%%%%%%%%%%%%%%%%%%%%%%
\begin{figure}
    \centering
    \includegraphics[scale=0.6]{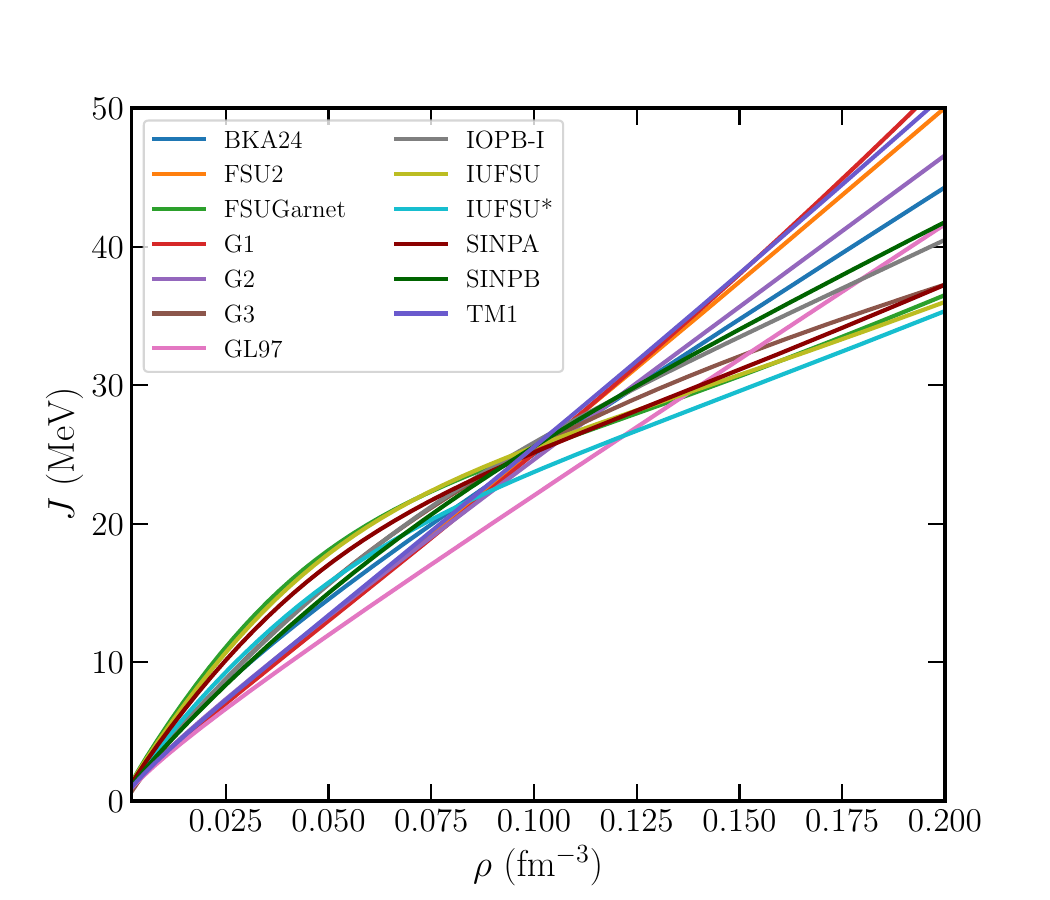}
    \caption{Symmetry energy of the models considered in Fig. \ref{fig:past_bar}.}
    \label{fig:symmetryenergy}
\end{figure}
%%%%%%%%%%%%%%%%%%%%%%%%%%%
Calculation of the inner crust composition is a problem of two-phase equilibrium, which is solved using suitable mechanical and dynamical equations \cite{Parmar_2022, Carreau_2019, Vishal_2021}. In such a system, the symmetry energy plays a deciding role \cite{Vishal_2021, Alam_2017} and is known to influence the inner crust EoS \cite{Pearson_2018}. Furthermore, with ever-improving astrophysical data, establishing available correlations among various nuclear matter and neutron star observables is highly desirable to constrain the equation of state. Nuclear matter properties such as symmetry energy, slope parameter, etc., are calculated at saturation density. These correlations are crucial to fine-tune the theoretical models. Since the relevant density range for crust properties of neutron stars lies below subsaturation density, i.e., below 0.1 fm$^{-3}$, 
one should not merely compare the crust properties of neutron stars with the saturation value of nuclear matter observables. To access the role of symmetry energy on crust EoS, Fig. \ref{fig:symmetryenergy} shows the density dependence of symmetry energy ($J$) for the parameter sets and the corresponding behaviour of equilibrium value of WS cell energy of the inner crust in Fig. \ref{fig:pasta_eos}. 

The density dependence of symmetry energy in the subsaturation density region seems to impact the WS cell energy directly. The parameter sets such as FSUGarnet, G3, IOPB-I, IUFSU, and IUFSU* show higher symmetry energy in the subsaturation density and hence higher crust-core transition density. These forces predict all five pasta phases. The parameter set BKA24, however, estimates lower symmetry energy yet predicts all the five pasta phases. The remaining forces, which estimate lower symmetry energy, estimate the possibility of only three pasta phases, i.e., sphere, cylinder, and slab, and lesser WS energy as shown in Fig. \ref{fig:pasta_eos}. It is relevant to mention that the behaviour of symmetry energy is different below and above the subsaturation density region, i.e., half the value of saturation density. Therefore, one must be cautious while analyzing the impact of symmetry energy on low-density EoS. Table \ref{tab:forceproperties} provides the values of $J$ and $L$ at saturation density and $\rho=0.05$fm$^{-3}$. Since the relative behaviour of symmetry energy among the considered force parameter somewhat remains the same below 0.075 fm$^{-3}$, therefore, the value of 0.05 fm$^{-3}$ is taken as a reference.
%%%%%%%%%%%%%%%%%%%%%%%%%%%
\begin{figure}
    \centering
    \includegraphics[width=0.75\textwidth]{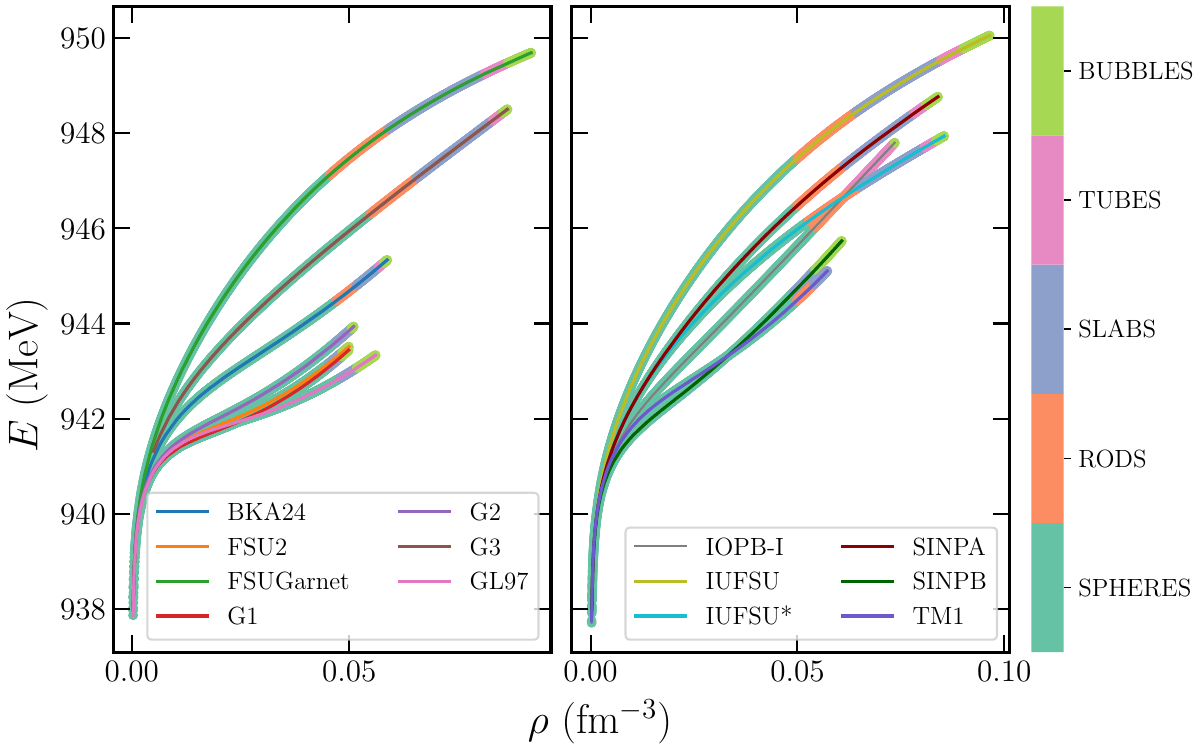}
    \caption{The equilibrium value of WS cell energy for various parameter sets considered in Fig. \ref{fig:past_bar} with the range of different pasta structures. }
    \label{fig:pasta_eos}
\end{figure}
%%%%%%%%%%%%%%%

\begin{figure}
    \centering
    \includegraphics[width=.65\textwidth]{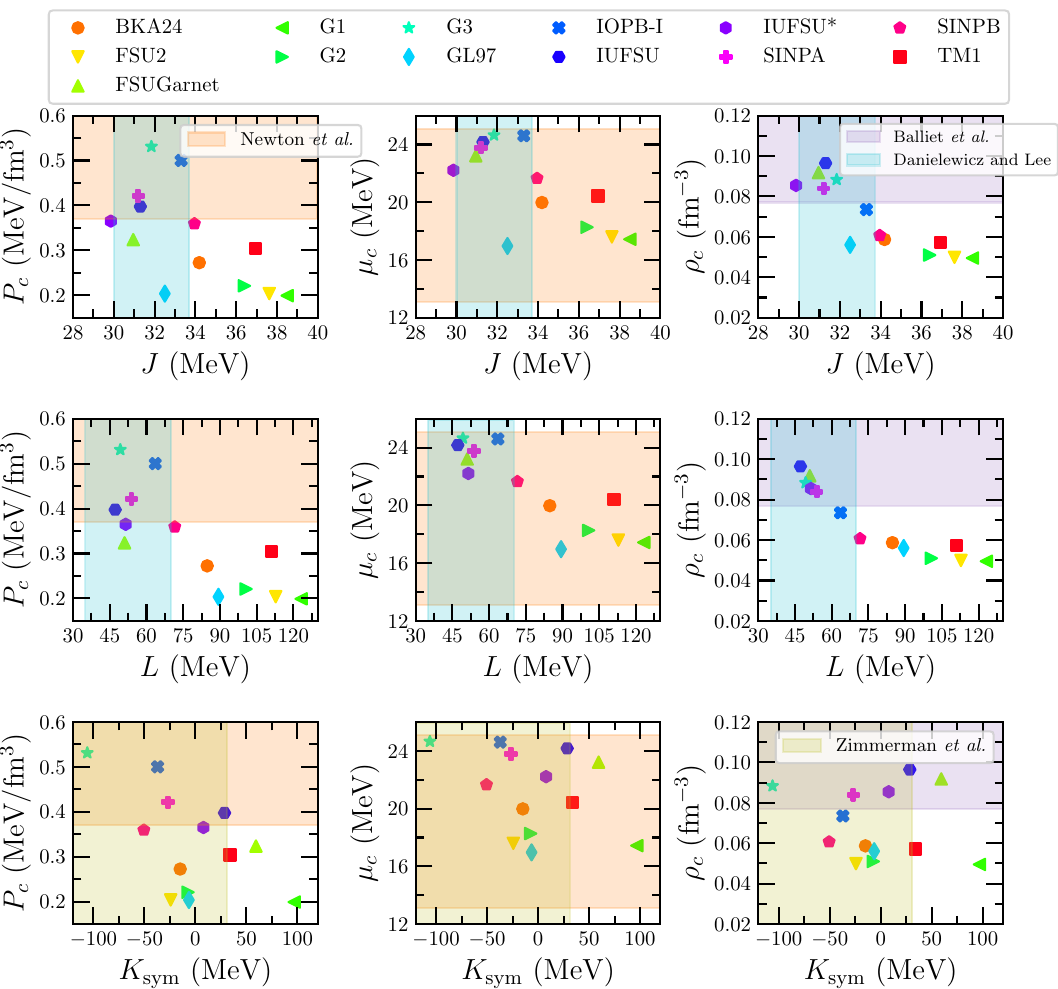}
    \caption{The crust-core transition pressure $P_c$, chemical potential $\mu_c$, and density $\rho_c$ as a function of symmetry energy $J$, slope parameter $L$ and $K_{\rm sym}$. The orange band represents the median range  obtained in Newton $\textit{et al.}$ \cite{Newton_2021} for the uniform Prior + PREX \cite{Adhikari_2021}  data while the purple band represents the uniform Prior + PNM band from the Balliet $\textit{et al.}$ \cite{Balliet_2021} for 95\% credible range. The vertical cyan band is for the empirical/experimental range of symmetry energy and its slope parameter constraints given By Danielwicz {\it et al.} \cite{Danielewicz_2014}. The olive vertical band represents the $K_{\rm sym}$ constraints by Zimmerman {\it et al.} \cite{Zimmerman_2020}.}
    \label{fig:pcc}
\end{figure}

As the density grows in the inner crust, the clusters' surface tension increases, and the system favours the homogeneous phase energetically. The transition from the heterogeneous crust to a homogeneous core is calculated where the energy of the WS cell becomes equal to the energy of the core,  $E_{\rm WS}(\rho_c)=E_{npe\mu}(\rho_c)$. However the crust-core boundary is not determined by the transition density ($\rho_c$), but instead by the transition pressure and chemical potential \cite{Balliet_2021}. The transition pressure controls the mass and moment of inertia of the crust (see Eqs. \ref{eq:pp} and \ref{eq:moic} ) while the chemical potential at crust-core transition determines the thickness of the crust and the pasta structures. Fig. \ref{fig:pcc} compares the transition pressure $P_c$, chemical potential $\mu_c$ and density $\rho_c$ as a function of symmetry energy $J$ and its higher order derivatives, slope parameter $L$ and curvature $K_{\rm sym}$ at the saturation density for various forces,  with the constraints obtained from the Bayesian inference analysis from the two separate studies of Newton \textit{et al.} \cite{Newton_2021} and Balliet \textit{et al.} \cite{Balliet_2021} which use  an extended Skyrme energy density functional within CLDM.
%%%%%%%%%%%%%%
%%%%%%%%%%%%%%
%%%%%%%%%%%%%%
\begin{figure}
    \centering
    \includegraphics[width=0.65\textwidth]{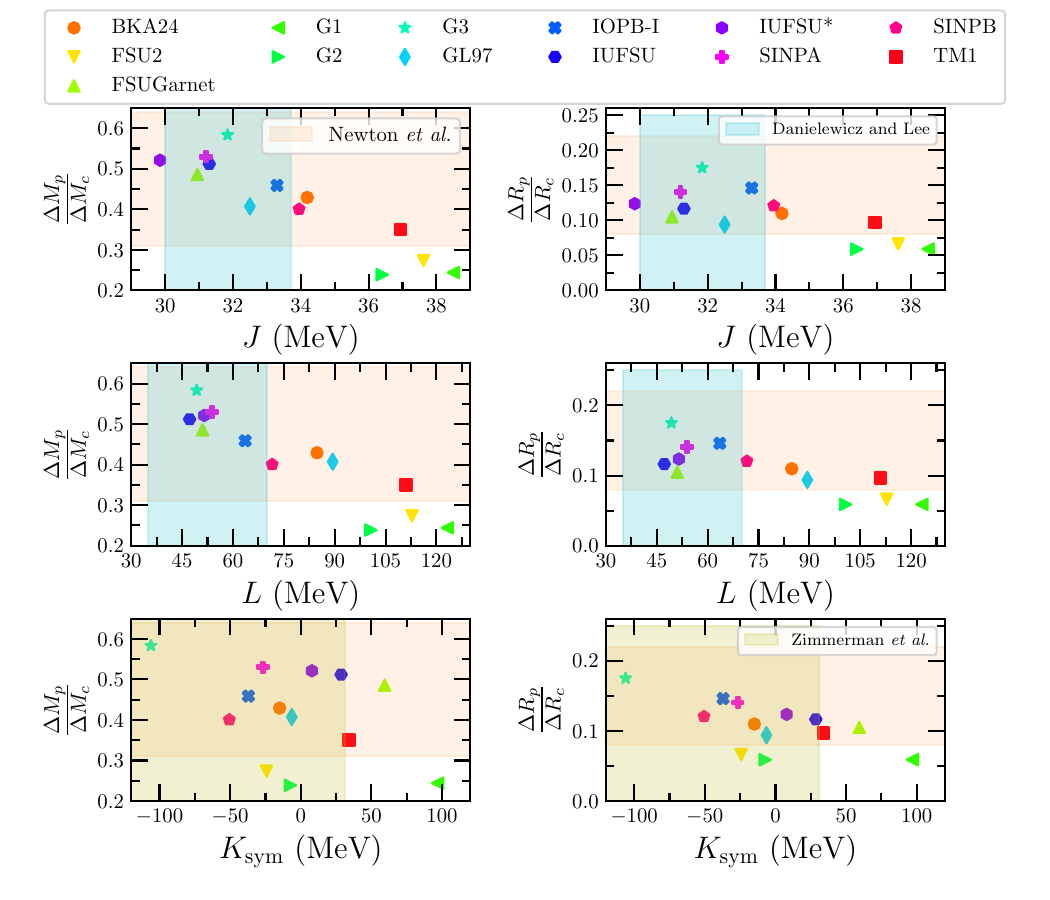}
    \caption{The mass/moment of inertia and thickness fractions of pasta as a function of symmetry energy $J$, slope parameter $L$ and $K_{\rm sym}$.}
    \label{fig:pasta_mass_thick}
\end{figure}
%%%%%%%%%%%%%%

The E-RMF models that satisfy the Newton $\textit{et al.}$ prior + PREX data are the ones that have a lower value of $J$ and $L$ in accordance with the isobaric analog states data \cite{Danielewicz_2014}. However, only  parameter sets SINPA, FSUGarnet, IUFSU, IUFSU$^*$ and TM1 satisfy a more stringent constraints on $P_c$  based on  Skins+ PNM data 
which  results in $P_c=0.38^{+0.08}_{-0.09}$. In contrast, all these models satisfy the prior + PNM constraint of Balliet \textit{et al.} \cite{Balliet_2021} which predict it to be $P_c= 0.49^{+0.27}_{-0.28}$ MeV fm$^{-3}$ on 95\% credible range. All the parameter sets estimate the transition chemical potential $\mu_c$  in agreement with the Newton \textit{et al.} \cite{Newton_2021}. At the same time, the models with lower symmetry energy do not obey the range of $\mu_c=14.7^{+4.7}_{-5.0}$ given by Balliet \textit{et al.} \cite{Balliet_2021}. For the transition density, only models IUFSU, IUFSU*, SINPA, G3, and FSUGarnet satisfy the available constraint from  Balliet \textit{et al.} \cite{Balliet_2021}. Furthermore, $P_c$, $\mu_c$, and $\rho_c$ seem to decrease with higher values of $J$, $L$ and $K_{\rm sym}$ advocating the role of symmetry energy on the crust parameters. The relationship of $K_{\rm sym}$ with $P_c$, $\mu_c$ and $\rho_c$ appears to have a large variance compared to the $J$ and $L$. It should be mentioned here that the transition density is almost half the value of saturation density where the respective values of $J$, $L$, and $K_{\rm sym}$ are calculated. Therefore, the above relationships should accompany the knowledge of symmetry energy in the subsaturation region \cite{Balliet_2021}.

%%%%%%%%%%%%%%%%%%%%%%%%%%%%%%%%%%%%%%%%%%%%%%%%%%%%
\section{Relative pasta layer thickness and mass}
\label{result:relativepasta}
%%%%%%%%%%%%%%%%%%%%%%%%%%%%%%%%%%%%%%%%%%%%%%%%%%%%
Various theoretical calculations predict that the pasta structures account for 15\% of the thickness of the crust and more than 50\% of its mass \cite{Newton_2021, Balliet_2021, Dinh_2021, Fabrizio_2014}. In view of this, following Ref. \cite{Lattimer_2007}, the mass and thickness of the nonspherical shapes using the E-RMF models is calculated and compared with the available theoretical range. The main ingredients are the chemical potential and pressure defined in section \ref{relativepasta}. In Fig. \ref{fig:pasta_mass_thick}, we show the relative mass and the thickness of the nonspherical shapes as a function of $J$, $L$, and $K_{\rm sym}$. All the models except G1, G2, and FSU2, which estimate a relatively larger value of symmetry energy and slope parameter, predict the nonspherical pasta mass and thickness within the range calculated by Newton $\textit{et al.}$ from PREX constraints. These are also consistent with the Skins+ PNM constraints of the  Newton $\textit{et al.}$ \cite{Newton_2021} ( $\frac{\Delta M_p}{\Delta M_c}= 0.49^{+0.06}_{-0.11}$ , $\frac{\Delta R_p}{\Delta R_c}= 0.132^{+0.023}_{-0.041}$), posterior estimations of Thi $\textit{et al.}$ \cite{Dinh_2021} ( $\frac{\Delta M_p}{\Delta M_c}= 0.485 \pm 0.138$, $\frac{\Delta R_p}{\Delta R_c}= 0.128 \pm 0.047$) using meta-model formalism \cite{Carreau_2019} and with the prior + PNM range of Balliet $\textit{et al.}$ \cite{Balliet_2021} ( $\frac{\Delta M_p}{\Delta M_c}= 0.62 ^{+0.03}_{-0.04}$  and $\frac{\Delta R_p}{\Delta R_c}= 0.29^{+0.04}_{-0.09}$).  Since the mass fraction is directly proportional to the amount of moment of inertia \cite{Lorenz_1993}, the behaviour of pasta mass also holds good for its moment of inertia content. Furthermore, the parameter sets with smaller $J$, $L$, and $K_{\rm sym}$  seem to give a larger mass and thickness of the pasta structure. A linear relationship between mass and thickness of pasta with $J$, $L$, and $K_{\rm sym}$ is also evident. 
%%%%%%%%%%%%%%
\begin{figure}
    \centering
    \includegraphics[scale=0.6]{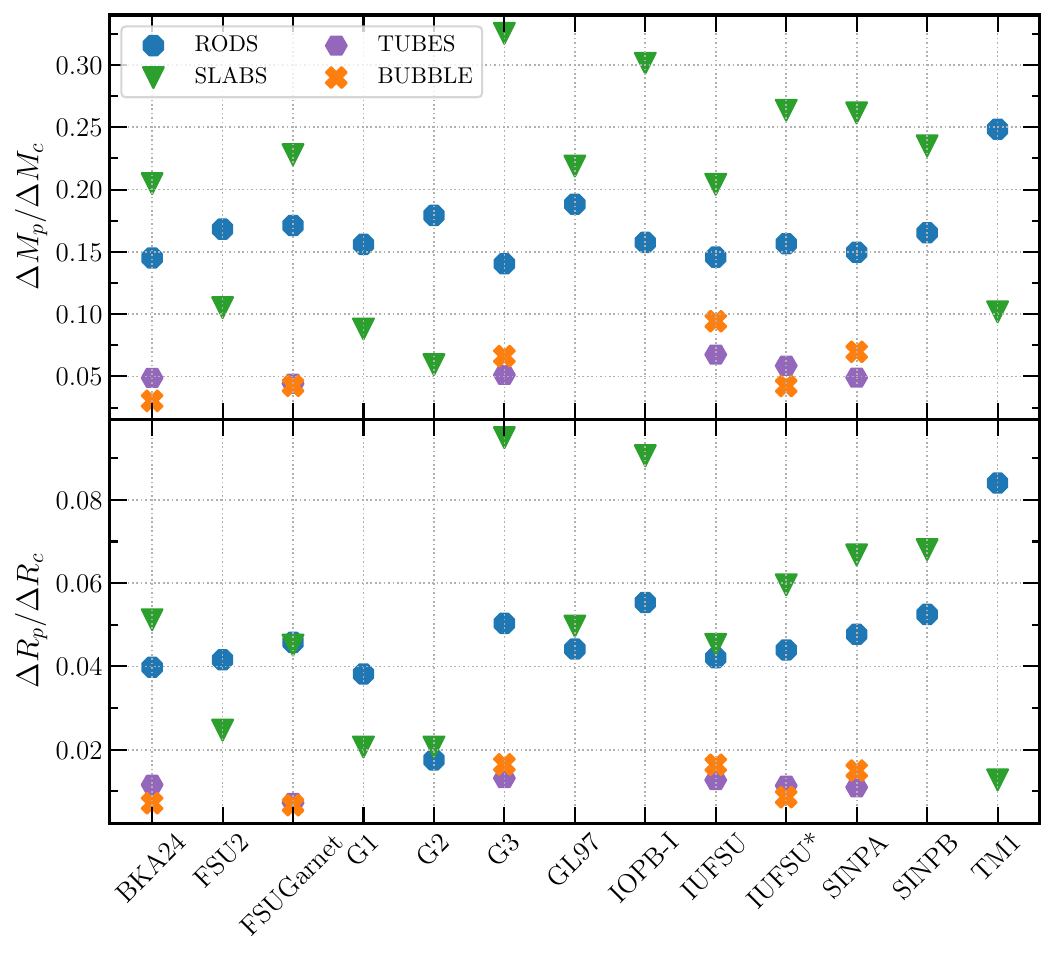}
    \caption{ Upper panel shows the relative mass of the different layers of pasta structures, and the lower panel shows the relative thickness compared to the total crust.}
    \label{fig:rel_mass_thick}
\end{figure}
%%%%%%%%%%%%

Fig. \ref{fig:rel_mass_thick} shows relative mass and thickness of different layer of pasta in the inner crust using the same method as for the total pasta content (Eqs.  \ref{eq:rr} and \ref{eq:pp}). In the present calculations of pasta phases, it is seen that all the models at least predict two nonspherical phases, namely, rods and slabs. The rod pasta phase has mass $\approx$ 15\% of the mass of the crust except for the TM1 set, which estimates its mass $\approx$ 25\%. The thickness of this phase is $\approx$ 4\% of the crust thickness. The parameter sets that predict the existence of only two nonspherical pasta phases before transiting into the homogeneous core have the mass and thickness of the slab phase lesser than the rod phase. The IOPB-I has an exception among these sets. It may be noted that  a third nonspherical tube phase for the IOPB-I set is visible but within a small density range, and hence is not considered (see Fig. \ref{fig:past_bar}). Once again, the symmetry energy seems to impact the relative amount of pasta structures. The parameter sets such as TM1, FSU2, G1, and G2 that have lower symmetry energy in the subsaturation density region predict the larger contribution of the rod phase compared to the slab phase. The remaining parameter sets predict the largest mass and thickness fraction for the slab phase. It accounts for $\approx$ 20\% of the crust mass and 5\% of the crust thickness. The G3 and IOPB-I sets estimate them as large as 30\% and 9\%, respectively. The tube and bubble phase has the smallest content in the inner crust. They account for about 5\% of the crust mass and 1\% of the thickness, subject to their occurrence.  

It is apparent that the existence of pasta structures in the inner crust is greatly influenced by the nuclear EoS. The density dependence of symmetry energy has a prominent role in determining their mass and thickness. To quantify the relationships discussed above,  a Pearson correlation analysis of various crust properties is carried out.
%%%%%%%%%%%%%%
\begin{figure}
    \centering
    \includegraphics[scale=0.5]{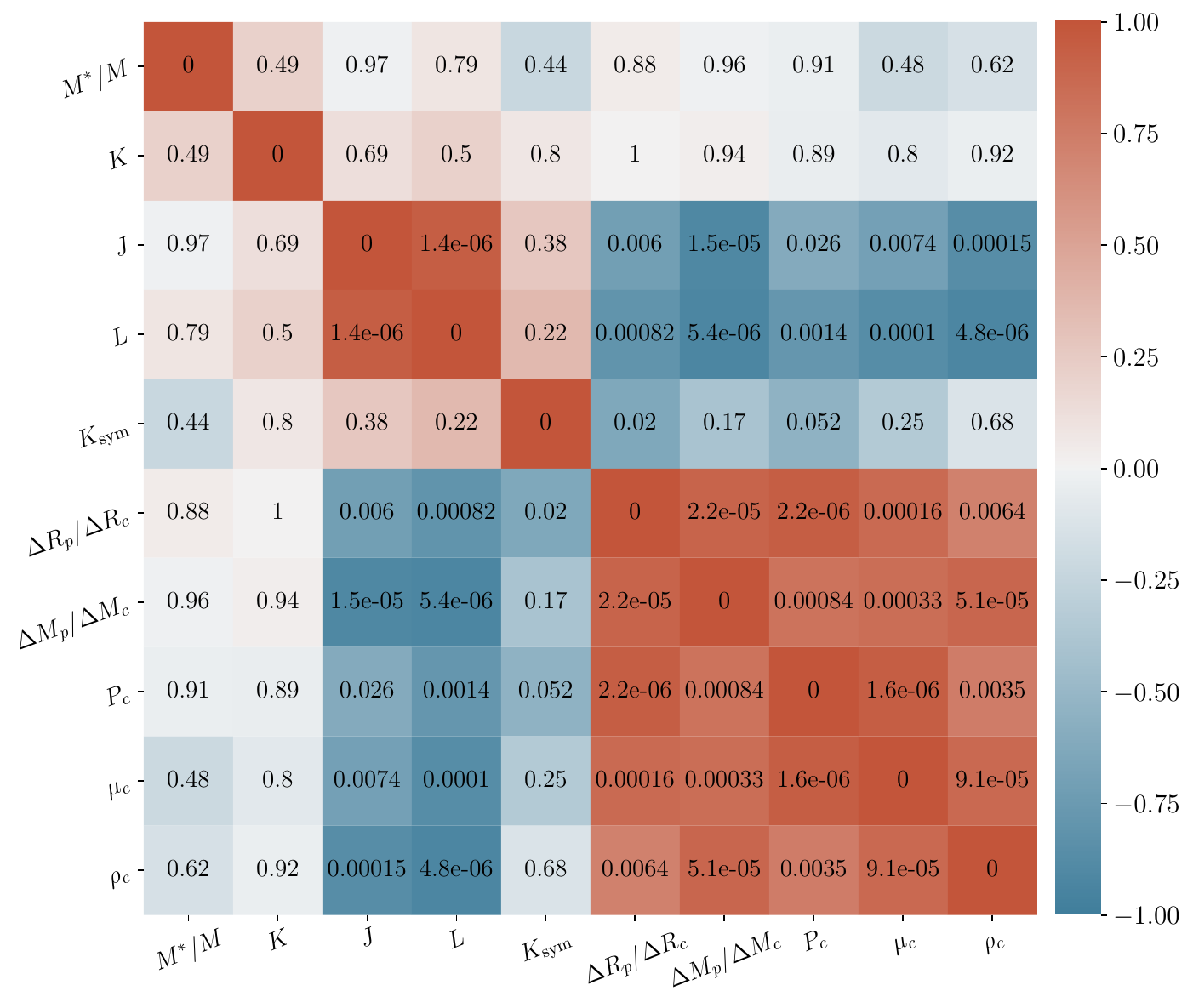}
    \caption{Correlation heat map of the bulk properties with the crustal properties and surface parameters. The color map signifies the strength of the correlation while the values represent the associated $p$-values.}
    \label{fig:corr}
\end{figure}
%%%%%%%%%%%%%%
Fig. \ref{fig:corr} shows the Pearson correlation matrix between the bulk properties, effective mass ($M^*/M$), incompressibility ($K$), symmetry energy ($J$), slope parameter ($L$) and curvature of symmetry energy ($K_{\rm sym}$) with crustal properties namely relative thickness ($\frac{\Delta R_p}{\Delta R_c}$) and mass of the pasta ($\frac{\Delta M_p}{\Delta M_c}$) along with the transition pressure ($P_c$), chemical potential ($\mu_c$) and density ($\rho_c$). The color shows the strength of the correlation while the values represent the statistical significance in the form of $p$-value or probability value \cite{james2013introduction}. A $p$-value signifies the statistical significance of the used statistics (here Pearson correlation), and a value less than 0.05/0.01 is generally considered statistically significant for a 95/99\% interval. It is seen that the bulk properties $M^*/M$ and $K$ do not correlate with the crustal properties. On the other hand, symmetry energy and slope parameter show a strong negative correlation with pasta mass and thickness along with the transition pressure, chemical potential, and density within a 95\% confidence interval. These relations are consistent with those obtained in previous studies \cite{Oyamatsu_2007, Newton_2012}. The $K_{\rm sym}$ shows some negative correlation with the relative thickness of the pasta. 

Additionally, the pasta's mass and thickness are strongly correlated with the transition pressure, chemical potential, and density. All of these relations, which are obtained within the E-RMF framework along with the CLDM formalism, are consistent with the recent work based on Bayesian inference of the neutron star crust \cite{Newton_2012, Balliet_2021, Dinh_2021}. Although these works are based on the relatively more straightforward nuclear interaction models as per the requirement of Bayesian analysis, they provide us with the relevant estimation of various crust properties. The E-RMF model considered in this work is all within reasonable agreement with the theoretical constraints and therefore suitable for further structural calculations of numerous neutron star properties such as superfluidity, conductivity, etc.    
%%%%%%%%%%%%%%%%%%%%%%%%%%%%%%%%%%%%%%%%%%%%%%%%%%%%%%%%%
\section{Shear modulus and torsional oscillation mode}
%%%%%%%%%%%%%%%%%%%%%%%%%%%%%%%%%%%%%%%%%%%%%%%%%%%%%%%%%
A magnetar, which is an exotic type of neutron star, is characterized by an extremely high magnetic field of the order of $10^{15}$G, which results in the powerful x-ray emission powered by the reconfiguration of the decaying field. The rapidly evolving field, when it strikes the solid crust, results in an associated starquake, detectable as quasiperiodic oscillations (QPOs) \cite{Steiner_2009, Thompson_2001, Israel_2005, Strohmayer_2005}. In this context, it becomes essential to understand the shear property of the crust. The shear modulus, which describes the  elastic response of the crust under the shear stress, leads to the shear oscillations.

Shear oscillations propagate within the crust of the star at a velocity known as the shear velocity ($V_s$). The composition of the crust is determined by its shear modulus and shear velocity, which in turn rely on the nuclear EoS and surface energy parametrization. This work use the Monte Carlo simulation results in the form of Eq. (\ref{eq:shearmodulus}) for the spherical portion of the inner crust. The elastic response of the nonspherical phase is not yet fully understood, but the crust's rigidity is expected to decrease and vanish at the crust-core boundary \cite{Gearheart_2011,Passamonti_2016}. To model the shear modulus in this region,  Eq. (\ref{eq:mubar}) is used. 
%%%%%%%%%%%%%%
\begin{figure}
    \centering
    \includegraphics[scale=0.7]{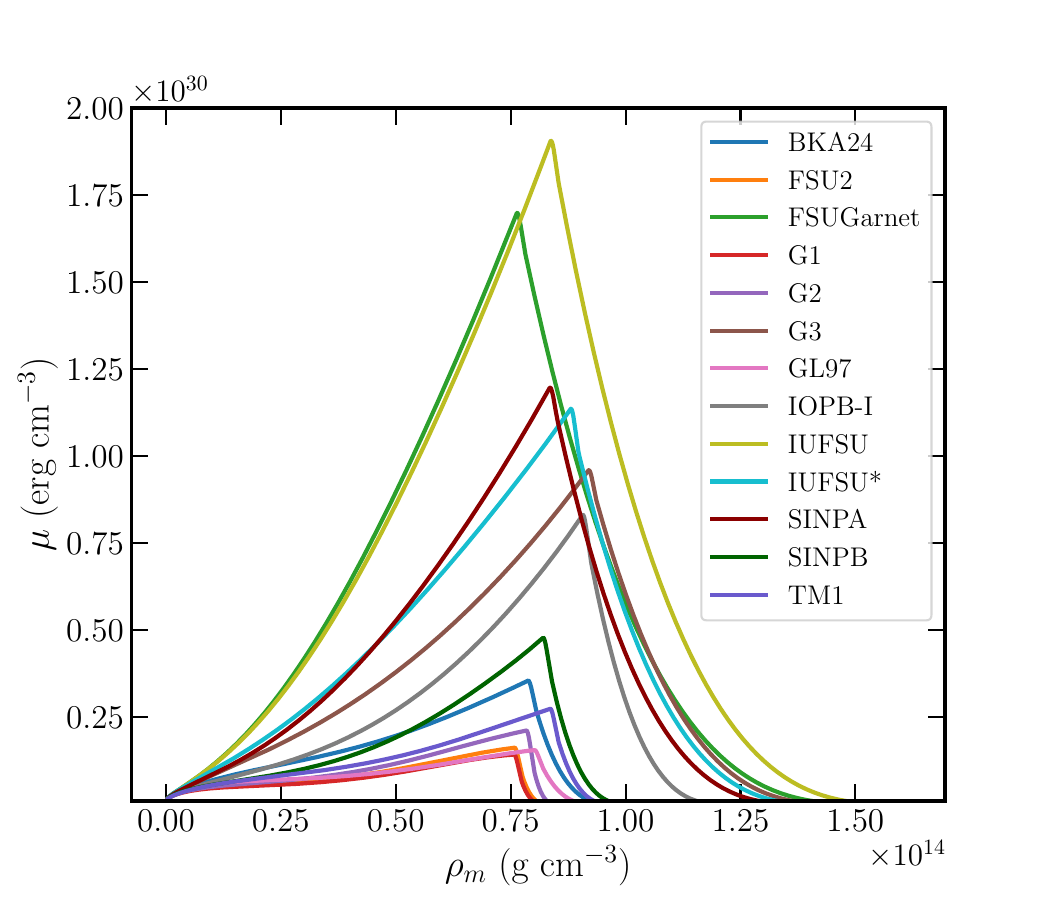}
    \caption{Shear modulus ($\mu$) of the inner crust for various E-RMF sets.}
    \label{fig:shear}
\end{figure}
%%%%%%%%%%%%%%

The complete behaviour of the shear modulus of the inner crust is shown in Fig. \ref{fig:shear}. As one moves deeper into the crust, the shear modulus increases monotonically until one reaches the density where the pasta phase appears. It then starts decreasing smoothly until the crust-core boundary and then vanishes. This behaviour directly results from our approximation of the shear modulus in the pasta phase region. There is a significant uncertainty among different models, which is the consequence of the inner crust composition predicted by these models. Since the density dependence of symmetry energy and slope parameter predominantly control the inner crust \cite{Parmar_2022},  its effect on the shear modulus are visible as well. In the subsaturation region, forces such as IUFSU, G3, BKA24, and FSUGarnet, which have a higher value of symmetry energy, estimate a larger shear stress value. 
%%%%%%%%%%%%%%
\begin{figure}
    \centering
    \includegraphics[scale=0.7]{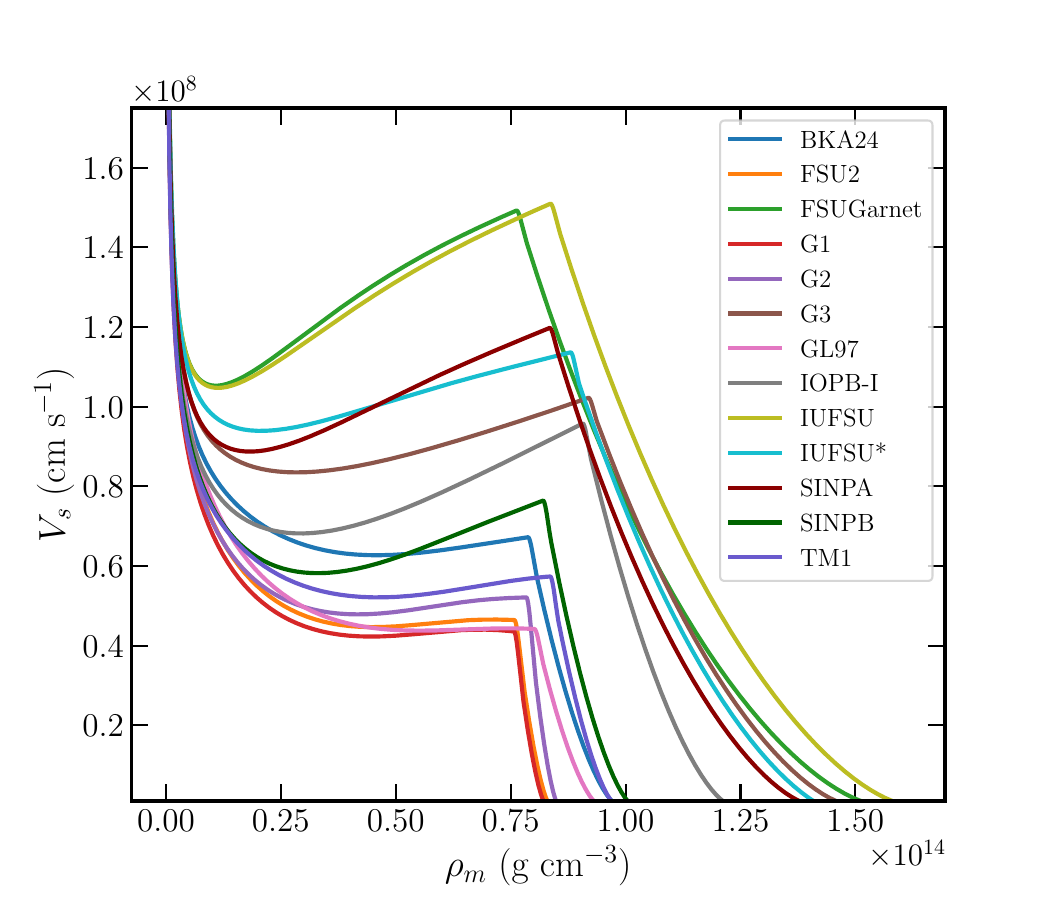}
    \caption{The shear velocity ($V_s$) as a function of the mass density for the various E-RMF models.}
    \label{fig:shear_speed}
\end{figure}
%%%%%%%%%%%%%%

Now the shear velocity  can be calculated using Eq.\ (\ref{eq:shearspeed}). In principle, the neutron superfluidity plays a crucial role in neutron star crust properties \cite{Haskell_2018}. The superfluid neutrons are unbound from the lattice moment and do not influence the shear modulus. However, Chamel  \cite{Chamel_2012} found that $\approx$ 90\% of the superfluid neutrons can be entrained to the lattice due to the Bragg scattering. However, this work considers the dynamical mass in Eq. (\ref{eq:shearspeed}) equal to its total mass density \cite{Steiner_2009} neglecting the effect of the superfluidity and entrainment effects. The calculated shear speed will then underestimate its value, but the qualitative nature will remain unaffected. Fig. \ref{fig:shear_speed} shows the behaviour of shear speed for the corresponding shear modulus in Fig. \ref{fig:shear}. The shear speed in the outer crust which is well established \cite{tews_2017} is not shown here. The shear speed increases with the increase in density. However, in the inner crust, it drops initially and then increases until the onset of the pasta structures. It decreases smoothly afterward and vanishes at the crust-core boundary. One can see that there is $\approx$ 3 fold difference between the lowest and highest value of shear speed among the parameter sets considered in this work. The dependence of shear velocity on the density also varies in a different way indicating the role of crust composition. 
%%%%%%%%%%%%%%
\begin{figure}
    \centering
    \includegraphics[scale=0.6]{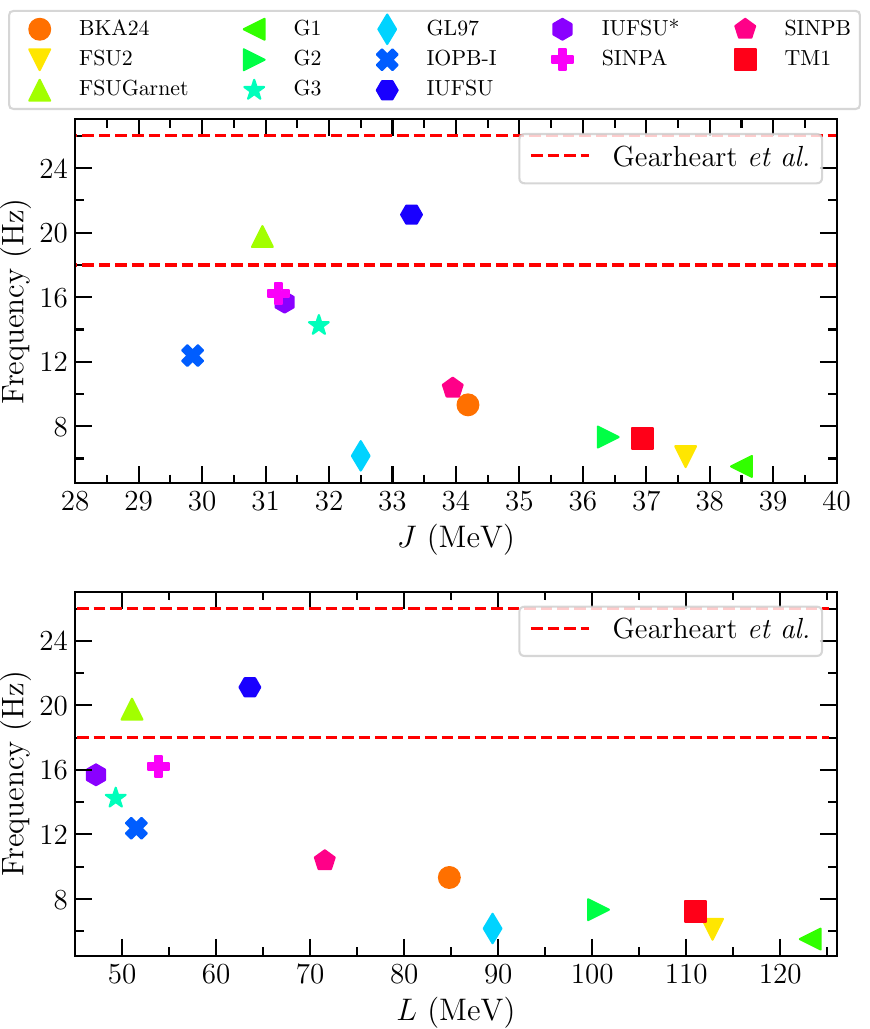}
    \caption{Frequency of fundamental torsional oscillation mode ($l=2$) in the crust for the maximum mass with $J$ and $L$. The two horizontal lines correspond to the observed value of 18 and 26 Hz. }
    \label{fig:freq}
\end{figure}
%%%%%%%%%%%%%%  
To approximately infer the fundamental torsional oscillation mode, the pasta shear modulus is assumed to be zero, considering the pasta as a liquid \cite{Gearheart_2011}. This means that the shear modulus and shear velocity in the solution of crustal shear perturbation equations (Eq. (\ref{eq:freq})) are calculated at the boundary between the phase of spherical nuclei and the pasta phases, i.e., $\rho=\rho_{ph}$. Fig. \ref{fig:freq} shows the calculated frequencies of fundamental oscillation mode ($l=2$) for the maximum mass as a function of $J$ and $L$ for the various E-RMF models along with the possible candidate of frequencies for the fundamental modes of QPOs: 18 Hz and 26 Hz \cite{Gearheart_2011, Israel_2005}. The fundamental frequency decreases with the symmetry energy and the slope parameter, which is consistent with the Refs. \cite{Gearheart_2011, Sotani_2013}. It is  seen that considering the pasta phase to be liquid and ignoring superfluid and entrainment effects, the fundamental mode frequency agrees with the observed QPOs from SGRs at low symmetry energy and slope parameter. Only FSUGarnet and IUFSU parameter sets match with the 18 Hz observed frequency. The frequencies also do not match with higher possible candidate frequencies of 28 Hz and  30 Hz \cite{Greif_2020, Israel_2005}. It is important to note that the current study employs a simplistic approach, represented by Eq. (\ref{eq:freq}), to calculate the fundamental torsional oscillation mode. Surprisingly, this approximation yields results comparable to those obtained from solving the full problem numerically, as demonstrated in \cite{Samuelsson_2006}. However, to comprehensively explain associated astrophysical observations, such as magneto-elastic oscillations \cite{Gabler_2013, Gabler_2011}, consistent calculations are required. Consequently, obtaining fully numerical solutions for the torsional mode can provide deeper insights into these phenomena. Moreover, recent work by Chamel et al. \cite{Chamel_2022} indicates that nuclear pasta structures may be much less abundant than previously believed, especially when considering shell and pairing effects. As a result, the frequencies calculated in this study could be considered as lower bounds for the fundamental frequency. Additionally, the pasta structures in the neutron star crust significantly influence the torsional mode. The frequency modes observed in QPOs can be utilized as asteroseismological sources to constrain the presence of pasta structures, in conjunction with other nuclear matter observables such as symmetry energy and slope parameter. Incorporating these factors into the analysis will help refine our understanding of the underlying physics and shed light on the intricate properties of neutron stars.

%It may be noted that considering pasta to be liquid and neglecting entrainment effects reduces the frequency by a factor of $\approx$ 3  \cite{Gearheart_2011}. 

%%%%%%%%%%%%%%%%%%%%%%%%%%%%%%%%%%%%%
\section{Neutron star observables}
%%%%%%%%%%%%%%%%%%%%%%%%%%%%%%%%%%%%%
%%%%%%%%%%%%%%
\begin{sidewaystable}\renewcommand{\arraystretch}{1.05}
\centering
\caption{The neutron star properties such as maximum mass ($M_{\rm max}$), radius corresponding to the maximum mass  ($R_{max}$), canonical radius ($R_{1.4}$), normalized maximum MI ($I_{max}$), normalized canonical MI ($I_{1.4}$), mass of the crust ($M_{\rm crust}$), thickness of the crust ($l_{crust}$), second Love number and dimensionless tidal deformability for canonical and maximum mass for 13 considered EoSs.}
\label{tab:NS_properties}
\renewcommand{\tabcolsep}{0.26cm}
\renewcommand{\arraystretch}{1.6}
\begin{tabular}{lllllllllllll}
\hline \hline
\begin{tabular}[c]{@{}l@{}}Parameter\\ sets\end{tabular} &
\begin{tabular}[c]{@{}l@{}}$M_{\rm max}$\\ ($M_\odot$)\end{tabular} &
\begin{tabular}[c]{@{}l@{}}$R_{\rm max}$\\  (km)\end{tabular} &
\begin{tabular}[c]{@{}l@{}}$R_{1.4}$\\ (km)\end{tabular} &
$I_{\rm max}$ &
$I_{1.4}$ &
$I_{\rm crust}/I$ &
\begin{tabular}[c]{@{}l@{}}$M_{\rm crust}$\\ ($M_\odot$)\end{tabular} &
\begin{tabular}[c]{@{}l@{}}$l_{\rm crust}$\\ (km)\end{tabular} &
$k_{2,1.4}$&
$\Lambda_{1.4}$ &
$k_{2,{\rm max}}$&
$\Lambda_{\rm max}$\\ \hline
BKA24&1.963 &11.61 &13.42 &0.401 &0.339 &0.0100  &0.008 &0.455 &0.0888 &681.25 &0.0307 &21.02
\\ 
FSU2&2.071 &12.12 &14.02 &0.405 &0.335 &0.0087 &0.008 &0.418 &0.0943 &899.64 &0.0302 &19.54
\\ 
FSUGarnet&2.065 &11.79 &13.19 &0.418 &0.343 &0.0100 &0.009 &0.542 &0.0889 &629.13 &0.0306 &17.54 \\ 
G1&2.159 &12.30 &14.15 &0.415 &0.331 &0.0084 &0.008 &0.413 &0.0922 &922.41 &0.0287 &16.34
\\ 
G2&1.937 &11.17 &13.27 &0.403 &0.333 &0.0069 &0.006 &0.378 &0.0819 &593.53 &0.0266 &16.11
\\ 
G3& 1.996 &10.95 &12.63 &0.425 &0.345 &0.0129 &0.011 &0.479 &0.0813 & 460.42 &0.0254 &11.98
\\ 
GL97&2.002 &10.81 &13.10 &0.423 &0.343 &0.0048  &0.004 &0.296 &0.0876 &596.43 &0.0221 &09.64
\\ 
IOPB-I&2.148 &11.96 &13.33 &0.428 &0.344 &0.0147 &0.014 &0.507 &0.0925 &686.49 &0.0292 &14.75
\\ 
IUFSU&1.939 &11.23 &12.61 &0.414 &0.351 &0.0110 &0.009 &0.510 &0.0871 &489.20 &0.0313 &19.10
\\ 
IUFSU*&1.959 &11.45 &12.92  &0.409 &0.347 &0.0114 &0.010 &0.526 &0.0880 &563.08 &0.0319 &20.66 \\ 
SINPA&2.000 &11.55 &12.93  &0.416  &0.349 &0.0136  &0.012 &0.515 &0.0908 &580.39 &0.0318 &19.36 \\ 
SINPB&1.993 &11.62 &13.16 &0.409 &0.342  &0.0128  &0.011  &0.486 &0.0881 &612.94 &0.0313 &19.98  \\ 
TM1&2.175 &12.36 &14.31 &0.415 &0.335 &0.0101 &0.009 & 0.444 &0.0979 &1037.5 &0.0285 & 15.97
\\ \hline \hline
\end{tabular}
\end{sidewaystable}
%%%%%%%%%%%%
Similar to the previous chapter, a complete neutron star is modelled by calculating the core EoS under the condition of charge neutrality \cite{Parmar_2022}, and $\beta$-equilibrium for each parameter set in Table \ref{tab:forceproperties}. The unified EoS in constructed by calculating the inner crust and core EoS using the same E-RMF parameter set along with the outer crust EoS discussed in Chapter \ref{chap5}. The unified EoSs are available  publicly in GitHub page\footnote{\url{https://github.com/hcdas/Unfied_pasta_eos}}. The unified treatment of each EoS ensures that the neutron star properties such as crust mass, thickness, the moment of inertia, etc., can be estimated and analyzed quite precisely. To calculate the neutron star observables, one can solve the TOV Eqs. (\ref{eq:pr} and \ref{eq:mr}) for a fixed central density to obtain the $M-R$ profile, second Love number, and dimensionless tidal deformability. The moment of inertia is calculated under the slow rotation approximation using Eq. (\ref{eq:moi}). The total crust mass and thickness is estimated by integrating the TOV Eqs. (\ref{eq:pr} and \ref{eq:mr}) from $R=0$ to $R=R_{\rm core}$, which depends on pressure as $P(R=R_{\rm core})=P_t$. Finally, the crustal moment of inertia is worked out using Eq. (\ref{eq:moic}). The detailed formalism of these quantities is provided in Refs. \cite{Lattimer_2000, Parmar_2022}. The mass and thickness of the crust for IOPB-I EoS are 0.013 $M_\odot$, and 0.490 km, respectively, without considering the pasta phase inside the crust (see Table 7 in Ref. \cite{Parmar_2022}). However, they are estimated to be 0.014 $M_\odot$ and 0.507 km, respectively, including the pasta structures. Hence, it is  noticed that the crustal mass doesn't change, but the crustal thickness increases slightly when one considers pasta phases inside the crust.

Table \ref{tab:NS_properties} gives the tabulated data for neutron star properties such as maximum mass ($M_{\rm max}$), radius corresponding to the maximum mass ($R_{\rm max}$), canonical radius ($R_{1.4}$), normalized maximum MI ($I_{\rm max}$), normalized canonical MI ($I_{1.4}$), mass of the crust ($M_{\rm crust}$), thickness of the crust ($l_{\rm crust}$), second Love number ($k_2$) and dimensionless tidal deformability ($\Lambda$)  for canonical and maximum mass for 13 considered EoSs. The maximum mass of all the sets reasonably satisfy the observational constraint  of massive pulsars such as PSR J0348+0432 ($M = 2.01\pm{0.04} \ M_\odot$) \cite{Antoniadis_2013} and PSR J0740+6620 ($M = 2.14_{-0.09}^{+0.10} \ M_\odot$) \cite{Cromartie_2019}. They are also  in accordance with the  radius constraints given by Miller {\it et al.} \cite{Miller_2019}, Riley {\it et al.} \cite{Riley_2019} and  PSR J0030+0451 with X-ray Multi-Mirror Newton for canonical star with $R_{1.4} = 12.35 \pm 0.75$ km \cite{Miller_2021}.

The normalized moment of inertia for slowly rotating neutron star is calculated for 13 EoSs. The numerical values are given in Table \ref{tab:NS_properties} both for the canonical and maximum mass star. There exists a Universal relation between the MI and the compactness of the star \cite{Lattimer_2005, Steiner_2016, LATTIMER_2016}. The value of $I_{\rm max}$ and $I_{1.4}$ for IOPB-I EoS was found to be 0.429 and 0.346, respectively,  without pasta phases in the previous chapter. By including the pasta phase, the values are slightly lesser and found to be 0.428 and 0.344, respectively. Similar cases are seen both for FSUGarnet and G3 EoSs. Hence, it is observed that the pasta phases don't significantly influence the moment of inertia of the star. However, the crustal moment of inertia ($I_{\rm crust}/I$) for maximum mass estimated from these EoSs are consistent with the fractional moment of inertia (FMI) observed from the 581 pulsar glitches catalog \cite{Espionza_2011, Parmar_2022}. One can also see that the mass of the crust ($M_{\rm crust}$) is equivalent to the crustal moment of inertia, advocating the importance of unified treatment of crust and core equation of state. 

The Love number and dimensionless tidal deformability for only quadrupole case ($l=2$) are calculated as described in Ref. \cite{Das_2022}. The numerical values are given in Table \ref{tab:NS_properties} for  considered EoSs. For a realistic star, the value of $k_2$ is 0.05--0.1 \cite{Hinderer_2008}. Our calculated results are well within this range. The constraint on $\Lambda_{1.4}$ given by LIGO/Virgo \cite{Abbott_2017, Abbott_2018} from the binary neutron star merger event GW170817 with, $\Lambda_{1.4}=190_{-70}^{+390}$. Only G3, IUFSU, and IUFSU* are within the GW170817 limit. It is also observed that the effects of pasta on both $k_2$ and $\Lambda$ are not significant as compared with only the spherical shape considered inside the crust. 

The relativistic nuclear models considered in this work suggest that $\approx$ 50\% of the crust mass and $\approx$ 15\% of the crust thickness is contained in the pasta structures. 
Since the entire crust itself comprises only 0.5-1\% of the neutron star mass and 5-10 \% of the radius, the pasta structures do not significantly impact the global properties of a neutron star such as maximum mass, the moment of inertia, Love number, dimensionless tidal deformability, etc. However, the pasta structure affects the microscopic properties of the neutron star, which essentially depend on the crust structure. The shear modulus, which determines the torsional oscillation mode of quasiperiodic oscillations (QPOs), is greatly influenced by the presence of pasta structures. The fractional crustal moment of inertia or mass is an important property to explain the pulsar glitches. The pasta content in the crust influences these properties by controlling the surface thickness. These structures also influence the magnetic field's decay rate, which explains the observed population of isolated X-ray pulsars \cite{Caplan_2017} and limits the maximum spin period of rotating neutron stars \cite{Pons_2013}. The properties such as viscosity, conductivity, neutrino cooling, etc., are also influenced by the nature of the structure present in the inner crust \cite{Caplan_2017}.
%%%%%%%%%%%%%%%%%%%%%
\section{Summary}

%%%%%%%%%%%%%%%%%%%%%
In summary, the present chapter investigates the existence of pasta structures in the inner crust of a neutron star employing the compressible liquid drop model along with the effective relativistic mean-field theory. It  considers three geometries: spherical, cylindrical, and planar, resulting in five configurations, namely sphere, rod, slab, tube, and bubble. The equilibrium configuration at a given baryon density is obtained by minimizing the energy of the five pasta structures. The main ingredient in calculating the inner crust is the proper treatment of the surface energy parametrization. 
   
The present  calculations have used 13 well-known parameter sets that satisfy the recent observational constraints on the maximum mass and radius of the neutron star. It constructs unified EoS for each of these sets to obtain the pasta and crustal properties consistently. The appearance of different pasta layers is model-dependent. The model dependency is attributed to the behaviour of symmetry energy in the subsaturation density region and the surface energy parametrization. A thicker crust favours the existence of more number of pasta layers in it. The pressure ($P_c$), chemical potential ($\mu_c$), and density ($\rho_c$) of the crust-core transition from the crust side is calculated and compared with the results from recent constraints proposed using Bayesian inference analyses \cite{Newton_2021, Balliet_2021}. The parameter sets with lower values of $J$, $L$, and $K_{\rm sym}$ seem to agree better with these theoretical constraints. 

It is seen that the ($P_c$) and ($\mu_c$) play a more critical role in determining the crust structure instead of ($\rho_c$). The mass and thickness of the total pasta layers in the inner crust are calculated using all the models considered in this work. The parameter sets with larger/smaller symmetry energy and slope parameter estimate thinner/thicker crust and thickness of the pasta structures. Alternatively, a larger negative/positive $K_{\rm sym}$ value corresponds to the thicker/thinner crust and pasta mass and thickness. The pasta mass and thickness are also in agreement with various theoretical constraints. Additionally, rod and slab configurations occupy the largest mass and thickness in the inner crust. The E-RMF models that predict the existence of only two nonspherical pasta phases before transiting into the homogeneous core have the mass and thickness of the slab phase lesser than the rod phase.
   
Quasiperiodic oscillations in soft gamma-ray repeaters are one of the observational means to constrain the inner crust structure and the amount of pasta structures in it. In view of this, the present chapter calculates the shear modulus and shear speed in the inner crust of a neutron star by using different methods for the spherical and pasta layers. These quantities are also model-dependent, and considerable uncertainty exists between them. It then considers the pasta layers to have zero shear modulus  to approximate the frequency of fundamental torsional oscillation mode in the crust for the maximum neutron star mass. The pasta structure significantly impacts the fundamental frequency mode. Out of 13 EoSs, only two parameter sets, FSUGarnet and IUFSU, agree with the 18Hz observational frequency. Finally, various neutron star properties for the constructed unified equation of states are estimated. The pasta phases do not impact the star's moment of inertia significantly. The fractional crustal moment of inertia ($I_{\rm crust}/I$) for maximum mass estimated from these EoSs are consistent with the pulsar glitch catalog. 
 
%%%%%%%%%%%%%%%%%%%%%

\clearpage
\addcontentsline{toc}{section}{Bibliography}
\printbibliography

%% file: Chapter_7/CHAP7.tex
\chapter{\label{chap7}  Magnetised neutron star crust}
\section{Introduction}
%%%%%%%%%%%%%%%%%%%%%%
In general, the global properties such as mass-radius profile, the moment of inertia, etc., of  neutron star are dictated by their core, where density reaches $\sim 10$ times the nuclear saturation density, and the matter is considered to be homogeneous \cite{Haenel_2007}. The core is covered by $\sim 1$ km thick heterogeneous crust characterized by fully ionized nuclei submerged in strongly degenerate electron gas known as the outer crust and the nuclear clusters (spherical in the shallower region and distorted (nuclear pasta) in dense regions) surrounded by electron and degenerate dripped neutron gas known as inner crust \cite{chamel2008physics}. This layer of the neutron star is of predominant area of curiosity to nuclear and astrophysicists as it acts as a unique exotic non-terrestrial laboratory to test theories of strong interaction and validate them using various observations. Recently it was shown that the crust plays an important role in stabilizing the magnetic field by solidification, which results in the development of elastic forces that consequently avoid the fast decay of the magnetic field \cite{Lander_2021}.  Therefore, an accurate description of neutron star crust in the presence of a magnetic field is essential to extract the core's properties and understand microscopic aspects of  the crust, such as cooling \cite{Ootes_2018}, entrainment,  quasiperiodic oscillations (QPOs) \cite{Stein_2016}, torsional vibrations, shattering \cite{Tsang_2009}, transport \cite{rezzolla2018physics} etc. The astrophysical phenomena related to the interaction and evolution of the magnetic field in the neutron star crust \cite{Baiko_2011, Konstantinos_2016, Sengo_2020} also make the study of neutron star crust in a magnetic environment highly desirable.

Most neutron star crust calculations in literature have been performed for an unmagnetised neutron star, and not much emphasis is given to the magnetised crust. Among a handful of studies that consider the magnetic field, the majority consider only the outer crust composition \cite{Basilico_2015, Chamel_2012, Chamel_2015, Arteaga_2011, Stein_2016}. The inner crust calculations are  either limited to the effect of magnetic field on the electrons  \cite{Mutafchieva_2019} or study  the crust-core transition properties employing Vlasov formalism for dynamical instability \cite{fang_2017, Fang_2017_1}. Only a few inner crust calculations have been performed using the self-consistent Thomas-Fermi approximation employing relativistic mean field theory \cite{Bao_2021, Nandi_2011, Lima_2013}, some of which \cite{Lima_2013} consider a fixed proton fraction instead of $\beta$-equilibrium in the inner crust. Therefore, the lack of comprehensive investigations of magnetised neutron crust in a unified manner using the realistic equation of state (EoS), which satisfies relevant nuclear matter and neutron star constraints, underscores the need for such analysis.

In this chapter, the aim is to investigate the possible changes in the neutron star crust's structure in presence of  magnetic field  and its EoS and study associated phenomena. 
To achieve this, the calculations of \textit{chapter 5 and 6} are extended for the case of magnetic fields by incorporating the magnetic field effects in the EoS. This will help  to understand the neutron star crust structure in a magnetic environment comprehensively and analyze the possible deviations as compared to the unmagnetised neutron star \cite{Xiaopeng_2022, Scurto_2023}. For the first time, to my knowledge, the CLDM method is used to estimate the crust structure of a magnetised neutron star. 

Similar to previous chapters, to calculate the equilibrium composition of the outer crust, the most recent experimental atomic mass evaluations AME2020 \cite{Huang_2021} in supplement with the theoretical calculations of Hartree-Fock-Bogoliubov (HFB) \cite{hfb2426, hfb14} and finite-range liquid-drop model (FRDM) \cite{MOLLER20161} are used. The inner crust structure is determined using six E-RMF parameter sets with varying saturation properties, namely: G3 \cite{Kumar_2017}, IOPB-I \cite{Kumar_2018}, FSUGarnet \cite{Chen_2015}, IUFSU \cite{Fattoyev_2010}, IUFSU$^*$ \cite{Fattoyev_2010}  and SINPB \cite{Mondal_2016}. Various inner crust properties, such as equilibrium composition, crust-core transition properties, pasta phase appearance, pasta mass, thickness, and frequency of QPOs in context to the soft gamma repeaters (SGRs), etc., are calculated for a magnetised neutron star. These properties play a central role in explaining various mechanisms of magnetar and pulsar activities which include transport of magnetic field lines (hall drift) \cite{Konstantinos_2016, Perna_2011}, sudden fractures in the crust due to accumulating crustal stress \cite{Beloborodov_2014}, neutron superfluidity that causes glitches in pulsars \cite{anderson1975pulsar}, transient heating of crust \cite{Li_2015}, etc. Furthermore, the present chapter also investigate the role of EoS on the crustal properties of neutron stars in a magnetised environment. The unified EoS at a given magnetic field, considering the same strength throughout the neutron star's interior is constructed. This work is restricted to the strength of the magnetic field under $B^* \le 5000$ to satisfy the assumptions made in \cite{Patra_2020} to calculate the neutron star structure using spherically symmetric treatment of the neutron star structure.

%%%%%%%%%%%%%%%%%%%%%%%%%%%%%%%%

\section{Outer crust}
\label{oc}
The equilibrium composition of the outer crust of a nonaccreting magnetised neutron star is determined by minimizing the Gibbs free energy given in Eq.\ \eqref{eq:gibbsminimization} at a fixed pressure. In principle,  the nuclear masses should modify in the presence of the magnetic field \cite{Arteaga_2011, Stein_2016, Basilico_2015}, which might affect the outer crust composition. However, a comprehensive mass table in the presence of a magnetic field is not yet available.  Moreover, Refs. \cite{Arteaga_2011, Stein_2016, Basilico_2015}  suggest that a field strength $> 10^{17} G$ is required to alter the nuclear ground state significantly, whereas, the highest observed field strength at the surface of the magnetar is $20 \cross 10^{14} G$ for SGR 1806-20 \cite{Olausen_2014} among  26 currently known magnetars. Therefore, the nuclear masses for the field-free case are used in this study, keeping the strength of magnetic field $\sim 10^{17}$ G.
%%%%%%%%%%%%%%%
\begin{figure}
  \centering
\subfloat[]{%
  \label{fig:fig1sub1}
  \includegraphics[width=0.49\linewidth]{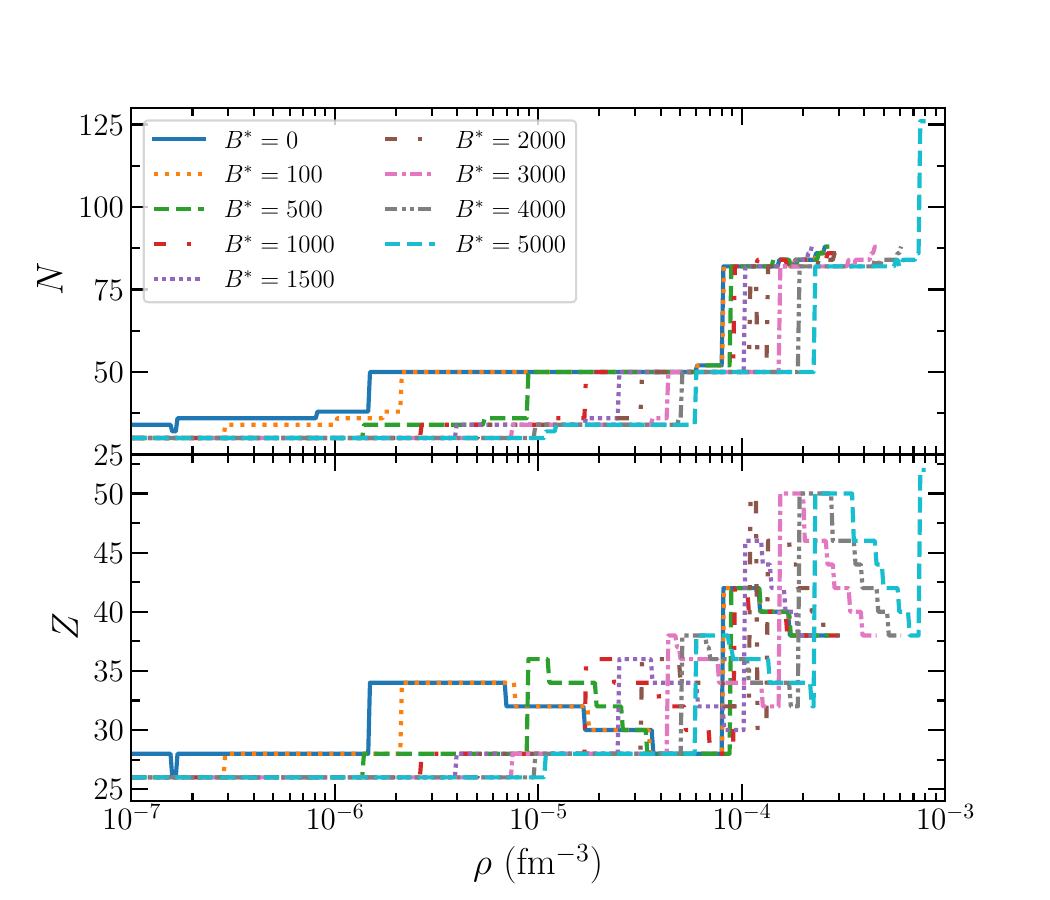}%
}\hfill
\subfloat[]{%
  \label{fig:fig1sub2}
  \includegraphics[width=0.49\linewidth]{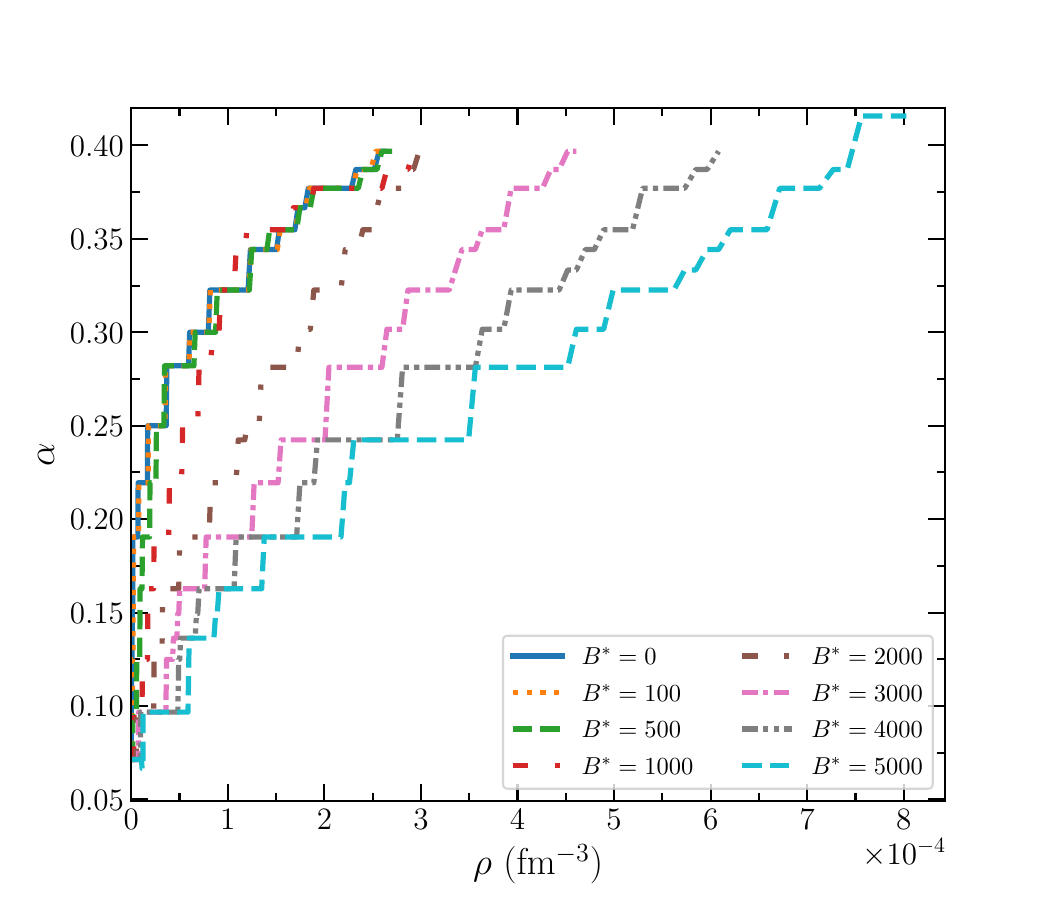}%
} 
\caption{(a) The charge/proton  number ($Z$) and neutron number ($N$) as a function of density in the outer crust of magnetised neutron star for various values of magnetic field strength $B^*$. The unknown masses are taken from the HFB26 mass model. (b) The variation of $\alpha=\frac{N-Z}{N+Z}$ as a function of density. }
\end{figure}
%%%%%%%%%%%%%

The equilibrium composition of the outer crust of a magnetised neutron star, with varying strengths of the magnetic field $B^*=B/B_c$, is depicted as a function of density in Fig. \ref{fig:fig1sub1}. The nuclear masses are  taken from  experimental AME2020 \cite{Huang_2021}, and HFB26 \cite{hfb2426} table. The Outer crust is stratified into various layers. For a weak magnetic field ($B^*\sim 10$), which is relevant for the pulsar \cite{becker2009neutron}, the composition remains similar to the field-free case. For the field strength $B^*>500$, the sequence of nuclei and the density at which they occur change significantly. The $Z=26$ and $N=30$ ($^{56}$Fe) plateau keeps extending with the increasing magnetic field and extends up to 1.0706E-5 fm$^{-3}$ for the field strength $B^*=5000$ as compared to the 4.9729E-9 fm$^{-3}$ for the field free case. The density at which the $N=50$ plateau appears (characteristic of the outer crust at $B^*=0$), increases monotonically with increasing field strength. This means that the nuclei become more and more symmetric with increasing magnetic fields at the same pressure. This is clear from Fig.\ \ref{fig:fig1sub2} where the asymmetry parameter $\alpha=\frac{N-Z}{N+Z}$ is plotted as a function of density. The isospin asymmetry increases almost linearly at higher magnetic field whereas an exponential behaviour is observed for the field-free case. However, the maximum $\alpha$ does not exceed $\sim 0.4$ for field strength as high as $B^*=5000$. This behaviour of the outer crust composition can be attributed to the EoS of the electron gas. With increasing field strength, the chemical potential or Fermi energy of the electrons decreases, which enforces a delay in the pressure where new nuclear species start appearing. 
%%%%%%%%%%%%%%%%%%%%%%%%
%%%%%%%%%%%%%%%
\begin{sidewaysfigure}
\centering
    \includegraphics[scale=0.6]{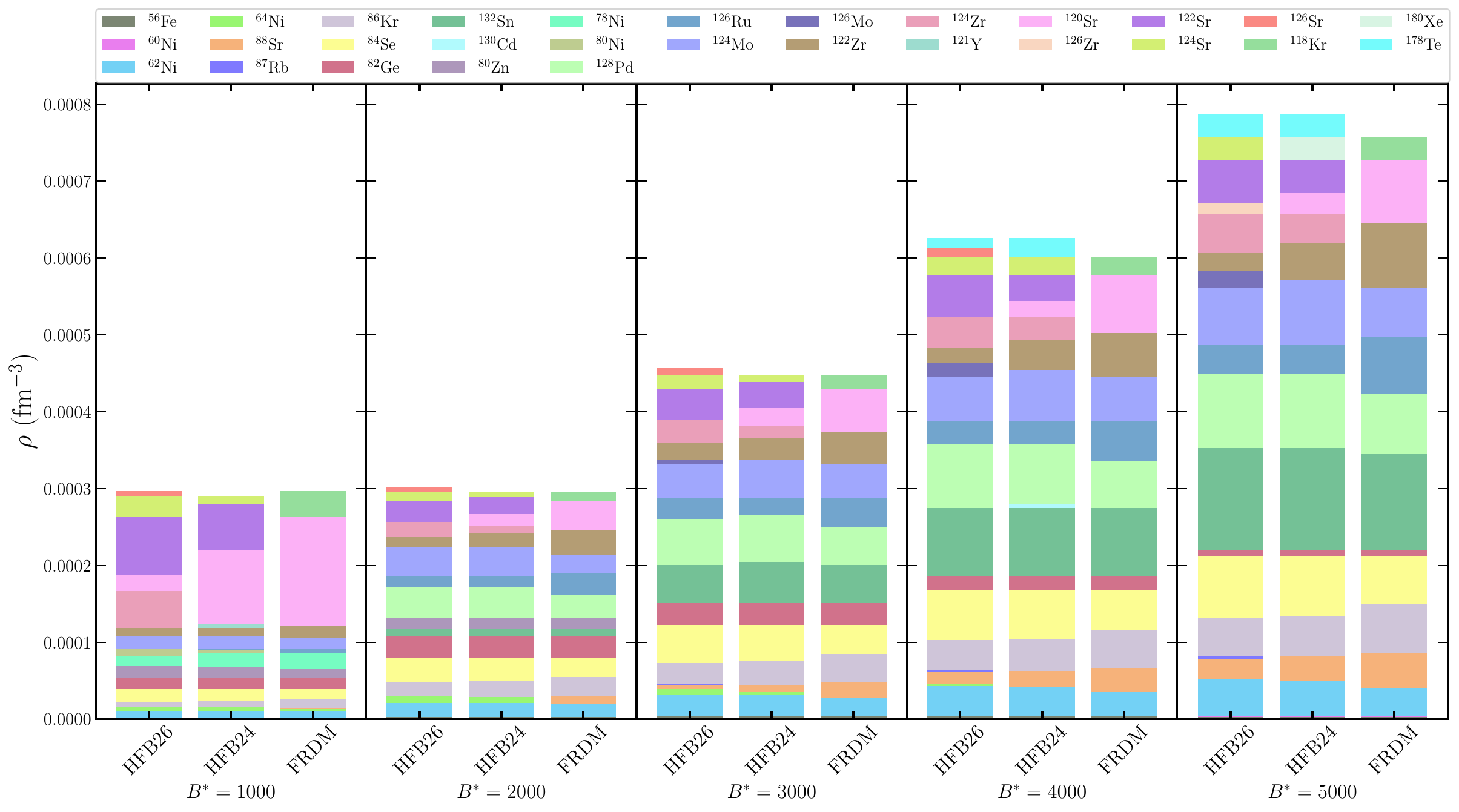}
    \caption{The arrangement of nuclei in the outer crust of a neutron star, sorted from low to high density and spanning various magnetic field strengths, is determined using a combination of experimental atomic mass evaluations from AME2020 \cite{Huang_2021} and theoretical models such as HFB24, HFB26 \cite{hfb2426}, and FRDM(2012) mass tables \cite{MOLLER20161} to estimate the masses of unknown nuclei.}
    \label{fig:nuc_seq}
\end{sidewaysfigure}

%%%%%%%%
%%%%%%%%
\begin{table}
\caption{The surface density of the outer crust ($\rho_{\rm surf}$) and neutron drip ($\rho_{\rm drip}$)  for the magnetic fields and mass models considered.}
\label{tab:oc_dens}
\renewcommand{\arraystretch}{1.5}
\resizebox{\textwidth}{!}{%
\begin{tabular}{llllllllllllllll}
\hline
\hline
$B^*$ & \multicolumn{3}{c}{1000} & \multicolumn{3}{c}{2000} & \multicolumn{3}{c}{3000} & \multicolumn{3}{c}{4000} & \multicolumn{3}{c}{5000} \\
\hline
$\rho$ (fm$^{-3}$)  & HFB26   & HFB24  & FRDM  & HFB26   & HFB24  & FRDM  & HFB26   & HFB24  & FRDM  & HFB26   & HFB24  & FRDM  & HFB26   & HFB24  & FRDM  \\
   \hline
 $\rho_{\rm surf}$ $\cross$ ($10^{-6}$) &1.0554  &1.0544 &1.0544 & 2.4708 & 2.4708 & 2.4708  &4.0537  & 4.0537  &4.0537 & 5.7897 &5.7897 &5.7897 &7.5783  &7.5783  & 7.5783  \\
 $\rho_{\rm drip}$ $\cross$ ($10^{-4}$)& 2.9791  &2.9207  & 2.9791 &3.0387& 2.9791  &   2.9791 &4.6057 &4.5154  &4.5154  & 6.3226 &6.3226  &  6.0771 & 7.9545  & 7.9545 &7.6456  \\
\hline \hline
\end{tabular}%
}
\end{table}
%%%%%%%
The qualitative observations in Fig.\ \ref{fig:fig1sub1} and \ref{fig:fig1sub2} are also supported by the nuclear mass model HFB24 \cite{hfb2426}, and FRDM \cite{MOLLER20161}. The detailed composition and EoS for these mass models, along with the unified magnetised EoSs, are provided in the GitHub link\footnote{\url{https://github.com/hcdas/Unified_mf_eos}}. As the deeper portion of the outer crust is determined using the mass excess from the theoretical mass models, there exists a model dependency of the sequence of the nuclear species \cite{Parmar_2022_1}. To investigate the model dependency in the presence of the magnetic field, Fig.\ \ref{fig:nuc_seq} shows the sequence of nuclear species along with the density of the surface ($\rho_{\rm surf}$), and neutron drip density ($\rho_{\rm drip}$) in Table \ref{tab:oc_dens}  for the HFB26, HFB24, and FRDM mass models at various magnetic field strengths. 

All the mass models predict approximately the same sequence of nuclei for a given magnetic field strength, except for the appearance/disappearance of certain nuclei. This shows that the outer crust of a neutron star is dominantly dependent on the structural effects (magic number of neutrons at $N=50, 82$) of the nucleus rather than the nuclear matter properties of the EoS with which their masses have been determined. Table \ref{tab:mass_model_sym} shows the symmetry energy ($J$) and slope parameter ($L$) of the mass models for reference. The outer crust preserves the $N=50 \, {\rm and} \, 80$ plateau for the magnetic field as high as $B^*=5000$. When comparing different mass models, it is seen that the FRDM estimates a constant $N=82$ nuclei while the HFB26 and HFB24 mass models deviate from it at higher density. The FRDM mass model estimates the presence of $^{118}$Kr at the neutron drip density for all the magnetic field strengths, while the HFB26 and HFB24 (having similar symmetry energy), estimate the different isotopes of Sr for $B^*$ up to $3000$ and $^{178}$Te at higher magnetic field strength.

Furthermore, as one increases the magnetic field strength, $^{132}$Sn appears in place of $^{80}$Zn. There is also a possibility of the existence of  $^{130}$Cd at higher magnetic field strength, although for a brief span. The odd number nucleus $^{121}$Y, observed for the field-free case using the HFB24 mass model, disappears at higher magnetic field strength.  For a particular value of the magnetic field, it is observed that the transition of one nucleus to another happens at a lower density for FRDM as compared to the HFB24, followed by HFB26 in the regions closer to neutron drip density (where the role of the theoretical mass model comes into play). This trend can be attributed to their decreasing slope parameter $L$.

%%%%%%%%%%%%%
\begin{table}
\centering
\caption{Symmetry energy ($J$) and slope parameter ($L$) coefficient for the HFB26, HFB24, and FRDM mass models.}
\label{tab:mass_model_sym}
\scalebox{1.0}{
\begin{tabular}{llll}
\hline
\hline
& HFB24 \cite{Pearson_2018}  & HFB26 \cite{Pearson_2018}  & FRDM \cite{Hiroyuki_frdmsym}  \\
\hline    
$J$ (MeV) & \hfill 30.0   & \hfill 30.0   &  \hfill 32.3  \\
$L$ (MeV) & \hfill 37.5   & \hfill 46.4   &  \hfill 53.5  \\
\hline

\end{tabular}%
}
\end{table}
%%%%%%%%%%%%

With an increase in magnetic field strength, the surface density of the outer crust of the neutron star increases as high as seven times when compared with the $B^*=1000$ and $B^*=5000$ case due to the magnetic condensation. This surface density is determined b on the basis of experimental mass of $^{56}$Fe and hence is the same for all the mass models. The neutron drip density increases exponentially with increasing magnetic field and becomes as high as three times compared to the $B^*=1000$. However, the neutron drip density exhibits an oscillatory quantum nature at a lower magnetic field arising from the Landau quantization of electrons resulting in the increase or decrease of neutron drip density. Similar behaviour was  demonstrated in \cite{Chamel_2015_landauquant}, which suggests the model independency of these oscillations, which occur for approximately $B^*<1300$. For higher magnetic field strengths, the electron EoS becomes strongly quantizing in the whole outer crust. A slight variation in the neutron drip density for various mass models at a given magnetic field strength is the consequence of the drip nuclei which these models predict. More neutron-rich (less bound) nuclei can sustain at lower pressure for a higher magnetic field. The asymmetry at the neutron drip density thus keeps increasing with increase in magnetic field strength. 
%%%%%%%%%%%%%%
\begin{figure}
    \centering
    \includegraphics[scale=0.5]{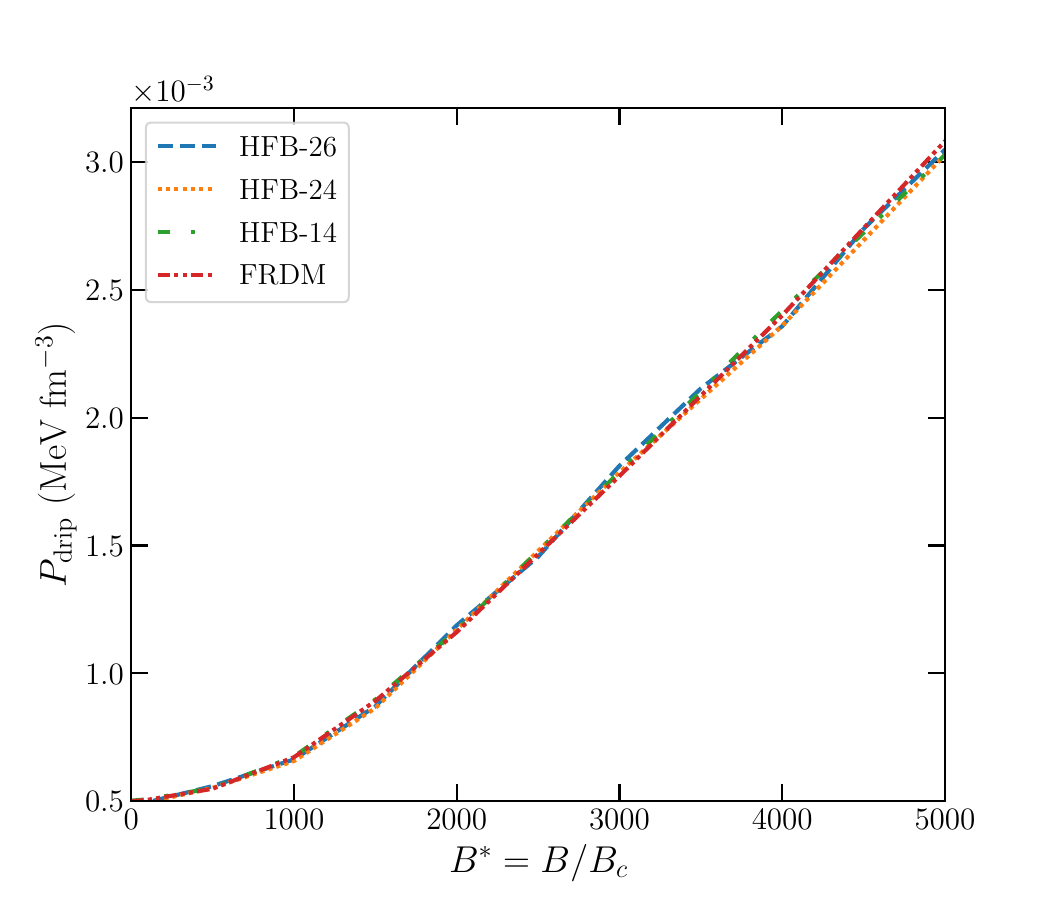}
    \caption{Transition pressure ($P_{\rm drip}$)  as a function of magnetic field strength.}
    \label{fig:pdrip_oc}
\end{figure}
%%%%%%%%%%%%%%

The transition pressure at the neutron drip point is more important than the transition density, as the former plays a direct role in calculating the mass of the crust \cite{Pearson_2018, Parmar_2022_1}. Fig.\ \ref{fig:pdrip_oc} shows the transition pressure for the various mass models: HFB14, HFB24, HFB26, and FRDM as a function of magnetic field strength. It increases linearly for a magnetic field greater than $\sim 1300$, which is the strongly quantizing regime (only the lowest Landau level $\nu=0$ is filled). Transition pressure increases almost six times for $B^*=5000$ as compared to the field-free case. A higher neutron drip transition pressure implies that the crustal mass of the outer crust will be larger for greater strength of the magnetic field. Furthermore, the neutron drip pressure as a function of magnetic field strength seems to be model-independent. Chamel {\it et al.} \cite{Chamel_2012} have determined analytic expression for the outer crust transition pressure ($P_{\rm drip}$) in the strongly quantizing regime (B$^* >$ 1300). Based on the evident model independency of the $P_{\rm drip}$ in Fig.\ \ref{fig:pdrip_oc}, the $P_{\rm drip}$ as a function of the magnetic field in both quantizing and non-quantizing regime can be written as

%%%%%%%%%%%%%%%%
\begin{align}
    P_{\rm drip}=&-1.81110^{-14}\cross B^{*3}+1.84910^{-10}\cross  B^{*2} \nonumber \\ 
    &+3.57810^{-8}\cross B^*+0.00049,    
\end{align}
%%%%%%%%%%%%%%%%
where the coefficients have relevant dimensions in the form of MeV fm$^{-3}$.
%%%%%%%%%%%%%%%
\begin{figure}
    \centering
    \includegraphics[scale=0.6]{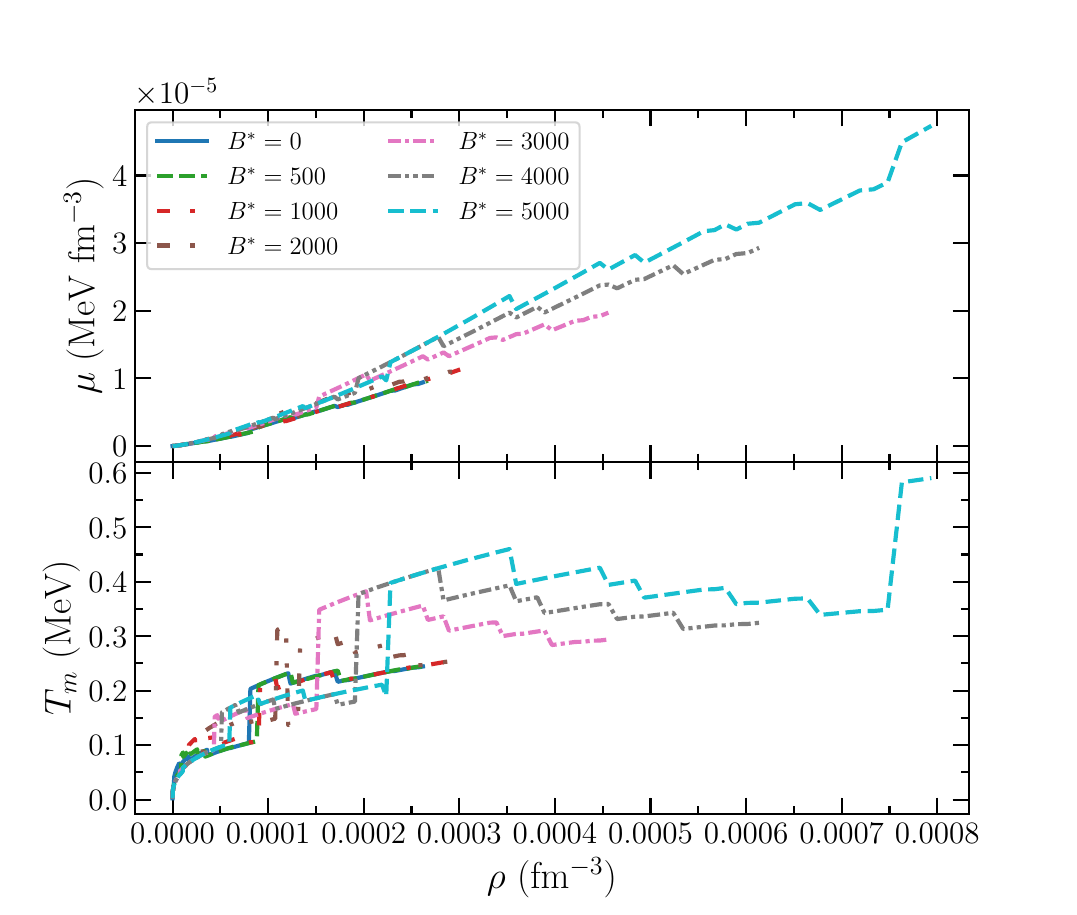}
    \caption {The variation of shear modulus $\mu$ and crystallization temperature $T_m$ in the outer crust of a neutron star at various magnetic field strengths. The theoretical mass model is HFB26.}
    \label{fig:oc_tmsm}
\end{figure}

Other essential aspects of the outer crust of a neutron star are its mechanical responses and melting temperature, which plays a prime role in describing properties such as crust failure \cite{Beloborodov_2014} and dynamics of the crust \cite{Newton_2012}. The mechanical properties are determined using its shear stress, which for a cold neutron star can be written  following Monte Carlo simulation \cite{Chugnov_2010} as
%%%%%%%%%%%%%%%%
\begin{equation}
    \mu=0.1106\Big(\frac{4 \pi}{3}\Big)^{1/3}A^{-4/3}\rho_i^{4/3}(1-X_n)^{4/3}(Ze)^2,
\end{equation}
%%%%%%%%%%%%%%%
where $\rho_i$ is the density of nuclei, and $X_n$ is the fraction of neutrons not confined to the nuclei. The melting or crystalline temperature, which defines the temperature at which the crystalline lattice converts to the gas of ions, is written in the one-component plasma (OCP) approximation as \cite{Newton_2012, Fatima_2020}
%%%%%%%%%%%%%%%%
\begin{equation}
    T_m= \frac{Z^2e^2}{k_b \Gamma_m}\Big(\frac{4 \pi}{3} \frac{\rho_i}{A} \Big)^{1/3}.
\end{equation}
%%%%%%%%%%%%%%%
Here, $\Gamma_m=175$ is the Coulomb coupling parameter at melting. Fig.\ \ref{fig:oc_tmsm} shows the variation of shear modulus ($\mu$) and melting temperature ($T_m$) in the outer crust at various magnetic field strengths. As magnetic field strength increases, there is a substantial enhancement of both shear modulus and melting temperature. The shear modulus increases up to four times for the field strength $B^*=5000$ compared to the field-free case. The melting temperature becomes as high as 0.6 MeV as opposed to $\sim 0.25$ MeV (0.01 MeV$=$ $1.16 \cross 10^8$K) resulting in considerable increment. The increase in the $\mu$ and $T_m$ is a consequence of the increase in the neutron drip density with a magnetic field that allows more neutron-rich nuclei to exist at lower pressure and density.  The results suggest that magnetic field might have profound implications in the transport, cooling, and magneto-rotational evolution of a neutron star as the outer crust structure principally drives them \cite{Fatima_2020}.
%%%%%%%%%%%%%%%
\begin{figure}
    \centering
    \includegraphics[scale=0.6]{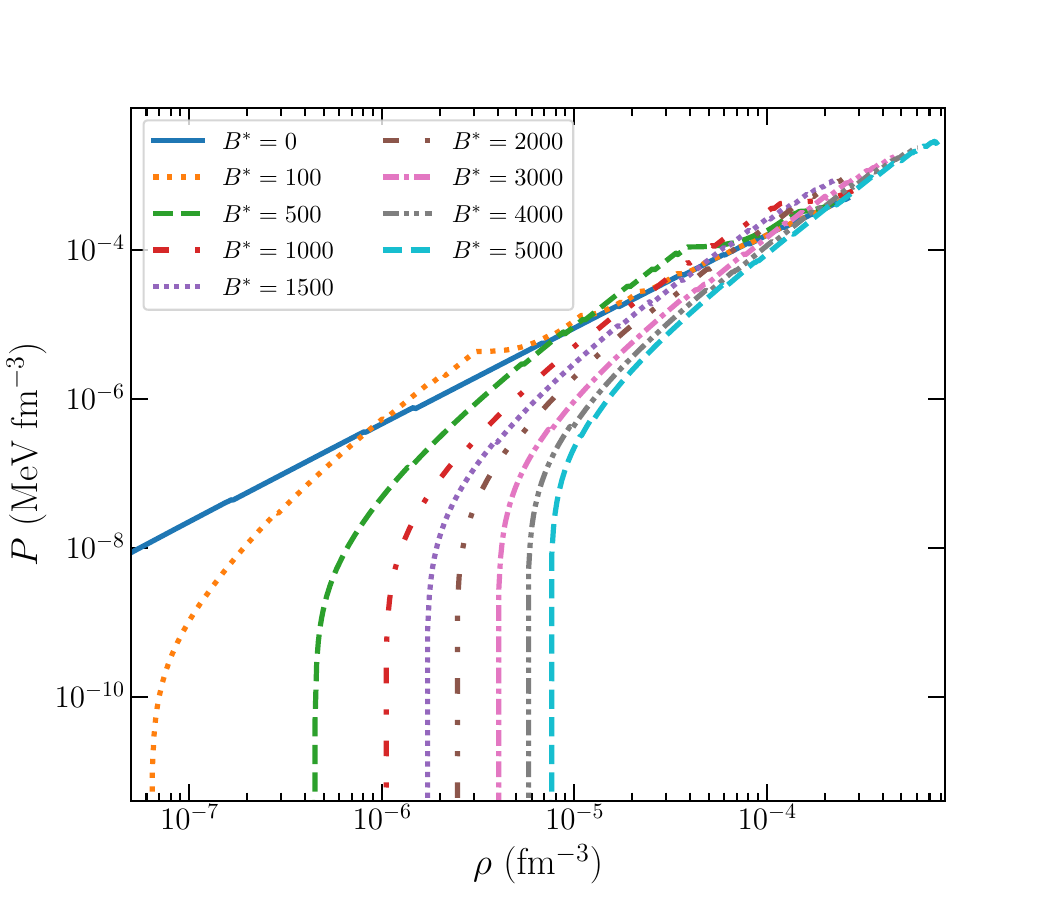}
    \caption{EoS of the outer crust of a neutron star at various magnetic field strengths. The theoretical mass model is taken as HFB26 with experimental evaluation of AME2020.}
    \label{fig:eos_oc}
\end{figure}
%%%%%%%%%%%%%%%

Fig.\ \ref{fig:eos_oc} presents the EoS of the outer crust of a cold nonaccreting neutron star at various magnetic field strengths using the HFB26 mass model. The EoS is significantly affected in the outer crust's shallower regions, where only a few Landau levels of electrons are filled. The density of the outer crust remains unchanged for a wide range of pressure, making the matter almost incompressible in the layers adjacent to the surface of the star. The composition in this region is essentially determined by the experimental evaluation and remains model-independent. As density grows, the EoS becomes similar to the field-free case due to the filling of the Landau levels. The discontinuity in the EoS for weaker field strength signifies that the lowest Landau level is fully occupied.
%%%%%%%%%%%%%%%%%%%%%%%%%%%%%%%%%%%%%%%%%%%%%%%%%%%%%%%%%%%%%%%%%%%%%%%%%%%%%%%%%%%%%%%%%%%%%%%%
\section{Inner crust}
%%%%%%%%%%%%%%%%%%%%%%%%
\begin{figure}
    \centering
    \includegraphics[scale=0.35]{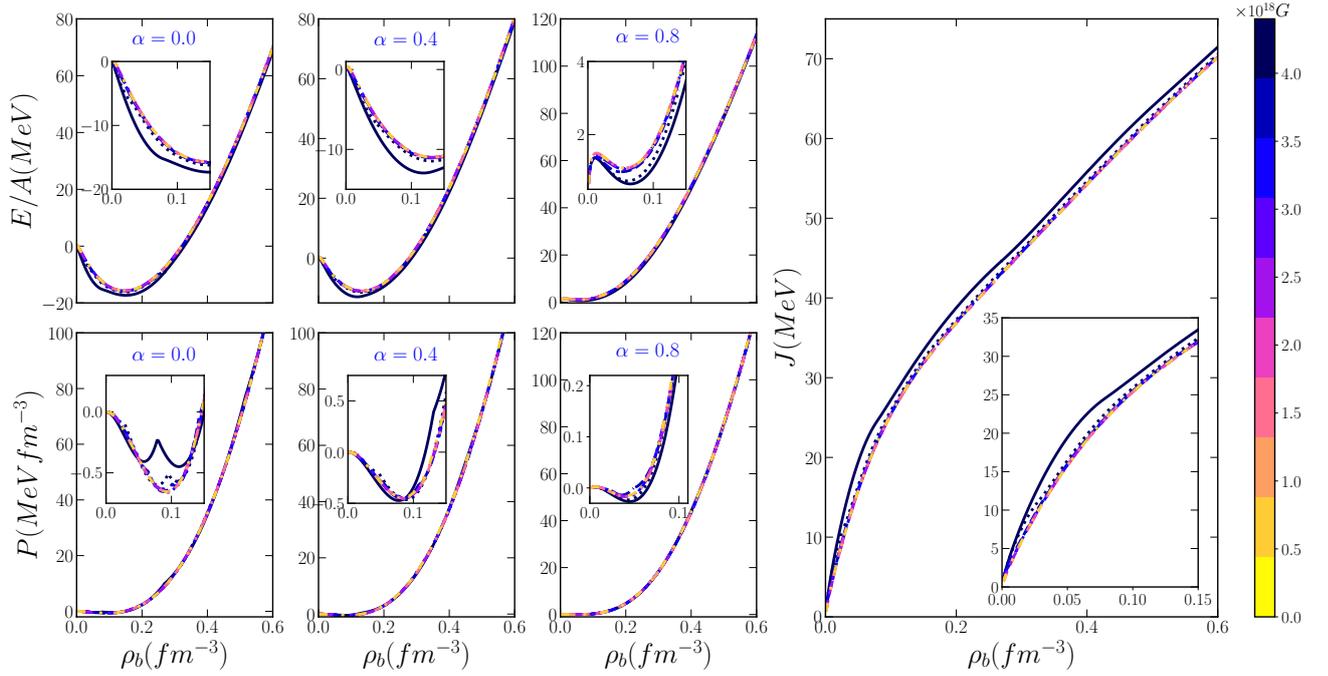}
    \caption{The binding energy ($E/A$), pressure ($P$) and symmetry energy ($J$) of homogeneous nuclear matter for various values of isospin asymmetry $\alpha=\frac{\rho_n-\rho_p}{\rho_n+\rho_p}$ at various magnetic field strengths (shown in the color bar). The parameter set taken is G3 \cite{Kumar_2017}.}
    \label{fig:bulk_eos}
\end{figure}

The calculation of the inner crust in the presence of a magnetic field is known to be tricky in the literature. Quantum oscillations possess various challenges to numerical solutions, especially in models where the energy minima becomes flat \cite{Pearson_2020, Mutafchieva_2019}. Furthermore, the small energy difference between various possible shapes (pasta structures) adds to the difficulty \cite{Parmar_2022_2}. Therefore, this work extends the simplistic CLDM formalism \cite{Parmar_2022_1, Carreau_2019, Newton_2013} of the inner crust to the finite magnetic field strength. It considers the WS cell to consist of a nuclear cluster in a BCC crystal surrounded by an ultrarelativistic electron gas along with the free neutron gas. Further, it does not take the effect of anomalous magnetic moment (AMM) on the EoS. The AMM for electron  is insignificant in comparison to the nucleon mass \cite{Broderick_2000}, whereas for baryons, it is observed that the AMM becomes important only for $B \ge 10^{18}$G \cite{Rabhi_2015, Rabhi_2008, Lima_2013, Bao_2021}. In addition, using the one-loop fermion self-energy, Ferrer {\it et al.} \cite{Ferrer_2015} suggest that AMM of charged particles does not affect the EoS significantly. However, AMM is an important factor for higher magnetic field strengths and play a significant role in  phenomena such as axion production and neutrino-antineutrino pair emission \cite{MARUYAMA2020135413, MARUYAMA2018160}.
%%%%%%%%%%%%%
\begin{table}[b]
\centering
\caption{Symmetry energy ($J$) and slope parameter ($L$) coefficient for the G3 \cite{Kumar_2017}, IOPB-I \cite{Kumar_2018}, FSUGarnet \cite{Chen_2015}, IUFSU \cite{Fattoyev_2010}, IUFSU$^*$ \cite{Fattoyev_2010}  and SINPB \cite{Mondal_2016} parameter sets at saturation density ($\rho_0$) and subsaturation density (0.05 fm$^{-3}$).}
\label{tab:rmf_sym}
\scalebox{1.0}{
\begin{tabular}{lllllll}
\hline
\hline
& FSUGarnet  & IUFSU  & IUFSU* & G3  & IOPB-I  & SINPB   \\
\hline    
$J_{\rho_0}$ (MeV) & \hfill 30.95   & \hfill 31.30  &  \hfill 29.85 & \hfill 31.84  & \hfill 33.30   &  \hfill 33.95 \\
$J_{0.05}$ (MeV) & \hfill 18.07   & \hfill 17.8 &  \hfill 15.73 & \hfill 15.66  & \hfill 15.6  &  \hfill 14.98 \\
$L_{\rho_0}$ (MeV) & \hfill 51.04   & \hfill 47.21&  \hfill 51.50 & \hfill 49.31   & \hfill 63.58  &  \hfill  71.55\\
$L_{0.05}$ (MeV) & \hfill 32.1058  & \hfill33.85  &  \hfill 32.26 & \hfill 36.781   & \hfill 37.2   &  \hfill 36.7  \\
\hline
\hline
\end{tabular}%
}
\end{table}
%%%%%%%%%%%%

For the calculation of inner crust composition, six E-RMF parameter sets are used for which the value of their symmetry energy and slope parameter are given in Table  \ref{tab:rmf_sym} at saturation density and subsaturation density (0.05 fm$^{-3}$), relevant for the inner crust. Fig.\ \ref{fig:bulk_eos} shows the variation of the binding energy $\big(\frac{E}{A}\big)$ and pressure as a function of baryon density ($\rho_b$) for various values of isospin--asymmetry at different magnetic field strength which is shown as a color bar. In addition,  the variation of the density-dependent symmetry energy is also shown. These calculations are performed for the parameter set G3 \cite{Kumar_2017}. The qualitative behaviour of other parameters remains the same, and their behaviour for the unmagnetised case has been well documented in the literature and earlier chapters. It is clear from Fig.\ \ref{fig:bulk_eos} that hadronic EoS is not significantly affected for B $<10^{17}G$, which is the case of this study. The small changes appear in the subsaturation density region, which is essential for the neutron star crust. At ultra-strong magnetic field strengths, the binding energy increases, making the system more and more bound. The  variation is more pronounced for symmetric matter as more charged particles are in the system. The kinks on the pressure (especially at low density)  appear due to the successive filling of Landau levels, which disappear at high densities because of the filling of more and more  Landau levels. This behaviour is analogous to free proton gas in a magnetic field \cite{Strickland_2012}.

The symmetry energy also does not change much for B $<10^{17}G$  and increases for very strong magnetic field strengths. Other nuclear matter properties, such as higher-order symmetry energy and incompressibility derivatives, also show no significant changes for B $<10^{17}G$. Therefore, the changes in the properties of the neutron star crust for the magnetic field case are predominantly due to the changes in electronic EoS, which in turn, changes the equilibrium compositions through the condition of $\beta$ equilibrium. However, this work includes the magnetic field on the hadronic EoS for the consistent and realistic calculations of various observables, which is absent in many crust calculations in literature  \cite{Nandi_2011}. Further, since the nuclear saturation energy does not change significantly until $B\sim10^{18}$ G \cite{Rabhi_2015}, the field free value of surface energy parameters (Eqs.\ \eqref{eq:sigma} and \eqref{eq:ecurv}) is used. These fits play a crucial role in the inner crust calculations using the CLDM formalism.  For details on the surface energy fits, please see \cite{Parmar_2022_1, Parmar_2022_1, Carreau_2019}.

Out of various bulk properties of nuclear matter, symmetry energy predominantly governs the properties of asymmetric systems, which include phase transition of asymmetric nuclear matter, crustal properties of the neutron star, neutron skin thickness, etc. \cite{Vishal_2021, Dutra_2021, Parmar_2022_1, Parmar_2022_2}. In the context of neutron star crust, where the typical baryon density is in the subsaturation region,  the uncertainties in the symmetry energy account for the variation in the crustal properties. The symmetry energy is not unique to a given E-RMF parameter set and is mostly governed by the  cross-coupling of $\rho$ and $\omega$  meson. Additionally, the behaviour of pure neutron matter (PNM) in the low-density regime is a crucial aspect of crustal properties, as the free neutron gas in the inner crust impacts the crust composition. The cross-coupling of the $\omega-\rho$ meson influences the PNM properties in the subsaturation density region \cite{Kumar_2017}, making it the most important factor determining the crust structure of a neutron star within the E-RMF framework. The E-RMF parameter sets used here are in harmony with the results obtained by various microscopic calculations for the PNM \cite{baldo_2008, Lovato_2022}. Furthermore, the six parameter sets differ in the way they behave in the low-density regime \cite{Parmar_2022_2} and, therefore, provide us with the flexibility to investigate the model dependency of our result and modification, if any, as compared to the zero magnetic field strength.

%We use six E-RMF parameter sets for which the value of their symmetry energy and slope parameter are given in Table  \ref{tab:rmf_sym} at saturation density and subsaturation density (0.05 fm$^{-3}$), relevant for the inner crust. These values are taken for the field-free case as they remain almost similar to the magnetic field strengths taken in this work. 

%%%%%%%%%%%%%%%
\begin{figure}
    \centering
    \includegraphics[scale=0.6]{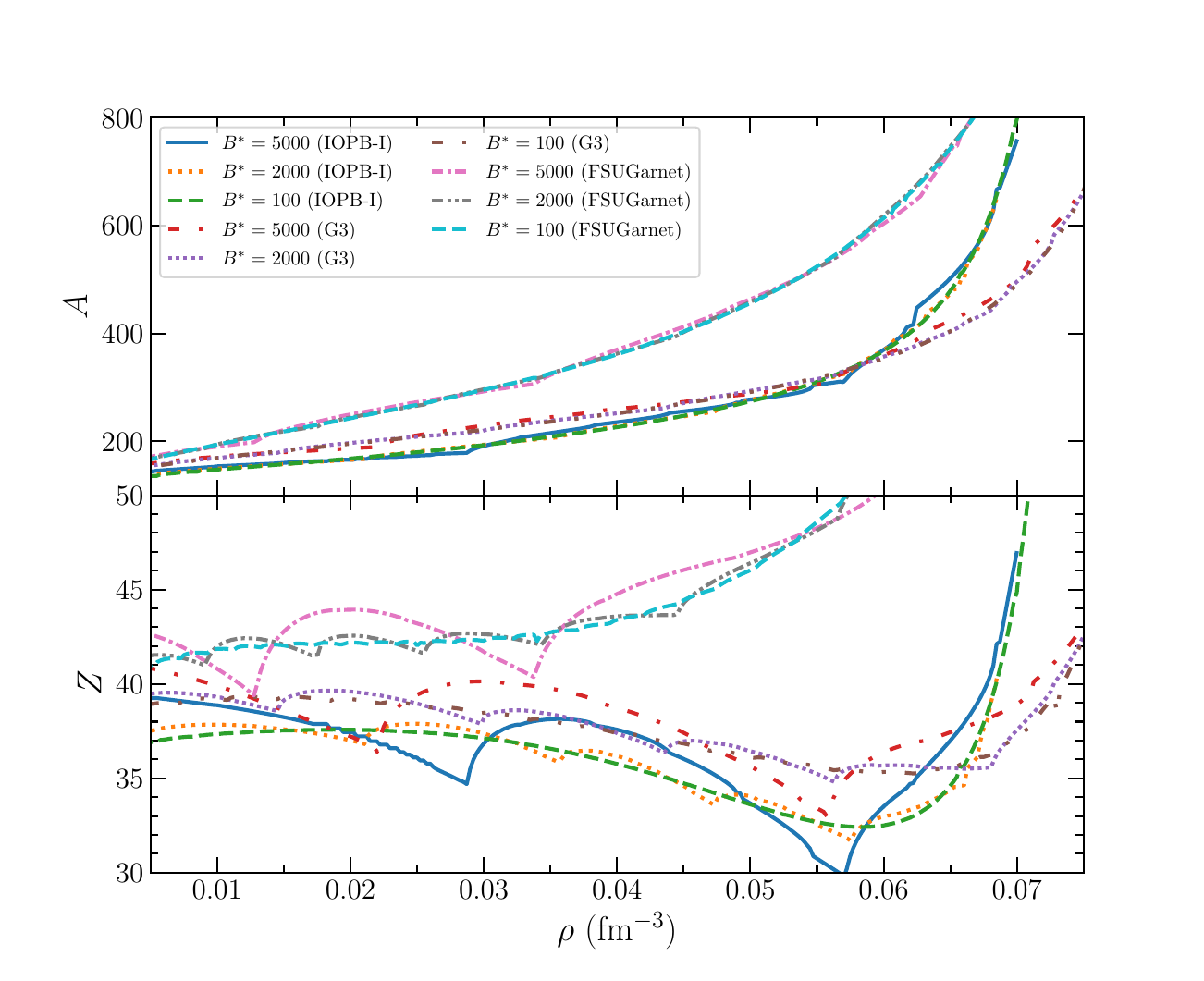}
    \caption{The distribution of the number of nucleons ($A$) in the cluster and charge number ($Z$) as a function of density in the inner crust of cold nonaccreting neutron star.}
    \label{fig:inner_crust_az}
\end{figure}
%%%%%%%%%%%%%%%

Fig.\ \ref{fig:inner_crust_az} shows the distribution of the total number of nucleons in the cluster and charge number as a function of density in the inner crust of a magnetised neutron star employing the CLDM formalism. For weak magnetic field strength, the atomic number distribution is similar to the field-free case (Fig. 4 of \cite{Parmar_2022_1}) as a large number of Landau levels are filled for electrons as well as protons. As the magnetic field strength increases, the atomic number distribution shows an oscillatory pattern due to the successive filling of electron Landau levels, also known as the De Haas–van Alphen effect \cite{chatterjee_2011}. For high magnetic field strength ($B^*=5000$), which becomes strongly quantizing for electrons, the oscillations typically represent the filling of $\nu=0,1,2$ Landau levels of electrons. However, the small oscillations in between the larger ones occur due to the filling of Landau levels of protons. Such behaviour is similar to one obtained in Lima {\it et al.} \cite{Lima_2013} for the Thomas-Fermi calculations (see Figs. 8 and 12 of Ref. \cite{Lima_2013}).

Furthermore, the effect of symmetry energy is also evident. The oscillations become broader for the IOPB-I parameter set with the least symmetry energy at subsaturation regions compared to the FSUGarnet with the highest symmetry energy. The difference occurs as nuclear matter EoS is affected due to the variation in electron chemical potential depending on corresponding  symmetry energy. The number of nucleons in the cluster does not change significantly for the magnetic field as high as $B^*=5000$ or $B=2.207 \cross 10^{17}$G. However, the oscillatory behaviour is similar to the distribution of atomic number. 
%%%%%%%%%%%%%%%
\begin{figure}
    \centering
    \includegraphics[scale=0.6]{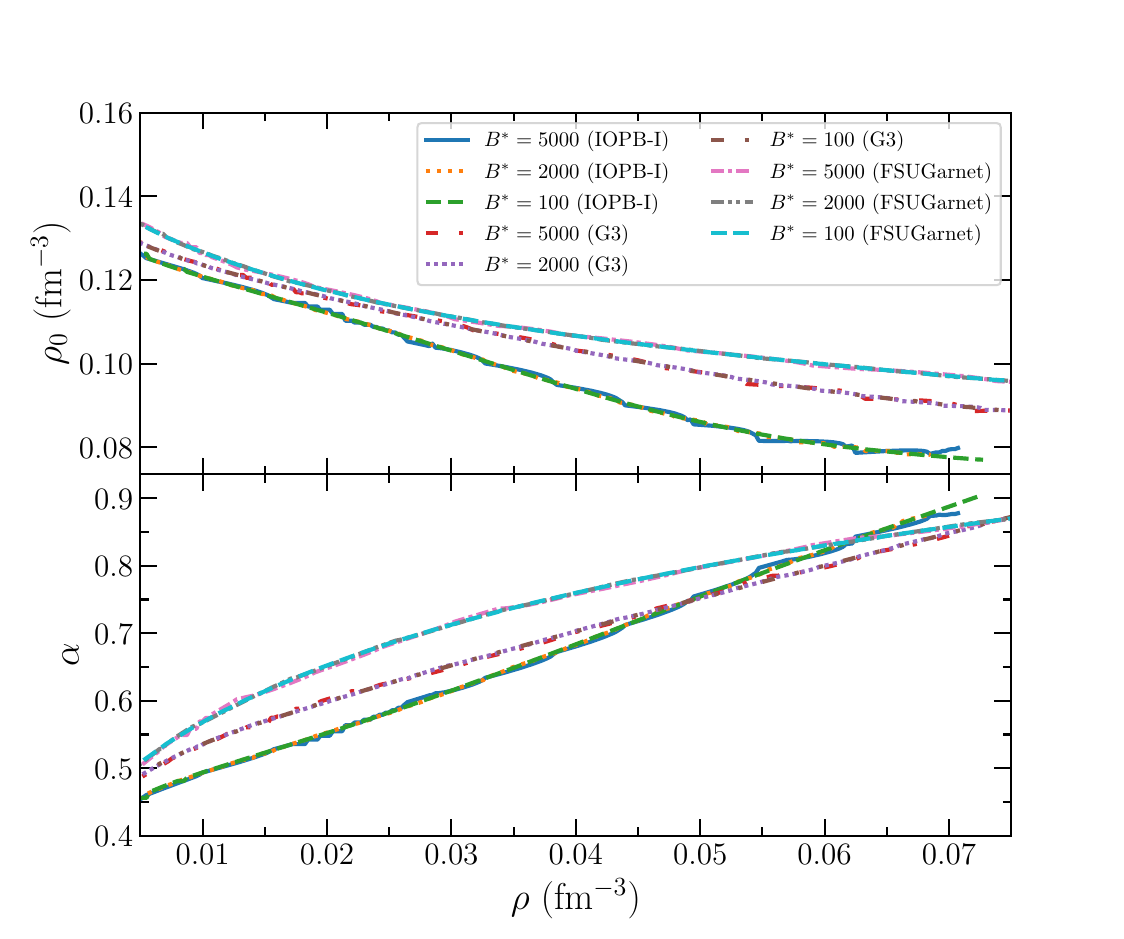}
    \caption{The average cluster density ($\rho_0$) and asymmetry in the cluster as a function of density in the inner crust for various values of magnetic field strength.}
    \label{fig:inner_crust_asy}
\end{figure}
%%%%%%%%%%%%%%

The average cluster density and the asymmetry ($\alpha$) in the cluster are shown in Fig.\ \ref{fig:inner_crust_asy} for various strengths of magnetic field. The average cluster density decreases while the asymmetry increases monotonically with inner crust density for all the magnetic field strengths, making the clusters more and more dilute and neutron-rich. The magnetic field seems to have a feeble impact on the cluster density and its asymmetry, as it does not change much except for the quantum oscillations arising due to the filling of Landau levels. A closer analysis, however, reveals that in the shallower regions ($\rho_0 \le 0.01$ fm$^{-3}$) of the inner crust, which is significantly affected due to magnetic field (only $\nu=0$ level is filled), the density of the cluster is larger for the higher magnetic field strength. The cluster becomes more symmetric with an increasing magnetic field in this density range. These results agree with Ref. \cite{Mutafchieva_2019}, which uses the extended Thomas Fermi method taking the magnetic field effects only on the electrons. For $\rho>0.01$ fm$^{-3}$, the behaviour of the cluster density and its asymmetry becomes comparable to the field-free case. The cluster cell size $r_c$  does not change much for the field strength $B^* \le 5000$. However, in the shallower region, it reduces as compared to the field-free case. Furthermore, the equilibrium composition of the inner crust changes significantly at very high  magnetic field strength, i.e., $B^* \ge 10000$. The presence of such a high magnetic field in the crust of magnetars has not yet been observed  \cite{Debarati_2015, Konstantinos_2016, DEXHEIMER2017487}. 

%%%%%%%%%%%%%%
\begin{figure}
    \centering
    \includegraphics[scale=0.6]{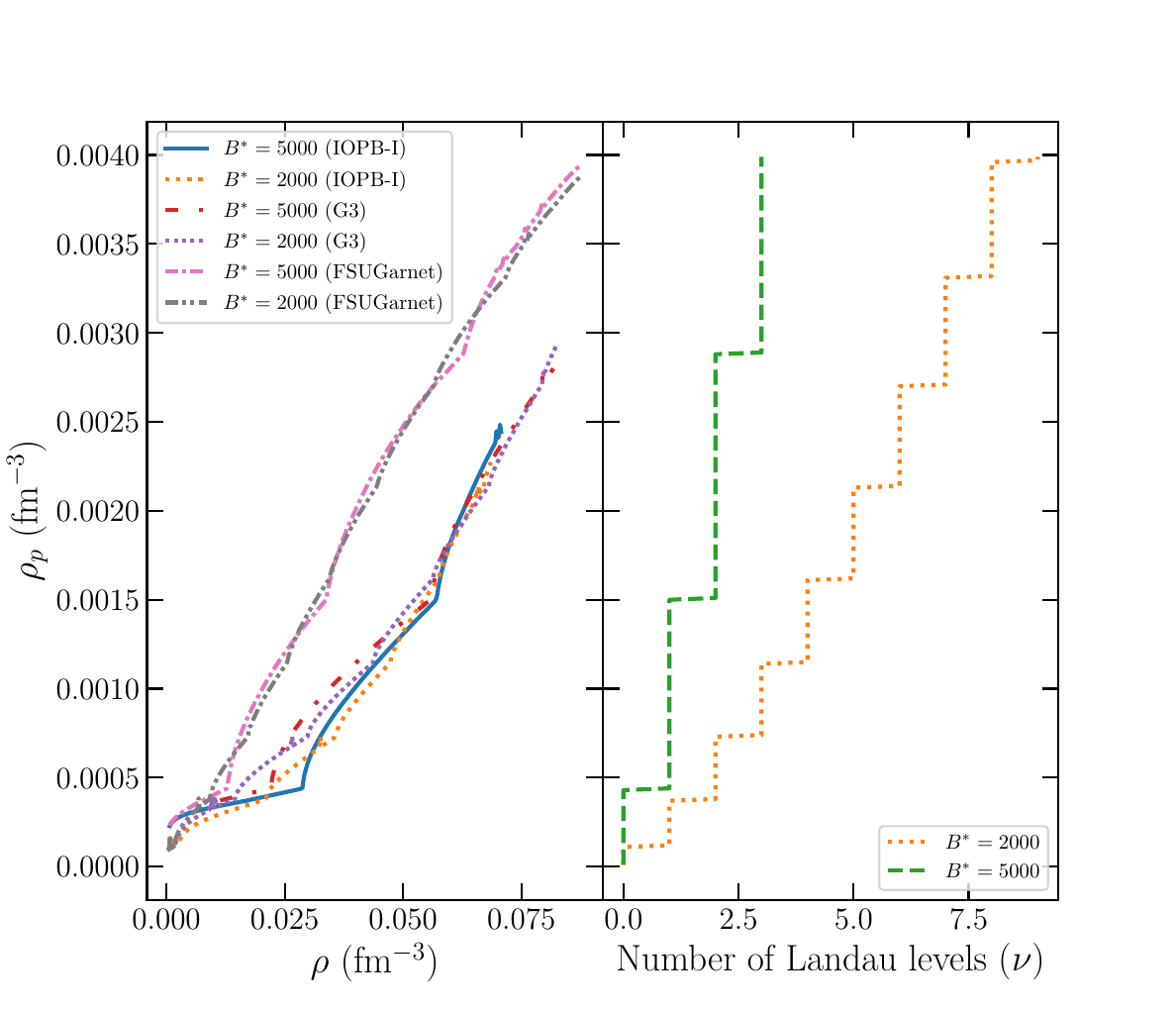}
    \caption{{\it Left:} Proton or electron density as a function of density in the inner crust. {\it Right:} The number of filled Landau levels of electrons at a given proton or electron density.}
    \label{fig:inner_crust_pd}
\end{figure}
%%%%%%%%%%%%
From the above analysis, it is clear that the magnetic field causes quantum oscillations in the inner crust composition, where electrons play  more crucial role as compared to  the baryons. For baryons, the critical magnetic field \Big($B^P_c=\big(\frac{M_p}{M_e}\big)^2B^e_c=1.487 \cross 10^{20}$ G\Big) is substantially greater than that of electrons. Since the magnetic field considered in this work and those observed in the neutron star's crust is much lower than the $B^P_c$, it does not significantly affect baryon properties. Fig.\ \ref{fig:inner_crust_pd} shows the proton density, which is equal to electron density in the inner crust of a neutron star along with the number of Landau levels filled by electrons. The proton density increases monotonically with the inner crust density. The equilibrium proton density at a given  density in the inner crust depends on the parameter set used. The FSUGarnet with the largest symmetry energy in the subsaturation region estimates a higher proton density than the G3 and IOPB-I sets, with lower symmetry energy at a particular magnetic field. However, the fluctuations in the proton density are guided by the filling of electron Landau levels. The discontinuity in the proton density arises where the filling of the subsequent Landau level takes place. This discontinuity occurs for the same proton density for all the parameter sets but at different inner crust densities. It is further observesd in the present calculations that the density of neutron gas is not significantly affected by the presence of the magnetic field. 
%%%%%%%%%%%%%%
\begin{figure}
    \centering
    \includegraphics[scale=0.6]{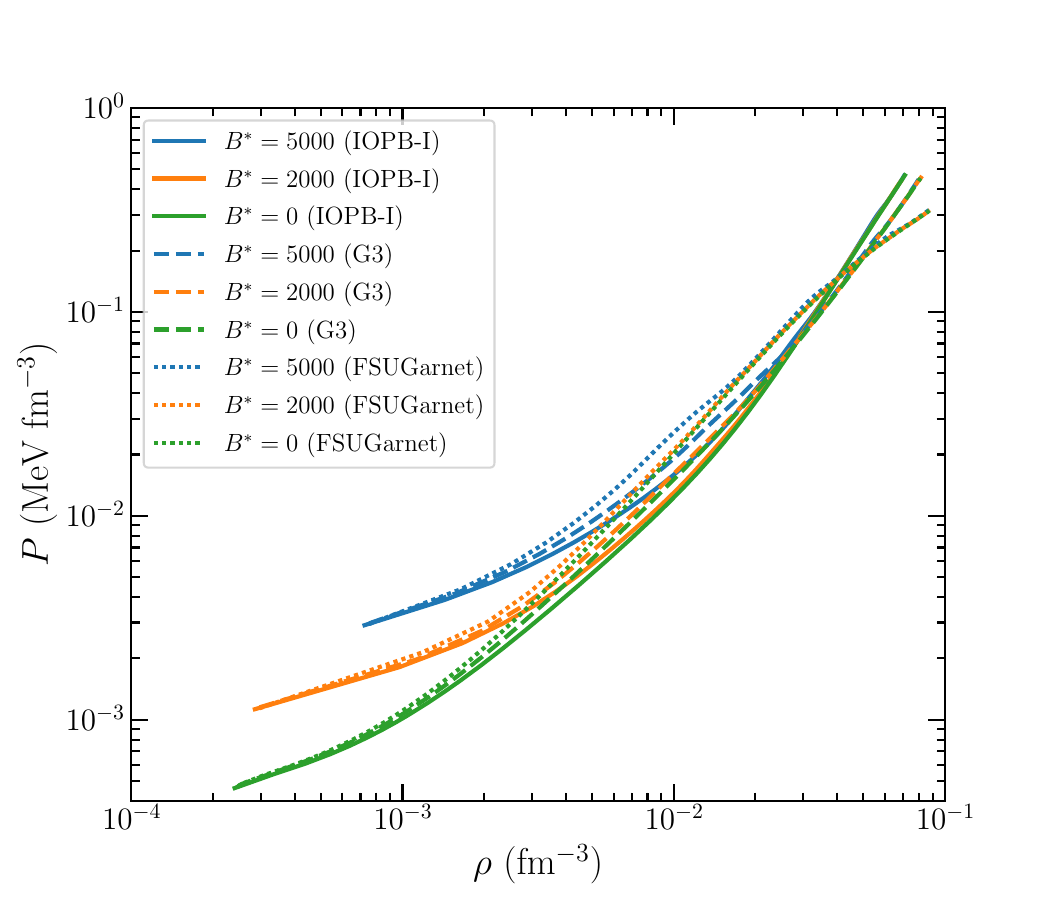}
    \caption{EoS of the inner crust of a magnetised neutron star using  IOPB-I, G3, and FSUGarnet E-RMF parameter sets.}
    \label{fig:inner_crust_eos}
\end{figure}
%%%%%%%%%%%%

The EoS of the inner crust at various magnetic fields for the E-RMF parameter sets FSUGarnet, G3, and IOPB-I are shown in Fig.\ \ref{fig:inner_crust_eos}. The EoS for the field-free case is also shown for comparison. It is seen that the magnetic field effects become prominent in the lower-density regions of the inner crust. As density approaches the crust-core transition density, magnetic field effects vanish as several Landau levels of electron and protons are filled, and EoSs imitates the field-free case. The inner crust experiences higher pressure at lower density for the larger magnetic field strength. The EoS preserves its dependence on the symmetry energy as for a given magnetic\label{ic} field strength, FSUGarnet shows the stiffest EoS followed by the G3 and IOPB-I parameter sets, a trend observed for the field free case \cite{Parmar_2022_1}. Comparing the results of EoS with switching on/off the magnetic field effects on the baryons reveals that the electrons play a far  critical role in the inner crust calculations than the protons. Furthermore, the above calculations are also performed for the IUFSU, IUFSU$^*$, and SINPB parameter sets, and it is seen that the qualitative results remain the same. 
%%%%%%%%%%%%%
\begin{table}
\centering
\caption{The density (fm$^{-3}$) of the onset of pasta structures in the inner crust of magnetised neutron star using the  G3 \cite{Kumar_2017}, IOPB-I \cite{Kumar_2018}, FSUGarnet \cite{Chen_2015}, IUFSU \cite{Fattoyev_2010}, IUFSU$^*$ \cite{Fattoyev_2010}  and SINPB \cite{Mondal_2016} parameter sets. Hom. (homogeneous)represents the crust-core transition density.}
\label{tab:pasta_density}
\scalebox{1.0}{
\begin{tabular}{lllllll}
\hline
\hline
                           &       & \multicolumn{5}{c}{Density (fm$^{-3}$)}    \\
                           \cline{3-7}
                           & B*    & Rod    & Slab   & Tube   & Bubble & Hom.    \\
\hline
\multirow{5}{*}{FSUGarnet} & 10000 & 0.0443 & 0.0583 & 0.0812 & 0.0857 & 0.0913 \\
                           & 5000  & 0.0453 & 0.0590 & 0.0811 & 0.0864 & 0.0925 \\
                           & 3000  & 0.0456 & 0.0591 & 0.0811 & 0.0864 & 0.0926 \\
                           & 1000  & 0.0456 & 0.0591 & 0.0810 & 0.0865 & 0.0926 \\
                           & 0     & 0.0456 & 0.0590 & 0.0810 & 0.0865 & 0.0918 \\
%\multicolumn{7}{c}{}    
&&&&&&\\
\multirow{5}{*}{IUFSU}     & 10000 & 0.0483 & 0.0634 & 0.0850 & 0.0896 & 0.0945 \\
                           & 5000  & 0.0497 & 0.0644 & 0.0846 & 0.0902 & 0.0960 \\
                           & 3000  & 0.0498 & 0.0643 & 0.0847 & 0.0900 & 0.0966 \\
                           & 1000  & 0.0500 & 0.0642 & 0.0847 & 0.0901 & 0.0965 \\
                           & 0     & 0.0499 & 0.0641 & 0.0847 & 0.0901 & 0.0965 \\
&&&&&&\\                           
\multirow{5}{*}{IUFSU*}    & 10000 & 0.0529 & 0.0632 & 0.0779 & 0.0824 & 0.0855 \\
                           & 5000  & 0.0526 & 0.0647 & 0.0807 & 0.0837 & 0.0855 \\
                           & 3000  & 0.0522 & 0.0652 & 0.0804 & 0.0836 & 0.0856 \\
                           & 1000  & 0.0526 & 0.0650 & 0.0807 & 0.0836 & 0.0858 \\
                           & 0     & 0.0525 & 0.0652 & 0.0807 & 0.0836 & 0.0858 \\
&&&&&&\\
\multirow{5}{*}{G3}        & 10000 & 0.0564 & 0.0631 & 0.0796 & 0.0834 & 0.0885 \\
                           & 5000  & 0.0559 & 0.0648 & 0.0833 & 0.0858 & 0.0896 \\
                           & 3000  & 0.0546 & 0.0657 & 0.0825 & 0.0854 & 0.0894 \\
                           & 1000  & 0.0549 & 0.0655 & 0.0830 & 0.0853 & 0.0889 \\
                           & 0     & 0.0551 & 0.0655 & 0.0830 & 0.0853 & 0.0889 \\
%\hline
&&&&&&\\
\multirow{5}{*}{IOPB-I}    & 10000 & 0.0548 & 0.0615 &        &        & 0.0712 \\
                           & 5000  & 0.0545 & 0.0617 &        &        & 0.0743 \\
                           & 3000  & 0.0542 & 0.0620 &        &        & 0.0737 \\
                           & 1000  & 0.0541 & 0.0619 &        &        & 0.0736 \\
                           & 0     & 0.0542 & 0.0618 &        &        & 0.0735 \\
&&&&&&\\
\multirow{5}{*}{SINPB}     & 10000 & 0.0499 & 0.0573 &        &        & 0.0600 \\
                           & 5000  & 0.0479 & 0.0542 &        &        & 0.0623 \\
                           & 3000  & 0.0488 & 0.0545 &        &        & 0.0620 \\
                           & 1000  & 0.0490 & 0.0542 &        &        & 0.0613 \\
                           & 0     & 0.0490 & 0.0543 &        &        & 0.0611\\
\hline
\hline

\end{tabular}%
}
\end{table}
%%%%%%%%%%%%

It has been previously shown and verified in our calculations that with an increase in magnetic field strength, the system's free energy decreases due to the Landau quantization \cite{Broderick_2000, Lima_2013}. Therefore, it may  impact the appearance of non-spherical shapes in the higher-density regime of the inner crust. To investigate this,  Table \ref{tab:pasta_density} shows the onset density of different pasta structures in the inner crust of a magnetised neutron star for various magnetic field strengths. The density at which different pasta structures appear does not change much for  $B^*=1000$. For $B^*>1000$, the onset densities changes, however, not significantly except for the crust-core transition density. These changes or small fluctuations in the onset density for different magnetic field strengths are not monotonic and essentially depend on the magnetic field strength and EoS. These results align with the self-consistent Thomas-Fermi approximation of Bao {\it et al.} \cite{Bao_2021}. The substantial changes can only be seen for magnetic field strength as high as $B^*=10000 = 4.414\cross10^{17}$ G. For $B^*=10000$, and the crust-core transition density decreases compared to the field-free case as the free energy decreases faster at higher density for higher magnetic fields. There is no  change in the number of pasta structures for a given E-RMF parameter set as compared to the field-free case, even for the magnetic field strength of $B^*=10000$.

The feeble changes in the pasta onset density and considerable change in the pressure because of the magnetic field effect in the inner crust prompted us to investigate the modifications which might occur in the mass and the thickness of pasta layers. The mass and thickness of the pasta structures are sensitively affected by the pressure and chemical potential as given by   Eqs.\ \eqref{eq:rr} and \eqref{eq:pp}  \cite{Newton_2021}.
 
%%%%%%%%%%%%%%
\begin{figure}
  \centering
\subfloat[]{%
  \label{fig:pasta_mass}
  \includegraphics[scale=0.55]{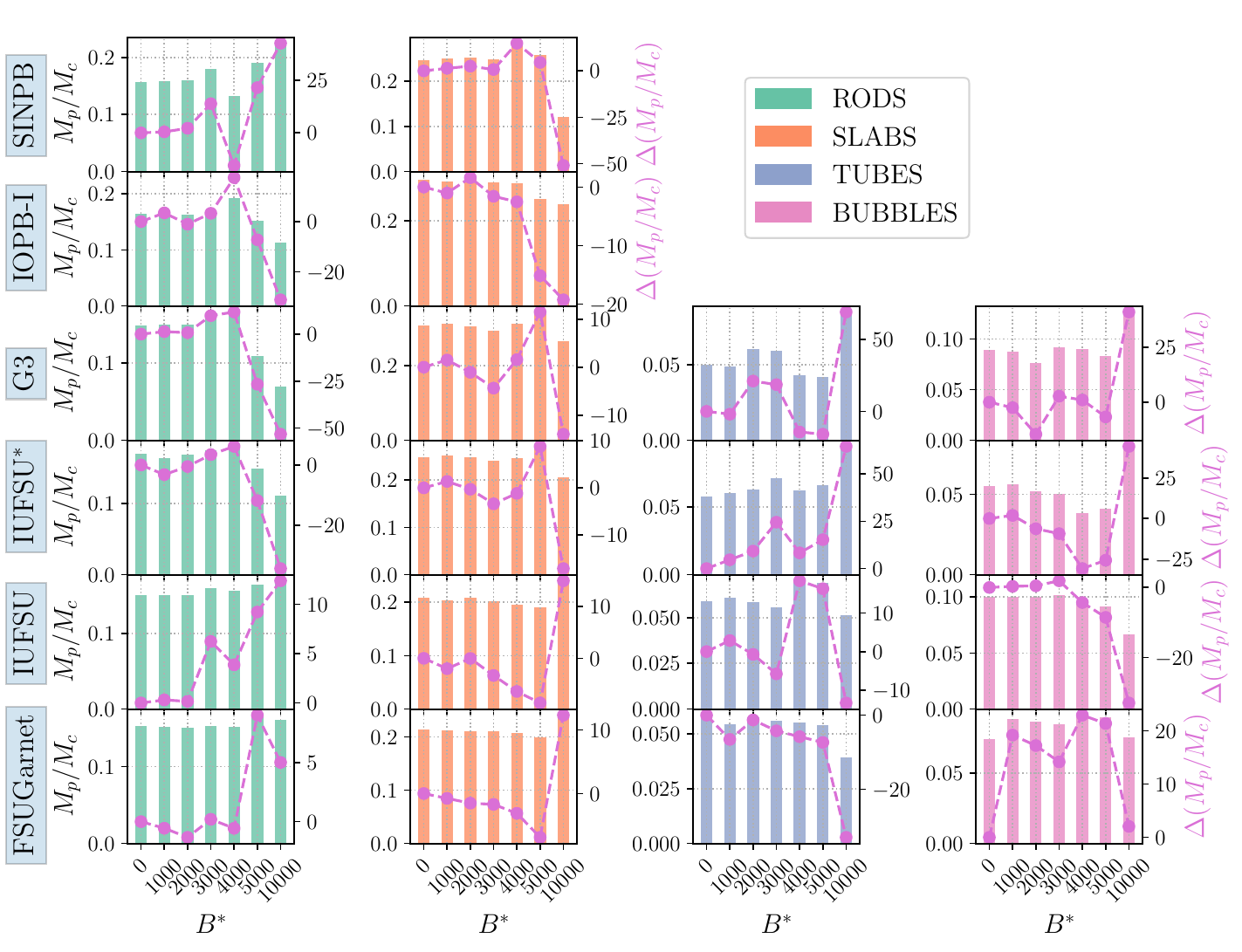}%
}\hfill
\subfloat[]{%
  \label{fig:pasta_thick}
  \includegraphics[scale=0.55]{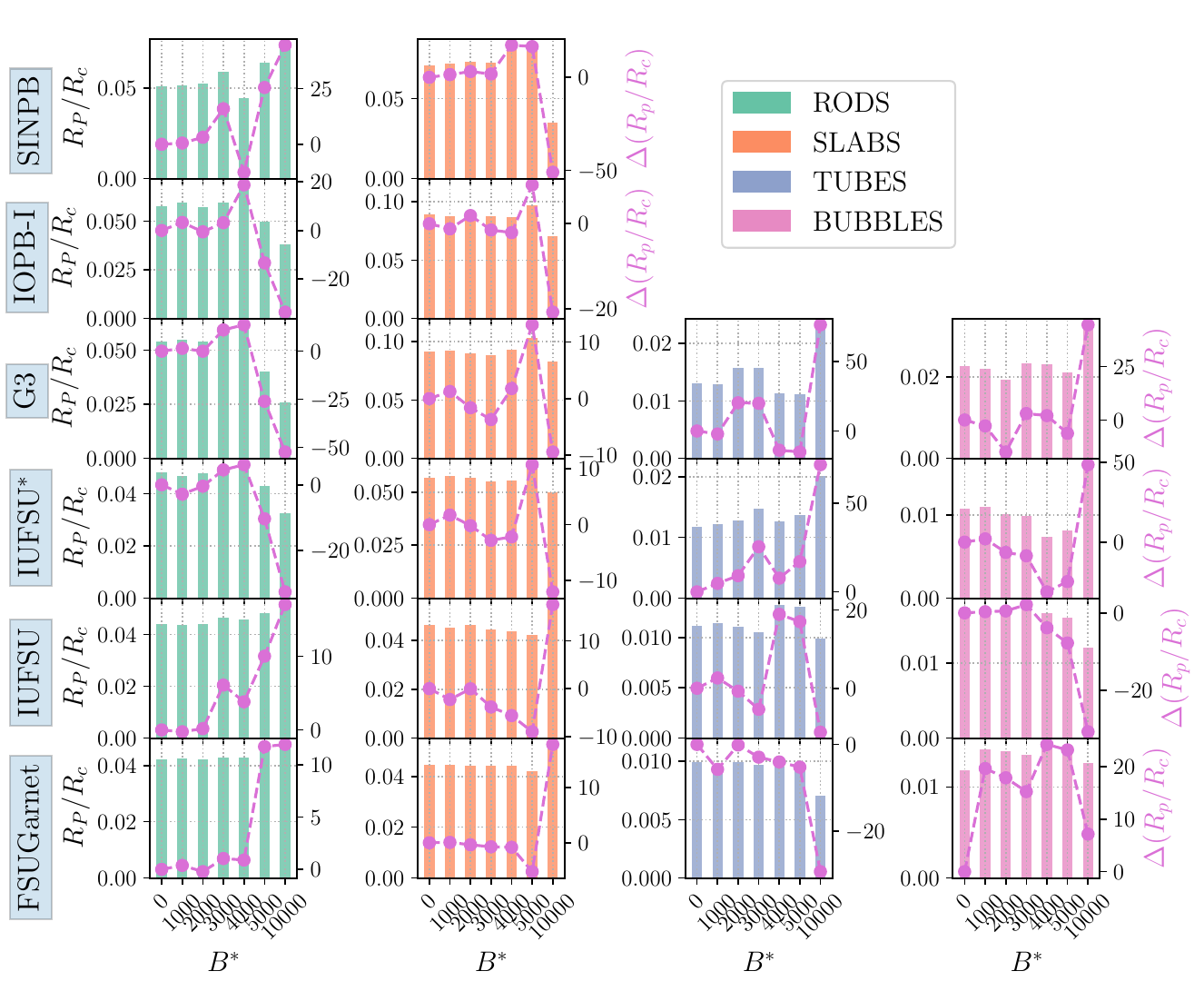}%
} 
\caption{(a) The relative mass and (b) thickness of the different layers of pasta structures at various magnetic field strengths for FSUGarnet, IUFSU, IUFSU$^*$, G3, IOPB-I, and SINPB E-RMF parameter sets. The left scale shows the absolute values, while the secondary right scale represents the percentage change with respect to the field-free case. }
\end{figure}
%%%%%%%%%%%%%

%%%%%%%%%%%%%%%
\begin{figure*}
    \centering
    \includegraphics[scale=0.6]{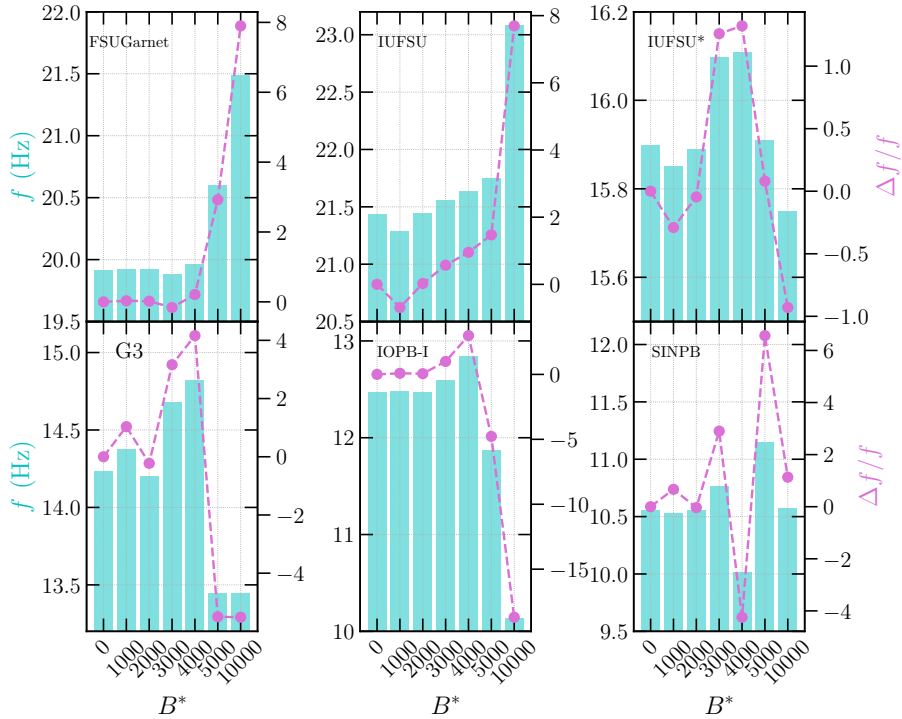}
    \caption{Frequency of fundamental torsional oscillation mode ($l=2$) in the crust of a magnetised neutron star for maximum mass at various magnetic field strengths.}
    \label{fig:freq}
\end{figure*}
%%%%%%%%%%%%%

Fig.\ \ref{fig:pasta_mass} and \ref{fig:pasta_thick} show the relative mass and thickness of different pasta structures, namely; rods, slabs, tubes, and bubbles for the  FSUGarnet, IUFSU, IUFSU$^*$, G3, IOPB-I, and SINPB parameter sets. The highest field strength is taken as $B^*=10000$ to investigate the thickness and mass for the upper limit. The E-RMF sets are plotted with increasing values of symmetry energy in the subsaturation density regions, with FSUGarnet having the highest symmetry energy and SINB the least. The relative mass and thickness of the different pasta layers are sensitive to a) the strength of the magnetic field and b) the EoS used. The changes with respect to the field-free case are similar to the mass and thickness of a particular pasta layer at a particular strength of the magnetic field. It is apparent that the relative mass and thickness of  pasta layers  fluctuate with the magnetic field  and do not behave in a particular fashion. One can see that deviations as high as ($25-30$) \% with respect to the field free case can be seen for magnetic field strength $B^*=10000$ while it remains $\sim (10-15)\ \%$ for $B^*=5000$. There is an interesting behaviour of EoS on the relative thickness and mass of pasta layers. The gross trend of fluctuations for a particular pasta layer changes when one compares different E-RMF calculations. The trend which is shown by E-RMF parameter sets having larger symmetry energy in the subsaturation density regions, namely: FSUGarnet and IUFSU, get reversed for higher magnetic field strength in comparison to IUFSU$^*$, G3, IOPB-I, and SINPB having comparatively lower symmetry energy. The large fluctuations have a profound implication for a neutron star  cooling period, where the magnetic field's decay occurs. The inner crust structure, therefore, can change considerably with the time period and magnetic field strength.

One of the significant characteristics of the inner crust is its shear modulus and shear speed which plays a crucial role in the crustal physics of neutron stars. These quantities are calculated using the E-RMF forces considered in the above analysis using the same method as in Ref. \cite{Parmar_2022_1} and observe that they also experience typical fluctuation due to the Landau quantization. Considering the pasta phases to have no shear modulus as prescribed in \cite{Parmar_2022_2}, the frequency of the fundamental torsional oscillation mode is computed using the approximate solution of crustal shear perturbation equation as given in Eq.\ \eqref{eq:freq} \cite{Gearheart_2011}.
%%%%%%%%%%%%%%%%
%\begin{equation}
%\label{eq:freq}
%    \omega_0^2 \approx \frac{e^{2\nu} V_s^2 (l-1)(l+2)}{2RR_c},
%\end{equation}
%%%%%%%%%%%%%%
Fig.\ \ref{fig:freq} shows the frequency variation of fundamental torsional oscillation mode ($l = 2$) in the crust of a magnetised neutron star using various E-RMF forces.  The shear modulus is calculated at the boundary of non-spherical structures in the inner crust. Out of all the forces, only FSUGarnet and IUFSU satisfy the possible candidates for the fundamental mode of QPOs: 18 Hz and 26 Hz \cite{Gearheart_2011}. The fundamental frequency also oscillates like other crustal properties of neutron stars with changing magnetic fields. For E-RMF force having a higher value of symmetry energy in the subsaturation region, the frequency tends to increase with the magnetic field, where it increases as high as 8\% as compared to the field-free case. Parameter sets IUFSU$^*$, G3 and IOPB-I also estimate an increase in frequency; however, a sharp dip for field strength higher than $B^*>4000$ can be seen for these sets. The frequency varies as low as 15 \% for the IOPB-I set. For SINPB having the least symmetry energy, more prominent fluctuations are observed. Therefore, the torsional oscillation mode frequency is sensitive to both the magnetic field strength and EoS (symmetry energy of the EoS in the subsaturtion region). These changes are significant in context to QPO, which are major asteroseismological sources to constrain neutron star crust properties such as pasta structures. 

It may be noted that this work do not consider possible deformation of the Wigner–Seitz (WS) cell due to the presence of magnetic field in this work. In literature, such attempts  are primarily based on Thomas-Fermi (TF) approximation with the modification that deformation of the cylindrical nature is introduced in the electronic distribution with spherically symmetric or deformed WS cell \cite{nag2010crustal, ghosh2011theoretical}. In such studies, a predetermined nature of deformation of the WS cell and the electronic gas distribution (cylindrical, prolate, etc.) is necessary, which might not be true always. The highly sophisticated molecular dynamics simulations of the present day can be used for such analysis. Moreover, solving Poisson's equations in the cylindrical coordinate needs a lot of approximation to handle the boundary value problems, such as using only the lowest Landau level, etc., and require a lot of computational power \cite{Avancini_2008}.

%%%%%%%%%%%%%%%%%%%%%%%%%%%%%%%%%%%%%%%%%%%%%%%%%%%%%
\section{Unified EoS and Neutron star observables}
%%%%%%%%%%%%%%%%%%%%%%%%%%%%%%%%%%%%%%%%%%%%%%%%%%%%%
\label{unified}
%%%%%%%%%%%%%%
\begin{figure}
    \centering
    \includegraphics[scale=0.7]{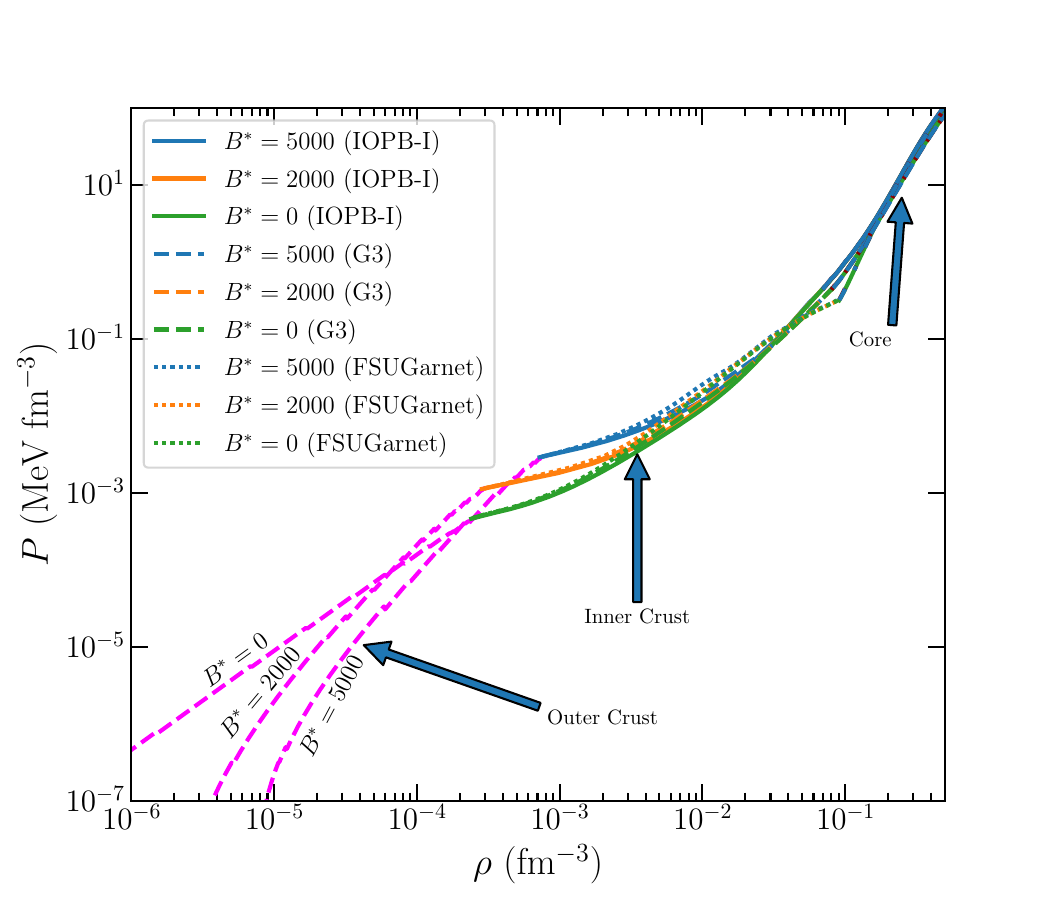}
    \caption{Unified  EoS for a magnetised neutron star at various magnetic field strengths.}
    \label{fig:unifid_eos}
\end{figure}
%%%%%%%%%%%%%
For completeness, a magnetised neutron star is modelled in a unified way from the surface to the core under the condition of $\beta$ equilibrium and charge neutrality. The core consists of neutrons, protons, electrons, and muons. The outer and inner crust EoS have been defined in the earlier section, whereas the EoS of the core is estimated using the same E-RMF set for which the inner crust is calculated. This procedure ensures the consistency between the crust and the core and helps to estimate the crustal parameters better. In general, the  density \cite{Debades_1997}, or chemical potential \cite{DEXHEIMER2017487} dependent magnetic field variation, are used for the neutron star calculation in literature. However, this work approximates the same magnetic field strength throughout the neutron star to calculate the properties such as outer and inner crust thickness, mass, etc. This approximation is taken for the simplicity of the calculations. It can be considered equivalent to the density/chemical potential dependent magnetic field profiles where the central field is quite low \cite{Patra_2020, Rather_2021}. In such circumstances, the EoS does not deviate much from the field-free case, and the magnetic field almost becomes constant for most parts of neutron star \cite{Patra_2020, Rather_2021}. Furthermore, the neutron star should be deformed due to the anisotropic pressure in the presence of magnetic field. However, this work considers the neutron star to be spherically symmetric, which allows us to use the Tolman-Oppenheimer-Volkoff (TOV) solution. This assumption holds good for the range of MF $10^{15}-10^{18}$ as the deformation from spherical symmetry turns out to be less than 1\% \cite{Patra_2020, Bordbar2022, Chu_2015}.

Fig.\ \ref{fig:unifid_eos} shows the unified EoS using the FSUGarnet, G3, and IOPB-I  parameters sets for a magnetised neutron star. The magnetic field effect becomes profound in the outer crust and shallow regions of the inner crust, whereas it vanishes in the core. The reason can be attributed to more Landau level filling at a large density, which makes the EoS similar to the field-free case.  A significant variation in the outer and inner crust EoS with changing magnetic fields means that the crust structure will be affected significantly, consequently impacting various microscopic properties of a neutron star such as elastic properties, cooling, bursts, etc. These conclusions remain valid even for more realistic magnetic field profiles. 
%%%%%%%%%%%%%%
\begin{sidewaystable}
\centering
\caption{The neutron star properties for canonical mass ($M=1.4M_\odot$), corresponding radius ($R_{\rm 1.4}$), radius of the outer crust ($R_{\rm oc}$), radius of the total crust ($R_{\rm c}$), mass of the outer crust ($M_{\rm oc}$), mass of the total crust ($M_{\rm crust}$), normalized maximum MI ($I_{\rm 1.4}$), fractional moment of inertia for outer crust ( $I_{ \rm oc}/I_{ \rm 1.4}$ ), fractional moment of inertia for total crust ( $I_{\rm c}/I_{\rm 1.4}$ ),   second Love number ($k_{2,1.4}$) and dimensionless tidal deformability ($\Lambda_{\rm max}$)  for various EoSs at various magnetic field strength.}
\label{tab:NS_properties}
\renewcommand{\tabcolsep}{0.2cm}
\renewcommand{\arraystretch}{1}
\begin{tabular}{llllllllllll}
\hline
\hline
\begin{tabular}[c]{@{}l@{}}Parameter\\ sets\end{tabular}         & B* & \begin{tabular}[c]{@{}l@{}}$R_{1.4}$\\ (km)\end{tabular}  & \begin{tabular}[c]{@{}l@{}}$R_{oc}$\\ (km)\end{tabular} & \begin{tabular}[c]{@{}l@{}}$R_{c}$\\ (km)\end{tabular} & \begin{tabular}[c]{@{}l@{}}$M_{\rm oc}$\\ ($M_\odot$)\end{tabular} & \begin{tabular}[c]{@{}l@{}}$M_{\rm c}$\\ ($M_\odot$)\end{tabular} & $I_{1.4}$  & $I_{oc}/I_{1.4}$ & $I_c/I_{1.4}$ & $k_{2,1.4}$  & $\lambda_{1.4}$ \\
\hline

\multirow{3}{*}{FSUGarnet} & 0  & 13.1935 & 0.5291   & 1.35600   & 0.00005   & 0.02473   & 0.34297 & 0.00007      & 0.03630      & 0.08891 & 629.131  \\
                        & 2000 & 12.9951 & 0.5269 & 1.20570 & 0.00011 & 0.02391 & 0.35245 & 0.00017 & 0.03456 & 0.09480 & 619.817 \\
                        & 5000 & 12.9881 & 0.5274 & 1.19900 & 0.00028 & 0.02424 & 0.35292 & 0.00044 & 0.03506 & 0.09514 & 620.265 \\
                        &      &          &         &         &         &         &         &         &         &         &           \\
\multirow{3}{*}{IUFSU}  & 0    & 12.6075 & 0.4728 & 1.17940 & 0.00004 & 0.02392 & 0.35138 & 0.00005 & 0.03507 & 0.08704 & 488.953 \\
                        & 2000 & 12.6153 & 0.4903 & 1.16400 & 0.00010 & 0.02563 & 0.35271 & 0.00015 & 0.03744 & 0.08844 & 498.451 \\
                        & 5000 & 12.6092 & 0.4906 & 1.15680 & 0.00025 & 0.02574 & 0.35314 & 0.00039 & 0.03764 & 0.08873 & 498.730 \\
                        &      &          &         &         &         &         &         &         &         &         &           \\
\multirow{3}{*}{IUFSU$^*$} & 0    & 12.9221 & 0.5025 & 1.23540 & 0.00004 & 0.02563 & 0.34704 & 0.00006 & 0.03780 & 0.08843 & 563.083 \\
                        & 2000 & 12.8100 & 0.5089 & 1.12460 & 0.00010 & 0.02605 & 0.35308 & 0.00016 & 0.03813 & 0.09230 & 561.581 \\
                        & 5000 & 12.8039 & 0.5098 & 1.12070 & 0.00027 & 0.02658 & 0.35350 & 0.00042 & 0.03891 & 0.09259 & 562.141 \\
                        &      &          &         &         &         &         &         &         &         &         &           \\
\multirow{3}{*}{G3}     & 0    & 12.6278 & 0.4753 & 1.26130 & 0.00004 & 0.03376 & 0.34453 & 0.00005 & 0.05035 & 0.08128 & 460.236 \\
                        & 2000 & 12.4931 & 0.4785 & 1.15580 & 0.00009 & 0.03257 & 0.35175 & 0.00014 & 0.04810 & 0.08546 & 458.863 \\
                        & 5000 & 12.4864 & 0.4800 & 1.15200 & 0.00024 & 0.03313 & 0.35220 & 0.00038 & 0.04895 & 0.08576 & 459.176 \\
                        &      &          &         &         &         &         &         &         &         &         &           \\
\multirow{3}{*}{IOPB-I} & 0    & 13.3316 & 0.5420 & 1.35420 & 0.00005 & 0.04004 & 0.34420 & 0.00007 & 0.05862 & 0.09244 & 686.640 \\
                        & 2000 & 13.3117 & 0.5580 & 1.33620 & 0.00012 & 0.03954 & 0.34491 & 0.00019 & 0.05791 & 0.09281 & 684.730 \\
                        & 5000 & 13.2987 & 0.5583 & 1.32950 & 0.00031 & 0.04055 & 0.34567 & 0.00050 & 0.05938 & 0.09335 & 685.145 \\
                        &      &          &         &         &         &         &         &         &         &         &           \\
\multirow{3}{*}{SINPB}  & 0    & 13.1576 & 0.5250 & 1.18490 & 0.00005 & 0.03053 & 0.34245 & 0.00006 & 0.04608 & 0.08814 & 612.943 \\
                        & 2000 & 13.2057 & 0.5481 & 1.23350 & 0.00012 & 0.02914 & 0.33945 & 0.00019 & 0.04401 & 0.08605 & 609.610 \\
                        & 5000 & 13.1297 & 0.5423 & 1.16630 & 0.00030 & 0.03024 & 0.34348 & 0.00048 & 0.04565 & 0.08865 & 610.223 \\
\hline
\hline
\end{tabular}%
\end{sidewaystable}
%%%%%%%%%%%%

Finally, let us calculate the global properties, such as mass, radius, etc., of a magnetised neutron star at a particular magnetic field using the TOV Eqs.\ \eqref{eq:pr} and \ref{eq:mr} considering the star to be spherically symmetric. This assumption remains valid for the strength of the magnetic field considered in this work \cite{Patra_2020}. It is  observed that the maximum mass and radius of the star do not change significantly up to $B^*=5000$ as the core EoS remains almost unaffected. The E-RMF parameter sets considered in this work reproduce the maximum mass and radius in agreement with the ($M=2.01^{+0.04}_{-0.04} M_\odot$) and ($M=2.14^{+0.10}_{-0.09}+M_\odot$) limit \cite{Antoniadis_2013, Cromartie_2019}.  To infer the effect of magnetic field on the crust mass and thickness, Table \ref{tab:NS_properties} shows neutron star properties for canonical mass ($M=1.4M_\odot$), corresponding radius ($R_{\rm 1.4}$), the radius of the outer crust ($R_{\rm oc}$), the radius of the total crust ($R_{\rm c}$), the mass of the outer crust ($M_{\rm oc}$), the mass of the total crust ($M_{\rm crust}$), normalized maximum MI ($I_{\rm 1.4}$), the fractional moment of inertia for the outer crust ( $I_{ \rm oc}/I_{ \rm 1.4}$ ), the fractional moment of inertia for the total crust ($I_{\rm c}/I_{\rm 1.4}$), second Love number ($k2_{1.4}$) and dimensionless tidal deformability ($\Lambda_{\rm 1.4}$) for various EoSs at various magnetic field strengths. Although the mass of the total crust remains almost similar with changing magnetic field strength, the outer crust mass increases $\sim 5-6$ times as compared to the field-free case. However, the thickness of the outer crust does not change much with magnetic field strength. The fractional crustal moment of inertia of the total crust also remains unaffected. In contrast, the outer crust moment of inertia increases $\sim 7-8$ times compared to the field-free case. The insensitiveness of the total crust mass, thickness, and crustal moment of inertia can be understood from Table \ref{tab:pasta_density} and Fig.\ \ref{fig:inner_crust_eos}, which show a minute change in crust-core transition density and pressure (and consequently chemical potential) which decides the mass and thickness of the crust. 

%%%%%%%%%%%%%%%%%%%%
\section{Summary}

%%%%%%%%%%%%%%%%%%%%
In summary, the present chapter investigates the impact of the magnetic field of the order of $\sim 10^{17}$G on the neutron star crust and associated phenomena. The outer crust structure predominantly depends on the structural effects in atomic nuclei and becomes increasingly symmetric as one increases the magnetic field strength. In the presence of magnetic field, more neutron-rich (less bound nuclei) can sustain at lower pressure in the outer crust due to the magnetic field. The surface density, neutron drip density, and neutron drip pressure increase substantially with an increase in the strength of the magnetic field. 

To calculate the equilibrium composition of the inner crust in the presence of the magnetic field, this work uses the CLDM formalism for the first time using the E-RMF framework for nuclear interaction. The equilibrium composition of the inner crust shows typical quantum fluctuations, which become prominent at a high magnetic field. In the shallow regions of the inner crust, the nuclear cluster becomes more dense and symmetric with an increase in magnetic field strength. The EoS of the electron plays the predominant role in infusing the magnetic field effects on inner crust EoS. In contrast, the symmetry energy of the nuclear matter EoS dictates the in-situ characteristics. The inner crust calculations using the CLDM formalism agree with the available self-consistent Thomas-Fermi calculations. 

It further investigates the possible change in the pasta structures in the presence of the magnetic field. The onset density of the various pasta structures fluctuates with the magnetic field strength. It results in substantial modifications in their mass and thickness, which ranges as high as 50\%. The frequency of the fundamental torsional oscillation mode is also investigated, and an increase/decrease of $\sim$ 5\% is seen depending upon the EoS and the magnetic field strength.

Finally, the unified EoSs of a magnetised neutron star are constructed by estimating the core EoS using the same E-RMF parameter as for the inner crust, ensuring consistency between both layers. The magnetic field impacts the outer crust and the shallower regions of the inner crust as the EoS of the electron gas becomes strongly quantizing in this region. As density increases, a large number of Landau level fillings lead to the core EoS imitating the field-free EoS. The outer crust mass and its fractional crustal moment of inertia increase substantially with an increasing magnetic field.  However, the total crust mass and thickness remain immune to the changing magnetic field strength. 

Although the present work considers magnetised neutron star crust to be composed of cold-catalyzed matter, it is unlikely for the crust, especially the outer crust, to be in complete thermodynamic equilibrium during its formation. Therefore, the structure of the neutron star crust in the presence of a magnetic field and some finite temperature becomes essential. This improvement deserves more investigation and will be carried out in future work to understand the neutron star structure holistically.

%%%%%%%%%%%%%%%%%%%%%%%%

\clearpage
\addcontentsline{toc}{section}{Bibliography}
\printbibliography
%%%%%%%%%%%%%%%%%%%%%%%%%%%%%%%%%%%%%%%%%%%%%%%%%%%%%%%%%%%%%%%%%%%%%%%%%%%%%%%%%

%% file: Summary/conclusion.tex
\chapter{ Summary and Outlook}
\label{chap8}
%%%%%%%%%%%%%%%%%%%%%%%%%%%%%%%%%%%%%%%%%%%%%
In the present thesis, the main focus is to examine the properties of nuclear matter as a function of various parameters such as density, pressure, temperature, isospin-asymmetry, magnetic field etc. Nuclear matter, in its three primary forms: \textbf{\textit{infinite nuclear matter, finite nuclei}} and \textbf{\textit{the neutron star}}, is considered in this thesis. To study these nuclear matter forms, this work uses the \textit{Effective Relativistic Mean Field Model} for the estimation of nuclear interactions. The main focus in reference to the infinite nuclear matter and finite nuclei are the investigations of their finite temperature properties. Various features such as liquid-gas phase transition, binodal, spinodal, EoS, limiting temperature of nuclei, etc. are investigated. On the other hand, this thesis provides a comprehensive analysis of the crustal properties of a neutron star in the magnetised as well as unmagnetised environments. The preeminent results obtained in the  thesis are enumerated below. 

\begin{enumerate}

    \item \textbf{\textit{Infinite nuclear matter:-}} The present thesis investigates various finite temperature properties of isospin symmetric and asymmetric nuclear matter over a wide range of density and pressure. The main focus is to understand the liquid-gas phase transition and its behaviour as a function of EoS and asymmetry. For the symmetric matter, the critical parameters, i.e. critical temperature ($T_c$), pressure ($P_c$) and density ($\rho_c$), are found to be model dependent. The  $T_c$ is one of the poorest constrained nuclear matter observables in theoretical and experimental estimations. Moreover, there are deviations in the theoretical and experimental estimates of the $T_c$.  The discrepancy in the experimental and theoretical values might be attributed to two main factors: ($i$) symmetric nuclear matter being an ideal system which is too difficult to simulate in experiments and ($ii$) finite size effect and very short time scale of multi fragment reactions which make it difficult to study thermodynamic equilibrium. While the vector-self coupling ($\zeta_0$) seems to impact the symmetric matter, the cross-coupling of $\omega-\rho$ meson ($\Lambda_\omega$) influences the properties of hot asymmetric nuclear matter. 

The temperature dependence of free nuclear symmetry energy ($F_{sym}$) and its higher-order derivatives are discussed. The  $F_{sym}$  increases with temperature at a given density due to a decrease in entropy density. It is observed that the thermal effects in E-RMF formalism depend mainly on the density dependence of Dirac effective mass. The Dirac effective mass is calculated self consistently, which depends on the  $\sigma$ and $\delta$ mesons.   A larger Dirac effective mass corresponds to larger thermal effects on the state variables. Furthermore, the change in magnitude of thermal contribution is principally attributed to their zero temperature variation, which results from the different nuclear matter observables such as incompressibility, symmetry energy and its higher order derivatives etc. The phase transition is studied for the asymmetric nuclear matter considering a two-component system with two conserved charges, i.e. Baryon number and isospin. A lower slope parameter value ($L_{sym}$) estimates the higher value of maximal asymmetry and critical pressure. A larger value of $\Lambda_\omega$ corresponds to the larger instability in asymmetric nuclear matter. Critical asymmetry is a quadratic function of temperature and exhibits different behaviour with change of temperature.

     \item \textbf{\textit{Hot finite nuclei:-}} In  HIC experiments and the low-density nuclear matter, such as in neutron star crust and supernovae, the nuclear matter is either in the form of finite nuclei or clusterised form. Therefore the investigation of the finite temperature properties of finite nuclei becomes essential. Compared to infinite matter, the inclusion of the Coulomb component in finite nuclei gives rise to instability, and results in the reduction of  critical temperature. To investigate this, the E-RMF framework is used to analyze the thermal properties of hot nuclei. The free energy of a nucleus is estimated by using temperature and density-dependent parameters of the liquid-drop model. The surface free energy is parametrized using two approaches based on the sharp interface of the liquid-gaseous phase and the semi-classical Seyler-Blanchard interaction. The thermal properties of the finite nuclear systems are influenced strongly by the effective mass and critical temperature ($T_c$)  of the E-RMF parameter sets employed. A larger effective mass corresponds to the higher excitation energy ($E^*$), level density ($a$), limiting temperature ($T_l$), and limiting excitation energy ($E^*_l$). The limiting temperature also depends on the behaviour of  EoS at subsaturation densities which helps to estimate the properties of surrounding nuclear gas. The effective mass shows a strong positive correlation with the critical parameters, namely ($T_c$,  $\rho_c$, $P_c$) and limiting temperature ($T_l$) of $^{208}Pb$ nucleus. On the other hand, the binding energy and saturation density act as independent parameters. 

       \item \textbf{\textit{Neutron star crust:-}} Neutron star is one of the prominent aspects of nuclear matter. Using E-RMF formalism, the present thesis provides a comprehensive analysis of neutron star crust structure and associated properties. The outer crust is treated within the well-known variational BPS formalism, while the structure of the inner crust is calculated using the compressible liquid drop model. This work uses the most recent atomic mass evaluation AME2020 and the highly precise microscopic HFB mass models, along with the experimental mass of available neutron-rich nuclei to find the equilibrium composition of the outer crust. For the inner crust, various E-RMF parameter sets are used to study the model dependence of the obtained results. 

The EoS and composition of the outer crust calculated using different mass models are compared, which points towards the persistent existence of $Z=28$ and $N= 50$ and $82$ nuclei. The majority of mass models predict the presence of even mass nuclei in the outer crust except for the HFB-14, which indicate a thin layer of $^{121}$Y at high pressure suggesting a possible ferromagnetic behaviour of neutron star. It is seen that the equilibrium configuration of the inner crust is strictly model-dependent and depends mainly on symmetry energy and slope parameter in the subsaturation density regime, and the surface energy parametrization. A higher value of symmetry energy or lower slope parameter results in larger mass and charge of the cluster. The crust-core transition density ($\rho_t$) and pressure ($P_t$) are calculated from the crust side, and it is seen that these values are sensitive to the isovector surface parameter $p$ and slope parameter $L$. In addition, this work provides unified EoS's, namely FSUGarnet-U, IOPB-I-U, and G3-U, which are available on Github.

Next, the influence of pasta structure in the inner crust of the neutron star is investigated. The calculations are done by using 13 well-known E-RMF parameter sets, that satisfy the recent observational constraints on the maximum mass and radius of the neutron star. Unified EoSs are constructed to obtain the pasta and crustal properties consistently. The appearance of different pasta layers is model-dependent. The model dependency is attributed to the behaviour of symmetry energy in the subsaturation density region and the surface energy parametrization. A thicker crust favours the existence of more number of pasta layers in it. The pressure ($P_c$), chemical potential ($\mu_c$), and density ($\rho_c$) of the crust-core transition from the crust side are calculated and compared with the results from recent constraints proposed using Bayesian inference analyses. The parameter set with lower values of $J$, $L$, and $K_{\rm sym}$ seem to give better agreement with these theoretical constraints. Fundamental frequency mode in context to quasiperiodic oscillations in soft gamma-ray repeaters is calculated, and  Out of 13 EoSs, only two parameter sets, FSUGarnet and IUFSU, agree with the 18Hz observational frequency.

Finally, the crustal properties of a  neutron star are investigated within the E-RMF framework in the presence of magnetic field strength $\sim 10^{17}$G. The outer crust structure predominantly depends on the structural effects in atomic nuclei and becomes increasingly symmetric as one increases the magnetic field strength. In the presence of magnetic field, more neutron-rich nuclei (less bound) can sustain at lower pressure in the outer crust due to the magnetic field. The surface density, neutron drip density, and neutron drip pressure increases substantially with an increase in the strength of the magnetic field. For the first time, this works uses CLDM formalism to calculate the inner crust structure of a magnetised neutron star. In the shallow regions of the inner crust, the nuclear cluster becomes more dense and becomes symmetric with an increase in magnetic field strength. The EoS of the electrons plays the predominant role in infusing the magnetic field effects on inner crust. In contrast, the symmetry energy of the nuclear matter EoS dictates the in-situ characteristics. The onset density of the various pasta structures fluctuates with the magnetic field strength. It results in substantial modifications in their mass and thickness, which ranges as high as 50\%. The frequency of the fundamental torsional oscillation mode is also investigated, and a change of $\sim$ 5\% is seen depending upon the choice of EoS and the magnetic field strength.

\end{enumerate}

%%%%%%%%%%%%%%%%%%%%%%%%%%%%%%%%%%%%%%%%%%%%%%%%%
\newpage
%%%%%%%%%%%%%%%%%%%%%%%%%%%%%%%%%%%%%%%%%%%%%%%%
\begin{center}
    \Large\textbf{Future scope}
\end{center}
\addcontentsline{toc}{chapter}{Future scope}
%%%%%%%%%%%%%%%%%%%%%%%%%%%%%%%%%%%%%%%%%%%%%%%%
One of the primary aims of nuclear physicists and astrophysicists is to put stringent constraints on nuclear matter observables and, consequently, on EoS. As we saw, the neutron star crust structure is essentially model dependent, where the symmetry energy plays the determining role. Therefore, the calculations open new doors to constrain the nuclear symmetry energy based on various Multi-messenger and Asteroseismological observations such as resonant shattering flare (RSF) from the binary neutron star inspiral. Various crust properties such as crust cooling in young neutron stars, which are not yet catalyzed and undergoing decay in the magnetic field, quasiperiodic oscillations, neutron superfluidity and associated entrainment effect etc., can be investigated to better ascertain the neutron star crust properties.  The clusterisation of nuclear matter is also expected to occur in the inner core of the neutron star, where there is a possibility of the quark-hadron phase transition. The possibility of such an exotic structure in the inner core of a neutron star has recently drawn particular interest in light of various astrophysical observations. In the literature, there are few unified EoS for magnetised EoS. The EoS presented in the present thesis can be used in various pulsar and magnetar structure simulations and their magnetic field evolution. 

While there are hundreds of available EoS for the cold catalyzed neutron star, the finite temperatures EoS's, which are relevant for the simulations of supernovae and neutron star merger events, are rare. The finite temperature calculations of the present thesis can be extended for the case of supernova matter. This will not only help in better estimations of neutron star structure and dynamics but will help us to understand various astrophysical phenomena such as nucleosynthesis in heavy star, the natural abundance of the elements etc. Furthermore, how the interior of a neutron star changes with temperature is one of the unsolved puzzles which can be brought up as an extension of the present thesis. 
%%%%%%%%%%%%%%%%%%%%%%%%%%%%%%%%%%%%%%%%%%%%%%%%%